\documentclass[10pt,onecolumn,twoside,a4paper,openany]{article}
\usepackage[dvips]{graphicx}
\usepackage{bbm}
\usepackage{amsfonts,amsmath,amssymb,graphics,psfrag}
\makeatletter
\@addtoreset{equation}{section}
\makeatother

\newcommand{\lL}{{}^{\scriptscriptstyle (\ell)}\!\!\:}

\newcommand{\nflop}[1]{\widehat{\!\widetilde{E}}_{#1}(S)}

\newcommand{\lp}{\ell_p}
\newcommand{\zu}{\!\!>}
\newcommand{\auf}{\;<} 
\newcommand{\sst}{\scriptscriptstyle}
\newcommand{\st}{\scriptstyle}

\newcommand{\op}[1]{\widehat{#1}}
\newcommand{\wt}[1]{\widetilde{#1}}
\newcommand{\bet}[2]{|\beta^{#1}\,{\st #2} \zu}
\newcommand{\betc}[2]{\auf\beta^{#1}\,{\st #2}|}
\newcommand{\alp}[3]{|\alpha^{\st #1}_{#2}\,{\st #3}\zu}
\newcommand{\alpc}[3]{\auf\alpha^{\st #1}_{#2}\,{\st #3}|}
\newcommand{\SixJ}[6]{\left\{\begin{array}{ccc} #1& #2& #3\\ #4 & #5 & #6\end{array}\right\}}

\DeclareMathOperator{\sgn}{sgn}
\DeclareMathOperator{\tr}{Tr}
\newtheorem{Theorem}{Theorem}[section]           

\newtheorem{Lemma}{Lemma}[section]
\newtheorem{Proof}{Proof}[section]
\def\be{\begin{equation}}
\def\ee{\end{equation}}
\def\ba{\begin{eqnarray}}
\def\ea{\end{eqnarray}}
\def\a{{\cal A}}
\def\ab{\overline{\a}}

\def\Nl{{\mathchoice
{\setbox0=\hbox{$\displaystyle\rm N$}\hbox{\hbox to0pt
{\kern0.4\wd0\vrule height0.9\ht0\hss}\box0}}
{\setbox0=\hbox{$\textstyle\rm N$}\hbox{\hbox to0pt
{\kern0.4\wd0\vrule height0.9\ht0\hss}\box0}}
{\setbox0=\hbox{$\scriptstyle\rm N$}\hbox{\hbox to0pt
{\kern0.4\wd0\vrule height0.9\ht0\hss}\box0}}
{\setbox0=\hbox{$\scriptscriptstyle\rm N$}\hbox{\hbox to0pt
{\kern0.4\wd0\vrule height0.9\ht0\hss}\box0}}}}
\def\Zl{{\mathchoice
{\setbox0=\hbox{$\displaystyle\rm Z$}\hbox{\hbox to0pt
{\kern0.4\wd0\vrule height0.9\ht0\hss}\box0}}
{\setbox0=\hbox{$\textstyle\rm Z$}\hbox{\hbox to0pt
{\kern0.4\wd0\vrule height0.9\ht0\hss}\box0}}
{\setbox0=\hbox{$\scriptstyle\rm Z$}\hbox{\hbox to0pt
{\kern0.4\wd0\vrule height0.9\ht0\hss}\box0}}
{\setbox0=\hbox{$\scriptscriptstyle\rm Z$}\hbox{\hbox to0pt
{\kern0.4\wd0\vrule height0.9\ht0\hss}\box0}}}}
\def\Ql{{\mathchoice
{\setbox0=\hbox{$\displaystyle\rm Q$}\hbox{\hbox to0pt
{\kern0.4\wd0\vrule height0.9\ht0\hss}\box0}}
{\setbox0=\hbox{$\textstyle\rm Q$}\hbox{\hbox to0pt
{\kern0.4\wd0\vrule height0.9\ht0\hss}\box0}}
{\setbox0=\hbox{$\scriptstyle\rm Q$}\hbox{\hbox to0pt
{\kern0.4\wd0\vrule height0.9\ht0\hss}\box0}}
{\setbox0=\hbox{$\scriptscriptstyle\rm Q$}\hbox{\hbox to0pt
{\kern0.4\wd0\vrule height0.9\ht0\hss}\box0}}}}
\def\Rl{{\mathchoice
{\setbox0=\hbox{$\displaystyle\rm R$}\hbox{\hbox to0pt
{\kern0.4\wd0\vrule height0.9\ht0\hss}\box0}}
{\setbox0=\hbox{$\textstyle\rm R$}\hbox{\hbox to0pt
{\kern0.4\wd0\vrule height0.9\ht0\hss}\box0}}
{\setbox0=\hbox{$\scriptstyle\rm R$}\hbox{\hbox to0pt
{\kern0.4\wd0\vrule height0.9\ht0\hss}\box0}}
{\setbox0=\hbox{$\scriptscriptstyle\rm R$}\hbox{\hbox to0pt
{\kern0.4\wd0\vrule height0.9\ht0\hss}\box0}}}}
\def\Cl{{\mathchoice
{\setbox0=\hbox{$\displaystyle\rm C$}\hbox{\hbox to0pt
{\kern0.4\wd0\vrule height0.9\ht0\hss}\box0}}
{\setbox0=\hbox{$\textstyle\rm C$}\hbox{\hbox to0pt
{\kern0.4\wd0\vrule height0.9\ht0\hss}\box0}}
{\setbox0=\hbox{$\scriptstyle\rm C$}\hbox{\hbox to0pt
{\kern0.4\wd0\vrule height0.9\ht0\hss}\box0}}
{\setbox0=\hbox{$\scriptscriptstyle\rm C$}\hbox{\hbox to0pt
{\kern0.4\wd0\vrule height0.9\ht0\hss}\box0}}}}
\def\Co{{\mathchoice
{\setbox0=\hbox{$\displaystyle\rm C$}\hbox{\hbox to0pt
{\kern0.4\wd0\vrule height0.9\ht0\hss}\box0}}
{\setbox0=\hbox{$\textstyle\rm C$}\hbox{\hbox to0pt
{\kern0.4\wd0\vrule height0.9\ht0\hss}\box0}}
{\setbox0=\hbox{$\scriptstyle\rm C$}\hbox{\hbox to0pt
{\kern0.4\wd0\vrule height0.9\ht0\hss}\box0}}
{\setbox0=\hbox{$\scriptscriptstyle\rm C$}\hbox{\hbox to0pt
{\kern0.4\wd0\vrule height0.9\ht0\hss}\box0}}}}
\def\Hl{{\mathchoice
{\setbox0=\hbox{$\displaystyle\rm H$}\hbox{\hbox to0pt
{\kern0.4\wd0\vrule height0.9\ht0\hss}\box0}}
{\setbox0=\hbox{$\textstyle\rm H$}\hbox{\hbox to0pt
{\kern0.4\wd0\vrule height0.9\ht0\hss}\box0}}
{\setbox0=\hbox{$\scriptstyle\rm H$}\hbox{\hbox to0pt
{\kern0.4\wd0\vrule height0.9\ht0\hss}\box0}}
{\setbox0=\hbox{$\scriptscriptstyle\rm H$}\hbox{\hbox to0pt
{\kern0.4\wd0\vrule height0.9\ht0\hss}\box0}}}}
\def\Ol{{\mathchoice
{\setbox0=\hbox{$\displaystyle\rm O$}\hbox{\hbox to0pt
{\kern0.4\wd0\vrule height0.9\ht0\hss}\box0}}
{\setbox0=\hbox{$\textstyle\rm O$}\hbox{\hbox to0pt
{\kern0.4\wd0\vrule height0.9\ht0\hss}\box0}}
{\setbox0=\hbox{$\scriptstyle\rm O$}\hbox{\hbox to0pt
{\kern0.4\wd0\vrule height0.9\ht0\hss}\box0}}
{\setbox0=\hbox{$\scriptscriptstyle\rm O$}\hbox{\hbox to0pt
{\kern0.4\wd0\vrule height0.9\ht0\hss}\box0}}}}
\title{Consistency Check on Volume and Triad Operator Quantisation\\
in\\ Loop Quantum Gravity II}
\author{
K. 
Giesel\thanks{Kristina.Giesel@aei.mpg.de, kgiesel@perimeterinstitute.ca}
~~ and ~
T. 
Thiemann\thanks{Thomas.Thiemann@aei.mpg.de, tthiemann@perimeterinstitute.ca}\\
\\
MPI f. Gravitationsphysik, Albert-Einstein-Institut, \\
           Am M\"uhlenberg 1, 14476 Potsdam, Germany\\
\\
and\\
\\
Perimeter Institute for Theoretical Physics, \\ 
31 Caroline Street N, Waterloo, ON N2L 2Y5, Canada}
\date{{\small Preprint AEI-2005-120}}

\begin{document}
\maketitle
\begin{abstract}
In this paper we provide the techniques and proofs for the resuls presented in our companion paper concerning the consistency check on volume and triad operator quantisation in Loop Quantum Gravity.
\end{abstract}
\newpage

\tableofcontents

\newpage
\section{Introduction}
\label{s1}                            %
In this paper we deliver the necessary techniques and proofs for the results discussed in our companion paper \cite{GT}. There a consistency check on the method of quantising triads by means of the so called Poisson bracket identity is performed. This identity allows us to replace triads by the Poisson bracket among the Ashtekar connection and the classical volume and places a prominent role in the dynamics of LQG \cite{2}. The consistency check is made by constructing an alternative flux operator based on the Poisson bracket identity whose action is then compared with the action of the usual flux operator, quantised in a standard way as a differential operator.
\newline
In particular we show that one must consider the electric field of LQG as a pseudo 2-form, since otherwise no consistent alternative flux operator can be obtained. Note, that classically, when taking the symplectic structure of LQG as fundamental the possibilities of taking the electric field as either a 2-form or a pseudo 2-form are equivalent. Furthermore a consistent alternative flux operator can only be achieved if one uses the volume operator introduced by Ashtekar and Lewandowski $\op{V}_{\sst AL}$ \cite{4}. The Rovelli-Smolin volume operator $\op{V}_{\sst RS}$ \cite{3} is inconsistent with the usual flux operator. The ambiguity of $\op{V}_{\sst AL}$ caused by regularisation can be uniquely fixed by this consistency check. Moreover since we apply the formula for matrix elements of the volume operator developed in \cite{14}, this formula is tested independently through our analysis here. Additionally, we could demonstrate that when considering higher representation weights than the fundamental one of $SU(2)$ for the holonmies involved in the alternative flux operator the results stay invariant. Hence, we get no ambiguities in the quantisation process.
Finally, the factor ordering of the alternative flux operator is unique if one insists on the principle of minimality.  
\newline\newline
These results show that instead of taking holonomies and fluxes as 
fundamental operators one could instead use holonomies and volumes as
fundamental operators. It also confirms that the method to quantise the 
triad developed in \cite{2} is mathematically consistent.\\
\\
This paper is organised as follows:\\
\\
In section two we review the regularisation and definition of the 
fundamental flux operator for the benefit of the reader and in order
to make the comparison with the alternative quantisation easier.

In section three we derive the classical expression for the alternative 
flux operator.

In section four we describe in detail the regularisation of the 
alternative flux operator and arrive at its explicit action on spin 
network functions.

In section five we draw first conclusions about and determine general 
properties of the expression obtained in section four.

In section six  we compute the full matrix elements of the alternative flux
operator.

In section seven we show that the chosen factor ordering is unique within
the minimalistic class of factor orderings mentioned above.

In section eight we compute the matrix elements of the fundamental flux 
operator.

In section nine we compare the two flux operators and discover that there 
is perfect match for any value of $\ell$ if and only if $C_{reg}=1/48$, if 
and only if the 
electric field is a pseudo two form and if and only if we use the AL 
volume operator. 

In section ten we rule out the RS volume operator explicitly. In particular, we stress that the fact that the RS volume operator is inconsistent could not have been guessed from the outset. The consistency check performed in this paper is non-trivial and should not be taken as criticism of the RS volume but rather as a mechanism to tighten the mathematical structure of LQG.

In section eleven we summarise and conclude.

Finally, in appendices A -- E we supply the detailed calculations and 
proofs for the claims that we have made in the main text.
\newpage
\section{Review of the Usual Flux Operator in LQG}%
\begin{figure}[hbt]
   \center
   \psfrag{t+}{$t=+\epsilon$}
   \psfrag{t=0}{$t=0$}
   \psfrag{t-}{$t=-\epsilon$}
   \psfrag{t}{$t$}
   \psfrag{S}{$S_t$}
   \includegraphics[height=6cm]{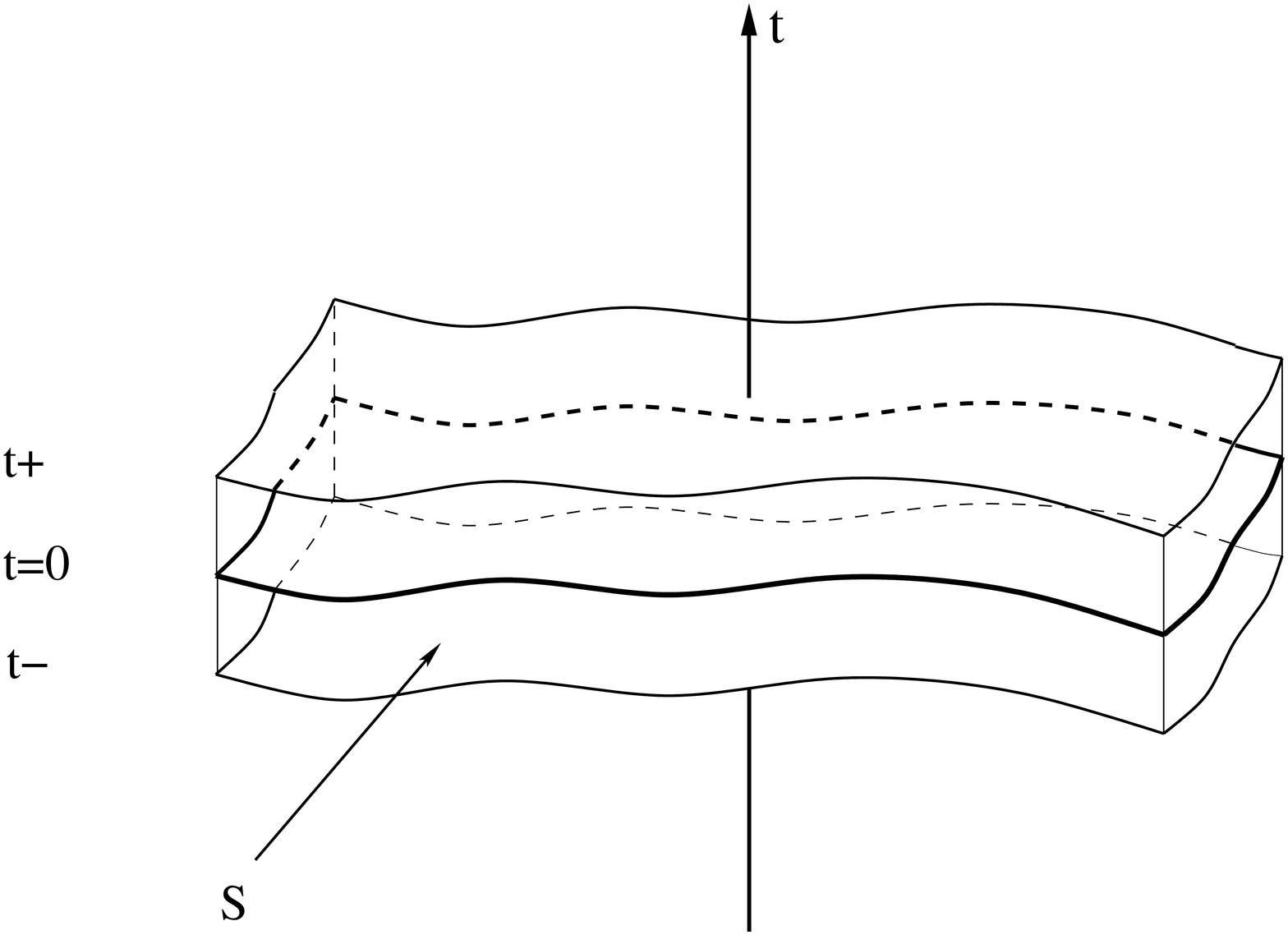}
   \caption{\label{Bild1}\small Smearing of the surface $S$ into the third dimension. We obtain an array of surfaces $S_t$ labelled by the parameter $t$ with $t\in\{-\epsilon,+\epsilon\}$. The original surface $S$ is associated with $t=0$.}
   \end{figure}

\begin{figure}[hbt]
   \center
   \psfrag{up}{up}
   \psfrag{down}{down}
   \psfrag{in}{in} 
   \psfrag{out}{out} 
   \includegraphics[height=4cm]{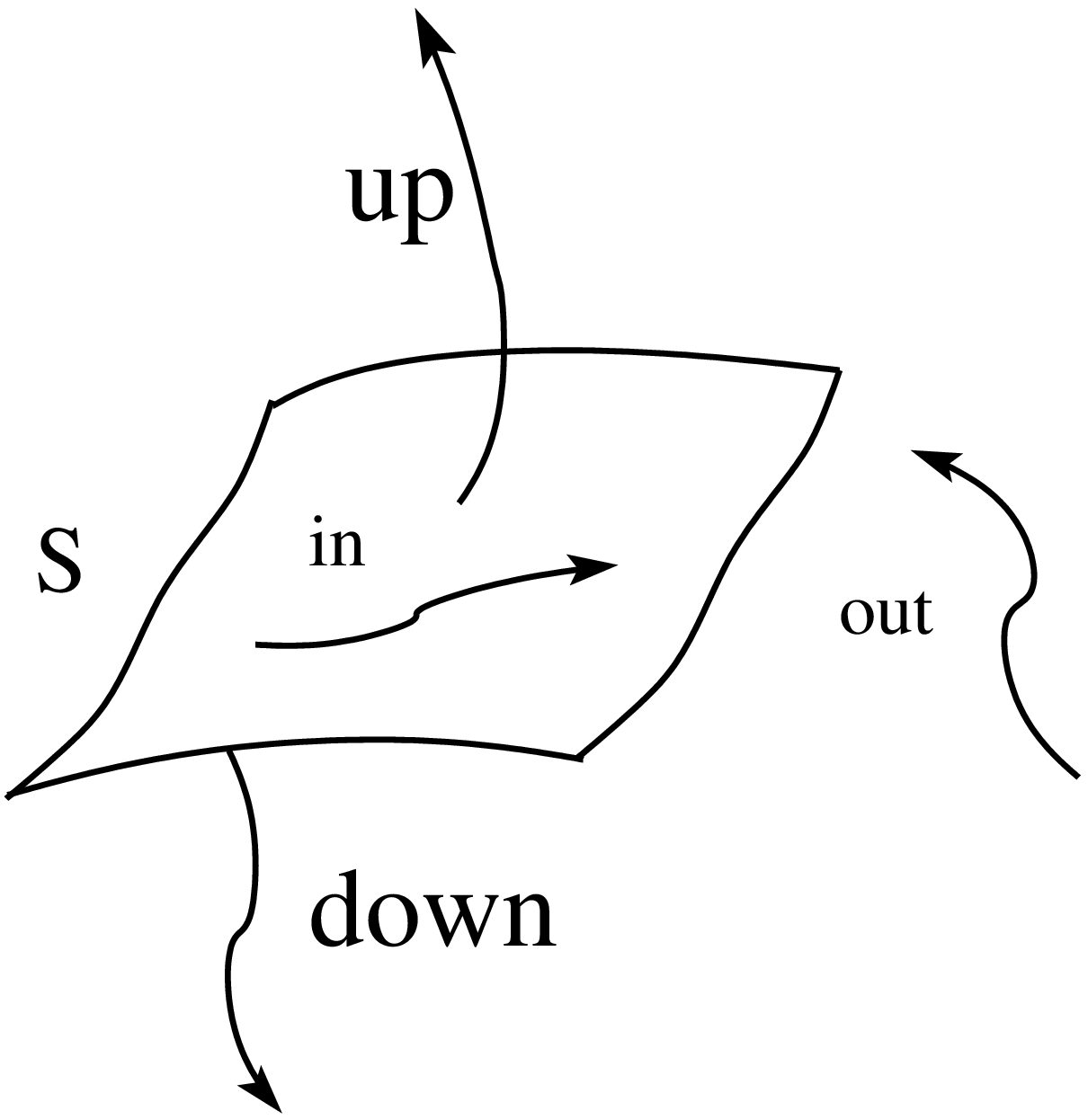}
   \caption{\label{Bild2}\small Edges of type up, down, in and out with respect to the surface $S$.}
 \end{figure}

The classical electric flux $E_k(S)$ through a surface $S$ in LQG is given by the integral of the densitised triad $E^a_k$ over a two surface $S$
\be
E_k(S)=\int\limits_{S}E^a_k\,n^{\st S}_a,
\ee
where $n^{\st S}_a$ is the conormal vector with respect to the surface $S$. 
In order to define a corresponding flux operator in the quantum theory, we have to consider  the Poisson bracket among the classical electric flux and an arbitrary cylindrical function $f_{\gamma}:G^{|E(\gamma)|}\to\Co$, where $G$ is the corresponding gauge group, namely $SU(2)$ in our case. 
\be
\left\{E_k(S),f_{\gamma}(\{h_e(A)\}_{e\in E(\gamma)})\right\}=\sum\limits_{e\in E(\gamma)}\left\{E^a_k,(h_e)_{\sst AB}\right\}\frac{\partial f_{\gamma}}{\partial(h_e)_{\sst AB}}.
\ee
We experience that the Poisson bracket among $E^a_k$ and $f_{\gamma}$ can be calculated whenever the Poisson bracket among $E^a_k$ and the holonomy $(h_e)_{\sst AB}$ is known. As the latter cannot be calculated on the manifold directly because terms including  distributions would appear, we have to regularise our electric flux and also the holonomy. Then we will investigate the regularised Poisson bracket, remove the regulator afterwards and hope that at the end of the day we will obtain a well defined operator. The regularisation can be implemented by smearing the two surface $S$ into the third dimension, shown in figure \ref{Bild1}, so that we get an array of surfaces $S_t$. The surface associated with $t=0$ is our original surface $S$. 
   \newline
We define our regularised classical flux as
\be
\label{regflux}
E^{\epsilon}_k(S):=\frac{1}{2\epsilon}\int\limits_{-\epsilon}^{+\epsilon}\,dt\,E_k(S_t).
\ee
The corresponding operator $\op{E}_k(S)$ in the quantum theory is then defined as
\be
\op{E}_k(S)f_{\gamma}:=i\hbar\lim\limits_{\epsilon\to {\st 0}}\left\{E^{\epsilon}_k(S),f_{\gamma}\right\}.
\ee
We have to derive the Poisson bracket among $E^{\epsilon}_k(S)$ and any 
possible cylindrical function $f_{\gamma}$. For this purpose, we can 
reduce the problem to investigating the Poisson bracket for any possible 
edge that is contained in the graph labelling the cylindrical function. 
The appearing edges can be classified as (i) up, (ii) down, (iii) in and 
(iv) out. Therefore, if we know the Poisson bracket for any of these types 
of edges, we will be able to derive the Poisson bracket among 
$E^{\epsilon}_k$ and any arbitrary $f_{\gamma}$. The calculation of the 
regularised Poisson bracket can be found e.g. in the 
second reference of \cite{1}. 
After having removed the regulator we end up with the following action of the flux operator on an arbitrary cylindrical function $f_{\gamma}$
\be
\op{E}_k(S)f_{\gamma}=\frac{i}{2}\,\lp^2\sum\limits_{e\in E(\gamma)}\epsilon(e,S)\left[\frac{\tau_k}{2}\right]_{AB}\frac{\partial f_{\gamma}(h_{e'})_{e'\in E(\gamma)}}{\partial (h_e)_{\sst AB}},
\ee
where $\tau_k$ is related to the Pauli matrices by $\tau_k:=-i\sigma_k$. The sum is taken over all edges of the  graph $\gamma$ associated with $f_{\gamma}$. The function $\epsilon(e,S)$ can take the values $\{-1,0,+1\}$ depending on the type of edge that is considered. It is +1 for edges of type up, -1 one for down and 0 for edges of type in or out (see figure \ref{Bild2}). 
\newline
If we introduce right invariant vector fields $X^e_k$, defined by $(X^e_k f)(h):=\frac{d}{dt}f(e^{t\tau_k}h)\Big|_{t=0}$, we can express the action of the flux operator by
\be
\op{E}_k(S)f_{\gamma}=\frac{i}{4}\,\lp^2\sum\limits_{e\in E(\gamma)}\epsilon(e,S)X^e_k\,f_{\gamma}.
\ee
 The right invariant vector fields fulfill the following commutator 
relations
\be
[X^r_e,X^s_e]=-2\epsilon_{rst}X^t_e.
\ee
By means of introducing the self-adjoint right invariant vector field $Y^k_e:=-\frac{i}{2}X^k_e$, we achieve commutator relations for $Y^k_e$ that are similar to the one of the angular momentum operators in quantum mechanics
\be
[Y^r_e,Y^s_e]=i\epsilon_{rst}Y^t_e.
\ee
Consequently, we can describe the action of $\op{E}_k(S)$ by the action of the self-adjoint right invariant vector field $Y^k_e$ on $f_{\gamma}$ 
\be
\label{flY}
\op{E}_k(S)f_{\gamma}=-\frac{1}{2}\lp^2\sum\limits_{e\in E(\gamma)}\epsilon(e,S)Y^k_e\,f_{\gamma}.
\ee
\section{Idea and Motivation of the Alternative Quantisation of the Flux Operator}%
\label{Idea}
Recall again the definition of the regularised classical flux $E^{\epsilon}_k(S)$ in eqn (\ref{regflux}). 
We take the Poisson bracket of the Ashtekar-connection $A^j_a$ and the densitised triad $E^b_k$   given by 
\be
\left\{A^j_a(x),E^b_k(y)\right\}=\delta^3(x,y)\delta^a_b\delta^k_j
\ee
as our fundamental starting point. If we use a canonical transformation in order to go from the ADM-formalism to the formulation in terms of Ashtekar varibales, we have two possibilities of choosing such a canonical transformation that both lead to the Poisson bracket above. These two possiblities are
\be
\begin{array}{llcl}
I & A^j_a=\Gamma^j_a +\gamma\sgn(\det(e))K^j_a &, &E^a_j=\frac{1}{2}\epsilon_{kst}\epsilon^{abc}e^s_b\,e^t_c\\
II & A^j_a=\Gamma^j_a + \gamma K^j_a &, & E^a_j=\frac{1}{2}\epsilon_{kst}\epsilon^{abc}e^s_b\,e^t_c\sgn(\det(e))
\end{array}
\ee
Here, $\Gamma^j_a$ is the SU(2)-spin connection, $K^j_a$ the extrinsic curvature and $\gamma$ the Imirzi-parameter.
Now the idea of defining an alternative regularised flux 
\be 
\label{altdtriad}
G{\epsilon}_k(S):=\frac{1}{2\epsilon}\int\limits_{-\epsilon}^{+\epsilon}\,dt\,\wt{E}_k(S_t)\, ;\quad\quad \wt{E}_k(S_t)=\int\limits_{S_t}E^a_k\,n^{\st S_t}_a
\ee
is to express the densitised triad $E^a_k$ in terms of the triads as above. Due to the two possibile canonical transformations, we have also two possibilities in defining an alternative densitised triad 
\be
\label{Eakdef}
E^a_k=\left\{\begin{array}{lcl} \det(e)e^a_k &=& \frac{1}{2}\epsilon_{kst}\epsilon^{abc}e^s_b\,e^t_c\\
                                         \sqrt{\det(q)}e^a_k &=& \frac{1}{2}\epsilon_{kst}\epsilon^{abc}\underbrace{\sgn(\det(e))}_{\displaystyle =:{\cal S}}e^s_b\,e^t_c\end{array}\right\}=:\left\{\begin{array}{c} E^{a,{\sst I}}_k \\ E^{a,{\sst II}}_k\end{array}\right\},
\ee
where $e^j_a$ is the cotriad related to the intrinsic metric as $q_{ab}=e^j_ae^j_b$. From now on we will use $E^{a,{\sst I}}_k$ and $E^{a,{\sst II}}_k$, respectively for the two cases.\\
So, instead of quantising the densitised triad directly, we could use the above classical identities, quantize them and check whether both quantisation procedures are consistent. The main difference between these two definitions is basically a signum factor which we will denote by ${\cal S}$. From the mathematical point of view  both definitions in eqn (\ref{altdtriad}) are equally viable, thus we will keep both possibilities and emphasise the differences that occur when we choose one or the other definition.
However, notice that case I leads to the anholonomic constraint $\det(E)\ge 0$ emphasised also in \cite{11} which already seems unlikely to be reproduced by quantising $E$ as a vector field on some space of connections. Notice that the precise distinction between case I and case II is often forgotten in the LQG literature where one treats $*E$ as a 2-form (I) when convinient and $*E$ as a vector density (II) when convinient. While this is classically immaterial as long as $\Sigma$ is orientable, we will see that in the Quantum theory this becomes crucial.  
\newline
If we parametrise the surface integral over $S_t$, we obtain for the alternative flux
\be
\label{altEk}
\wt{E}_k(S_t)=\left\{\begin{array}{lcrl}\int\limits_{S_t}d^2u\;\epsilon_{kst}\left[e^s_b(X(u))X^b_{,u_{\sst 3}}(u)\right]\left[e^t_c(X(u))X^c_{,u_{\sst 4}}(u)\right]&\quad ,E^{a,{\sst I}}_k=\det(e)e^a_k\\
\int\limits_{S_t}d^2u\;\epsilon_{kst}\left[e^s_b(X(u))X^b_{,u_{\sst 3}}(u)\right]{\cal S}\left[e^t_c(X(u))X^c_{,u_{\sst 4}}(u)\right]&\quad ,E^{a,{\sst II}}_k=\sqrt{\det(q)}e^a_k\end{array}\right\},
\ee                 
where we used the expression of the conormal vector  $n^{\st S_t}_a=\epsilon_{aqr}X^q_{,u_{\sst 3}}(u)X^r_{,u_{\sst 4}}(u)$ associated with the surface $S_t$ in terms of an arbitrary embedding $X:(-\frac{1}{2},+\frac{1}{2})^2\to S;(u_{\sst 3},u_{\sst 4})\mapsto X(u_{\sst 3},u_{\sst 4})$ 
\section{Construction of the Alternative Flux Operator}%
Our strategy in quantising the alternative expression of the electric flux will be as follows:
 First of all we  express the triads such as $e^s_b$ in eqn (\ref{altEk}) in terms of the Poisson bracket among the components of the connection $A^s_b$ and the Volume $V(R)$, given by $\{A^s_b,V(R)\}$. Here $V(R)=\int\limits_{R}d^3x\sqrt{\det q}$ is the volume of the region $R$. This kind of quantisation procedure was first introduced in \cite{2} in order to derive a well defined expression for the Hamiltonian Constraint in the quantum theory and is used in various applications in LQG nowadays. By comparing the action of the alternative flux operator with the one for the usual flux operator later on, we are able to verify whether this particular way of quantising leads to the correct and expected result. Therefore, this can be seen as an independent check of this particular way of quantisation. As a second step, we want to replace the connection by holonomies, for which well defined operators exist. For this reason we will have to partition each surface $S_t$ and consider the limit where the partition gets finer and finer. This will be explained in more in detail later. Before we apply canonical quantisation and replace the Poisson brackets by the corresponding commutators, we want to 
get various issues out of the way.
\subsection{Replacement of the Triads by means of the Poisson bracket}%
As before  we want to derive the relation between the Poisson bracket $\{A^s_b,V(R)\}$  and the cotriads for both expression of $E^a_k$ in eqn (\ref{altEk}). The explicit definition of the densitised triad $E^a_k$ in terms of the $e^a_k$ enters the calculation. Thus it is not surprising that the final result is different for the two cases
\be
\left\{A^s_b,V(R)\right\}=\left\{\begin{array}{lcl}-\frac{\kappa}{2}{\cal S}e^s_b&,& \quad E^{a,{\sst I}}_k=\frac{1}{2}\epsilon_{\sst kst}\epsilon^{\sst abc}e^s_be^t_c\\
                                                    -\frac{\kappa}{2}e^s_b&,& \quad E^{a,\sst II}_k=\frac{1}{2}\epsilon_{\sst kst}\epsilon^{\sst abc}{\cal S}e^s_be^t_c
\end{array}\right\}.
\ee 
By using the above identity and inserting it into eqn (\ref{altEk}) we get
\be
\wt{E}_k(S_t)=\left\{\begin{array}{lcll}\frac{4}{\kappa^2}\int\limits_{S_t}d^2u\;\epsilon_{kst}\left\{A^s_b(X(u))X^b_{,u_{\sst 3}}(u),V(R)\right\}\left\{A^t_c(X(u))X^c_{,u_{\sst 4}}(u),V(R)\right\}&\quad ,E^{a,{\sst I}}_k=\det(e)e^a_k\\
\frac{4}{\kappa^2}\int\limits_{S_t}d^2u\;\epsilon_{kst}\left\{A^s_b(X(u))X^b_{,u_{\sst 3}}(u),V(R)\right\}{\cal S}\left\{A^t_c(X(u))X^c_{,u_{\sst 4}}(u),V(R)\right\}&\quad ,E^{a,{\sst II}}_k=\sqrt{\det(q)}e^a_k\end{array}\right\}. 
\ee 
Here $V(R)$ is any region containing $\cup S_t$, $t\in [-\epsilon,\epsilon]$ and we used ${\cal S}{\cal S}\in\{0,+1\}$ and thus could completely neglect the signum factor in the case of $E^a_k=\det(e)e^a_k$, because classically the 3-metric is nondegenerate, hence ${\cal S}^2=1$.
At this stage we already see that the main difference between the two expressions of $E^a_k$ is whether we will have a signum factor in the final (classical) expression or not. Exactly this feature will be very important in the quantum expression, because the action of the corresponding operator differs remarkably if the operator contains a corresponding signum operator or if it does not.
\subsection{Replacement of the Connections by Holonomies}%
\begin{figure}[hbt]
   \center
    \psfrag{t}{$t$}
   \psfrag{n}{$\vec{n}^{\sst S_t}$}
   \psfrag{e'}{${\st \epsilon'}$}
   \psfrag{v}{${\st v(}{\sst \Box}{\st )}$}
   \psfrag{e3}{${\st e}_{\sst 3}{\st({\sst \Box})}$}
   \psfrag{e4}{${\st e}_{\sst 4}{\st ({\sst \Box})}$}
   \psfrag{PofS}{${\cal P}_t$ of $S_t$}
   \includegraphics[height=7cm]{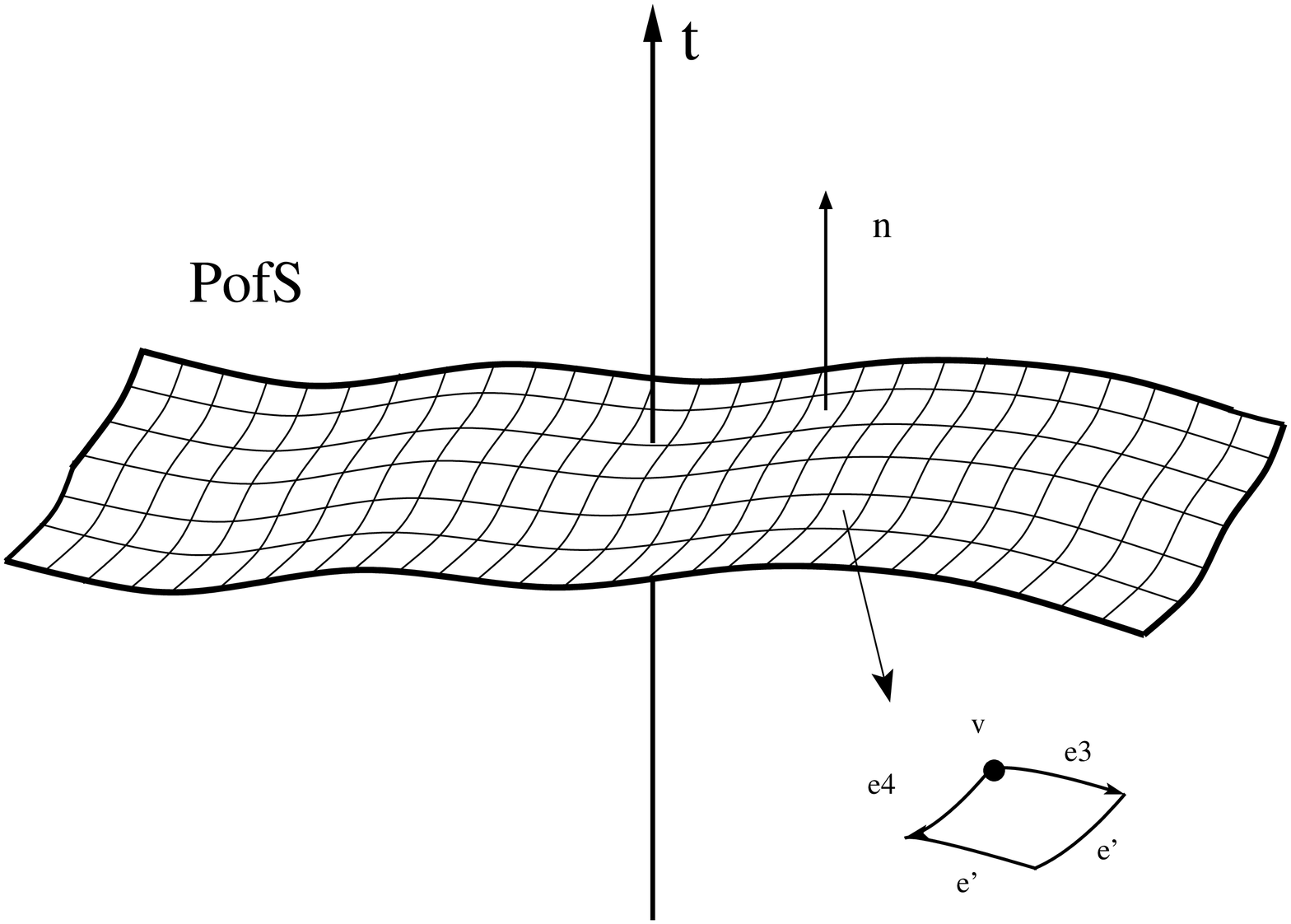}
   \caption{\label{Bild3}\small Partition ${\cal P}_t$ of the surface $S_t$ into small squares with an parameter edge length $\epsilon'$.}
    \end{figure}

Our main aim is to express 
the components of the connections $A^s_b(X_S(u))$ in terms of holonomies 
 for which well defined operators on the quantum level are known. For this 
 reason we partition each surface $S_t$ into small squares with an 
 parameter edge length $\epsilon'$ as shown in figure \ref{Bild3}. We can 
 therefore express the integral over $S_t$ as the sum over the integrals 
 over all small squares in the limit where the partition gets 
infinitesimally small. Consequently, we can rewrite eqn (\ref{altEk}) as
\be
\label{Ekpart}
\wt{E}_k(S_t)=\left\{\begin{array}{lcll}\lim\limits_{{\cal P}_t\to S_t}\sum\limits_{\sst {\Box}\in {\cal P}_t}\frac{4}{\kappa^2}\epsilon_{kst}\left\{A^s_{\sst 3}({\sst {\Box}}),V(R_{v({\sst \Box})})\right\}\left\{A^t_{\sst 4}({\sst {\Box}}),V(R_{v({\sst \Box})})\right\}&\quad ,E^{a,{\sst I}}_k=\det(e)e^a_k\\
\lim\limits_{{\cal P}_t\to S_t}\sum\limits_{\sst {\Box}\in {\cal P}_t}\frac{4}{\kappa^2}\epsilon_{kst}\left\{A^s_{\sst 3}({\sst {\Box}}),V(R_{v({\sst \Box})})\right\}{\cal S}\left\{A^t_{\sst 4}({\sst {\Box}}),V(R_{v({\sst \Box})})\right\}&\quad ,E^{a,{\sst II}}_k=\sqrt{\det(q)}e^a_k\end{array}\right\},
\ee 
 where we introduced the notation $A^s_{\st I}({\sst 
 {\Box}})=\int\limits_{e_{\sst I}({\sst {\Box}})}A^s\, ,I=3,4$ for the 
 integral over the connection along the edge $e_{\sst I}({\sst {\Box}})$ 
 of $\sst {\Box}$. Here $R_{v_{\sst \Box}}$ is any region containing the 
 point $e_3({\sst \Box}) \cap e_4({\sst \Box})$ and in the limit 
$\epsilon'\to 0$ also  $R_{v_{\sst \Box}}\to v({\sst \Box})$.
\newline
 If we choose $\epsilon'$ small enough, we can use the following approximation
\be
\label{happrox1}
\left\{A^s_{\st I}({\sst {\Box}}),V(R_{v({\sst \Box})})\right\}\frac{\tau_s}{2}+o(\epsilon'^2)=+h_{e_{\sst I}}\left\{h^{-1}_{e_{\sst I}},V(R_{v({\sst \Box})})\right\}.
\ee
The above equation holds for holonomies in the spin-$\frac{1}{2}$-representation. We would like to generalise this relation to the case of holonomies with an arbitrary weight $\ell$ in order to construct an operator that could contain arbitrary spin representations. This could be useful in the sense that we are then able to analyse whether the result of our alternative flux operator is sensitive to the chosen weight. That is, we investigate the effect of this particular kind of factor ordering ambiguity in the classical limit. The generalisation of eqn (\ref{happrox1}) is straight forward and leads to
\be
\left\{A^s_{\st I}({\sst {\Box}}),V(R_{v({\sst \Box})})\right\}\frac{1}{2}\pi_{\ell}(\tau_s)+o(\epsilon'^2)=+\pi_{\ell}(h_{e_{\sst I}})\left\{\pi_{\ell}(h^{-1}_{e_{\sst I}}),V(R_{v({\sst \Box})})\right\},
\ee
where we denote a representation with weight $\ell$ by $\pi_{\ell}$.
By choosing $\epsilon'$ small enough, we are allowed to replace the Poisson brackets $\{A^s_{\sst 3}({\sst {\Box}}),V(R_{v({\sst \Box})})\}$ and $\{A^t_{\sst 4}({\sst {\Box}}),V(R_{v({\sst \Box})})\}$, respectively, by Poisson brackets including holonomies. Thus the basis of the alternative flux operator will be the following classical identity
\ba
\label{clid}
\lL\wt{E}^{\sst I/II}_k(S_t)&=&\lim_{{\cal P}_t\to S_t}\sum\limits_{\sst {\Box}\in{\cal P}_t}\epsilon_{kst}\frac{4}{\kappa^2}\left\{A^s_{\sst 3}({\sst {\Box}}),V(R_{v({\sst \Box})})\right\}{\cal S}\left\{A^t_{\sst 4}({\sst {\Box}}),V(R_{v({\sst \Box})})\right\}\nonumber\\    
&=&\lim_{{\cal P}_t\to S_t}\sum\limits_{\sst {\Box}\in{\cal P}_t}\frac{16}{\kappa^2}\frac{1}{\frac{4}{3}\ell(\ell+1)(2\ell+1)}\nonumber \\
&&\quad\quad\quad\quad
\tr\left(\pi_{\ell}(h_{e_{\sst 3}({\sst {\Box}})})\left\{\pi_{\ell}(h^{-1}_{e_{\sst 3}({\sst {\Box}})}),V(R_{v({\sst \Box})})\right\}\pi_{\ell}(\tau_k)\fbox{${\cal S}$}\pi_{\ell}(h_{e_{\sst 4}({\sst {\Box}})})\left\{\pi_{\ell}(h^{-1}_{e_{\sst 4}({\sst {\Box}})}),V(R_{v({\sst \Box})})\right\}\right).
\ea
The box around the signum factor ${\cal S}$ indicates that it is not contained in the equation if we choose $E^{a,{\sst I}}$, but occurs when we use $E^{a,{\sst II}}_k$. If one wants to show the  correctness of the above identity, one has to use the following indentity $\tr\left(\pi_{\ell}(\tau_s)\pi_{\ell}(\tau_k)\pi_{\ell}(\tau_t)\right)=-\frac{4}{3}\ell(\ell+1)(2\ell+1)\epsilon_{skt}$ which is derived in 
appendix \ref{Taus}.
\newline
Hence, we managed to derive an alternative expression for the flux operator on the classical level which we are able to quantise by means of well known operators
\ba
\label{alflres}
\lL\wt{E}^{\sst I/II}_k(S_t)&=&\lim_{{\cal P}_t\to S_t}\sum\limits_{\sst 
{\Box}\in{\cal P}_t}\frac{16}{\kappa^2}\frac{1}{\frac{4}{3}\ell(\ell+1)
(2\ell+1)}\\
&&\quad\quad\quad\quad
\tr\left(\pi_{\ell}(h_{e_{\sst 3}({\sst {\Box}})})\left\{\pi_{\ell}(h^{-1}_{e_{\sst 3}({\sst {\Box}})}),V(R_{v({\sst \Box})})\right\}\pi_{\ell}(\tau_k)\,\fbox{${\cal S}$}\,\pi_{\ell}(h_{e_{\sst 4}({\sst {\Box}})})\left\{\pi_{\ell}(h^{-1}_{e_{\sst 4}({\sst {\Box}})}),V(R_{v({\sst \Box})})\right\}\right)
\nonumber
\ea
From now on we will neglect the dependence of the edges $e_{\sst I}({\sst {\Box}})$ on the particular point $P_{\sst I}({\sst {\Box}})$ in order to keep the expressions clearer. 
\subsection{Notion of Convergence and Factor Ordering}  
\label{factord}
Let us now discuss in which sense the limit $\epsilon\to 0$ is to be
understood. First of all we formally have for any spin network state $T_s$
\be
\lL\hat{G{\epsilon}_k} T_s:=\frac{1}{2\epsilon}
\int\limits_{-\epsilon}^\epsilon  \sum_{s'}
\auf T_{s'}\,|\,\lL\widehat{\wt{E}}_k(S_t)\,|\, T_s\zu T_{s'}
\ee
where we sum over all spin network labels $s'$ (resolution of unity).
Notice that the sum $\sum_{s'}$ must be taken under the integral as 
otherwise the result would automatically be zero. Moreover, notice that
for each $t$ the number of $s'$ contributing is finite. 
Next we have 
\be
\auf T_{s'}\,|\,\lL\widehat{\wt{E}}_k(S_t) \,|\, T_s\zu=\lim_{{\cal 
P}_t\to S_t}
\sum_{\Box\in {\cal P}_t}  
\auf T_{s'}\,|\, \lL\widehat{\wt{E}}_k(\Box)\,|\, T_s\zu  
\ee
In order to simplify the notation, let us assume that for all $t$
the limit $\epsilon'\to 0$ implies ${\cal P}_t\to S_t$ while the 
parameter area of the squares within the partitions decay to zero as 
$(\epsilon')^2$. Then we can combine the two formulas and write
\be
\lL\hat{\tilde{E}}^{\epsilon,\epsilon'}_k T_s:=\frac{1}{2\epsilon}
\int\limits_{-\epsilon}^\epsilon dt\sum_{s'} \sum_{\Box\in {\cal P}_t}
\auf T_{s'}\,|\,\lL\widehat{\wt{E}}_k(\Box)\,|\, T_s\zu T_{s'}
\ee
It is easy to see that the Hilbert norm of this object vanishes
with respect to the Hilbert 
space ${\cal H}_{Kin}=L_2(\ab,d\mu_{AL})$ of LQG where $\ab$ is the 
Ashtekar -- Isham space of {\it generalized} connections and $\mu_{AL}$ is 
the 
Ashtekar -- Lewandowski measure. Basically 
this happens because the norm squared involves a double integral over 
$t,t'$ while the integrand has support only on the measure zero subset
$t=t'$.
Hence, we cannot use the strong operator topology as a notion of 
convergence. The same applies to the weak operator topology. Rather, we 
will use the same notion of convergence as the one 
that has been used for the fundamental flux operator: Given a point 
$A\in \a$ in the space of {\it smooth} connections, we may evaluate the 
above expression at $A$ and obtain a function on $\a$
\be
[\lL\hat{\tilde{E}}^{\epsilon,\epsilon'}_k T_s](A):=\frac{1}{2\epsilon}
\int_{-\epsilon}^\epsilon dt\sum_{s'} \sum_{\Box\in {\cal P}_t}
\auf T_{s'}\,|\, \lL\widehat{\wt{E}}_k(\Box)\,|\, T_s\zu  T_{s'}(A)
\ee
We now take the limit $\epsilon'\to 0$ {\it before} the limit 
$\epsilon \to 0$ in the following sense: We say that 
\be
\lim\limits_{\epsilon\to 0} \lim\limits_{\epsilon'\to 0} 
\lL\hat{\tilde{E}}_k^{\epsilon,\epsilon'}(S)=
\lL\widehat{\wt{E}}_k(S)
\ee
provided that for any $A\in \a$ and any spin network label $s$
\be
\lim\limits_{\epsilon\to 0} \lim\limits_{\epsilon'\to 0} 
|[\lL\hat{\tilde{E}}^{\epsilon,\epsilon'}_k(S) \; T_s](A)
-[\lL\widehat{\wt{E}}_k(S) T_s](A)| =0
\ee
Notice that the limit is pointwise in $A,s$ and not uniform. Notice 
also that this is a limit from the space of operators on the space of 
functions of smooth connections to operators on ${\cal H}_{Kin}$ and 
not a convergence of operators on ${\cal H}_{Kin}$.\\
\newline\newline
With these preparations out of the way we may now draw already some 
first conclusions about the action of the final operator  
$\lL\hat{\wt{E}}_k(S)$. We may assume without loss of generality that 
both graphs $\gamma=\gamma(s),\;\gamma'=\gamma(s')$ underlying 
$\auf T_{s'}\,|\,\lL\widehat{\wt{E}}_k(\Box)\,|\, T_s\zu$ 
are adapted to $S$ in 
the sense that each of their edges has well defined type with respect to 
$S$. If an edge $e$ is of type up or down respectively then 
$S_t\cap e\not=\emptyset$ only for $t\ge 0$ or $t\le 0$ respectively. 
If $e$ is of type in or out respectively then for sufficiently small 
$\epsilon$ 
we have $S_t\cap e\not=\emptyset$ only for $t=0$ or for no $t$ at all 
respectively. Now consider $\auf T_{s'}\,|\,\lL\widehat{\wt{E}}_k(\Box)\,|\, T_s\zu$ at 
finite $\epsilon'$. Since $\widehat{\tilde{E}}_k(\Box)$ involves the volume 
operator which has non -- trivial action only when the state on which it 
acts has at least one at least trivalent vertex, no matter whether we use 
the RS or AL volume operator, it easily follows that 
$[h_{e_I(\Box)}^{-1},\hat{V}_{v(\Box)}],\;I=3,4$ annihilates $T_s$ unless 
$e_I(\Box),\gamma$ intersect each other. Hence, in order to obtain a 
non -- vanishing contribution at all, we must refine ${\cal P}_t$ 
in such a way each edge $e\in E(\gamma)$ intersecting $S_t$ at all does so 
by intersecting with at least one of the $e_I(\Box)$ for some $I\in 
\{3,4\}$ and at least one $\Box \in {\cal P}_t$. 
Making use of the fact 
that classically the limit ${\cal P}_t \to S_t$ is independent of the 
refinement we refine ${\cal P}_t$ graph dependently by demanding that 
eventually $e\cap S_t$ coincides with precisely one of the $v(\Box)$
if $e$ is of type up or down respectively and $t\ge 0$ or $t\le 0$ 
respectively. This is motivated by the fact otherwise no such edge would 
contribute if we use the AL version. If $e$ is of type in and $t=0$ then 
the number
of intersections of $e$ with the $e_I(\Box)$ necessarily diverges as
$\epsilon'\to 0$. However, if we use the AL volume, all these 
contributions vanish because, in order that its action be non trivial, 
it needs non -- coplanar vertices, except 
if $v(\Box)$ coincides with an endpoint of $e$ where there might be 
additional edges of $e$ adjacent which are transversal to $S$. If we use 
the RS volume then all these intersections contribute and the 
sum over $\Box$ diverges for suitable $s'$ as $\epsilon'\to 0$. 
However, since we perform the integral over $t$ before taking 
$\epsilon'\to 0$ and  
the support of the integrand for type in edges consists
of the measure zero set $t=0$, the contribution vanishes, again 
no matter whether we use the RS or AL volume.
\begin{figure}[hbt]
\center
    \psfrag{n}{$\vec{n}^{\sst S_t}$}
   \psfrag{e'}{$\epsilon'$}
   \psfrag{v}{$v({\sst \Box})$}
   \psfrag{e1}{$e_{\sst 1}({\sst \Box})$}
   \psfrag{e2}{$e_{\sst 2}({\sst \Box})$}
   \psfrag{e3}{$e_{\sst 3}({\sst \Box})$}
   \psfrag{e4}{$e_{\sst 4}({\sst \Box})$}
\includegraphics[height=5cm]{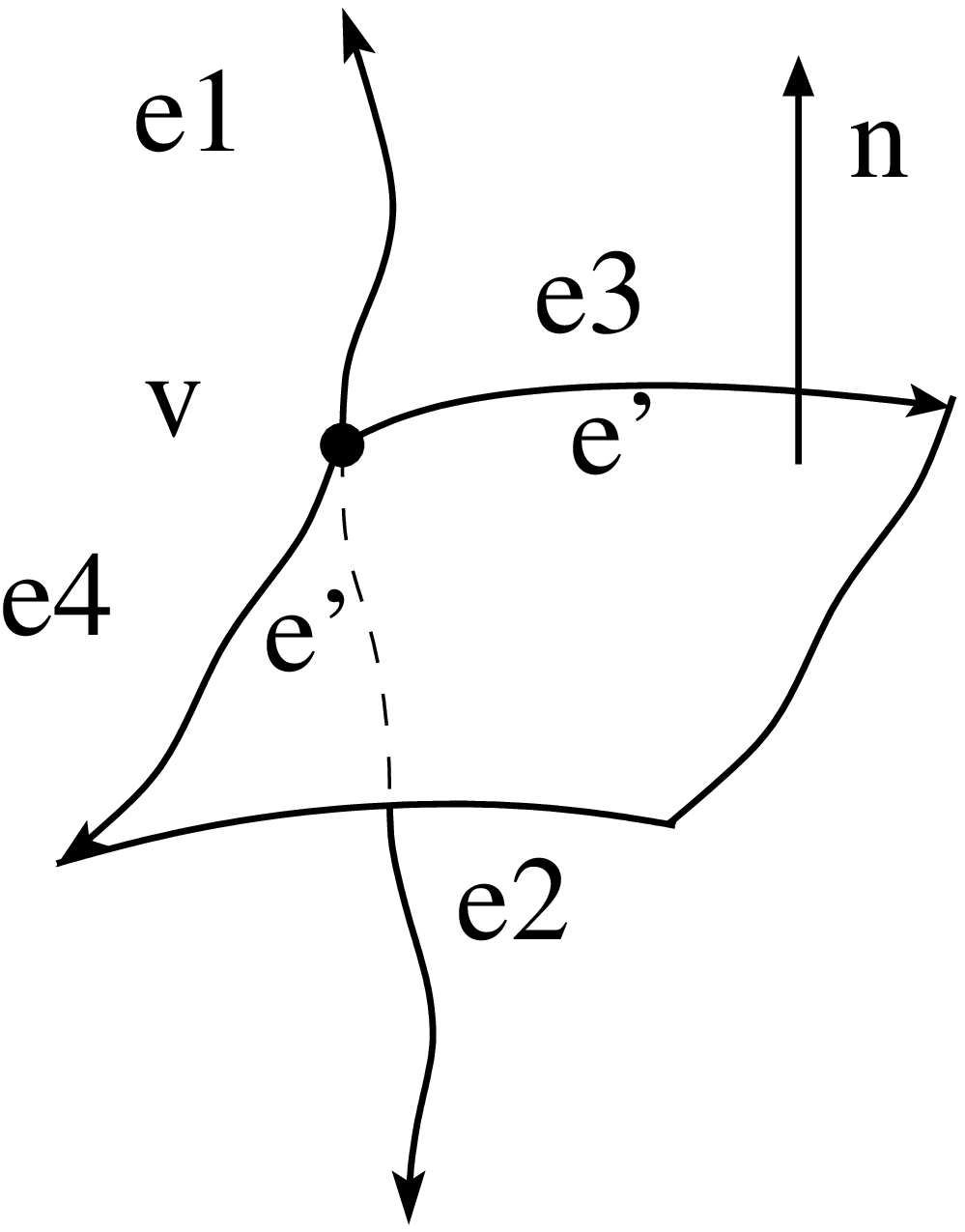}
   \caption{\small An non-vaishing contribution to $\auf T_{s'}\,|\,\lL\widehat{\wt{E}}_k(\Box)\,|\, T_s\zu$ can only be achieved if $T_s$ contains edges of type up and/or down, respectively with respect to the surface $S_t$. Moreover, the edges $e_{\sst 3}({\sst \Box}),e_{\sst 4}({\sst \Box})$ have to be attachted to $T_s$ in this specific way.}      
   \end{figure} 
\newline         
We conclude that for both versions of the volume operator only edges of 
type up or down will contribute, exactly as for the fundamental flux 
operator. However, for the AL volume the required ordering is more 
restrictive because there must be terms with both edges $e_3(\Box), 
e_4(\Box)$ to the right of $\hat{V}_{v(\Box)}$. For the RS volume there 
are more possibilities available which we will discuss in a later part of 
the paper. 
\newline
Now let us derive which $s'$ contribute to  
$\auf T_{s'}\,|\,\lL\widehat{\wt{E}}_k(\Box)\,|\, T_s\zu$ for given $s$ and $\Box\in S_t$.
We may restrict to edges of type up or down as just discussed.
The factors $\pi_{\ell}(h_{e_I(\Box)}),\; \pi_{\ell}(h_{e_I(\Box)})^{-1}$
involved could a priori change the graph $\gamma$ by adding the edge 
$e_I(\Box)$ with spin $J=0,1,..,2\ell$. 
However, the operator $\lL\hat{\wt{E}}_k(\Box)$ is invariant under gauge 
transformations at the endpoints of the $e_I(\Box)$ by construction, hence
we must necessarily have $J=0$. Thus, even at finite $\epsilon'$ the 
operator $\lL\widehat{\wt{E}}_k(\Box)$ does not change the range of 
the graph $\gamma$. Hence, the only difference between 
$s',s$ is that $\gamma'=\gamma$ but the edge $e\in E(\gamma')$ appears 
splitted into $e^t_1, e^t_2$ with $e=(e^t_2)^{-1} \circ (e^t_1)$ and 
$e^t_1\cap e^t_2=v(\Box)=e\cap S_t$. Notice also that with $dt$ measure 
one the point $S_t\cap e$ is an interior point of $e$. This is important 
because the contribution of $\auf T_{s'}\,|\, \lL\widehat{\wt{E}}_k(\Box)\,|\, T_s\zu$
for $\Box\in S$ differs from that for $\Box \in S_t,\;t\not=0$ because 
in the former case $v(\Box)$ maybe a vertex of higher valence than four.
\newline\newline
Finally notice that 
$\lL\widehat{\wt{E}}_k(\Box) |T_s\zu$ transforms in the spin one 
representation at $v(\Box)$ because $T_s$ is gauge invariant there.
Hence $T_{s'}$ bust have a spin one intertwiner at $v(\Box)$.

What happens now when we take the limit as discussed is the following:
For each value of $t$ the sum over $\Box$ can be replaced by a finite 
number of terms, one for each $e\in E(\gamma)$ of type up or down and taking the
limit $\epsilon'\to 0$ becomes trivial. Next, for each 
value of $t$ and each edge $e\in E(\gamma)$ there will be a finite 
number of states $T_{s'_{e,t}}$ which contribute to the sum over $s'$ and 
which are mutually orthogonal for different $e,t$. 
The numbers 
$\auf T_{s'_{e,t}}\,|\,\lL\widehat{\wt{E}}_k(\Box)\,|\, T_s\zu$ do not 
depend on $t$ (thanks of the diffeomorphism invariance of the measure), 
however
the states $T_{s'_{e,t}}$ do. Fortunately, considered as functions of 
smooth connections, the limit $\epsilon \to 0$ converges and results in 
states $T_{s'_e}$ where $\gamma(s'_e)=\gamma$ not only have the same range 
but also the same edge sets. Then $s,s'_e$ differ only by the intertwiner 
at the point $v=b(e)$.
\subsection{Classical Identity} %
Collecting all the arguments of the discussion of the last section, we end up with the following ordering of the classical terms\footnote{In the case of $\op{V}_{\sst RS}$ exist more than this symmetric factor ordering. We will discuss this aspect later in the paper.}
\ba
\label{Classfinal}
 \lL\wt{E}^{\sst I/II}_k(S_t)&=&-\lim_{{\cal P}_t\to S_t}\sum\limits_{\sst 
 {\Box}\in{\cal P}_t}\frac{16}{\kappa^2}\frac{1}{\frac{4}{3}\ell(\ell+1)
(2\ell+1)}\pi_{\ell}(\tau_k)_{\sst CB}\\
&&\quad\quad\quad\quad
\pi_{\ell}(h_{e_{\sst 4}})_{\sst CD}\left\{\pi_{\ell}(h^{-1}_{e_{\sst 3}}),V(R_{v({\sst \Box})})\right\}_{\sst AB}\,\fbox{${\cal S}$}\,\left\{V(R_{v({\sst \Box})}),\pi_{\ell}(h^{-1}_{e_{\sst 4}})\right\}_{\sst DE}\pi_{\ell}(h_{e_{\sst 3}})_{\sst EA},
\nonumber
\ea
where the indices $\{ A,B,C,D,E\}\in\{-\ell,...,+\ell\}$.
\newline
Since the Poisson bracket in eqn (\ref{Classfinal}) inlcudes the classical Volume $V(R_{v({\sst \Box})})$, the coresponding alternative flux operator contains the volume operator $\op{V}$. As mentioned in the introduction, in LQG exist two different volume operators, $\op{V}_{\sst RS}$ and $\op{V}_{\sst AL}$. Thus, for each case I and II respectively, we have 2 different alternative flux operators depending on the choice of $\op{V}_{\sst RS}$ and $\op{V}_{\sst AL}$ respectively. Hence, after canonical quantisation, we end up with four different versions of the alternative flux operator. For these four operators, we use the following notation
\ba
 \op{\wt{E}}^{\sst I}_k(S_t)&\longrightarrow&\op{\wt{E}}^{\sst I,AL}_k(S_t),\quad\op{\wt{E}}^{\sst I,RS}_k(S_t)\nonumber\\
 \op{\wt{E}}^{\sst II}_k(S_t)&\longrightarrow&\op{\wt{E}}^{\sst II,AL}_k(S_t),\quad\op{\wt{E}}^{\sst II,RS}_k(S_t)\\
 \ea
 Before, we want to apply canonical quantisation on the classical identity in eqn (\ref{Classfinal}) we want to discuss the two volume operators $\op{V}_{\sst RS},\op{V}_{\sst AL}$ in more detail.
 \subsection{The two Volume Operators of LQG}
 \label{TwoVs}  %
 \subsubsection{The Volume Operator $\op{V}_{\sst RS}$ of Rovelli and Smolin}
 The idea that the volume operator acts only on vertices of a given graph was first mentioned in \cite{15}. The  first version of a volume operator can be found in \cite{3} and is given by
 \ba
\label{Voldef}
\op{V}(R)_{\gamma}&=&\int\limits_{R}\,d^3p\op{V}(p)_{\gamma}\nonumber\\
\op{V}(p)_{\gamma}&=&\lp^3\sum\limits_{v\in V(\gamma)}\delta^{(3)}(p,v)\op{V}_{v,\gamma}\nonumber\\
\op{V}^{\sst RS}_{v,\gamma}&=&\sum\limits_{I,J,K}\sqrt{
\Big| \frac{i}{8} C_{reg}
\epsilon_{ijk}X^i_{e_{\sst I}}X^j_{e_{\sst J}}X^k_{e_{\sst K}}\Big|}.
\ea
Here we sum over all triples of edges  at the vertex $v\in V(\gamma)$ of a given graph $\gamma$. $\op{V}_{\sst RS}$ is not sensitive to the orientation of the edges, thus also linearly dependent triples have to be considered in the sum. Moreover, we introduced a constant $C_{reg}\in\Rl$ that we will keep arbitrary for the moment and that is basically fixed by the particular regularisation scheme one chooses. 
As for the usual flux operator, we want to express $\op{V}$ in terms of self-adjoint vector fields $Y_e^k:=-\frac{i}{2}X_e^k$. Hence, we have
\be
\label{XandJ}
\epsilon_{ijk}X^i_{e_{\sst I}}X^j_{e_{\sst J}}X^k_{e_{\sst K}}=-8i\epsilon_{ijk}Y^i_{e_{\sst I}}Y^j_{e_{\sst J}}Y^k_{e_{\sst K}}.
\ee
and thus
\be
\op{V}^{\sst RS}_{v,\gamma}=\sum\limits_{I,J,K}\sqrt{\Big|iC_{reg}\epsilon_{ijk}Y^i_{e_{\sst I}}
Y^j_{e_{\sst J}}Y^k_{e_{\sst K}}\Big|}.
\ee
In order to selecet the gauge invariant states properly, we have to express our abstract angular momentum states in terms of the recoupling basis. The following identity \cite{10} holds
\be
\label{qijkDef}
\frac{1}{8}\epsilon_{ijk}X^i_{e_{\sst I}}X^j_{e_{\sst J}}X^k_{e_{\sst K}}=\frac{1}{4}[Y^2_{\sst IJ},Y^2_{\sst JK}]=:\frac{1}{4}q_{\sst IJK}^{\sst Y},
\ee
where $Y_{\sst IJ}:=Y_{\sst I}+Y_{\sst J}$. Consequently, we get 
\ba
\label{RSVqijk}
\op{V}(R)_{\gamma}^{\st Y,{\sst RS}}|\,J\,M\,;\,M'\zu&=&\lp^3\sum\limits_{v\in 
V(\gamma)\cap R}\sum\limits_{I<J<K}
\underbrace{\sqrt{\Big|\frac{3!i}{4} C_{reg}\op{q}_{\sst IJK}^{\sst Y}\Big|}}_{\displaystyle \op{V}^{\sst RS}_{v,\gamma}}\;\;|\,J\,M\,;\,M'\zu.
\ea
 The additional factor of $3!$ is due to the fact that we sum only over 
ordered triples $I<J<K$ now.  
The way  to calculate eigenstates and eigenvalues of $\op{V}$ is as follows. 
Let us introduce the operator $\op{Q}^{\sst Y,RS}_{v,{\sst IJK}}$ as
\be 
\label{RSQ}
\op{Q}^{\sst Y,RS}_{v,{\sst IJK}}:=\lp^6\frac{3!i}{4}C_{reg}\op{q}_{\sst IJK}^{\sst Y}
\ee
As a first step we have to calculate the eigenvalues and corresponding eigenstates for $\op{Q}^{\sst Y,RS}_{v,{\sst IJK}}$. If for example $|\phi\zu$ is an eigenstate of $\op{Q}^{\sst Y,RS}_{v,{\sst IJK}}$ with corresponding eigenvalue $\lambda$, then we obtain $\op{V}|\phi\zu=\sqrt{|\lambda|}|\phi>$. Consequently, we see that while $\op{Q}^{\sst Y,RS}_{v,{\sst IJK}}$ can have positive and negative eigenvalues, $\op{V}$ has only positive ones. Furthermore, if we consider the eigenvalues $\pm\lambda$ of $\op{Q}^{\sst Y,RS}_{v,{\sst IJK}}$ and the corresponding eigenstate $|\phi_{+\lambda}\zu,|\phi_{-\lambda}\zu$, we notice that these eigenvalues will be degenerate
  in the case of the operator $\op{V}$, as 
 $\sqrt{|+\lambda|}=\sqrt{|-\lambda|}$.
 \subsubsection{The Volume Operator $\op{V}_{\sst AL}$ of Ashtekar and Lewandowski}  %
  Another version of the volume operator which differs by the chosen regularisation scheme was defined in \cite{4}
  \ba
\label{Vqijk}
\op{V}(R)_{\gamma}^{\st Y,{\sst AL}}|\,J\,M\,;\,M'\zu&=&\lp^3\sum\limits_{v\in 
V(\gamma)\cap R}
\underbrace{\sqrt{\Big|\frac{3!i}{4} C_{reg}\sum\limits_{I<J<K}\epsilon(e_{\sst I},e_{\sst J},e_{\sst K})\,\op{q}_{\sst IJK}^{\sst Y}\Big|}}_{\displaystyle \op{V}^{\sst AL}_{v,\gamma}}\;\;|\,J\,M\,;\,M'\zu.
\ea
The major difference between $\op{V}_{\sst AL}$ and $\op{V}_{\sst RS}$ is the factor $\epsilon(e_{\sst I},e_{\sst J},e_{\sst K})$ that is sensitive to the orientation of the tangent vectors of the edges $\{e_{\sst I},e_{\sst J},e_{\sst K}\}$. $\epsilon(e_{\sst I},e_{\sst J},e_{\sst K})$ is $+1$ for right handed, $-1$ for left handed and $0$ for linearly dependent triples of edges. In the case of $\op{V}_{\sst AL}$ it is convinient to introduce an operator $\op{Q}^{\sst Y,AL}_v$ that is defined as the expression that appears inside the absolute value under the square root in $\op{V}^{\sst AL}_{v,\gamma}$
\be 
\op{Q}^{\sst Y,AL}_v:=\lp^6\frac{3!i}{4}C_{reg}\sum\limits_{I<J<K}\epsilon(e_{\sst I},e_{\sst J},e_{\sst K})\,\op{q}_{\sst IJK}^{\sst Y}
\ee
\newline
By comparing eqn (\ref{RSVqijk}) with (\ref{Vqijk}) we notice that another difference between $\op{V}_{\sst RS}$ and $\op{V}_{\sst AL}$ is the fact that for the first one, we have to sum over the triples of edges outside the square root, while for the latter one, we sum inside the absolute value under the square root.
Besides the difference of the sign factor, the difference in the summation will play an important role later on. Notice that one arrives at (\ref{Vqijk}) also from a usual point splitting regularisation \cite{10}.
\subsection{Canonical Quantisation}   %
 Usually the densitised triads, appearing in the classical flux $E_k(S)$ 
 are quantised as differential operators, while holonomies are quantised 
 as multiplication operators. If we choose the alternative expression 
 $\wt{E}_k(S)$ we will instead get the scalar volume $\op{V}$ and the so 
 called signum $\op{\cal S}$ operator into our quantised expression.  The 
 properties of this $\op{\cal S}$ will be explained in more detail below. 
 Moreover, we have to replace Poisson brackets by commutators, following 
 the replacement rule $\{.\,,\,.\}\rightarrow (1/i\hbar)[.\,,\,.]$. In 
 order to simplify the following calculations, we
  want to  achieve a form of the operator such that on the left hand side 
 only inverses of the holonomies appear while right beneath the product of 
 operators $\op{V}\fbox{$\op{\cal S}$}\op{V}$ only holonomies appear. 
 Thus, we make use of the identities
$
\op{\pi}_{\ell}(h^{-1}_{e_{\sst I}})_{\sst AB}=\pi_{\ell}(\epsilon)_{\sst AC}\pi_{\ell}(\epsilon)_{\sst BD}\op{\pi}_{\ell}(h_{e_{\sst I}})_{\sst DC}$  and
$\op{\pi}_{\ell}(h_{e_{\sst I}})_{\sst AB}=\pi_{\ell}(\epsilon)_{\sst CA}\pi_{\ell}(\epsilon)_{\sst DB}\op{\pi}_{\ell}(h^{-1}_{e_{\sst I}})_{\sst DC}$, 
where $\pi_{\ell}(\epsilon)$ stands for the $\epsilon_{\sst AB}$ of SU(2) in a higher representation with weight $\ell$. The explicit form can be derived from eqn (\ref{FormSexl}) in appendix and is given by $\pi_{\ell}(\epsilon)_{\sst AB}=(-1)^{\ell{\st -A}}\delta_{\sst A+B,0}$. Clearly, we want the total operator to be self-adjoint, so we will calculate the adjoint of $\lL\op{\wt{E}}_k(S_t)$ and define the total and final operator to be $\lL\op{\wt{E}}_{k,{\sst tot}}(S_t)=\frac{1}{2}(\op{\wt{E}}_k(S_t)+\op{\wt{E^{\dagger}}}_k(S_t))$ that is self-adjoint by construction. Hence, the final operator for $\op{V}_{\sst RS}$ which we will use through the calculation of this paper is given by
 \ba
\label{RSEktot}
\lL\op{\wt{E}}^{\sst I/II,RS}_{k,{\sst tot}}(S_t)&=&\lim_{{\cal P}_t\to S_t}
\sum\limits_{\sst {\Box}\in{\cal P}_t}
\frac{8\,\lp^{-4}(-1)^{2\ell}}{\frac{4}{3}\ell(\ell+1)
(2\ell+1)}\pi_{\ell}(\tau_k)_{\sst CB}\pi_{\ell}(\epsilon)_{\sst EI}
\nonumber \\
&&\quad\quad\quad\quad
\Big\{+\pi_{\ell}(\epsilon)_{\sst FC}\left[\op{\pi}_{\ell}(h_{e_{\sst 4}})_{\sst FG}\right]^{\dagger}\left[\left[\op{\pi}_{\ell}(h_{e_{\sst 3}})_{\sst BA}\right]^{\dagger},\op{V}_{\sst RS}\right]\fbox{$\op{\cal S}$}\left[\op{V}_{\sst RS},\op{\pi}_{\ell}(h_{e_{\sst 4}})_{\sst IG}\right]\op{\pi}_{\ell}(h_{e_{\sst 3}})_{\sst EA}\nonumber\\
&&\quad\quad\quad\quad\quad
 -\pi_{\ell}(\epsilon)_{\sst FB}\left[\op{\pi}_{\ell}(h_{e_{\sst 3}})_{\sst IG}\right]^{\dagger}\left[\left[\op{\pi}_{\ell}(h_{e_{\sst 4}})_{\sst EA}\right]^{\dagger},\op{V}_{\sst RS}\right]\fbox{$\op{\cal S}$}\left[\op{V}_{\sst RS},\op{\pi}_{\ell}(h_{e_{\sst 3}})_{\sst FG}\right]\op{\pi}_{\ell}(h_{e_{\sst 4}})_{\sst CA}\Big\},
\ea
whereby we used the identity $\pi_{\ell}(h^{-1}_{e_{\sst I}})_{\sst AB}=\left[\pi_{\ell}(h_{e_{\sst I}})_{\sst BA}\right]^{\dagger}$, the definition of the Planck length $\lp^{-4}:=(\hbar\kappa)^{-2}$,  and  additionally,   $\pi_{\ell}(\epsilon)_{\sst GD}\pi_{\ell}(\epsilon)_{\sst DH}=(-1)^{2\ell}\delta_{\sst G,H}$.
\newline
 Considering the operator $\op{V}_{\sst AL}$, we know that for each commutator only one term will contribute, because otherwise we cannot construct linearly independet triples of edges since $\{e_{\sst 1},e_{\sst 2},e_{\sst 3/4}\}$ are linearly dependent. Therefore in the case of $\op{V}_{\sst AL}$ we obtain the following final expression 
\ba
\label{Ektot}
\lL\op{\wt{E}}^{\sst I/II,AL}_{k,{\sst tot}}(S_t)&=&\lim_{{\cal P}_t\to S_t}
\sum\limits_{\sst {\Box}\in{\cal P}_t}
\frac{8\,\lp^{-4}(-1)^{2\ell}}{\frac{4}{3}\ell(\ell+1)
(2\ell+1)}\pi_{\ell}(\tau_k)_{\sst CB}\pi_{\ell}(\epsilon)_{\sst EI}
\nonumber \\
&&\quad\quad\quad\quad
\Big\{+\pi_{\ell}(\epsilon)_{\sst FC}\left[\op{\pi}_{\ell}(h_{e_{\sst 4}})_{\sst FG}\right]^{\dagger}\left[\op{\pi}_{\ell}(h_{e_{\sst 3}})_{\sst BA}\right]^{\dagger}\op{V}_{\sst AL}\fbox{$\op{\cal S}$}\op{V}_{\sst AL}\op{\pi}_{\ell}(h_{e_{\sst 4}})_{\sst IG}\op{\pi}_{\ell}(h_{e_{\sst 3}})_{\sst EA}\nonumber\\
&&\quad\quad\quad\quad\quad
 -\pi_{\ell}(\epsilon)_{\sst FB}\left[\op{\pi}_{\ell}(h_{e_{\sst 4}})_{\sst IG}\right]^{\dagger}\left[\op{\pi}_{\ell}(h_{e_{\sst 3}})_{\sst EA}\right]^{\dagger}\op{V}_{\sst AL}\fbox{$\op{\cal S}$}\op{V}_{\sst AL}\op{\pi}_{\ell}(h_{e_{\sst 4}})_{\sst FG}\op{\pi}_{\ell}(h_{e_{\sst 3}})_{\sst CA}\Big\}.
\ea
Here again for case II the signum operator is inlcuded, whereas in case I it is not.
\newline
By looking at the equation above, we see that the operator contains a lot of sums, so it does not seem to be that trivial to actually compute expectation values. However, we will show in the next section how we can  use the given structure of the operator and derive some properties from it that will simplify the summation and therefore the calculation of expectation values. 
\section{General Properties of the Operator $\lL\op{\wt{E}}_{k,{\sst tot}}(S_t)$}%
In this section we will discuss some general properties of the alternative flux operator. Since this properties are valid independent of the choice of $\op{V}_{\sst AL}$ or $\op{V}_{\sst RS}$ we will drop this labeling of the volume operator here. If not explicitly mentioned otherwise these properties also hold independetly from the fact whether we are considering case I or case II respectively. Thus, we will only talk about the operator $\lL\op{\wt{E}}_{k,{\sst tot}}(S_t)$.
\subsection{Correspondence Between the AL and the Abstract Angular Momentum System  Hilbert Space}%
Going back to the action of the usual flux operator in eqn (\ref{flY}) we 
see that the action of the flux operator can be expressed in terms of  
selfadjoint right invariant vector fields $Y^k_e$. The same is true for 
the volume operator appearing in the new alternative flux operator. Since we would like to utilise the technology of Clebsch-Gordan coefficents (CGC), $6j$-symbols and the like in order to calculate matrix elements of these operators with respect to spin network states, we will discuss in detail how the AL-Hilbert space and the abstract angular momentum system Hilbert space are related.  
\newline\newline
Consider the explicit expression for the matrix elements of the 
unitary transformation matrix $[\pi_j(g)]_{mn}$ for the components 
$\psi_m$
of a totally symmetric symmetric spinor of rank $2j$ under $SU(2)$ 
gauge transformations reviewed in appendix 
E, that is, $\psi'_m=\sum\limits_{n=-j}^j\;[\pi_j(g)]_{mn}\;\psi_n$. By 
elementary linear algebra, the unitary representation $g\mapsto U(g)$
of $SU(2)$ on the linear span of the standard angular momentum states 
$|jm>$ is obtained by transposition, i.e. $U(g)\;|j\,m>= 
\sum_{n=-j}^j\;[\pi_j(g)]_{nm}\;|j\,n>$. To see this, it is enough 
to check that the standard angular momentum operators $J^k$ when written 
in terms of ladder operators have the same action as the infinitesimal 
generators of the one parameter groups $t\mapsto U(\exp(it \tau_k/2))$.
(Recall that $i\tau_k=\sigma_k$ are the Pauli matrices.) Explicitly we 
find 
$$
J^k |j\,m>= +\sum_n \frac{i}{2} [\pi_j(\tau_k)]_{nm} |j\,n> 
$$
where $\pi_j(\tau_k)$ are the matrices derived in appendix E. 

Now consider the functions 
\be
<h_e|j\,m>_{m'}:=\sqrt{2j+1} [\pi_j(h_e)]_{mm'},
\ee
where $h_e$ denotes the holonomy along some edge $e$. For fixed $m'$ they 
are orthonormal just as the $|j\,m>$. Moreover, the operators 
$Y_e^k:=-i X_e^k/2$, where $X_e^k$ are the right invariant vector 
fields on $SU(2)$, satisfy the same algebra as the $J^k$. Let us drop 
the label $e$ for the purposes of this paragraph. From the 
explicit representation of the gauge transformation on the 
$<h|j\,m>_{m'}$ given by $V(g) <h|j\,m>_{m'}=<g\,h|j\,m>_{m'}=
[\pi_j(g)]_{mn} <h|jn>_{m'}$ we can explicitly calculate that the 
$Y^k$ are the infinitesimal generators of the one parameter groups
$t\mapsto V(\exp(i t \tau_k/2))$, explicitly
\be
Y^k |j\,m>_{m'}= -\sum_n \frac{i}{2} [\pi_j(\tau_k)]_{mn} |j\,n>_{m'} 
\ee
It is instructive to verify the angular momentum algebra for $J^k,Y^k$.
\newline
The fluxes are expressed in terms of the 
$Y_e^k$ and the spin network states are expressed in terms of the 
$|j\,m>^e_{m'}$ (the superscript $e$ reminds of the edge to which the state
$|j\,m>_{m'}$ is associated to). In order to write these in terms of 
$J^k$ and $|j\,m>$ we must determine the unitary operator 
\be
W:\;\;{\cal H}^{jm'} \to {\cal H}^j_{m'};\;\; W |j\,m\,;\,m'>=\sum_n W_{jmn} 
|j\,n>_{m'}
\ee
such that $WJ^k W^{-1} = Y^k$. Here ${\cal H}^{jm'}$ is the 
linear span of abstract angular momentum eigenstates $|j\,m\,;\,m'>$ 
which for fixed $m'\in \{-j,-j+1,..,j\}$ are just the $|jm>$ 
with additional label $m'$ while ${\cal 
H}^j_{m'}$ is the linear span of the spin network states 
$|j\,m>_{m'}$ and $W_{jmn}$ is a unitary matrix. 

It is not difficult to see from the above formulae that 
$W_{jmn}=[\pi_j(\epsilon)]_{mn}$ where 
$\epsilon=-\tau_2$. Therefore 
\be
\label{Wmap}
W |j\,m\,;\,m'>=[\pi_j(\epsilon)]_{mn} |j\,n>_{m'} \;\;
\Leftrightarrow 
W^{-1} |j\,m>_{m'}=[\pi_j(\epsilon^{-1})]_{mn} |j\,n\,;\,m'>
\ee
and we will make frequent use of the identities 
$\epsilon^{-1}=\epsilon^T=-\epsilon$, $\epsilon g^T \epsilon^T=g^{-1}$
valid for any $g\in SL(2,\Cl)$ such as $g=\tau_k$ and 
$\tau_k^{-1}=-\tau_k=\overline{\tau_k}^T$.

Now in order to use these identities, consider some spin network states 
$T_{\tilde{\gamma},\vec{\tilde{j}},\vec{\tilde{m}},\vec{\tilde{m}}'},
T_{\gamma,\vec{j},\vec{m},\vec{m}'}$ and some operator $\hat{O}_Y$ which 
we think of as a function in the operators $Y_e^k$. Then by unitarity
\be
\begin{array}{ll}
<T_{\tilde{\gamma},\vec{\tilde{j}},\vec{\tilde{m}},\vec{\tilde{m}}'}\,|\,
\hat{O}_Y\,|\, 
T_{\gamma,\vec{j},\vec{m},\vec{m}'}>_{SNF}
=&\\ 
\sum\limits_{\vec{\tilde{n}},\vec{n}} 
\prod\limits_{\tilde{e}\in E(\tilde{\gamma})} \; 
[\pi_{j_{\tilde{e}}}(\epsilon^{-1})]_{\tilde{m}_{\tilde{e}}
\tilde{n}_{\tilde{e}}}
\;
\prod\limits_{e\in E(\gamma)} \; 
[\pi_{j_e}(\epsilon^{-1})]_{m_e n_e}
\;
<T'_{\tilde{\gamma},\vec{\tilde{j}},\vec{\tilde{n}},\vec{\tilde{m}}'}\,|\,
\hat{O}_J\,|\, 
T'_{\gamma,\vec{j},\vec{n},\vec{m}'}>_{ABS}&
\end{array}
\ee
where $SNF$ stands for the spin network Hilbert space and 
$ABS$ for the abstract angular momentum system Hilbert space. 
We use the following notation. Whenever we address SNF we call them $T$ and express them in terms of $|jm>_{m'}$. In contrast if we refer to states in the abstract angular momentum system Hilbert space, we use the notation $T'$ for the abstract angular momentum system functions which result from $T$ upon 
substituting $|j\,m>_{m'}$ by $|j\,m\,;\,m'>$.
 The operator $\hat{O}_J$ is the 
same as $\hat{O}_Y$ just that $Y_e^k$ is everywhere replaced by $J_e^k$
\newline\newline
 The discussion above shows that we have to map the holonomies 
$\pi_{\ell}(h)_{\st AB}$ in the alternative flux operator in eqn 
(\ref{Ektot}) into the abstract angular momentum system Hilbert space via 
the unitary map $W$ in eqn (\ref{Wmap}) in order to apply technical tools 
of usual angular momentum recoupling theory. 
Thus, if we apply the unitary map $W$ (summation convention is assumed)
 \be
W\op\pi_{\ell}(h)_{\sst AB}=\frac{\pi_{\ell}(\epsilon^{-1})_{\st AC}}{\sqrt{2\ell+1}}\,\auf\,h\,|\,\ell\,{\st C}\,{\st ;}\,{\st B}\zu,
\ee
 use the fact that $\pi_{\ell}(\tau_k)_{\st BC}=-\left[\pi_{\ell}(\epsilon)\pi_{\ell}(\tau_k)\pi_{\ell}(\epsilon^{-1})\right]_{\st CB}$ and the following properties of  $\pi_{\ell}(\epsilon^{-1})$
\be
\pi_{\ell}(\epsilon^{-1})_{\st AB}=(-1)^{2\ell}\pi_{\ell}(\epsilon)_{\st AB}\quad\quad
\pi_{\ell}(\epsilon)_{\st AB}\pi_{\ell}(\epsilon)_{\st BC}=(-1)^{2\ell}\delta_{\st AC}\quad\quad \pi_{\ell}(\epsilon)_{\st AB}=(-1)^{2\ell}\pi_{\ell}(\epsilon)_{\st BA}
\ee 
that can easily be derived from the explicit expression of $\pi_{\ell}(\epsilon)_{\st AB}$, we end up with
\ba
\label{Ektcoup}
\lL\op{\wt{E}}_{k,{\sst tot}}(S_t)&=&
-\lim_{{\cal P}_t\to S_t}\sum\limits_{\sst {\Box}\in{\cal P}_t}\frac{8\,\lp^{-4}(-1)^{2\ell}}{\frac{4}{3}\ell(\ell+1)(2\ell+1)}\frac{1}{(2\ell+1)^2}\pi_{\ell}(\tau_k)_{\sst CB}\pi_{\ell}(\epsilon)_{\sst EI}\nonumber \\
&&\Big\{+\pi_{\ell}(\epsilon)_{\sst FC}\Big(\otimes\auf_{e_{\sst 3}}\ell\,{\st B}\, {\st ;}\,{\st A}|\otimes\auf_{ e_{\sst 4}}\ell\,{\st F}\, {\st ;}\,{\st G}|\,\op{O}_{\sst 1}\,|\,\ell\,{\st I}\, {\st ;}\,{\st G}\zu_{e_{\sst 4}}\otimes|\,\ell\, {\st E}\, {\st ;}\,{\st A}\zu_{e_{\sst 3}}\otimes\Big)\nonumber\\
&&\quad -\pi_{\ell}(\epsilon)_{\sst FB}\Big(\otimes\auf_{e_{\sst 3}}\ell\,{\st E}\, {\st ;}\,{\st A}|\otimes\auf_{e_{\sst 4}}\ell\,{\st I}\, {\st ;}\,{\st G}|\,\op{O}_{\sst 2}\,|\,\ell\,{\st F}\, {\st ;}\,{\st G}\zu_{e_{\sst 4}}\otimes|\,\ell\,{\st C}\, {\st ;}\,{\st A}\zu_{e_{\sst 3}}\otimes\Big)\Big\}.
\ea
The definition of the operators $\op{O}_{\sst 1}$ and $\op{O}_{\sst 2}$ in the four different cases
 are shown in eqn (\ref{O1O2}). We introduced the notation $V_{q_{\sst IJK}}$ in the RS case meaning that only the contribution of the triple  $\{e_{\sst I},e_{\sst J},e_{\sst K}\}$ is taken into account. Why $\op{O}_{\sst 1,2}$ have this particular in the case of RS will is explained more in detail in appendix  \ref{RSOp}.Basically, the structure displayed is due to the various contributions from the four terms involved in the product of 2 commutators in eqn (\ref{RSEktot}).
 \ba
 \label{O1O2}
 O_{\sst 1}^{\sst I,AL}&=&\op{V}_{\sst AL}^2\nonumber\\
O_{\sst 1}^{\sst I,RS}&=&\op{V}_{\sst RS}^2+\op{V}_{q_{\sst 124}}\op{V}_{q_{\sst 123}}-\op{V}_{q_ {\sst 124}}\op{V}_{\sst RS}-\op{V}_{\sst RS}\op{V}_{q_{\sst 123}}\nonumber\\
O_{\sst 2}^{\sst I,AL}&=&\op{V}_{\sst AL}^2\nonumber\\
O_{\sst 2}^{\sst I,RS}&=&\op{V}_{\sst RS}^2+\op{V}_{q_{\sst 123}}\op{V}_{q_{\sst 124}}-\op{V}_{q_ {\sst 123}}\op{V}_{\sst RS}-\op{V}_{\sst RS}\op{V}_{q_{\sst 124}}\nonumber\\ 
O_{\sst 1}^{\sst II,AL}&=&\op{V}_{\sst AL}\op{\cal S}\op{V}_{\sst AL}\nonumber\\
O_{\sst 1}^{\sst II,RS}&=&\op{V}_{\sst RS}\op{\cal S}\op{V}_{\sst RS}+\op{V}_{q_{\sst 124}}\op{\cal S}\op{V}_{q_{\sst 123}}-\op{V}_{q_ {\sst 124}}\op{\cal S}\op{V}_{\sst RS}-\op{V}_{\sst RS}\op{\cal S}\op{V}_{q_{\sst 123}}\nonumber\\ 
O_{\sst 2}^{\sst II,AL}&=&\op{V}_{\sst AL}\op{\cal S}\op{V}_{\sst AL}\nonumber\\
O_{\sst 2}^{\sst II,RS}&=&\op{V}_{\sst RS}\op{\cal S}\op{V}_{\sst RS}+\op{V}_{q_{\sst 123}}\op{\cal S}\op{V}_{q_{\sst 124}}-\op{V}_{q_ {\sst 123}}\op{\cal S}\op{V}_{\sst RS}-\op{V}_{\sst RS}\op{\cal S}\op{V}_{q_{\sst 124}}
\ea
Recall from the discussion in section 
 \ref{factord} that the action of both operators 
$\op{E}_k(S),\;\lL\nflop{k}$ on any SNF was totally 
 determined by its action on single edges of type up, down, in and out, 
 and that the latter two were annihilated by this operator. 
The surface $\Box$ which intersects an edge $e$ of type up or down 
necessarily transversally splits $e$ as $e=e_2(\Box)^{-1}\circ 
e_1(\Box)$
where $e_1(\Box),\; e_2(\Box)$ is of type up or down with 
respect to $\Box$ (or $S_t$) respectively if $e$ is of type up with 
respect to $S$ 
and conversely if $e$ is of type down. Notice that $e_I(\Box),\;I=1,2$
inherit from $e$ the same spin label $j$ coupling to total spin $j_{12}$ 
at the point $v(\Box)=e_1(\Box)\cap e_2(\Box)$.
\begin{figure}[hbt]
\center
    \psfrag{e'}{$\epsilon'$}
   \psfrag{v}{$v({\sst \Box})$}
   \psfrag{e1}{$e_{\sst 1}({\sst \Box})$}
   \psfrag{e2}{$e_{\sst 2}({\sst \Box})$}
   \psfrag{j1}{$j,n_{\sst 1}$}
   \psfrag{j2}{$j,n_{\sst 2}$}
   \psfrag{j12}{$j_{\sst 12},n_{\sst 12}$}
   \psfrag{beta}{$\bet{j_{\sst 12}}{n_{\sst 12}}$}
\includegraphics[height=5cm]{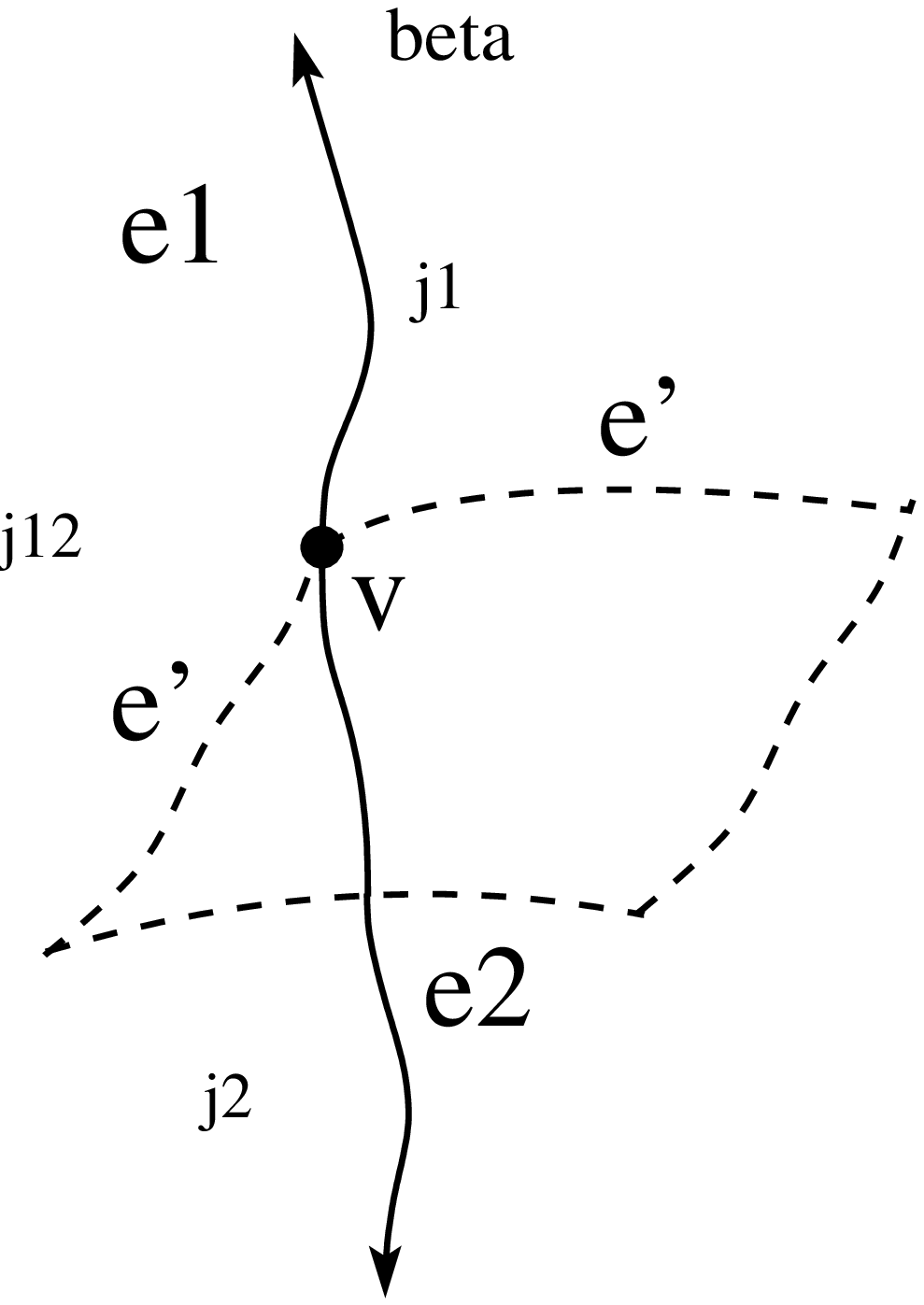}
   \caption{\label{Bild5}\small SNF $\bet{j_{\sst 12}}{n_{\sst 12}}$ that consists of two edges, whereby one is of type up and the other of type down with respect to the surface $S_t$. These two edges carry both a spin label $j$ and couple at the vertex $v({\sst \Box})$ to an resulting angular momentum $j_{\sst 12}$.}        
   \end{figure}     
 \newline 
As the operators $\op{O}_{\sst 1}$ and $\op{O}_{\sst 2}$ in eqn \ref{Ektcoup} contain the volume operator $\op{V}(R_{v({\sst \Box})})$, at some point we will have to calculate matrix elements of $\op{V}$. With this in mind it is advisable to work in the so called recoupling basis right from the beginning, because the formula for matrix elements of $\op{V}$ derived in \cite{14} applies only to states in that particular basis\footnote{Recall that in the tensor basis a state is characterised by the spin labels $j_i$ and the magnetic quantum numbers $m_i$ and an additional label $m_i'$ that are attached to the edges $e_i$ of a particular vertex of the corresponding graph $\gamma$. We express a given SNF in this basis by tensor products between states $|j\, m_i\zu_{m_i'}^e$ multiplied by corresponding intertwiners. In contrast, in the recoupling basis states are characterised by the total angular momentum $J$, the total magnetic quantum number $M$ to which the edges couple at a particular vertex of the graph $\gamma$ and  the value of the intermediate couplings. In order to know what kind of intermediate couplings are possible, we have to fix an order in which we want to couple the edges associated at one particular vertex from the very beginning. Then the intermediate couplings  $a_i$ are successively defined by  $a_{i+1}:=\{|a_i-j_{i+1}|,...,a_i+j_{i+1}\}$ with $a_{\sst 1}:=j_{\sst 1}$. If we choose a different order of coupling, we will end up  with a different recoupling scheme, where these two recoupling schemes are related by so called $3nj-$symbols. (For a brief introduction to recoupling theory see for example \cite{10,14}.)}.
 The particular SNF we want to work with can be characterised in the 
 recoupling basis by its total angular momentum $j_{\sst 12}$ and its 
 magnetic quantum number $n_{\sst 12}$ (and two additional labels 
 $m'_{\sst 1},m'_{\sst 2}$) since the first intermediate coupling $a_{\sst 
 1}$ is equivalent to the spin label of the first edge which is fixed and 
 $j$ in our case. Therefore, we will call those states $|\beta^{j_{\sst 
12}}\,n_{\sst 12}\zu_{m'_{\sst 1},m'_{\sst 2}}:=|a_{\sst 1}=j\,a_{\sst 2}=j_{\sst 12}\, n_{\sst 12}\,\zu_{m'_{\sst 1}\,m'_{\sst 2}}$ where 
 $n_{\sst 12}\in\{-j_{\sst 12},...,j_{\sst 12}\}$ and $m'_{\sst 
 1},m'_{\sst 2}$ can be treated as  additional indices unimportant for the 
 recoupling procedure. This means to a fixed choice of $j_{\sst 12}$ we 
 have $(2j_{\sst 12}+1)(2n_{\sst 12}+1)$ orthogonal states 
 $|\beta^{j_{\sst 12}}\,n_{\sst 12}\zu_{m'_{\sst 1},m'_{\sst 2}}$ being a 
 basis of the Hilbert space for this particular value of $j_{\sst 12}$. 
This SNF is also shown in figure \ref{Bild5}. 
\newline
As before we want to map the SNF $|\beta^{j_{\sst 12}}\,n_{\sst 12}\zu_{m'_{\sst 1},m'_{\sst 2}}$ and the operators $\op{O}^{\sst Y}_{\sst 1/2}$ into the abstract angular momentum system Hilbert space
\ba
W|\beta^{j_{\sst 12}}\,n_{\sst 12}\zu_{m'_{\sst 1},m'_{\sst 2}}&=&
\sum\limits_{m_{\sst 12}}\pi_{j_{\sst 12}}(\epsilon^{-1})_{n_{\sst 12}m_{\sst 12}}|\beta^{j_{\sst 12}}\,m_{\sst 12}\,;\,m'_{\sst 1},m'_{\sst 2}\zu
\ea
Consequently, the map $W$ has the following effect on the matrix element of $\lL\op{\wt{E}}_{k,{\sst tot}}(S_t)$
\be
\begin{array}{l}
\label{Mtrafo}
_{\wt{m}'_{\sst 1},\wt{m}'_{\sst 2}}\auf 
\beta^{\tilde{j}_{\sst 12}}\,\wt{n}_{\sst 12}\,|\lL\hat{\tilde{E^{\st Y}}}_{k,{\sst 
tot}}(S_t)|\,\beta^{j_{\sst 12}}\,n_{\sst 12}\zu_{m'_{\sst 1},
m'_{\sst 2}}\nonumber\\
=\sum\limits_{m_{\sst 12},\wt{m}_{\sst 12}}\pi_{\tilde{j}_{\sst 
12}}(\epsilon^{-1})_{\tilde{n}_{\sst 12}\wt{m}_{\sst 12}}\pi_{j_{\sst 12}}(\epsilon^{-1})_{n_{\sst 12}m_{\sst 12}}
\auf \beta^{\tilde{j}_{\sst 12}}\,\wt{m}_{\sst 12}\,;
\wt{m}'_{\sst 1}\,\wt{m}'_{\sst 2}\,|\lL\hat{\tilde{E^{\st J}}}_{k,{\sst 
tot}}(S_t)|\,\beta^{j_{\sst 12}}\,m_{\sst 12}\,;\,m'_{\sst 1}\,\,m'_{\sst 2}\zu.
\end{array}
\ee
where for reasons of clarity we denoted by the superscripts $Y,J$
respectively the same algebraic expression in terms of the $Y,J$ operators
respectively. In what follows we will drop this label and it will be
understood that we will be working in the abstract angular momentum space
only.
The same transformation applies to the matrix element of the usual flux 
 operator. Since the inverse of the matrices $\pi_{j_{\sst 
 12}}(\epsilon^{-1})$ exists, we can conclude that in order to show that 
the matrix element of the usual flux operator and the one of  
 $\lL\op{\wt{E}}_{k,{\sst tot}}(S_t)$ are identical, we only have to show 
 that after taking the limits $\lim_{\epsilon\to 0}\lim_{\epsilon'\to 0}$ 
the matrix element 
 $\auf \beta^{\tilde{j}_{\sst 12}}\,\wt{m}_{\sst 12}\,;m'_{\sst 
1}\,m'_{\sst 2}\,|\lL\op{\wt{E}}_{k,{\sst tot}}(S_t)|\,\beta^{j_{\sst 12}}\,
 m_{\sst 12}\,;\,m'_{\sst 1}\,\,m'_{\sst 2}\zu$ agrees with the matrix 
 element of the usual flux operator $\auf \beta^{\tilde{j}_{\sst 
 12}}\,\wt{m}_{\sst 12}\,;m'_{\sst 1}\,m'_{\sst 
2}\,|\,\op{E}_{k}(S)|\,\beta^{j_{\sst 12}}\,m_{\sst 12}\,;\,m'_{\sst 1}\,\,m'_{\sst 2}\zu$
 for every possible value of $\wt{m}_{\sst 12},m_{\sst 12}$. Thus, we do not have to consider the two additional  $\pi_{j_{\sst 12}}(\epsilon^{-1})$.
\newline
 Note that also the explit value 
 of the matrix element of the usual flux operator will be contracted by 
these $\pi_{j}(\epsilon^{-1})$. Considering gauge-invariant states 
($j=0$) only, $\pi_{0}(\epsilon^{-1})$=1  is only a single number. 
Thus, if one would work with gauge-invariant operators only, all 
$\pi_{j}(\epsilon^{-1})$ would drop out in eqn \ref{Mtrafo}.
\newline
For the further calculation of $\auf \beta^{\tilde{j}_{\sst 12}}\,\wt{m}_{\sst 12}\,;\wt{m}'_{\sst 1}\,\wt{m}'_{\sst 2}\,|\lL\op{\wt{E}}_{k,{\sst tot}}(S_t)|\,\beta^{j_{\sst 12}}\,m_{\sst 12}\,;\,m'_{\sst 1}\,\,m'_{\sst 2}\zu$ we will introduce the following abbreviations
\ba
\auf \beta^{\tilde{j}_{\sst 12}}\,\wt{m}_{\sst 12}\,;m'_{\sst 1}\,
m'_{\sst 2}\,|&:=&\auf \beta^{\tilde{j}_{\sst 12}}\,\wt{m}_{\sst 12}|\nonumber\\
|\,\beta^{j_{\sst 12}}\,m_{\sst 12}\,;\,m'_{\sst 1}\,\,m'_{\sst 2}\zu&:=&|\,\beta^{j_{\sst 12}}\,m_{\sst 12}\zu
\ea
\subsection{The Explicit Action of $\lL\op{\wt{E}}_{k,{\sst tot}}(S_t)$}%
\begin{figure}[bht]
\center
   \psfrag{n}{$\vec{n}^{\sst S_t}$}
   \psfrag{e'}{$\epsilon'$}
   \psfrag{v}{$v({\sst \Box})$}
   \psfrag{e1}{$e_{\sst 1}({\sst \Box})$}
   \psfrag{e2}{$e_{\sst 2}({\sst \Box})$}
   \psfrag{e3}{$e_{\sst 3}({\sst \Box})$}
   \psfrag{e4}{$e_{\sst 4}({\sst \Box})$}
   \psfrag{j1}{$j$}
   \psfrag{j2}{$j$}
   \psfrag{j12}{$j_{\sst 12},n_{\sst 12}$}
   \psfrag{beta}{$\bet{j_{\sst 12}}{n_{\sst 12}}$}
   \psfrag{alpha}{$\alp{J}{i}{\st M}$}
   \psfrag{J}{$J,M$}
   \psfrag{ell}{$\ell$}
   \psfrag{OP(E)}{$\lL\op{\wt{E}}_{k,{\sst tot}}(S_t)\bet{j_{\sst 12}}{n_{\sst 12}}$}      
\includegraphics[height=6cm]{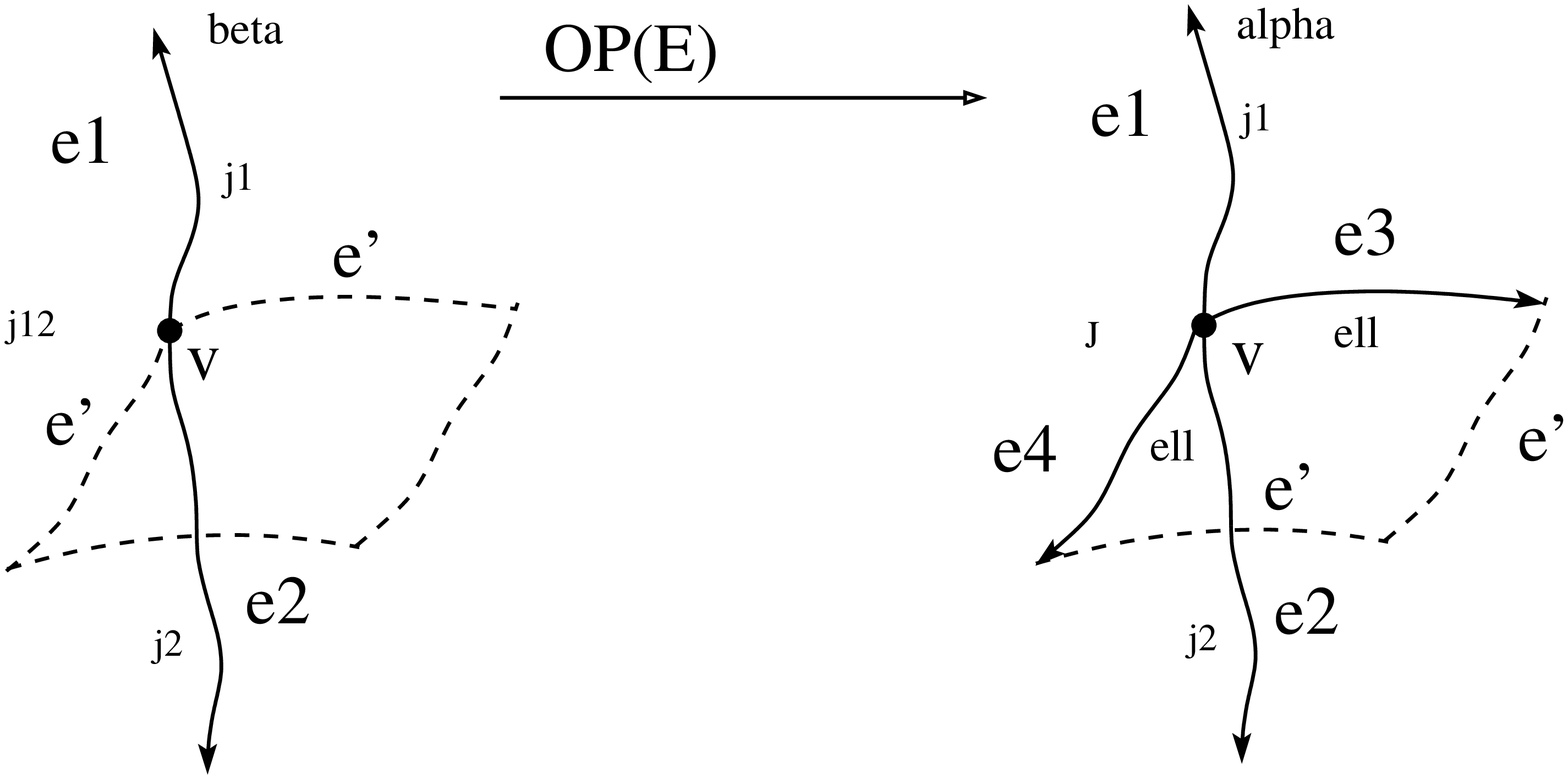}
   \caption{\label{Bild6}\small The SNF $\bet{j_{\sst 12}}{n_{\sst 12}}$ is transformed into an new SNF $\alp{J}{i}{\st M}$ by the action of $\lL\op{\wt{E}}_{k,{\sst tot}}(S_t)$.}     
   \end{figure}

\label{ExpAction}                                                       %
If the operator acts on such a state  $\bet{j_{\sst 12}}{n_{\sst 12}}_{m'_{\sst 1},m'_{\sst 2}}$  it will basically add two additional edges $e_{\sst 3}$ and $e_{\sst 4}$ to the SNF. These edges lie in the surface $S_t$ as can be seen in figure \ref{Bild6}. Consequently, applying the operator to the states $\bet{j_{\sst 12}}{n_{\sst 12}}$  means nothing else than coupling the two additional edges $e_{\sst 3},e_ {\sst 4}$ to the already existing edges $e_{\sst 1},e_{\sst 2}$ and constructing a new SNF with four edges that we will call $\alp{J}{i}{\st M}$. We label these new states $\alp{J}{i}{\st M}$ again by their resulting total angular momentum $J$ and their corresponding magnetic quantum number $M$. The two additional edges both carry a spin label $\ell$. 
 These states $\alp{J}{i}{\st M}$ include three intermediate couplings $a_{\sst 1},a_{\sst 2},a_{\sst 3}$ and $a_{\sst 4}$ is equal to the total angular momentum $J$. In contrast to $\bet{j_{\sst 12}}{m_{\sst 12}}$ we need an additional index $i$ here for distinguishing all possible states $\alp{J}{i}{\st M}$, because it will be the case that for a particular value of $J$ several values of intermediate couplings $a_{\sst 2},a_{\sst 3}$ exist. (This becomes clearer when we explicitly describe the set of states that belongs to a particular total angular momentum $J$ and that build a basis of the corresponding Hilbert space.) Therefore the action of $\lL\op{\wt{E}}_{k,{\sst tot}}(S_t)$ can be expressed in terms of the recoupling basis states $\alp{J}{i}{M}$, where the expansion coefficients are the corresponding CGC.
 \newline
 Therefore the action and consequently the matrix element of $\lL\op{\wt{E}}_{k,{\sst tot}}(S_t)$ can be described by the following expression
\ba
\label{act2Ek}
\lefteqn{\auf \beta^{\tilde{j}_{\sst 12}}\,\wt{m}_{\sst 12}\,|\lL\op{\wt{E}}_{k,{\sst tot}}(S_t)|\,\beta^{j_{\sst 12}}\,m_{\sst 12}\zu}\nonumber\\
&=&
-\lim_{{\cal P}_t\to S_t}\sum\limits_{\sst {\Box}\in{\cal P}_t}\frac{8\,\lp^{-4}(-1)^{2\ell}}{\frac{4}{3}\ell(\ell+1)(2\ell+1)}\frac{1}{(2\ell+1)^2}\nonumber\\
&&\sum\limits_{\sst A,B,C,E,F,G=-\ell}^{+\ell}\Big\{\pi_{\ell}(\tau_k)_{\sst CB}\pi_{\ell}(\epsilon)_{\sst E-E}
\sum\limits_{\wt{J}=|\wt{a}_{\sst 3}-\ell|}^{\wt{a}_{\sst 3}+\ell}\sum\limits_{J=|a_{\sst 3}-\ell|}^{a_{\sst 3}+\ell}\sum\limits_{\wt{a}_{\sst 3}=|\tilde{j}_{\sst 12}-\ell|}^{\tilde{j}_{\sst 12}+\ell}\sum\limits_{a_{\sst 3}=|j_{\sst 12}-\ell|}^{j_{\sst 12}+\ell}\delta_{\st \wt{J},J}\nonumber\\
&&\Big[+\pi_{\ell}(\epsilon)_{\sst FC}
\auf \tilde{j}_{\sst 12}\, \wt{m}_{\sst 12}\, {\st ;}\,\ell\, {\st B}|\wt{a}_{\sst 3}\, \wt{m}_{\sst 12}{\st +B}\zu 
\auf \wt{a}_{\sst 3}\, \wt{m}_{\sst 12}{\st +B}\, {\st ;}\,\ell\,{\st F}|\wt{J}\,\wt{m}_{\sst 12}{\st +B+F}\zu\nonumber\\
&&\hspace{1.8cm}
\auf j_{\sst 12}\, m_{\sst 12}\, {\st ;}\,\ell\, {\st E}\,|\,a_{\sst 3}\, m_{\sst 12}{\st {\st +E}}\zu  
\auf a_{\sst 3}\, m_{\sst 12}{\st +E}\, {\st ;}\,\ell\, -E\,|\,J\, m_{\sst 12}\zu\delta_{\wt{m}_{\sst 12}{\sst +F+B},m_{\sst 12}}\nonumber\\
&&\hspace{1.8cm}
\auf \alpha^{\st \wt{J}}_{\wt{i}}\, {\st M}=\wt{m}_{\sst 12}{\st +B+F}\, {\st ;}\, \wt{m}'_{\sst 1}\,\wt{m}'_{\sst 2}\,{\st A\,G}\,|\,\op{O}_{\sst 1}\,|\,\alpha^{\st J}_i\,{\st M}=m_{\sst 12}\, {\st ;}\, m'_{\sst 1}\,m'_{\sst 2}\,{\st A\,G}\zu\nonumber\\ 
&&\quad
-\pi_{\ell}(\epsilon)_{\sst FB}
\auf \tilde{j}_{\sst 12}\, \wt{m}_{\sst 12}\, {\st ;}\,\ell\, {\st E}\,|\,\wt{a}_{\sst 3}\, \wt{m}_{\sst 12}{\st +E}\zu 
\auf \wt{a}_{\sst 3}\, \wt{m}_{\sst 12}{\st +E}\, {\st ;}\,\ell\, {\st -E}\,|\,\wt{J}\wt{m}_{\sst 12}\zu\nonumber\\
&&\hspace{1.9cm}
\auf j_{\sst 12}\, m_{\sst 12}\, {\st ;}\,\ell\, {\st C}\,|\,a_{\sst 3}\, m_{\sst 12}{\st +C}\zu  
\auf a_{\sst 3}\, m_{\sst 12}{\st +C}\, {\st ;}\,\ell\, {\st F}|J\, m_{\sst 12}{\st +C+F}\zu\delta_{m_{\sst 12}{\sst +C+F},\wt{m}_{\sst 12}}\nonumber\\
&&\hspace{1.9cm}
\auf \alpha^{\st \wt{J}}_{\wt{i}}\,{\st M}=\wt{m}_{\sst 12}\, {\st ;}\, \wt{m}'_{\sst 1}\,\wt{m}'_{\sst 2}\,{\st A\,G}\,|\,\op{O}_{\sst 2}\,|\,\alpha^{\st J}_i\,{\st M}=m_{\sst 12}{\st+C+F}\, {\st ;}\, m'_{\sst 1}\,m'_{\sst 2}\,{\st \,A\,G}\zu\Big]\Big\}.
\ea
Here $\auf j_1\,m_{\sst 1}\, {\st ;}\,j_2\,m_2\,|\,J\,M\zu$ denotes the CGC that describes the coupling of the angular momentum $j_1$ and $j_2$ with magnetic quantum numbers $m_{\sst 1},m_2$ to a resulting angular momentum $J$ with magnetic quantum number $M$.
 \newline
Since, the states $|\alpha^{\st J}_i\,{\st M}\zu$ for different angular momentum and different magnetic quantum numbers are orthogonal to each other, meaning $\auf\alpha^{\st \wt{J}}_{\wt{i}}\,\wt{\st M}\,|\,\alpha^{\st J}_i\,{\st M}\zu=\delta_{\sst \wt{J},J}\delta_{\sst \wt{M},M}\auf\alpha^{\st J}_{\wt{i}}\,{\st M}\,|\,\alpha^{\st J}_i\,{\st M}\zu$ and the operator $\op{O}$ leaves $J$ and $M$ invariant, we replaced $\wt{J}$ and $\wt{M}$ by $J$ and $M$ and added the necessary $\delta$-function $\delta_{\st \wt{J},J}$. Furthermore, we used the definition $\pi_{\ell}(\epsilon)_{\sst EI}=(-1)^{\ell{\st -E}}\delta_{\sst E+I,0}$ and substituted $I$ by $-E$ in the whole equation.
This restriction of $I$ together with the constraint that $|\alpha^{\st J}_i\,{\st M}\zu$ and $\auf\alpha^{\st J}_{\tilde{i}}\,{\st M}|$ must have the same magnetic quantum number leads to two other $\delta$-functions including $m_{\sst 12}$ and $\wt{m}_{\sst 12}$.
Although the $\delta$-functions above will surely simplify the summation, we still have 11 sums in total and some even depend on each other. Especially the summation over $\wt{J}$ and $J$ contains a lot of terms. But fortunately due to the structure of the operator we can reduce these sums.
\begin{Theorem}
\label{theorem1}
\end{Theorem}
The resulting angular momentum $J$ and $\wt{J}$ of the states $|\alpha^{\st J}_i\,{\st M}\zu$ and $\auf\alpha^{\st \wt{J}}_{\wt{i}}\,{\st M}|$ that do contribute to the matrix element 
$\auf \beta^{\tilde{j}_{\sst 12}}\,\wt{m}_{\sst 12}\,|\, \lL\op{\wt{E}}_{k,{\sst tot}}(S_t)\,|\,\beta^{j_{\sst 12}}\,m_{\sst 12}\zu$
 are only $j_{\sst 12}$ and $\tilde{j}_{\sst 12}$ of the incoming states.
\newline
More precisely, the only contribution to $\lL\op{\wt{E}}_{k}(S_t)$ is the angular  momentum $J=j_{\sst 12}$, while the only contribution to $\lL\op{\wt{E^{\dagger}}}_{k}(S_t)$ is  $\wt{J}=\tilde{j}_{\sst 12}$. 
\newline
(Recall that the first term of the sum in eqn (\ref{act2Ek}) is caused by $\lL\op{\wt{E}}_{k}(S_t)$ and the second and negative part belongs to $\lL\op{\wt{E}^{\dagger}}_{k}(S_t)$.)

\begin{Proof}\end{Proof}
First of all we will prove the following lemma. Afterwards  we will use it so as to be able to prove the theorem just stated.
\setcounter{Lemma}{1}
\begin{Lemma}
\label{lemma1}
\ba
\label{Lem1eqn}
\lefteqn{\sum\limits_{E=-\ell}^{+\ell}\pi_{\ell}(\epsilon)_{\st E-E}\auf j_{\sst 12}\,m_{\sst 12}\, {\st ;}\,\ell\,{\st E}\,|\,a_{\sst 3}\,m_{\sst 12}{\st +E}\zu\auf a_{\sst 3}\,m_{\sst 12}{\st +E}\, {\st ;}\, \ell {\st -E}\,|\,J\,m_{\sst 12}\zu}
\nonumber\\
&=&(-1)^{{\st -}j_{\sst 12}{\st -}\ell{\st -}3a_{\sst 3}}\frac{\sqrt{2a_{\sst 3}+1}}{\sqrt{2j_{\sst 12}+1}}\delta_{J,j_{\sst 12}}\Big(\delta_{m_{\sst 12},-j_{\sst 12}}+\delta_{m_{\sst 12},-j_{\sst 12}+1}+...+\delta_{m_{\sst 12},j_{\sst 12}}\Big).
\ea
\end{Lemma}
The proof of lemma \ref{lemma1} is shown in appendix \ref{Beweis}.
\newline\newline
We use lemma (\ref{lemma1}) for performing the sum over $E$ in eqn (\ref{act2Ek}).
The summation over $\wt{J}$ and $J$ contains only one term now, so we can easily carry out these two sums. Moreover, since the operator $\op{O}$ does not change the $m'$-indices,  we can trivially sum over the indices $A,G$. This leads to an additional factor of $(2\ell+1)^2$. Accordingly, the final version of the matrix element of the operator $\lL\op{\wt{E}}_{k,{\sst tot}}(S_t)$ with which we will start in the next section is
\ba
\label{act5Ek}
\lefteqn{\auf \beta^{\tilde{j}_{\sst 12}}\,\wt{m}_{\sst 12}\,|\lL\op{\wt{E}}_{k,{\sst tot}}(S_t)|\,\beta^{j_{\sst 12}}m_{\sst 12}\zu}\nonumber\\
&=&
-\lim_{{\cal P}_t\to S_t}\sum\limits_{\sst {\Box}\in{\cal P}_t}\frac{8\,\lp^{-4}}{\frac{4}{3}\ell(\ell+1)(2\ell+1)}
\sum\limits_{\sst B,C,F=-\ell}^{+\ell}\Big\{\pi_{\ell}(\tau_k)_{\sst CB}
\sum\limits_{\wt{a}_{\sst 3}=|\tilde{j}_{\sst 12}-\ell|}^{\tilde{j}_{\sst 12}+\ell}\sum\limits_{a_{\sst 3}=|j_{\sst 12}-\ell|}^{j_{\sst 12}+\ell}\nonumber\\
&&\Big[+(-1)^{\st -F}\delta_{\sst F+C,0}(-1)^{{\st -}j_{\sst 12}{\st -}3a_{\sst 3}}\frac{\sqrt{2a_{\sst 3}+1}}{\sqrt{2j_{\sst 12}+1}}\delta_{\wt{m}_{\sst 12}{\sst +F+B},m_{\sst 12}}\nonumber\\ 
&&\hspace{2.2cm}
\auf \tilde{j}_{\sst 12}\, \wt{m}_{\sst 12}\, {\st ;}\,\ell\, {\st B}\,|\,\wt{a}_{3}\,\wt{m}_{\sst 12}{\st +B}\zu 
\auf \wt{a}_{\sst 3}\, \wt{m}_{\sst 12}{\st +B}\, {\st ;}\,\ell\, {\st F}\,|\, j_{\sst 12}\,\wt{m}_{\sst 12}{\st +B+F}\zu\nonumber\\
&&\hspace{2.2cm}
\auf \alpha^{j_{\sst 12}}_{\wt{i}}\,{\st M}=\wt{m}_{\sst 12}{\st +B+F}\, {\st ;}\, \wt{m}'_{\sst 1}\,\wt{m}'_{\sst 2}\,|\,\op{O}_{\sst 1}\,|\,\alpha^{j_{\sst 12}}_i\,{\st M}=m_{\sst 12}\, {\st ;}\, m'_{\sst 1}\,m'_{\sst 2}\zu\nonumber\\
&&
-(-1)^{\st -F}\delta_{\sst F+B,0}
(-1)^{{\st -}\tilde{j}_{\sst 12}{\st -}3\wt{a}_{\sst 3}}\frac{\sqrt{2\wt{a}_{\sst 3}+1}}{\sqrt{2\tilde{j}_{\sst 12}+1}}\delta_{m_{\sst 12}{\sst +C+F},\wt{m}_{\sst 12}}\nonumber\\
&&\hspace{2.2cm}
\auf j_{\sst 12}\, m_{\sst 12}\, {\st ;}\,\ell\, {\st C}\,|\,a_{\sst 3}\, m_{\sst 12}{\st +C}\zu  
\auf a_{\sst 3}\, m_{\sst 12}{\st +F}\, {\st ;}\,\ell\, {\st C}\,|\,\tilde{j}_{\sst 12}\, m_{\sst 12}{\st +C+F}\zu\nonumber\\
&&\hspace{2.2cm}
\auf \alpha^{\st \tilde{j}_{\sst 12}}_{\wt{i}}\,{\st M}=\wt{m}_{\sst 12}\, {\st ;}\, \wt{m}'_{\sst 1}\,\wt{m}'_{\sst 2}\,|\,\op{O}_{\sst 2}\,|\,\alpha^{\tilde{j}_{\sst 12}}_i\,{\st M}=m_{\sst 12}{\st+C+F}\, {\st ;}\, m'_{\sst 1}\,m'_{\sst 2}\zu\Big]\Big\}.
\ea 
where we used $\delta_{J,\wt{J}}\delta_{J,j_{\sst 12}}=\delta_{\wt{J},j_{\sst 12}}$ and $(-1)^{4\ell}=+1$. We omitted the sum over the $\delta$-function acting on the magnetic quantum number $m_{\sst 12}$ and $\wt{m}_{\sst 12}$ respectively (see lemma \ref{lemma1}). This is possible as long as we keep in mind that the action of the operator $\lL\op{\wt{E}}_{k,{\sst tot}}(S_t)$ is identical for each fixed $m_{\sst 12}$ and $\wt{m}_{\sst 12}$ of the states $|\beta^{j_{\sst 12}}\,m_{\sst 12}\zu$ and $\auf\beta^{\tilde{j}_{\sst 12}}\,\wt{m}_{\sst 12}|$.\newline
\newline
 However, by simply looking at eqn (\ref{act5Ek}) we see that only the resulting angular momentum $J=\wt{J}=j_{\sst 12},\tilde{j}_{\sst 12}$ contribute to the matrix element $\auf \beta^{\tilde{j}_{\sst 12}}\,\wt{m}_{\sst 12}\,|\lL\op{\wt{E}}_{k,{\sst tot}}(S_t)|\,\beta^{j_{\sst 12}}\,m_{\sst 12}\zu$.\newline
Consequently, we have proven theorem (\ref{theorem1}).$\quad\quad{\sst\blacksquare}$
\newline
We can read off from eqn (\ref{act5Ek}) that we  have already managed to reduce the number of summations down to 5 just by investigating the physical properties of $\lL\nflop{k}$.
\subsection{Behaviour of $\lL\nflop{k,tot}$ Under Gauge Transformations}%
\label{gaugetrans}                                                      %
Now we will take a closer look at the behaviour of  $\lL\nflop{k}$ under gauge transformations and see that this will constrain the possible values of $\tilde{j}_{\sst 12}$. Applying a gauge transformation on eqn (\ref{clid}) under which $h_{e_{\sst I}}$ transforms as $h_{e_{\sst I}}^g\rightarrow g(b(e_{\sst I}))g^{-1}(f(e_{\sst I}))$ with $b(e_{\sst I})$ and $f(e_{\sst I})$ being the beginning and the final point of the edge $e_{\sst I}$, we obtain
\ba
\left[\lL\wt{E}_k(S_t)_{k,{\sst tot }}\right]^g&=&\lim_{{\cal P}_t\to S_t}\sum\limits_{\sst {\Box}\in{\cal P}_t}\frac{16}{\kappa^2}\frac{1}{\frac{4}{3}\ell(\ell+1)(2\ell+1)}\nonumber \\
&&
\tr\left(\pi_{\ell}(h_{e_{\sst 3}})\left\{\pi_{\ell}(h^{-1}_{e_{\sst 3}}),V(R_{v({\sst \Box})})\right\}g^{-1}(b(e))\pi_{\ell}(\tau_k)g(b(e))\fbox{${\cal S}$}\pi_{\ell}(h_{e_{\sst 4}})\left\{\pi_{\ell}(h^{-1}_{e_{\sst 4}}),V(R_{v({\sst \Box})})\right\}\right).\nonumber\\
&&
\ea
Thus the classical expression transforms in the spin-1-representation, due to the term $g^{-1}(b(e))\pi_{\ell}(\tau_k)g(b(e))$. Consequently, we know that if we applied the corresponding operator on an incoming state $\bet{j_{\sst 12}}{m_{\sst 12}}$, the action of $\lL\nflop{k}$ would change the intertwiner at the vertex $v(\gamma)$ by $0,\pm 1$. Therefore, if we consider matrix elements of the kind $\betc{\tilde{j}_{\sst 12}}{\wt{m}_{\sst 12}}\,\lL\nflop{k}\,\bet{j_{\sst 12}}{m_{\sst 12}}$ the only non-vanishing values for $\tilde{j}_{\sst 12}$ are $\tilde{j}_{\sst 12}=j_{\sst 12},j_{\sst 12}\pm 1$. In the specific case where $j_{\sst 12}=0$, $\tilde{j}_{\sst 12}$ can only take the value $\tilde{j}_{\sst 12}=j_{\sst 12}+1$. Of course, we only want to consider incoming states that are physically relevant. Therefore we have to choose an incoming state $\bet{j_{\sst 12}}{m_{\sst 12}}$ with a total angular momentum $j_{\sst 12}=0$ in order to ensure that this state is gauge invariant.
Hence, the transformation property of  $\lL\nflop{k}$ leads to the restriction of $\tilde{j}_{\sst 12}=1$. Therefore, by means of theorem (\ref{theorem1}), the only total angular momentum $J$ of the states $|\alpha^{\st J}_i\,{\st M}\zu$ that contribute to the matrix element of  $\lL\op{\wt{E}}_{k,{\sst tot}}(S_t)$ are $J=0,1$. Therefore, eqn (\ref{act5Ek}) can be rewritten, according to our particular choices of $j_{\sst 12}=0$ and $\tilde{j}_{\sst 12}=1$, as
\ba
\label{act6Ek}
\lefteqn{\auf \beta^{\st 1}\,\wt{m}_{\sst 12}\,|\lL\op{\wt{E}}_{k,{\sst tot}}(S_t)|\,\beta^{\st 0}\,m_{\sst 12}\zu}\nonumber\\
&=&-\lim_{{\cal P}_t\to S_t}\sum\limits_{\sst {\Box}\in{\cal P}_t}\frac{8\,\lp^{-4}}{\frac{4}{3}\ell(\ell+1)(2\ell+1)}
\sum\limits_{\sst B,C,F=-\ell}^{+\ell}\Big\{\pi_{\ell}(\tau_k)_{\sst CB}
\sum\limits_{\wt{a}_{\sst 3}=|1-\ell|}^{1+\ell}\sum\limits_{a_{\sst 3}=|-\ell|}^{+\ell}\nonumber\\
&&\Big[+(-1)^{\st -F}\delta_{\sst F+C,0}(-1)^{{\st -}3a_{\sst 3}}\sqrt{2a_{\sst 3}+1}
\delta_{\wt{m}_{\sst 12}{\sst +F+B},m_{\sst 12}}\nonumber\\
&&\hspace{2.2cm}
\auf 1\, \wt{m}_{\sst 12}\, {\st ;}\,\ell\, {\st B}\,|\,\wt{a}_{\sst 3}\, \wt{m}_{\sst 12}{\st +B}\zu 
\auf \wt{a}_{\sst 3}\, \wt{m}_{\sst 12}{\st +B}\, {\st ;}\,\ell\, {\st F}\,|\,\,0\,\wt{m}_{\sst 12}{\st +B+F}\zu\nonumber\\
&&\hspace{2.2cm}
\auf \alpha^{\st 0}_{\wt{i}}\,{\st M}=\wt{m}_{\sst 12}{\st +B+F}\, {\st ;} \wt{m}'_{\sst 1}\,\wt{m}'_{\sst 2}\,|\,\op{O}_{\sst 1}\,|\,\alpha^{\st 0}_i\,{\st M}=m_{\sst 12}\, {\st ;}\, m'_{\sst 1}\,m'_{\sst 2}\zu\nonumber\\
&&\quad
-(-1)^{\st -F}\delta_{\sst F+B,0}
(-1)^{{\st -}1{\st -}3\wt{a}_{\sst 3}}\frac{\sqrt{2\wt{a}_{\sst 3}+1}}{\sqrt{3}}
\delta_{m_{\sst 12}{\sst +C+F},\wt{m}_{\sst 12}}\nonumber\\
&&\hspace{2.2cm}
\auf 0\, m_{\sst 12}\, {\st ;}\,\ell\, {\st C}\,|\,a_{\sst 3}\, m_{\sst 12}{\st +C}\zu  
\auf a_{\sst 3}\, m_{\sst 12}{\st +C}\, {\st ;}\ell\, {\st F}\,|\,\,1\, m_{\sst 12}{\st +C+F}\zu\nonumber\\
&&\hspace{2.2cm}
\auf \alpha^{\st 1}_{\wt{i}}\,{\st M}=\wt{m}_{\sst 12}\, {\st ;} \wt{m}'_{\sst 1}\,\wt{m}'_{\sst 2}\,|\,\op{O}_{\sst 2}\,|\,\alpha^{1}_i\,{\st M}=m_{\sst 12}{\st+C+F}\, {\st ;}\, m'_{\sst 1}\,m'_{\sst 2}\zu\Big]\Big\},
\ea
where $\wt{m}_{\sst 12}=\{-1,0,1\}$ and $m_{\sst 12}=0$ is the only possible value of the magnetic quantum number for $|\beta^{\st 0}_{m_{\sst 12}}\zu$. 
\newline
In the next section we will calculate the matrix elements $\auf \beta^{\st 1}\,\wt{m}_{\sst 12}\,|\lL\op{\wt{E}}_{k,{\sst tot}}(S_t)|\,\beta^{\st 0}\,m_{\sst 12}\zu$ of all four versions $\lL\op{\tilde{E}}^{\sst I,AL}_{k,{\sst tot}}(S_t),\lL\op{\tilde{E}}^{\sst I,RS}_{k,{\sst tot}}(S_t),\lL\op{\tilde{E}}^{\sst II,AL}_{k,{\sst tot}}(S_t),\lL\op{\tilde{E}}^{\sst II,RS}_{k,{\sst tot}}(S_t)$ of the new flux operator.
\section{Matrix Elements of the New Flux Operator $\lL\op{\wt{E}}_{k,{\sst tot}}(S_t)$}%
\label{Matrixelements}                                                                 %
 Before we explicitly calculate the necessary matrix elements of $\op{O}_{\sst 1},\op{O}_{\sst 2}$, the question arises, what are the matrix elements that we need, or rather what kind of matrix elements will appear in the recoupling procedure of eqn (\ref{act6Ek}). 
As the action of $\op{V}$ and accordingly also the action of  $\op{q}_{\st IJK}$  leaves the total angular momentum $J$ of a state $|\alpha^{\st J}_i\,{\st M}\zu$ invariant, the whole matrix that includes the elements of all  possible values of $J$ belonging to a particular choice of $j_{\sst 12}$ and  $\tilde{j}_{\sst 12}$ would be divided into orthogonal submatrices for each fixed total angular momentum $J$. Consequently, we can actually calculate the eigenvalues and eigenstates separately for every possible value of $J$.
Hence, in our case we should take a detailed look at the corresponding 
Hilbert spaces of $J=0,1$. Similarly to $\bet{j_{\sst 12}}{m_{\sst 12}}$ the spin label of $e_{\sst 1}$ and $e_{\sst 2}$ of $\alp{J}{i}{M}$ are identical ($j_{\sst 1}=j_{\sst 2}=j$). Therefore, we already know that $a_{\sst 2}=j\otimes j\in\{0,+1,...,2j\}=j_{\sst 12}$ can only be an integer. Hence, a basis of the Hilbert space belonging to $J=0$ is given by
\ba
\label{HJ0}
\alp{0}{1}{M}&:=&|\,a_{\sst 1}=j\,a_{\sst 2}=0\,a_{\sst 3}=\ell\,J=0\zu\nonumber\\
\alp{0}{2}{M}&:=&|\,a_{\sst 1}=j\,a_{\sst 2}=1\,a_{\sst 3}=\ell\,J=0\zu\nonumber\\
\alp{0}{3}{M}&:=&|\,a_{\sst 1}=j\,a_{\sst 2}=2\,a_{\sst 3}=\ell\,J=0\zu\nonumber\\
...&&\nonumber\\
\alp{0}{2j+1}{M}&:=&|\,a_{\sst 1}=j\,a_{\sst 2}=2j\,a_{\sst 3}=\ell\,J=0\zu
\ea 
Here, the only possible value for $a_{\sst 3}$  is $a_{\sst 3}=\ell$, because otherwise $a_{\sst 3}$ and $j_{\sst 4}=\ell$ could not couple to a resulting angular momentum $J=0$. Furthermore, we have assumed that the condition $a_{\sst 2}\le 2\ell$ has to be fulfilled to ensure that a resulting total angular momentum of $J=0$ can be achieved. If this is not the case, the number of states reduces down to the number of states where the condition $a_{\sst 2}\le 2\ell$ is still true. \footnote{Consequently, only for large enough $\ell$  the Hilbert space belonging to a zero total angular momentum will be $(2j+1)$-dimensional, e.g. for the simplest case $\ell=\frac{1}{2}$ it is only two dimensional.}
Fortunately, we will not have to calculate matrix elements of all possible combination of states. In our case, we  already know that $\tilde{j}_{\sst 12}=1$ and $j_{\sst 12}=0$. This is equivalent to $\wt{a}_{\sst 2}=1$ and $a_{\sst 2}=0$ and we realise that we only have to calculate the matrix element $\alpc{0}{2}{M}\,\op{q}_{\sst 134}\,\alp{0}{1}{M}$ here. 
\newline
The transformation properties of the operator $\lL\nflop{k}$, discussed in section \ref{gaugetrans}, led us to this restriction $\tilde{j}_{\sst 12}=1$. Even if we had not at all worried about any transformation properties of the operator before, we see at this point by simply looking at eqn (\ref{RSq134}) that all other possible matrix elements $\alpc{0}{i}{M}\,\op{q}_{\sst 134}\,\alp{0}{1}{M}$ where $i>2$ will vanish anyway. This is due to the fact that for $i>2$ $\Delta a_{\sst 2}:=|\wt{a}_{\sst 2}-a_{\sst 2}|>1$. In this case the 6j-symbols in eqn (\ref{RSq134}) in the last bracket will be zero and this makes the whole matrix element vanish. Summarising, if we start with a gauge invariant state $\bet{0}{0}$ there exists only one non-vanishing matrix element for the case $J=0$ which is $\alpc{0}{2}{M}\,\op{q}_{\sst 134}\,\alp{0}{1}{M}$ in our notation.
\newline
Let us analyse the case of a total angular momentum $J=1$ now. In this case we have three different values of the intermediate coupling $a_{\sst 3}=\{\ell-1,\ell,\ell+1\}$ to ensure that a total angular momentum of $J=1$ can be achieved. Hence, a basis of the corresponding Hilbert space is given by
\ba
\label{HJ1}
\alp{1}{1}{M}&:=&|\,a_{\sst 1}=j\,a_{\sst 2}=0\,a_{\sst 3}=\ell\,J=1\zu\nonumber\\
\alp{1}{2}{M}&:=&|\,a_{\sst 1}=j\,a_{\sst 2}=1\,a_{\sst 3}=\ell-1\,J=1\zu\nonumber\\
\alp{1}{3}{M}&:=&|\,a_{\sst 1}=j\,a_{\sst 2}=1\,a_{\sst 3}=\ell\,J=1\zu\nonumber\\
\alp{1}{4}{M}&:=&|\,a_{\sst 1}=j\,a_{\sst 2}=1\,a_{\sst 3}=\ell+1\,J=1\zu\nonumber\\
...&&\nonumber\\
\alp{1}{6j-1}{M}&:=&|\,a_{\sst 1}=j\,a_{\sst 2}=2j\,a_{\sst 3}=\ell-1\,J=1\zu\nonumber\\
\alp{1}{6j}{M}&:=&|\,a_{\sst 1}=j\,a_{\sst 2}=2j\,a_{\sst 3}=\ell\,J=1\zu\nonumber\\
\alp{1}{6j+1}{M}&:=&|\,a_{\sst 1}=j\,a_{\sst 2}=2j\,a_{\sst 3}=\ell+1\,J=1\zu\nonumber\\
\ea 
Here the condition on $a_{\sst 2}$ and $\ell$ is $a_{\sst 2}\le 2\ell+1$. Notice that in the special and simplest case where $\ell=\frac{1}{2}$ the intermediate coupling $a_{\sst 3}=\ell-\frac{1}{2}$ is not sensible, therefore this state has to be dropped  here and the Hilbert space includes only $5\times 3=15$ states. Again, due to the  construction of the operator $\lL\nflop{k}$, we only have to consider the matrix elements with $\wt{a}_{\sst 2}=1,a_{\sst 2}=0$ and these are precisely $\alpc{1}{i}{M}\,\op{q}_{\sst 134}\,\alp{1}{1}{M}$ where $i=2,3,4$. Hence, we see that for $J=1$  three different matrix elements will contribute to the final result. Similarly to the case $J=0$ all matrix elements $\alpc{1}{i}{M}\,\op{q}_{\sst 134}\,\alp{1}{1}{M}$ for $i>4$ vanish, because then $\Delta a_{\sst 2}:=|\wt{a}_{\sst 2}-a_{\sst 2}|>1$.
\newline
We will now go back to eqn (\ref{act6Ek}) and apply our new results. Furthermore, the discussion above showed that in the first term of eqn (\ref{act6Ek}) the only possible value for $a_{\sst 3},\wt{a}_{\sst 3}$ is $\ell$ (case $J=0$). In the second term $a_{\sst 3}=\ell$ is still valid, but here $\wt{a}_{\sst 3}$ can take the values  $\wt{a}_{\sst 3}=\{\ell-1,\ell,\ell+1\}$. Therefore eqn (\ref{act6Ek}) simplifies to
\ba
\label{act7Ek}
\lefteqn{\auf \beta^{\st 1}\,\wt{m}_{\sst 12}\,|\lL\op{\wt{E}}_{k,{\sst tot}}(S_t)|\,\beta^{\st 0}\,{\st 0}\zu}\nonumber\\
&=&-\lim_{{\cal P}_t\to S_t}\sum\limits_{\sst {\Box}\in{\cal P}_t}\frac{8\,\lp^{-4}(-1)^{3\ell}}{\frac{4}{3}\ell(\ell+1)(2\ell+1)}
\sum\limits_{\sst B,C,F=-\ell}^{+\ell}\Big\{\pi_{\ell}(\tau_k)_{\sst CB}\nonumber\\
&&\Big[+(-1)^{\st -F}\delta_{\sst F+C,0}\sqrt{2\ell+1}\delta_{\wt{m}_{\sst 12}{\st +B+F,0}}\nonumber\\
&&\hspace{2.2cm}
\auf 1\, \wt{m}_{\sst 12}\, {\st ;}\,\ell\, {\st B}\,|\,\ell\, \wt{m}_{\sst 12}{\st +B}\zu 
\auf \ell\, \wt{m}_{\sst 12}{\st +B}\, {\st ;}\,\ell\, {\st F}\,|\,\,0\,0\zu
 \nonumber\\
&&\hspace{2.2cm}
\auf \alpha^{\st 0}_{2}\,{\st M}=\wt{m}_{\sst 12}{\st +B+F}\, {\st ;}\, \wt{m}'_{\sst 1}\,\wt{m}'_{\sst 2}\,|\,\op{O}_{\sst 1}\,|\,\alpha^{\st 0}_1\,{\st M}=0\, {\st ;}\, m'_{\sst 1}\,m'_{\sst 2}\zu\nonumber\\
&&
-(-1)^{\st -F}\delta_{\sst F+B,0}
\delta_{{\sst C+F},\wt{m}_{\sst 12}}\nonumber\\
&&\hspace{2.2cm}\auf 0\, 0\, {\st ;}\ell\, {\st C}|\ell\, {\st C}\zu  
\auf \ell\, {\st C}\, {\st ;}\ell\, {\st F}|\,1\, {\st C+F}\zu\nonumber\\
&&\hspace{2.2cm}
\Big[+\frac{\sqrt{2\ell-1}}{\sqrt{3}}
\auf \alpha^{\st 1}_{2}\,{\st M}=\wt{m}_{\sst 12}\, {\st ;} \wt{m}'_{\sst 1}\,\wt{m}'_{\sst 2}\,|\,\op{O}_{\sst 2}\,|\alpha^{1}_1\,{\st M}={\st C+F}\, {\st ;}\, m'_{\sst 1}\,m'_{\sst 2}\zu\nonumber\\
&&\,\,\,\,\hspace{2.2cm}
-\frac{\sqrt{2\ell+1}}{\sqrt{3}}
\auf \alpha^{\st 1}_{3}\,{\st M}=\wt{m}_{\sst 12}\, {\st ;} \wt{m}'_{\sst 1}\,\wt{m}'_{\sst 2}\,|\,\op{O}_{\sst 2}\,|\alpha^{1}_1\,{\st M}={\st C+F}\, {\st ;}\, m'_{\sst 1}\,m'_{\sst 2}\zu\nonumber\\
&&\,\,\,\,\hspace{2.2cm}
+\frac{\sqrt{2\ell+3}}{\sqrt{3}}
\auf \alpha^{\st 1}_{4}\,{\st M}=\wt{m}_{\sst 12}\, {\st ;} \wt{m}'_{\sst 1}\,\wt{m}'_{\sst 2}\,|\,\op{O}_{\sst 2}\,|\alpha^{1}_1\,{\st M}={\st C+F}\, {\st ;}\, m'_{\sst 1}\,m'_{\sst 2}\zu\Big]\Big\}.
\ea
\subsection{Matrix Elements of $\op{Q}^{\sst AL}_v$ and $\op{Q}^{\sst RS}_{v,\sst IJK}$}%
In order to calculate the matrix elements of $\op{O}_{\sst 1},\op{O}_{\sst 2}$ we have to calaculate the matrix elements of $\op{Q}^{\sst AL}_v$ and $\op{Q}^{\sst RS}_{v,\sst IJK}$ as an intermediate step. Thus, we will discuss this calculation first before we talk about the four different cases separately.
\subsubsection{Matrix Elements of $\op{Q}^{\sst AL}_v$} %
First, we have to apply the map $W$ in eqn (\ref{Wmap}) to $\op{Q}^{\sst Y,AL}_v$ since  we 
 need the corresponding operator in the abstract angular momentum system Hilbert space depending on $J$ 
\be
\label{Qdef} 
\op{Q}^{\sst J,AL}_v:=\lp^6\frac{3!i}{4}C_{reg}\sum\limits_{I<J<K}\epsilon(e_{\sst I},e_{\sst J},e_{\sst K})\,\op{q}_{\sst IJK}^{\sst J},
\ee
whereby $\op{q}_{\sst IJK}^{\sst J}$ results from $\op{q}_{\sst IJK}^{\sst Y}$ upon replacing $Y^k_e$ everywhere by $J^k_e$. From now on we will neglect the explicit label $J$ for $\op{Q}^{\sst J,AL}_v$ and keep in mind that we are working in the abstract angular momentum system Hilbert space.
Our SNF under consideration $\alp{J}{i}{M}$ contains two linearly independent triples constructed from the edges $\{e_{\sst 1},e_{\sst 3},e_{\sst 4}\}$ and $\{e_{\sst 2},e_{\sst 3},e_{\sst 4}\}$ for which the signum factor is non-vanishing. Here we split an edge of type up or down as $e=e^{-1}_{\sst 2}\circ e_{\sst 1}$ and then $\epsilon(e_{\sst 1},e_{\sst 3},e_{\sst 4})=-\epsilon(e_{\sst 2},e_{\sst 3},e_{\sst 4})=\pm 1$ for edges of type up and down respectively. Hence, we have
\be
\label{RSQdef} 
\op{Q}^{\sst J,AL}_v:=\lp^6\frac{3!i}{4}C_{reg}\left(\op{q}_{\sst 134}^{\sst J} - \op{q}_{\sst 234}^{\sst J}\right)
\ee
We notice that the matrix elements of $\op{Q}^{\sst AL}_v$ apart from the constant prefactor $\lp^6\frac{3!i}{4}C_{reg}$ are basically equal to the matrix elements of $\op{q}_{\sst IJK}^{\sst J}$.
In \cite{14} a general formula for the matrix element $\alpc{J}{i}{M}\,\op{q}_{\sst IJK}\,\alp{J}{\tilde{i}}{M}$ for an arbitrary n-valent vertex was derived (eqn (47) in \cite{14}). We can use this result in order to get the desired matrix element of $\op{q}_{\sst 134}$ and $\op{q}_{\sst 234}$. We obtain
\ba
\label{RSq134}
\lefteqn{\alpc{J}{i}{M}\,\op{q}_{\sst 134}\,\alp{J}{\tilde{i}}{M}} \\
&=&\frac{1}{4}(-1)^{+2j+\ell+J}\sqrt{2j(2j+1)(2j+2)}\left[2\ell(2\ell+1)(2\ell+2)\right]^{\frac{3}{2}}
\sqrt{(2a_{\sst 2}+1)(2\wt{a}_{\sst 2}+1)}\sqrt{(2a_{\sst 3}+1)(2\wt{a}_{\sst 3}+1)}\nonumber\\
&&\hspace{-0.5cm}
\SixJ{j}{j}{a_{\sst 2}}{1}{\wt{a}_{\sst 2}}{j}
\SixJ{J}{\ell}{a_{\sst 3}}{1}{\wt{a}_{\sst 3}}{\ell}
\Big[(-1)^{\wt{a}_{\sst 3}+\wt{a}_{\sst 2}}\SixJ{\wt{a}_{\sst 2}}{\wt{a}_{\sst 3}}{\ell}{1}{\ell}{a_{\sst 3}}
                                           \SixJ{a_{\sst 3}}{\ell}{\wt{a}_{\sst 2}}{1}{a_{\sst 2}}{\ell}
-(-1)^{a_{\sst 3}+a_{\sst 2}}\SixJ{a_{\sst 2}}{a_{\sst 3}}{\ell}{1}{\ell}{\wt{a}_{\sst 3}}
                                           \SixJ{\wt{a}_{\sst 3}}{\ell}{a_{\sst 2}}{1}{\wt{a}_{\sst 2}}{\ell}\!\!\!\Big]
\nonumber
\ea 
\ba
\label{RSq234}
\lefteqn{\alpc{J}{i}{M}\,q_{\sst 234}\,\alp{J}{\tilde{i}}{M}}\nonumber\\
&=&+\frac{1}{4}(-1)^{+2j+\ell+{\st J}}
\sqrt{2j(2j+1)(2j+2)}[2\ell(2\ell+1)(2\ell+2)]^{\frac{3}{2}}
\sqrt{(2\wt{a}_{\sst 2}+1)(2a_{\sst 2}+1)}\sqrt{2(a_{\sst 3}+1)(2\wt{a}_{\sst 3}+1)}\nonumber\\
&&
\SixJ{j}{j}{a_{\sst 2}}{1}{\wt{a}_{\sst 2}}{j}
\SixJ{{\st J}}{\ell}{a_{\sst 3}}{1}{\wt{a}_{\sst 3}}{\ell}
\left[(-1)^{a_{\sst 2}+\wt{a}_{\sst 3}}
\SixJ{a{\sst 3}}{\ell}{a_{\sst 2}}{1}{\wt{a}_{\sst 2}}{\ell}
\SixJ{\wt{a}_{\sst 2}}{\ell}{a_{\sst 3}}{1}{\wt{a}_{\sst 3}}{\ell}
-(-1)^{\wt{a}_{\sst 2}+a_{\sst 3}}
\SixJ{\wt{a}{\sst 3}}{\ell}{a_{\sst 2}}{1}{\wt{a}_{\sst 2}}{\ell}
\SixJ{a_{\sst 2}}{\ell}{a_{\sst 3}}{1}{\wt{a}_{\sst 3}}{\ell}\right]\nonumber\\
\ea
The explicit derivation can be found in appendix \ref{Derqijk}. Here we already used that $j_{\sst 1}=j_{\sst 2}=j$, $j_{\sst 3}=j_{\sst 4}=\ell$ and $a_{\sst 4}=J$ and $\SixJ{a}{b}{c}{d}{e}{f}$ are the 6j-symbols defined in eqn (120) in \cite{14}.  
\subsubsection{Matrix Elements of $\op{Q}^{\sst RS}_{v,{\sst IJK}}$} %
If we consider the operator $\op{Q}^{\sst RS}_{v,{\sst IJK}}$ we also have to consider linearly dependent triples. Therefore also the triples $\{e_{\sst 1},e_{\sst 2},e_{\sst 3}\}$ and $\{e_{\sst 1},e_{\sst 2},e_{\sst 4}\}$ will contribute. Since the sum over the triples is positioned outside the square root and the abolute value in the case of $RS$ (see eqn (\ref{RSVqijk}) for details), we moreover have to deal with four separated operators, namely $\op{Q}^{\sst RS}_{v,{\sst 134}}$,$\op{Q}^{\sst RS}_{v,{\sst 234}}$,$\op{Q}^{\sst RS}_{v,{\sst 123}}$,$\op{Q}^{\sst RS}_{v,{\sst 124}}$. From eqn (\ref{RSQ}) we can read off that the matrix element of $\op{Q}^{\sst RS}_{v,{\sst IJK}}$ is derived from the matrix element of $\op{q}_{\sst IJK}$ multiplied by the constant $\lp^6\frac{3!i}{4}C_{reg}$. Thus, here we also need the matrix elements of $\op{q}_{\sst 123}$ and $\op{q}_{\sst 124}$ which are presented below
\ba
\label{RSq123}
\lefteqn{\alpc{J}{i}{M}\,\op{q}_{\sst 123}\,\alp{J}{\tilde{i}}{M}}\nonumber\\
&=&
+\frac{1}{2}(-1)^{+2j+\ell+1}(-1)^{\wt{a}_{\sst 2}-a_{\sst2}+a_{\sst 3}}
X(j,\ell)^{\frac{1}{2}}A(a_{\sst 2},\wt{a}_{\sst 2})
\SixJ{j}{j}{a_{\sst 2}}{1}{\wt{a}_{\sst 2}}{j}
\SixJ{a_{\sst 3}}{\ell}{a_{\sst 2}}{1}{\wt{a}_{\sst 2}}{\ell}
\Big[a_{\sst 2}(a_{\sst 2}-1) - \wt{a}_{\sst 2}(\wt{a}_{\sst 2}-1)\Big]\delta_{a_{\sst 3},\wt{a}_{\sst 3}}\nonumber\\
\ea
and
\ba
\label{RSq124}
\lefteqn{\alpc{J}{i}{M}\,\op{q}_{\sst 124}\,\alp{J}{\tilde{i}}{M}}\nonumber\\
&=&
+\frac{1}{2}(-1)^{+2j+{\st J}}X(j,\ell)^{\frac{1}{2}}A(a_{\sst 2},\wt{a}_{\sst 2})A(a_{\sst 3},\wt{a}_{\sst 3})
\SixJ{j}{j}{a_{\sst 2}}{1}{\wt{a}_{\sst 2}}{j}\SixJ{\ell}{a_{\sst 2}}{a_{\sst 3}}{1}{\wt{a}_{\sst 2}}{\wt{a}_{\sst 3}}
\SixJ{a_{\sst 4}}{\ell}{a_{\sst 3}}{1}{\wt{a}_{\sst 3}}{\ell}\Big[a_{\sst 2}(a_{\sst 2}-1)-\wt{a}_{\sst 2}(\wt{a}_{\sst 2}+1)\Big].\nonumber\\
\ea
\subsection{Case $\lL\op{\tilde{E}}^{\sst I,AL}_{k,{\sst tot}}(S_t)$ i.e. $E^{a,{\sst I}}_k=\det(e)e^a_k$ and $\op{V}_{\sst AL}$}
\label{OV2}
If we consider the case of $\lL\op{\tilde{E}}^{\sst I,AL}_{k,{\sst tot}}(S_t)$, the operators $\op{O}_{\sst 1},\op{O}_{\sst 2}$ in eqn (\ref{act7Ek}) are $\op{O}_{\sst 1}=\op{O}_{\sst 2}=\op{V }_{\sst AL}^2$.
Going back to eqn (\ref{RSVqijk}) and (\ref{Qdef}), we see that 
$\op{V}^2_{\sst AL}=|\op{Q}^{\sst AL}_v|$. Consequently, the task of calculating matrix elements of $\op{V}^2_{\sst AL}$ can be treated in the following way. As a first step we compute the eigenvalues $\lambda^{\sst Q}_j$ and eigenstates $\{\vec{e}_j\}$ of $\op{Q}^{\sst AL}_v$. Afterwards we expand the matrix elements of $\op{V}^2_{\sst AL}$ in terms of the eigenvectors of $\op{Q}^{\sst AL}_v$
\be
 \alpc{\wt{J}}{\tilde{i}}{\wt{M}}\,\op{V}^2_{\sst AL}\,\alp{J}{i}{M}=\sum\limits_j
 \left|\lambda^{\sst Q}_j\right|\alpc{\wt{J}}{\tilde{i}}{\wt{M}}\,\vec{e}_j\zu\auf\vec{e}_j\,\alp{J}{i}{M},
\ee
whereby we used that $\op{V}^2_{\sst AL}$ and $\op{Q}^{\sst AL}_v$ have the same eigenvectors and if $\lambda^{\sst Q }_j$ is an eigenvalue of $\op{Q}^{\sst AL}_v$, so is $|\lambda^{\sst Q }_j|$ an eigenvalue of $\op{V}^2_{\sst AL}$.
\newline
The four matrix elements of $\op{V }^2_{\sst AL}$ that occur in eqn (\ref{act7Ek}) are $\alpc{0}{2}{0}\,\op{V}_{\sst AL}^2\alp{0}{1}{0}$ and $\alpc{1}{i}{M}\,\op{V}_{\sst AL}^2\alp{1}{1}{M}$ where $i=2,3,4$.
As the operator $\op{V}^2_{\sst AL}$ does not change the total angular momentum $J$ and magnetic quantum number $M$ of the states $\alp{J}{i}{M}$ and moreover the Hilbert spaces belonging to different $J$ are orthogonal to each other, we can calculate the cases of $J=0$ and $J=1$ separately.
 Since these Hilbert spaces for arbitrary spin $\ell$ of the edges $e_{\sst 3},e_{\sst 4}$ in general are $(2j+1)-$ and $(6j+1)\times 3-$dimensional for $J=0$ and $J=1$, respectively (see also eqn (\ref{HJ0}) and (\ref{HJ1}) for this ) this is a lot of work that has to be done. The diagonalisation of $\op{Q}^{\sst AL}_v$ for the two most easiest cases $\ell=0.5,1$, where the dimension of the Hilbert spaces in these cases is so small that we were still able to calculate the eigensystems of $\op{Q}^{\sst AL}_v$ analytically, can be found in appendix \ref{CaseV2}. Applying the eigenvector expansion, we obtain the following matrix elements\footnote{Note, that in the case $\ell=1/2$ the state $\alp{1}{1}{M}$ does not exist (see eqn (\ref{HJ1}) for the definition of $\alp{1}{1}{M}$. That is the reason why we do not have to consider this particular matrix element.} for $\op{V}^2_{\sst AL}$ 
 \begin{center}
\begin{tabular}{|l|l|}\hline
{$\ell=0.5$} &{$\ell=1$} \\ \hline\hline
$\alpc{0}{2}{0}\,\op{V}^2_{\sst AL}\alp{0}{1}{0}=0$&$\alpc{0}{2}{0}\,\op{V}^2_{\sst AL}\alp{0}{1}{0}=0$ \\ \hline
 & $\alpc{1}{2}{M}\,\op{V}^2_{\sst AL}\alp{1}{1}{M}=0$\\ \hline
$\alpc{1}{3}{M}\,\op{V}^2_{\sst AL}\alp{1}{1}{M}=0$&$\alpc{1}{3}{M}\,\op{V}^2_{\sst AL}\alp{1}{1}{M}=0$\\ \hline
$\alpc{1}{4}{M}\,\op{V}^2_{\sst AL}\alp{1}{1}{M}=0$&$\alpc{1}{4}{M}\,\op{V}^2_{\sst AL}\alp{1}{1}{M}=0$\\ \hline
\end{tabular}
\end{center}
Surprisingly, all matrix elements turned out to be identical to zero. Therefore the operator $\lL\op{\tilde{E}}^{\sst I,AL}_{k,{\sst tot}}(S_t)$, at least for the spin labels $\ell=0.5,1$, becomes the zero operator! Consequently,it is not consistent with the usual flux operator $\op{E}_k(S_t)$ which is definitely not the zero operator.
\subsection{Case $\lL\op{\tilde{E}}^{\sst I,RS}_{k,{\sst tot}}(S_t)$ i.e. $E^{a,{\sst I}}_k=\det(e)e^a_k$ and $\op{V}_{\sst RS}$}
\label{RSOV2}
 As pointed out before we have to take into account the linearly dependent triples. The total $\op{V}_{\sst RS}$ is then given by
 \be
 \label{VRStot}
 \op{V}_{\sst RS}=\op{V}_{q_{\sst 134}}+\op{V}_{q_{\sst 234}}+\op{V}_{q_{\sst 123}}+\op{V}_{q_{\sst 124}},
 \ee
 whereby for each $\op{V}_{q_{\sst IJK}}$ the operator identity $\op{V}_{q_{\sst IJK}}=\sqrt{|Q^{\sst RS}_{v,{\sst IJK}}|}$ holds. If we consider the expression of $\op{V}_{\sst RS}$ in eqn (\ref{VRStot}) together with the definition of the operators $\op{O}_{\sst 1},\op{O}_{\sst 2}$ in eqn (\ref{O1O2}), we can rewrite the operators $\op{O}_{\sst 1},\op{O}_{\sst 2}$ in the following way
 \ba
 \label{C1RSO1O2}
 \op{O}_{\sst 1}^{\sst I,RS}&=&\op{V}_{q_{\sst 134}}^2+\op{V}_{q_{\sst 234}}^2+\op{V}_{q_{\sst 134}}\op{V}_{q_{\sst 234}}+\op{V}_{q_{\sst 234}}\op{V}_{q_{\sst 134}}+\op{V}_{q_{\sst 134}}\op{V}_{q_{\sst 123}}+\op{V}_{q_{\sst 124}}\op{V}_{q_{\sst 134}}+\op{V}_{q_{\sst 234}}\op{V}_{q_{\sst 123}}+\op{V}_{q_{\sst 124}}\op{V}_{q_{\sst 234}}+\op{V}_{q_{\sst 124}}\op{V}_{q_{\sst 123}}\nonumber\\
 \op{O}_{\sst 2}^{\sst I,RS}&=&\op{V}_{q_{\sst 134}}^2+\op{V}_{q_{\sst 234}}^2+\op{V}_{q_{\sst 234}}\op{V}_{q_{\sst 134}}+\op{V}_{q_{\sst 134}}\op{V}_{q_{\sst 234}}+\op{V}_{q_{\sst 123}}\op{V}_{q_{\sst 134}}+\op{V}_{q_{\sst 134}}\op{V}_{q_{\sst 124}}+\op{V}_{q_{\sst 123}}\op{V}_{q_{\sst 234}}+\op{V}_{q_{\sst 234}}\op{V}_{q_{\sst 124}}+\op{V}_{q_{\sst 123}}\op{V}_{q_{\sst 124}}
 \ea
  Similar to $\op{V}_{\sst AL}$ we are restricted to the spin labels $\ell=0.5,1$ of the additional edges $e_{\sst 3},e_{\sst 4}$, because for higher spin labels the matrices of $Q^{\sst RS}_{v,{\sst IJK}}$ cannot be diagonalised analytically anymore. Using the operator identity $\op{V}_{q_{\sst IJK}}=\sqrt{\left|Q^{\sst RS}_{v,{\sst IJK}}\right|}$, we can, as before, expand each $\op{V}_{q_{\sst IJK}}$ in terms of the eigenvectors of $Q^{\sst RS}_{v,{\sst IJK}}$ and use that if $\lambda^{\sst Q}_j$ is an eigenvalue of $Q^{\sst RS}_{v,{\sst IJK}}$, then $\sqrt{|\lambda^{\sst Q}_j|}$ is also an eigenvalue of $\op{V}_{q_{\sst IJK}}$
 \be
 \alpc{J}{\tilde{i}}{M}\,\op{V}_{q_{\sst IJK}}\,\alp{J}{i}{M}=\sum\limits_j\sqrt{|\lambda^{\sst Q}_j|}\alpc{J}{\tilde{i}}{M}\,\vec{e}_j\zu\auf\vec{e}_j\,\alp{J}{i}{M}
 \ee
  The detailed calculations of the matrix elements of $\op{O}_{\sst 1},\op{O}_{\sst 2}$ can be found in appendix section \ref{RSOp}. Here we will list only the final results. The matrix elements that are included in $\lL\op{\tilde{E}}^{\sst I,RS}_{k,{\sst tot}}(S_t)$ are  precisely $\alpc{0}{2}{0}\,\op{O}^{I,\sst RS}_{\sst 1}\,\alp{0}{1}{0}$ and $\alpc{1}{i}{M}\,\op{O}^{I,\sst RS}_{\sst 2}\,\alp{1}{1}{M}$ where $i=3,4$ for $\ell=0.5$ and $i=2,3,4$ if $\ell=1$, respectively. We get 
 \newline
 \begin{center}
\begin{tabular}{|l|l|}\hline
{$\ell=0.5$} &{$\ell=1$} \\ \hline\hline
$\alpc{\st 0}{2}{\st 0}\,\op{O}^{I,\sst RS}_{\sst 1}\,\alp{\st 0}{1}{\st 0}=0$&$\alpc{\st 0}{2}{\st 0}\,\op{O}^{I,\sst RS}_{\sst 1}\,\alp{\st 0}{1}{\st 0}=0$\\ \hline
&$\alpc{\st 1}{2}{\st M}\,\op{O}^{I,\sst RS}_{\sst 2}\,\alp{\st 1}{1}{\st M}=0$\\ \hline 
$\alpc{\st 1}{3}{\st M}\,\op{O}^{I,\sst RS}_{\sst 2}\,\alp{\st 1}{1}{\st M}=0$&$\alpc{\st 1}{3}{\st M}\,\op{O}^{I,\sst RS}_{\sst 2}\,\alp{\st 1}{1}{\st M}=0$\\ \hline 
$\alpc{\st 1}{4}{\st M}\,\op{O}^{I,\sst RS}_{\sst 2}\,\alp{\st 1}{1}{\st M}=0$&$\alpc{\st 1}{4}{\st M}\,\op{O}^{I,\sst RS}_{\sst 2}\,\alp{\st 1}{1}{\st M}=0$\\ \hline
\end{tabular}
\end{center}
Consequently, similarly to our calculations before with $V^2_{\sst AL}$, we obtain only vanishing matrix elements of $\op{O}^{I,\sst RS}_{\sst 1},\op{O}^{I,\sst RS}_{\sst 2}$. Thus the matrix element $\betc{1}{\wt{m}_{\sst 12}}\,\lL\op{\tilde{E}}^{\sst I,RS}_{k,{\sst tot}}(S_t)\,\bet{0}{0}$ is zero as well. Consequently, analogous to the case of $\op{V}_{\sst AL}$ $\lL\op{\tilde{E}}^{\sst I,RS}_{k,{\sst tot}}(S_t)$ becomes the zero operator.
\newline
It is true that due to the absence of the factor $\epsilon(e_{\sst 
I},e_{\sst J},e_{\sst K})$ other orderings for the RS volume operator are 
available in which not $V^2_{\sst  RS}$ but rather two fators of $V_{\sst RS}$ sandwiched 
between holonomies appear and such orderings could potentially lead 
to non vanishing matrix elements. Unfortunately, all these orderings also 
lead to identically vanishing matrix elements as we prove explicitly in 
appendix \ref{RSOp}.
\subsection{Summarising the Results of Case I}  %
The analysis of the last two section showed that either the operator $\lL\op{\tilde{E}}^{\sst I,AL}_{k,{\sst tot}}(S_t)$ nor the operator $\lL\op{\tilde{E}}^{\sst I,RS}_{k,{\sst tot}}(S_t)$ are consistent with the usual flux operator, because both of them are the zero operator. This is due to the fact that all matrix elments of the operators $\op{O}_{\sst 1},\op{O}_{\sst 2}$ that occur in eqn (\ref{act7Ek}) vanish. Since the action on an arbitrary SNF can be determined from the matrix element $\betc{1}{\wt{m}_{\sst 12}}\,\lL\op{\tilde{E}}^{\sst I,AL/RS}_{k,{\sst tot}}(S_t)\,\bet{0}{0}$, we know that the vanishing of this matrix element is equivalent to the fact that $\lL\op{\tilde{E}}^{\sst I,AL/RS}_{k,{\sst tot}}(S_t)$ becomes the zero operator. For this reason we can conclude, at least in the cases where we choose $\ell=0.5,1$, that the choice of $E^{a,{\sst I}}_k(S_t)=\det(e)e^a_k$ does not lead to an alternative flux operator that is consistent with the usual one.
 To rule out the choice $E^a_k(S_t)=\det(e)e^a_k$ completely, we need to investigate the matrix element for arbitrary representation weights $\ell$. 
For higher values of $\ell$ the calculation cannot be done analytically any 
more simply due to the fact that the roots of the characteristic 
polynomial of 
Hermitean matrices of the form $Q=iA,\;A^T=-A$ can be found by quadratures 
in general only up to rank nine. However, the results for $\ell=0.5,1$ 
indicate that there is an abstract reason which leads to the vanishing 
of the matrix elements for {\it any} $\ell$. We were not able to find such 
an abstract argument yet. However, even if that was not the case
and there would be a range of values for $\ell$ for which not all of the 
matrix elements would vanish, it would be awkward that the classical
theory is independent of $\ell$ while the quantum theory strongly depends on 
$\ell$ even in the correspondence limit of large $j$. 
\subsection{Case $\lL\op{\tilde{E}}^{\sst II,AL}_{k,{\sst tot}}(S_t)$ i.e. $E^{a,{\sst II}}_k={\cal S}\det(e)e^a_k$ and $\op{V}_{\sst AL}$}
\label{QVSV}
Considering the case of the operator $\lL\op{\tilde{E}}^{\sst II,AL}_{k,{\sst tot}}(S_t)$, we can read of from eqn \ref{O1O2} the expressions $\op{O}_{\sst 1}=\op{V}_{\sst AL}\op{{\cal S}}\op{V}_{\sst AL}=\op{O}_{\sst 2}$. Hence, again we have to compute special matrix elements of the operators $\op{O}_{\sst 1},\op{O}_{\sst 2}$. Since the signum operator $\op{{\cal S}}$ that corresponds to the classical expression ${\cal S}:=\sgn(\det(e))$ does not exist in the literature so far, we will in detail explain how the operator $\op{{\cal S}}$ has to be understood.
\subsubsection{The Sign Operator $\op{{\cal S}}$}%
\label{SignOP}
We are dealing now with case II meaning that the densitised triad is given by $E^{a,{\sst II}}_k={\cal S}\det(e)e^a_k$, where ${\cal S}:=\det(e)$. Applying the determinant onto $E^{a,{\sst II}}_k$, we get 
\be
\label{DefdetE}
\det(E)=\sgn(\det(e))\det(q)\quad\mbox{with}\quad \det(q)=[\det(e)]^2\ge0.
\ee
Therefore, we obtain
\be
\sgn(\det(E))=\sgn(\det(e))={\cal S}. 
\ee
In the following we want to show that ${\cal S}=\sgn(\det(E))$ can be identified with the signum of the expression inside the absolute value under the square roots in the definition of the volume. For this purpose let us first discuss this issue on the classical level and afterwards go back into the quantum theory and see how the corresponding operator $\op{{\cal S}}$ is connected with the operator $\op{Q}^{\sst AL}_v$ in eqn ({\ref{Qdef}).
\newline
In order to do this let us consider eqn (\ref{Classfinal}). This equation contains the classical volume
$V(R_{v({\sst \Box})})$ where $R_{v({\sst \Box})}$ denotes a region centred around the vertex $v({\sst \Box})$.
\newline
The volume of such a cube is given by
\be
\label{VBox}
V(R_{v({\sst \Box})})=\int\limits_{R_{v({\sst \Box})}}\sqrt{\det(q)}d^3x=
\int\limits_{R_{v({\sst \Box})}}\sqrt{|\det(E)|}d^3x,
\ee 
where we used $\det(q)=|\det(E)|$ from eqn (\ref{DefdetE}). Introducing a parametrisation of the cube now,  we end up with
\be
V(R_{v({\sst \Box})})=\int\limits_{[-\frac{\epsilon'}{2},+\frac{\epsilon'}{2}]^3}\left|\frac{\partial X^{\st I}(u)}{\partial u_{\st J}}\right|\sqrt{|\det(E)(u)|}d^3u
=\int\limits_{[-\frac{\epsilon'}{2},+\frac{\epsilon'}{2}]^3}\left|\det(X)\right|\sqrt{|\det(E)(u)|}d^3u.
\ee
In order to be able to carry out the integral  we choose the cube $R_{v({\sst \Box})}$ small enough and thus, the volume can be approximated by
\be
\label{ResVBox}
V(R_{v({\sst \Box})})\approx\epsilon'^3\left|\det(\frac{\partial X}{\partial u})(v)\right|\sqrt{|\det(E)(v)|}.
\ee
Using the definition of $\det(E)=\frac{1}{3!}\epsilon_{abc}\epsilon^{jkl}E^a_jE^b_kE^c_l$, we can rewrite  eqn (\ref{VBox}) as
\be
V(R_{v({\sst \Box})})=\int\limits_{\sst \Box}\sqrt{\Big|\frac{1}{3!}\epsilon_{abc}\epsilon^{jkl}E^a_jE^b_kE^c_l\Big|}d^3x
\ee
If we again choose $R_{v({\sst \Box})}$ small enough and define the square surfaces of the cube as $S^{\sst I}$, we can re-express the volume integral over the densitised triads in terms of their corresponding electric fluxes through the surfaces $S^{\sst I}$
\ba
\label{Vflux}
V(R_{v({\sst \Box})})&\approx&\sqrt{\Big|\frac{1}{3!}\epsilon_{\sst IJK}\epsilon^{jkl}E_j(S^{\sst I})E_k(S^{\sst J})E_l(S^{\sst K})\Big|}.
\ea
The flux through a particular surfaces $S^{\sst I}$ is defined as
\be
\label{VFlux}
E_j(S^{\sst I})=\int\limits_{S^{\sst I}}E^a_j n^{S^{\sst I}}_a \quad\quad n^{S^{\sst I}}_a=\frac{1}{2}\epsilon^{\sst IJK}\epsilon_{abc}X^b_{,u_{\sst J}}X^c_{,u_{\sst
K}}\Big|_{n^{\sst I}=0}.
\ee
Here $n^{S^{\sst I}}_a$ denotes the conormal vector associated with the surface $S^{\sst I}$. 
Regarding eqn (\ref{Vflux}) we realise that inside the absolute value in eqn (\ref{Vflux}) appears exactly the definition of $\det(E_j(S^{\sst I}))$. Therefore we get
\be
\label{VdetES}
V(R_{v({\sst \Box})})\approx\sqrt{\Big|\det(E_j(S^{\sst I}))\Big|}.
\ee
On the other hand, by taking advantage of the fact that the surfaces $S^{\sst I}$ are small enough so that the integral can be approximated by the value at the vertex times the size of the surface itself, we obtain for $\det(E_j(S^{\sst I}))$
\ba
\label{detES}
\det(E_j(S^{\sst I}))&\approx &\det(E^a_j(v)n^{S^{\sst I}}_a(v)\epsilon'^2)\nonumber\\
&=&\det(E^a_j(v))\det(n^{S^{\sst I}}_a(v))\epsilon'^6\nonumber\\
&=&\det(E(v))\det(n^{S^{\sst I}}_a(v))\epsilon'^6.
\ea
If we consider the definition of the normal vector in eqn (\ref{VFlux}),we can show the following identity
\ba
\label{detn}
n^{S^{\sst I}}_a&=&\det(X)X_a^{S^{\sst I}}\nonumber\\
\det(n^{S^{\sst I}}_a)&=&\det(X)^3\det(X^{-1})=\frac{\det(X)^3}{\det(X)}=\det(X)^2.
\ea
Inserting eqn (\ref{detn}) back into eqn (\ref{detES}) we have
\be
\label{detES2}
\det(E_j(S^{\sst I}))\approx \det(E(v))[\det(X(v))]^2\epsilon'^6
\ee
 and can conclude that eqn (\ref{VdetES}) is consistent with the usual definition of the volume in eqn (\ref{ResVBox}). 
\newline
Since we want to identify ${\cal S}:=\sgn(\det(E))$ with the signum that appears inside the absolute value under the square root in the definition of the volume, we can read off from eqn (\ref{VdetES}), that we still have to show $\sgn(\det(E))=\sgn(\det(E_j(S^{\sst I})))$. However, this can be done by means of eqn (\ref{detES2})
\ba
\sgn(\det(E_j(S^{\sst I})))&\approx &\sgn(\det(E(v))[\det(X(v))]^2\epsilon'^6)\nonumber\\
&=&\sgn(\det(E(v)))\sgn([\det(X(v))]^2)\sgn(\epsilon'^6)\nonumber\\ 
&=&\sgn(\det(E(v))).
\ea
Consequently, we can identify ${\cal S}$ with the signum that appears inside the absolute value under the square root in the definition of the volume $V$ in the classical theory, because it was precisely the expression $\det(E_j(S_{\st I}))$ that was used in the construction of the volume operator, defined as the square root of absolute value of $\det(E)$. In the quantum theory, we introduced the operator $\op{Q}$ in eqn (\ref{Qdef}), which is basically the expression inside the absolute value in the definition of the volume operator. Hence, it can be seen as the squared version of the volume operator that additionally contains  information about the signum of the expression inside the absolute values. Consequently, we can identify the operator $\op{Q}_{v}^{\sst AL}$ with $\op{Q}^{\sst AL}_v=\op{V}_{\sst AL}\op{\cal S}\op{V}_{\sst AL}$. 
Now we will be left with the task to calculate particular matrix elements for $\op{Q}^{\sst AL}_v$ which can be done by means of the formula derived in \cite{14}.
\newline
In order to apply the the operator 
$\op{\cal S}$ onto states expressed in terms of abstract angular momentum states, we have to use the $W$-map defined in eqn (\ref{Wmap}). Classically it is the signum 
of $\det(E)$ which is quantized by smearing the $E^a_j$ with surfaces 
upon which we obtain fluxes. Using that $\det((E_j(S^I)))\approx
[\det((\partial X^a/\partial u^I))]^2 \det(E^a_j)$ as the surfaces 
shrink to a point $v$ as we saw above,
the signum of $\det(E)$ is the signum of the determinant of the fluxes 
which in turn gives the operator $\hat{Q}_v$ which is related to 
$\hat{V}_v$ by $\hat{V}_v=\sqrt{|\hat{Q}_v|}$. Now $\op{Q}^{\sst AL}_v$ is given by
\be
\op{Q}^{\sst Y,AL}_v=C_{reg} \sum_{I,J,K} \epsilon(e_I,e_J,e_K) \epsilon_{ijk} 
(i \ell_p^2 X_{e_I}^i)
(i \ell_p^2 X_{e_J}^j)
(i \ell_p^2 X_{e_K}^k)
=-8C_{reg} \ell_p^6\sum_{I,J,K} \epsilon(e_I,e_J,e_K) \epsilon_{ijk} 
Y_{e_I}^i Y_{e_J}^j Y_{e_K}^k,
\ee
because $\op{E}_j(S)=i\ell_p^2 \sum_e \sigma(e,S) X_e^j$. Applying the map
$W$  then simply transforms the $Y$ into the $J$. Due to the global minus 
sign  in the above equation, we will obtain a global minus sign in front 
of the whole operator $\lL\op{\tilde{E}}^{\sst II,AL}_{k,{\sst tot}}(S_t)$. Thus, the 
minus sign in eqn (\ref{act7Ek}) gets cancelled. 
\subsubsection{Matrix Elements of $\op{O}_{\sst 1},\op{O}_{\sst 2}$ in the Case of $\lL\op{\tilde{E}}^{\sst II,AL}_{k,{\sst tot}}(S_t)$}
\label{ALMEO}
For $\lL\op{\tilde{E}}^{\sst II,AL}_{k,{\sst tot}}(S_t)$ the operators $\op{O}_{\sst 1},\op{O}_{\sst 2}=\op{V}_{\sst AL}\op{\cal S}\op{V}_{\sst AL}$. We showed in the last section, where $\op{\cal S}$ was introduced, the following operator identity $\op{Q}^{\sst AL}_v=\op{V}_{\sst AL}\op{\cal S}\op{V}_{\sst AL}$. 
Therefore calculating matrix elements of $\op{O}_{\sst 1/2}$ is equivalent to calculate matrix elements of $\op{Q}^{\sst AL}_v$. Hence, in order to get the matrix element for $\lL\op{\tilde{E}}^{\sst II,AL}_{k,{\sst tot}}(S_t)$, we need to compute the matrix elements $\alpc{0}{2}{0}\,\op{Q}^{\sst AL}_v\alp{0}{1}{0}$ and $\alpc{1}{i}{M}\,\op{Q}^{\sst AL}_v\alp{1}{1}{M}$ with $i=2,3,4$. And now one big advantage of the occurance of the sign operator $\op{\cal S}$ can be observed. In case I when we were forced to compute particular matrix elements of $\op{V}_{\sst AL}^2$ we had to calculate the whole eigensystem of $\op{Q}^{\sst AL}_v$  as a first step in order to use an eigenstate expansion for the matrix elements of $\op{V}_{\sst AL}^2$. Here, since $\lL\op{\tilde{E}}^{\sst II,AL}_{k,{\sst tot}}(S_t)$ includes matrix elements of $\op{Q}^{\sst AL}_v$ , we can use the formula derived in \cite{14} to get $\alpc{0}{2}{0}\,\op{Q}^{\sst AL}_v\alp{0}{1}{0}$ and $\alpc{1}{i}{M}\,\op{Q}^{\sst AL}_v\alp{1}{1}{M}$ and no involved diagonalisation of $\op{Q}^{\sst AL}_v$ is needed anymore.
\newline
 Moreover, we are only considering matrix elements with $\alp{1}{1}{M}$ as an incoming state $\alp{1}{1}{M}$. This state has the property that the intermediate coupling $a_{\st 2}$ of the edges $e_{\sst 1},e_{\sst 2}$ is zero. Thus, we have $J_{e_{\sst 1}}=-J_{e_{\sst 2}}$ and therefore obtain in these cases $\op{q}_{\sst 134}=-\op{q}_{\sst 234}$. Hence we only have to deal with one of the triples. So, in our special case we get  
\be
\label{defQ}
\op{Q}^{\sst AL}_v=\lp^6\frac{3!i}{4}C_{reg}\big(\epsilon(e_{\sst 1},e_{\sst 3},e_{\sst 4})\op{q}_{\sst 134}+\epsilon(e_{\sst 2},e_{\sst 3},e_{\sst 4})\op{q}_{\sst 234}\big)=\sigma\lp^6\frac{3!i}{2}C_{reg}\,\op{q}_{\sst 134},
\ee
 where we introduced $\sigma=+1$ for edges of type up and $\sigma=-1$ for 
 edges of type down. Moreover, we have chosen to take $\op{q}_{\sst 134}$ 
without loss of generality.
In the following calculation we will consider the case of an up edge, so we choose $\sigma=+1$. The whole calculation is analogous for an edge of type down with the only difference that all subsequent formulae have to be multiplied by a factor of $-1$.
Taking  formulae\footnote{This formula was originally derived for gauge invariant SNF only, but can easily be extended to gauge variant states with a total angular momentum different from zero\cite{CommJB}.} for the matrix elements of $\op{q}_{\sst 134},\op{q}_{\sst 234}$ in eqn (\ref{RSq134}),(\ref{RSq234}) and obtain the following result
\ba
\label{Resq134}
\alpc{0}{2}{M}\,\op{q}_{\sst 134}\,\alp{0}{1}{M}&=&\frac{4}{\sqrt{3}}\sqrt{j(j+1)}\sqrt{\ell(\ell+1)}\nonumber\\
\alpc{1}{2}{M}\,\op{q}_{\sst 134}\,\alp{1}{1}{M}&=&\frac{4}{\sqrt{3}}\sqrt{j(j+1)}\frac{\sqrt{(\ell+1)^3(2\ell-1)}}{\sqrt{\ell(2\ell+1)}}
\nonumber\\
\alpc{1}{3}{M}\,\op{q}_{\sst 134}\,\alp{1}{1}{M}&=&\frac{4}{\sqrt{3}}\sqrt{j(j+1)}\frac{(\ell(\ell+1)-1)}{\sqrt{\ell(\ell+1)}}\nonumber\\
\alpc{1}{4}{M}\,\op{q}_{\sst 134}\,\alp{1}{1}{M}&=&\frac{4}{\sqrt{3}}\sqrt{j(j+1)}\frac{\sqrt{\ell^3(2\ell+3)}}{\sqrt{(\ell+1)(2\ell+1)}}
\ea
The matrix elements do not depend on the magnetic quantum number $M$ and are therefore identical for any chosen value of $M$. From eqn (\ref{defQ}) we can read off that the matrix elements of $\op{Q}^{\sst AL}_v$ are given by eqn (\ref{Resq134}) multiplied by a factor of $\left(i\lp^6\frac{3!i}{2}C_{reg}\right)$.  Quite promising at this stage is the fact that the $j$ and $\ell$ dependence of the matrix elements factorises, because it might be a slight indication that the whole $\ell$ dependence will cancel exactly in the end.
With the result of the matrix elements we can go ahead in computing the matrix element of $\lL\op{\tilde{E}}^{\sst II,AL}_{k,{\sst tot}}(S_t)$ by inserting the matrix elements above into eqn (\ref{act7Ek}).
\subsubsection{Explicit Calculation of the Matrix Elements of $\lL\op{\tilde{E}}^{\sst II,AL}_{k,{\sst tot}}(S_t)$}
\label{ExCalc}
Multiplying the matrix elements in eqn (\ref{Resq134}) by the necessary factor of $\left(i\lp^6\frac{3!i}{2}C_{reg}\right)$, inserting them into eqn (\ref{act7Ek}) and taking into account the global factor of $-1$ due to the $W-$map of $\op{\cal S}$, we obtain
\ba
\label{act9Ek}
\lefteqn{\auf \beta^{\st 1}\,\wt{m}_{\sst 12}\,|\lL\op{\wt{E}}_{k,{\sst tot}}(S_t)|\,\beta^{\st 0}\,{\st 0}\zu}\nonumber\\
&=&
\lim_{{\cal P}_t\to S_t}\sum\limits_{\sst {\Box}\in{\cal P}_t}\frac{8\,\lp^{2}C_{reg}}{\frac{4}{3}\ell(\ell+1)(2\ell+1)}\frac{(-1)^{3\ell}3!2i}{\sqrt{3}}\sqrt{j(j+1)}\nonumber\\
&&\Big[+\sum\limits_{\sst B=-\ell}^{+\ell}\Big\{\pi_{\ell}(\tau_k)_{\sst B(\wt{m}_{12}+B)}(-1)^{{\st B+}\wt{m}_{\sst 12}}\sqrt{2\ell+1}\sqrt{\ell(\ell+1)}
\auf 1\, \wt{m}_{\sst 12}\, {\st ;}\ell\, {\st B}|\,\ell\, \wt{m}_{\sst 12}{\st +B}\zu 
\auf \ell\, \wt{m}_{\sst 12}{\st +B}\, {\st ;}\ell\, {\st -}(\wt{m}_{\sst 12}{\st +B})|\,0\,0\zu
\Big\}\nonumber\\
&&\;\;-\sum\limits_{\sst C=-\ell}^{+\ell}\Big\{\pi_{\ell}(\tau_k)_{\sst (C-\wt{m}_{12})C}(-1)^{{\st C-}\wt{m}_{\sst 12}}
\auf 0\, 0\, {\st ;}\ell\, \wt{m}_{\sst  12}{\st C}|\ell\, \wt{m}_{\sst  12}{\st C}\zu  
\auf \ell\, \wt{m}_{\sst  12}{\st C}\, {\st ;}\ell\, \wt{m}_{\sst 12}{\st -C}|\,1\, \wt{m}_{\st 12}\zu\nonumber\\
&&\quad\quad\quad\quad\quad
\Big(\frac{\sqrt{2\ell-1}}{\sqrt{3}}
\frac{\sqrt{(\ell+1)^3(2\ell-1)}}{\sqrt{\ell(2\ell+1)}}
-\frac{\sqrt{2\ell+1}}{\sqrt{3}}
\frac{(\ell(\ell+1)-1)}{\sqrt{\ell(\ell+1)}}
+\frac{\sqrt{2\ell+3}}{\sqrt{3}}
\frac{\sqrt{\ell^3(2\ell+3)}}{\sqrt{(\ell+1)(2\ell+1)}}\Big)\Big\}\Big].
\ea
where we put a global factor of $(\lp^6\frac{3!i}{2}C_{reg})\frac{(-1)^{3\ell}4}{\sqrt{3}}\sqrt{j(j+1)}$ in front of the summation. 
In order to get rid of the $\delta-$functions, we performed the sum over the indices $C,F$ in the first term and the sum over $B,F,$ in the last term. Hence, only one summation is left. Compared to  our starting point eqn (\ref{act9Ek}), eqn (\ref{act2Ek}) has become effectively simplified. Nevertheless, for carrying out the last sum, we have to insert the explicit expressions for the remaining CGC. They are given by
\ba
\label{CGCall}
\auf1\,\wt{m}_{\sst 12}\, {\st ;}\,\ell{\st B}\,|\,\ell\,\wt{m}_{\sst 12}{\st +B}\zu&=&
\frac{1}{\sqrt{\ell(\ell+1)}}\left\{\begin{array}{l}-\frac{1}{\sqrt{2}}\sqrt{\ell(\ell+1)-{\st B(B-1)}}\,\delta_{\wt{m}_{\sst 12},-1}\vspace{0.2cm}\\  -B\,\delta_{\wt{m}_{\sst 12},0}\vspace{0.2cm}\\ +\frac{1}{\sqrt{2}}\sqrt{\ell(\ell+1)-{\st B(B+1)}}\,\delta_{\wt{m}_{\sst 12},+1}\end{array}\right\}\nonumber\\
\auf\ell\,\wt{m}_{\sst 12}{\st +B}\, {\st ;}\,\ell\,-(\wt{m}_{\sst 12}{\st +B})\,|\,0\,0\zu&=&\frac{(-1)^{\ell{\st -B+}{\wt{m}_{\sst 12}}}}{\sqrt{2\ell+1}}\nonumber\\
\auf0\,0\, {\st ;}\,\ell\,{\st C}\,|\,\ell\,{\st C}\zu&=&1\nonumber\\
\auf\ell\,{\st C}\, {\st ;}\,\ell\,\wt{m}_{\sst 12}{\st -C}\,|\,1\,\wt{m}_{\sst12}\zu&=&
\frac{(-1)^{\ell{\st -C}}\sqrt{3}}{\sqrt{\ell(\ell+1)(2\ell+1)}}\left\{\begin{array}{l}-\frac{1}{\sqrt{2}}\sqrt{\ell(\ell+1)-{\st C(C+1)}}\delta_{\wt{m}_{\sst 12},-1}\vspace{0.2cm}\\  +C\,\delta_{\wt{m}_{\sst 12},0}\vspace{0.2cm}\\+\frac{1}{\sqrt{2}}\sqrt{\ell(\ell+1)-{\st C(C-1)}}\,\delta_{\wt{m}_{\sst 12},+1}\end{array}\right\}.
\ea
If we insert these CGC into eqn (\ref{act9Ek}), we will get an additional factor of $(-1)^{\ell}$ which, combined with the already existing factor of $(-1)^{3\ell}$, leads to a total of $(-1)^{4\ell}=+1$ and can therefore be neglected. Furthermore, the factors $(-1)^{\st B}$ and $(-1)^{\st C}$ are cancelled by the corresponding inverse factors included in the CGC in eqn (\ref{CGCall}). Hence, we get
\ba
\label{act10Ek}
\lefteqn{\auf \beta^{\st 1}\,\wt{m}_{\sst 12}\,|\lL\op{\wt{E}}_{k,{\sst tot}}(S_t)|\,\beta^{\st 0}\,{\st 0}\zu}\nonumber\\
&=&\lim_{{\cal P}_t\to S_t}\sum\limits_{\sst {\Box}\in{\cal P}_t}\frac{8\,\lp^{2}C_{reg}}{\frac{4}{3}\ell(\ell+1)(2\ell+1)}\frac{3!2i}{\sqrt{3}}\sqrt{j(j+1)}\nonumber\\
&&\Big[+\sum\limits_{\sst B=-\ell}^{+\ell}\Big\{\pi_{\ell}(\tau_k)_{\sst B(\wt{m}_{12}+B)}
\left\{\begin{array}{l}-\frac{1}{\sqrt{2}}\sqrt{\ell(\ell+1)-{\st B(B-1)}}\,\delta_{\wt{m}_{\sst 12},-1}\vspace{0.2cm}\\ -B\,\delta_{\wt{m}_{\sst 12},0}\vspace{0.2cm}\\+\frac{1}{\sqrt{2}}\sqrt{\ell(\ell+1)-{\st B(B+1)}}\,\delta_{\wt{m}_{\sst 12},+1}\end{array}\right\}
\Big\}\nonumber\\
&&\;\;\;-\sum\limits_{\sst C=-\ell}^{+\ell}\Big\{\pi_{\ell}(\tau_k)_{\sst (C-\wt{m}_{12})C}
\left\{\begin{array}{l}+\frac{1}{\sqrt{2}}\sqrt{\ell(\ell+1)-{\st C(C+1)}}\delta_{\wt{m}_{\sst 12},-1}\vspace{0.2cm}\\  +C\,\delta_{\wt{m}_{\sst 12},0}\vspace{0.2cm}\\-\frac{1}{\sqrt{2}}\sqrt{\ell(\ell+1)-{\st C(C-1)}}\,\delta_{\wt{m}_{\sst 12},+1}\end{array}\right\}\nonumber\\
&&\quad\quad\quad\quad
\left[\frac{(2\ell-1)}{(2\ell+1)}\frac{(\ell+1)}{\ell}
-\Big(1-\frac{1}{\ell(\ell+1)}\Big)
+\frac{(2\ell+3)}{(2\ell+1)}\frac{\ell}{(\ell+1)}\right]\Big\}\Big].
\ea
Here we have used $(-1)^{2m_{\sst 12}}=+1$ in the first term, absorbed the factor of $(-1)^{m_{\sst 12}}$ in a change of sign in the CGC for $\wt{m}_{\sst 12}=\pm1$ and combined and cancelled square roots where appropriate. Fortunately, the expression in the square bracket in the second sum is identical to one, so eqn (\ref{act10Ek}) simplifies to
\ba
\label{act11Ek}
\lefteqn{\auf \beta^{\st 1}\,\wt{m}_{\sst 12}\,|\lL\op{\wt{E}}_{k,{\sst tot}}(S_t)|\,\beta^{\st 0}\,{\st 0}\zu}\nonumber\\
&=&\lim_{{\cal P}_t\to S_t}\sum\limits_{\sst {\Box}\in{\cal P}_t}\frac{3!16i\,\lp^{2}C_{reg}\sqrt{j(j+1)}}{\sqrt{3}\frac{4}{3}\ell(\ell+1)(2\ell+1)}\nonumber\\
&&\hspace{-0.4cm}\left[
\sum\limits_{\sst B=-\ell}^{+\ell}\pi_{\ell}(\tau_k)_{\sst B(\wt{m}_{12}+B)}
\left\{\begin{array}{l}-\frac{1}{\sqrt{2}}\sqrt{\ell(\ell+1)-{\st B(B-1)}}\,\delta_{\wt{m}_{\sst 12},-1}\vspace{0.2cm}\\ -B\,\delta_{\wt{m}_{\sst 12},0}\vspace{0.2cm}\\+\frac{1}{\sqrt{2}}\sqrt{\ell(\ell+1)-{\st B(B+1)}}\,\delta_{\wt{m}_{\sst 12},+1}\end{array}\right\}
+\sum\limits_{\sst C=-\ell}^{+\ell}\pi_{\ell}(\tau_k)_{\sst (C-\wt{m}_{12})C}
\left\{\begin{array}{c}-\frac{1}{\sqrt{2}}\sqrt{\ell(\ell+1)-{\st C(C+1)}}\delta_{\wt{m}_{\sst 12},-1}\\  -C\,\delta_{\wt{m}_{\sst 12},0}\\+\frac{1}{\sqrt{2}}\sqrt{\ell(\ell+1)-{\st C(C-1)}}\,\delta_{\wt{m}_{\sst 12},+1}\end{array}\right\}\right],\nonumber\\
\ea
where we absorbed the minus sign in front of the second sum into the GCG.
The tau-matrices for an arbitrary SU(2)-representation with weight $\ell$ are 
derived in appendix \ref{Taus}
\begin{eqnarray}
\label{1Taugeneral}
\pi_\ell(\tau_1)_{mn}&=&-i\sqrt{\ell(\ell+1)-m(m-1)}\,\delta_{m-n,1}-
i\sqrt{\ell(\ell+1)-m(m+1)}\,\delta_{m-n,-1}\nonumber\\
\pi_\ell(\tau_2)_{mn}&=&\sqrt{\ell(\ell+1)-m(m+1)}\,\delta_{m-n,-1}-
\sqrt{\ell(\ell+1)-m(m-1)}\,\delta_{m-n,1} \nonumber\\
\pi_\ell(\tau_3)_{mn}&=&-2\,i\,m\,\delta_{m-n,0}.
\end{eqnarray}
Taking a closer look at the structure of these tau-matrices, we realise that a different choice of $\wt{m}_{\sst 12}$ in eqn (\ref{act11Ek}) projects onto 
different tau-matrices, e.g. only $\pi_{\ell}(\tau_3)$ will contribute to the case $\wt{m}_{\sst 12}=0$, while in the case $\wt{m}_{\sst 12}=\pm 1$ only $\pi_{\ell}(\tau_1)$ and $\pi_{\ell}(\tau_2)$ have to be considered. Formulating this fact in terms of $\delta-$functions and using the explicit expressions for the tau-matrices in eqn (\ref{1Taugeneral}), we obtain
\ba
\label{act13Ek}
\auf \beta^{\st 1}\,\wt{m}_{\sst 12}\,|\lL\op{\wt{E}}_{k,{\sst tot}}(S_t)|\,\beta^{\st 0}\,{\st 0}\zu
&=&\lim_{{\cal P}_t\to S_t}\sum\limits_{\sst {\Box}\in{\cal P}_t}
\frac{3!16i\,\lp^2C_{reg}}{\sqrt{3}}\sqrt{j(j+1)}
\left\{\begin{array}{l}-\frac{1}{\sqrt{2}}\delta_{\wt{m}_{\sst 12},-1}\left\{-i\delta_{k{\sst,1}}+\delta_{k{\sst ,2}}\right\}\vspace{0.2cm}\\ +i\delta_{\wt{m}_{\sst 12},0}\delta_{k{\sst ,3}}\vspace{0.2cm}\\+\frac{1}{\sqrt{2}}\delta_{\wt{m}_{\sst 12},1}\left\{-i\delta_{k{\sst ,1}}-\delta_{k{\sst ,2}}\right\}\end{array}\right\},
\ea
whereby we used $\sum\limits_{B=-\ell}^{\ell}B^2=\frac{1}{3}\ell(\ell+1)(2\ell+1)$.
\newline\newline
Now we want to take the limit $\lim_{\epsilon\to 0}\lim_{\epsilon'\to 0}$ The discussion in section \ref{factord} showed that taking  the $\lim_{\epsilon'\to 0}$ (that is equivalent to 
$\lim_{{\cal P}_t\to S_t}$) is trivial and taking the $\lim_{\epsilon\to 0}$ leads to an additional overall factor of $1/2$. So, when calculating the action of the alternative flux operator on the state $\bet{0}{0}$, we use the expansion
\be
\lL\op{\wt{E}}_{k,{\sst tot}}(S)\bet{0}{0}=
\sum\limits_{\wt{m}_{\sst 12}=-1}^{+1}\betc{1}{\wt{m}_{\sst 12}}\,\lL\op{\wt{E}}_{k,{\sst tot}}(S)\,\bet{0}{0}\bet{1}{\wt{m}_{\sst 12}}
\ee
and end up with the final result
\ba
\label{NEkres}
\lL\op{\wt{E}}^{\sst II,AL}_{1,{\sst tot}}(S)\bet{0}{0}&=&
-\frac{3!8\,\lp^2C_{reg}}{\sqrt{6}}\sqrt{j(j+1)}\big\{\bet{1}{-1}-\bet{1}{+1}\Big\}\nonumber\\
\lL\op{\wt{E}}^{\sst II,AL}_{2,{\sst tot}}(S)\bet{0}{0}&=&
-\frac{3!8i\,\lp^2C_{reg}}{\sqrt{6}}\sqrt{j(j+1)}\big\{\bet{1}{-1}+\bet{1}{+1}\big\}\nonumber\\
\lL\op{\wt{E}}^{\sst II,AL}_{3,{\sst tot}}(S)\bet{0}{0}&=&
-\frac{3!8\,\lp^2C_{reg}}{\sqrt{3}}\sqrt{j(j+1)}\,\bet{1}{0}.
\ea
Remarkably, in the final result the $\ell$-dependence drops out completly.
\subsection{Case $\lL\op{\tilde{E}}^{\sst II,RS}_{k,{\sst tot}}(S_t)$ i.e. $E^{a,{\sst II}}_k={\cal S}\det(e)e^a_k$ and $\op{V}_{\sst RS}$}%
\label{RSVSV}
In this case the operators $\op{O}_{\sst 1},\op{O}_{\sst 2}$ have the following form
\ba
\label{C2RSO1O2}
\op{O}_{\sst 1}&=&+\op{V}_{q_{\sst 134}}\op{\cal S}\op{V}_{q_{\sst 134}}+\op{V}_{q_{\sst 234}}\op{\cal S}\op{V}_{q_{\sst 234}}+\op{V}_{q_{\sst 134}}\op{\cal S}\op{V}_{q_{\sst 234}}+\op{V}_{q_{\sst 234}}\op{\cal S}\op{V}_{q_{\sst 134}}+\op{V}_{q_{\sst 134}}\op{\cal S}\op{V}_{q_{\sst 123}}+\op{V}_{q_{\sst 124}}\op{\cal S}\op{V}_{q_{\sst 134}}+\op{V}_{q_{\sst 234}}\op{\cal S}\op{V}_{q_{\sst 123}}\nonumber\\
&&+\op{V}_{q_{\sst 124}}\op{\cal S}\op{V}_{q_{\sst 234}}+\op{V}_{q_{\sst 124}}\op{\cal S}\op{V}_{q_{\sst 123}}\nonumber\\
\op{O}_{\sst 2}&=&+\op{V}_{q_{\sst 134}}\op{\cal S}\op{V}_{q_{\sst 134}}+\op{V}_{q_{\sst 234}}\op{\cal S}\op{V}_{q_{\sst 234}}+\op{V}_{q_{\sst 234}}\op{\cal S}\op{V}_{q_{\sst 134}}+\op{V}_{q_{\sst 134}}\op{\cal S}\op{V}_{q_{\sst 234}}+\op{V}_{q_{\sst 123}}\op{\cal S}\op{V}_{q_{\sst 134}}+\op{V}_{q_{\sst 134}}\op{\cal S}\op{V}_{q_{\sst 124}}+\op{V}_{q_{\sst 123}}\op{\cal S}\op{V}_{q_{\sst 234}}\nonumber\\
&&+\op{V}_{q_{\sst 234}}\op{\cal S}\op{V}_{q_{\sst 124}}+\op{V}_{q_{\sst 123}}\op{\cal S}\op{V}_{q_{\sst 124}}
\ea
But, before continuing we want to discuss some difficulties that occur if one uses the volume operator $\op{V}_{\sst RS}$ in this case.
\subsubsection{Problems with the Sign Operator $\op{{\cal S}}$ in the case of RS}
\label{Problems}       %
When we introduced the quantisation of ${\cal S}\rightarrow\op{\cal S}$ in section \ref{SignOP}, we realised that $\op{\cal S}$ has a precise relation to the operator $\op{Q}^{\sst AL}_v$ i.e. $\op{Q}^{\sst AL}_v=\op{V}_{\sst AL}\op{\cal S }\op{V}_{\sst AL}$. However, this was possible, because $\op{V}_{\sst AL}$ sums over the triples inside the absolute value under the square root ( see eqn (\ref{Vqijk}) ). In constrast, $\op{V}_{\sst RS}$, defined in eqn (\ref{RSVqijk}) consists of a sum of single square roots. 
Consequently, we are not able to repeat the calculations done in section 
\ref{SignOP} if we choose $\op{V}_{\sst RS}$, because there is no 
possible origin for a sign.  This means, there exists no signum operator $\op{\cal S}$ that is quantised in the same way $\op{V}_{\sst RS}$ is quantised. Accordingly, in a strict sense the operator $\lL\op{\tilde{E}}^{\sst II,RS}_{k,{\sst tot}}(S_t)$ does not exist, because $\op{\cal S}$ can not be implemented in 
the quantum theory just using the regularization that leads 
to $\op{V}_{\sst RS}$.
The conclusion is that $\lL\op{\tilde{E}}^{\sst II,RS}_{k,{\sst tot}}(S_t)$ is inconsistent with the usual flux operator. In retrospect there is a simple argument why the only possibility $\lL\op{\tilde{E}}^{\sst I,RS}_{k,{\sst tot}}(S_t)$ (since $\lL\op{\tilde{E}}^{\sst II,RS}_{k,{\sst tot}}(S_t)$ does not exist) is ruled out $\op{V}_{\sst RS}$ without further calculation:
Namely, the lack 
 of a factor of orientation in $\op{V}_{\sst RS}$, like $\epsilon(e_{\sst 
I},e_{\sst J},e_{\sst K})$ in $\op{V}_{\sst AL}$, leads to 
the following basic disagreement with the usual flux operator. 
 Suppose we had chosen the orientation of the surface $S$ in the 
opposite way. Then the type of the edge $e$ switches between up and down 
and similarly for $e_{\sst 1},e_{\sst 2}$. Then, the result of the usual flux 
 operator would differ by a minus sign. In the case of $\op{V}_{\sst AL}$ 
 we would get this minus sign as well due to $\epsilon(e_{\sst I},e_{\sst 
J},e_{\sst K})$, whereas a change of the orientation of 
$e_{\sst 1},e_{\sst 2}$ would not modify the result of the alternative flux 
operator if we used $\op{V}_{\sst RS}$ instead, because it is not 
sensitive to the orientation of the edges. 
\newline
\newline
A way out would be to use the somehow 'artificial' construction $\op{V}_{\sst RS }\op{\cal S}_{\sst AL}\op{V}_{\sst RS}$, where $\op{\cal S}$ denotes the signum operator $\op{\cal S}$ introduced in section \ref{SignOP}. We attached  the label $AL$ to it in order to emphasize that its quantisation is in agreement with $\op{V}_{\sst AL}$. This is artificial for the following reason. Suppose we have a classical quantity $A:=\det(E)$ and two different functions $f_1:=\sqrt{|A|}$ and $f_2:=\sgn(A)$. If we want to quantise the functions $f_1$ and $f_2$, we do this with the help of the corresponding operator $\op{A}$ and obtain due to the spectral theorem  $\op{f}_1=\sqrt{|\op{A}|}$ and $\op{f}_2=\sgn(\op{A})$. The product of operators $\op{V}_{\sst RS }\op{\cal S}_{\sst AL}\op{V}_{\sst RS}$ rather corresponds to $\op{g}_1=\op{A}^{\prime}$ and $\op{g}_2=\sgn(\op{A})$, because $\op{V}_{\sst RS}$ is quantised with a different regularisation scheme than $\op{S}$ is. This would only be justified if $\sqrt{|\op{A}|}$ and $\op{A}^{\prime}$ would agree semiclassically.  However they do not: If we compare the expressions for $V_{\sst AL}$ and $V_{\sst RS}$ 
then, schematically,
they are related in the following way when restricted to a vertex: 
$\op{V}_{v,\sst AL}=|\frac{3!i}{4}C_{reg}\sum\limits_{\sst I<J<K}\epsilon(e_{\sst 
I},e_{\sst J},e_{\sst K})\op{q}_{\sst IJK}|^{1/2}$ while 
$\op{V}_{v,\sst RS}=\sum\limits_{\sst I<J<K} |\frac{3!i}{4}C_{reg}\op{q}_{\sst IJK}|^{1/2}$. It is clear that apart from the sign
$\epsilon(e_{\sst I},e_{\sst J},e_{\sst K})$ the two operators can agree at most on states where
only one of the $\op{q}_{\sst IJK}$ is non vanishing (three or four valent graphs)
simply because $\sqrt{|a+b|}\not=\sqrt{|a|}+\sqrt{|b|}$ for generic real
numbers $a,b$.
\subsubsection{Matrix Elements of $\op{O}_{\sst 1},\op{O}_{\sst 2}$ in the Case of $\lL\op{\tilde{E}}^{\sst II,RS}_{k,{\sst tot}}(S_t)$}  %
Nevertheless, we can analyse whether $\lL\op{\tilde{E}}^{\sst II,RS}_{k,{\sst tot}}(S_t)$ inlcuding $\op{V}_{\sst RS }\op{\cal S}_{\sst AL}\op{V}_{\sst RS}$ is consistent with the usual flux operator $\op{E}_k(S)$.\newline
In the case of $\lL\op{\tilde{E}}^{\sst II,AL}_{k,{\sst tot}}(S_t)$ no diagonalisation of the $\op{Q}^{\sst AL}_v$ matrices was necessary because of the operator identification $\op{Q}^{\sst AL}_v=\op{V}_{\sst AL}\op{\cal S }\op{V}_{\sst AL}$. Since this is not possible for $\lL\op{\tilde{E}}^{\sst II,RS}_{k,{\sst tot}}(S_t)$ we have to diagonalise the $\op{Q}^{\sst RS}_{v,{\sst IJK}}$ in order to get the eigenvalues and eigenvectors. Then we can compute the matrix elments for instance $\alpc{0}{2}{M}\,\op{O}_{1}\,\alp{0}{1}{M}$ by an eigenvector expansion for each operator contained in $\op{O}_1$
\ba
\alpc{0}{2}{M}\,\op{V}_{q_{\sst IJK}}\op{\cal S}\op{V}_{q_{\sst \tilde{I}\tilde{J}\tilde{K}}}\,\alp{0}{1}{M}&=&
\sum\limits_{|\alpha'\zu,|\alpha''\zu}
\alpc{0}{2}{M}\,\op{V}_{q_{\sst IJK}}\,|\alpha'\zu\auf\alpha'|\,\op{\cal S}\,|\alpha''\zu\auf\alpha''|\,\op{V}_{q_{\sst \tilde{I}\tilde{J}\tilde{K}}}\,\alp{0}{1}{M}\nonumber\\
&=&
\sum\limits_{|\alpha'\zu,|\alpha''\zu}\sum\limits_{k,k',k''}
\alpc{0}{2}{M}\,\op{V}_{q_{\sst IJK}}\,|\,\vec{e}_k\zu\auf\vec{e}_k\,|\,\alpha'\zu\auf\alpha'|\,\op{\cal S}\,|\vec{e}_{k'}\zu\auf\vec{e}_{k'}\,|\,\alpha''\zu\nonumber\\
&&\hspace{+2.2cm}\auf\alpha''|\,\op{V}_{q_{\sst \tilde{I}\tilde{J}\tilde{K}}}\,|\,\vec{e}_{k''}\zu\auf\vec{e}_{k''}\,\alp{0}{1}{M},
\ea
whereby $|\vec{e}_k\zu$ are the eigenvectors of the corresponding operators and $|\alpha'\zu$  are all states belonging to the Hilbert space ${\cal H}^{\sst J}$. We calculated the matrix elements $\alpc{0}{2}{M}\,\op{O}_{1}\,\alp{0}{1}{M}$, $\alpc{1}{3}{M}\,\op{O}_{2}\,\alp{1}{1}{M}$, $\alpc{1}{4}{M}\,\op{O}_{2}\,\alp{1}{1}{M}$ that occur in eqn (\ref{act7Ek}) for $\ell=0.5$. The details can be found in  appendix  \ref{RSOp}. The results are shown below
\ba
\label{RSMEII}
\alpc{0}{2}{M}\,\op{O}^{\sst RS}_{1}\,\alp{0}{1}{M}&=&C_1(\ell)\alpc{0}{2}{M}\,\op{O}^{\sst AL}_{1}\,\alp{0}{1}{M}\nonumber\\
\alpc{1}{3}{M}\,\op{O}^{\sst RS}_{2}\,\alp{1}{1}{M}&=&C_3(j,\ell)\alpc{1}{3}{M}\,\op{O}^{\sst AL}_{2}\,\alp{1}{1}{M}\nonumber\\
\alpc{1}{4}{M}\,\op{O}^{\sst RS}_{2}\,\alp{1}{1}{M}&=&C_4(j,\ell)\alpc{1}{4}{M}\,\op{O}^{\sst AL}_{2}\,\alp{1}{1}{M}
\ea
Here $C_1(\ell),C_i(j,\ell)\in\Rl$ and the explicit expression can be found in eqn (\ref{C1}) and eqn (\ref{C3C4}). Furthermore we expressed the matrix elements in terms of the associated AL-matrix elements, because the whole calculation has already been done for $\lL\op{\tilde{E}}^{\sst II,AL}_{k,{\sst tot}}(S_t)$ and therefore in this way of writing we can easily note where differences occur. 
$C_i(j,\ell)$ are  real constants whose values depend on the explicit value of the spin labels $j$ and $\ell$. In the case of $\lL\op{\tilde{E}}^{\sst II,AL}_{k,{\sst tot}}(S_t)$ we could see that the whole dependence on the spin label $\ell$ drops out in the final result. Hence, if in the case of $\lL\op{\tilde{E}}^{\sst II,RS}_{k,{\sst tot}}(S_t)$ we obtain not exactly the same matrix elements as for $\lL\op{\tilde{E}}^{\sst II,AL}_{k,{\sst tot}}(S_t)$, we already know that the $\ell$-dependence will not be canceled in the final result here.
The $j-$ dependence is basically caused by terms proportional to $(\sqrt{j(j+1)+c})(\sqrt{j(j+1)})^{-1}$, whereby $c\in\Nl$. Thus semiclassically, i.e. in the limit of large $j$, the numerator and the denominator become equal and accordingly the $j$-dependence vanishes, $C_i(j,\ell)\to C_i(\ell)$. By reinserting the matrix elements from eqn (\ref{RSMEII}) into eqn (\ref{act7Ek}) and repeat all the steps of the former $\lL\op{\tilde{E}}^{\sst II,RS}_{k,{\sst tot}}(S_t)$ calculation for $\ell=0.5$, we end up with
\ba
\label{RSNEkres}
\lL\op{\wt{E}}^{\sst II,RS}_{1,{\sst tot}}(S)\bet{0}{0}&=&
-\frac{\lp^2}{\sqrt{6}}C(j,\ell)\,C_{reg}\sqrt{j(j+1)}\big\{\bet{1}{-1}-\bet{1}{+1}\Big\}\nonumber\\
\lL\op{\wt{E}}^{\sst II,RS}_{2,{\sst tot}}(S)\bet{0}{0}&=&
-\frac{\lp^2}{\sqrt{6}}C(j,\ell)\,C_{reg}\sqrt{j(j+1)}\big\{\bet{1}{-1}+\bet{1}{+1}\big\}\nonumber\\
\lL\op{\wt{E}}^{\sst II,RS}_{3,{\sst tot}}(S)\bet{0}{0}&=&
-\frac{\lp^2}{\sqrt{3}}C(j,\ell)\,C_{reg}\sqrt{j(j+1)}\,\bet{1}{0}.
\ea
Here, $C(j,\ell)\in\Rl$ with $C(j,\ell)\to C(\ell)$ semiclassically and
\be
\label{Clhalf}
C(\ell)=\left[C_1(\frac{1}{2})-C_3(\frac{1}{2})\left(1-\frac{1}{\ell(\ell+1)}\right)+C_4(\frac{1}{2})\left(\frac{(2\ell+3)}{(2\ell+1)}
\frac{\ell}{(\ell+1)}\right)\right]_{\ell=0.5}.
\ee
The functions $C_1(\ell),C_2(\ell),C_3(\ell),C_4(\ell)$ can be computed analytically only for $\ell=0.5,1$. Note, that $C_2(\ell)$ is zero for $\ell=1/2$ since the state $\alp{1}{2}{M}$ does not exist for $\ell=0.5$ (see eqn (\ref{Hj05J1})). For this reason it does not occur in eqn (\ref{Clhalf}).  However, suppose we would know this constants the precise $\ell$-depence of $C(\ell)$ would be
\be
C(\ell)=C_1(\ell)+C_2(\ell)\left(\frac{(2\ell-1)}{(2\ell+1)}\frac{(\ell+1)}{\ell}\right)-C_3(\ell)\left(1-\frac{1}{\ell(\ell+1)}\right)+C_4(\ell)\left(\frac{(2\ell+3)}{(2\ell+1)}\frac{\ell}{(\ell+1)}\right)
\ee
Since the $\ell$-dependence of $C_i(\ell)$ should be result from the $\ell$-dependence of $\op{Q}_{v,{\sst IJK}}^{\sst RS}$ which is non-trivial in general, it is very unlikely that the whole $\ell$-dependence is canceled for arbitrary $\ell$ as in the case of $\lL\op{\tilde{E}}^{\sst II,AL}_{k,{\sst tot}}(S_t)$, where $C_i({\ell})=1$ for $i=1,2,3,4$.
\newline 
 Thus, we conclude that the volume operator introduced by Rovelli and 
Smolin 
 is not appropriate to reproduce the result of the usual flux operator 
 $\op{E}_k(S_t)$ and can therefore not be used to construct the alternative 
 flux operator. In other words, the RS-operator is inconsistent with the 
fundamental flux operator on which it is based.
\subsection{Summarising the Results of Case II}  %
 Considering the operator $\lL\op{\tilde{E}}^{\sst II,AL}_{k,{\sst tot}}(S_t)$ the operators  $\op{O}_1,\op{O}_2$  whose matrix elemnts are included in eqn (\ref{act7Ek}) are given by $\op{O}_1=\op{O}_2=\op{V}_{\sst AL}\op{\cal S}\op{V}_{\sst AL}$. Thus we have to implement the signum operator ${\cal S}=\sgn(\det(e))\to\op{\cal S}$ on the quantum level. In section \ref{SignOP} we showed in detail that $\op{\cal S}$ has a well defined relation with $\op{Q}^{\sst AL}_v$, in particular $\op{\cal S}=\sgn(\op{Q}^{\sst AL}_v)$. This relation is equivalent to the operator identity $\op{Q}^{\sst AL}_v=\op{V}_{\sst AL}\op{\cal S}\op{V}_{\sst AL}$. Consequently, it remarkably turned out that the operators  $\op{O}_1,\op{O}_2$ are identical to the operator $\op{Q}^{\sst AL}_v$ in the case of $\lL\op{\tilde{E}}^{\sst II,AL}_{k,{\sst tot}}(S_t)$. Along with this comes the nice side effect that thus a diagonalisation of the operator $\op{Q}^{\sst AL}_v$ is no longer necessary since now the matrix elements of $\op{Q}^{\sst AL}_v$ instead of matrix elements of $\op{V}_{\sst AL}$ contribute to the calculation. Therefore, we can apply the general formula for matrix elements of $\op{Q}^{\sst AL}_v$ derived in \cite{14}, even for arbitrary spin labels $\ell$, and we are done. The expression for the matrix elements of $\op{Q}^{\sst AL}_v$ is given in section \ref{ALMEO} in eqn (\ref{Resq134}). By reinserting these matrix elements into eqn (\ref{act7Ek}) and follow the intermediate steps discussed in section \ref{ExCalc}, we end up with the final result in eqn (\ref{NEkres}). For $\lL\op{\tilde{E}}^{\sst II,AL}_{k,{\sst tot}}(S_t)$ the whole dependence on the spin label $\ell$ that is associated with the two additional edges $e_{\sst 3},e_{\sst 4}$ drops out in the final result. Hence, the result is independent of the chosen respresentation of the holonomies in the alternative flux operator $\lL\op{\tilde{E}}^{\sst II,AL}_{k,{\sst tot}}(S_t)$. 
 \newline
 \newline
 In the case of $\lL\op{\tilde{E}}^{\sst II,RS}_{k,{\sst tot}}(S_t)$ the operators $\op{O}_1,\op{O}_2$ have quite lengthly expressions that can be found in eqn (\ref{C2RSO1O2}). $\op{O}_1,\op{O}_2$ are both given by a sum of operators that have the form $\op{V}_{q_{\sst IJK}}\op{\cal S}\op{V}_{q_{\sst \tilde{I}\tilde{J}\tilde{K}}}$, whereby $\op{V}_{q_{\sst IJK}}$ denotes the operator $\op{V}_{\sst RS}$ when only the contribution of the tripel $\{e_{\sst I},e_{\sst J},e_{\sst K}\}$ is considered. In contrast to $\op{Q}^{\sst AL}_v$, for $\sum\limits_{\sst IJK}\op{Q}^{\sst RS}_{v,{\sst IJK}}$ no relation with the signum operator $\op{\cal S}$ can be derived. This fact is dealt with in section \ref{Problems}. Consequently, it is impossible to quantise $\op{\cal S}$ in an analogous way as $\op{V}_{\sst RS}$ is quantised. This is a big difference to $\op{V}_{\sst AL}$ where $\op{\cal S}$ and $\op{V}_{\sst AL}$ could be quantised in the same manner. Therefore, the operator $\lL\op{\tilde{E}}^{\sst II,RS}_{k,{\sst tot}}(S_t)$ cannot be defined rigorously since $\op{\cal S}$ does not exist for $\op{V}_{\sst RS}$. Thus, the operator $\lL\op{\tilde{E}}^{\sst II,RS}_{k,{\sst tot}}(S_t)$ is inconsistent with the usual flux operator $\op{E}_k(S_t)$.
\newline
\newline
Nevertheless, we analysed the artificial construction $\op{V}_{\sst RS}\op{\cal S}\op{V}_{\sst AL}$ for $\lL\op{\tilde{E}}^{\sst II,RS}_{k,{\sst tot}}(S_t)$. It is artificial, because the operator $\op{V}_{\sst RS}$ an the operator $\op{\cal S}$ are quantised with respect to different regularisation schemes and are not semiclassically consistent with each other. The results are shown in eqn (\ref{RSNEkres}).
\newline
In the next section we will calculate the matrix element of the usual flux operator $\op{E}_k(S)$ in order to compare it with the results in eqn (\ref{NEkres}) and eqn (\ref{RSNEkres}) respectively afterwards.
 \section{Matrix Elements of the Usual Flux Operator $\op{E}_k(S)$}%
In this section we will calculate the action of the usual flux operator on our SNF $\bet{0}{0}$ that was used through all the calculations of the alternative flux operator before\footnote{Notice that it so happens that for $\op{O}=\op{V}\op{S}\op{V}=\op{Q}$ an explicit diagonalisation
of $\op{Q}$ is not necessary so we may refrain from using the recoupling basis
and can work directly in the tensor basis. The associated calculations
are of a similar
length but sidestep the use of CGC's and hence may be used as an
independent check of our result. We did this and the result completely
agrees with the recoupling basis calculation. However, for $\op{O}=\op{V}^2\neq\op{Q}$
it is necessary to diagonalise $\op{Q}$ and the use of the recoupling basis
becomes calculationally mandatory, which is why we have done all
calculations in this paper in the recoupling basis.}. If we want to use the technical tools of angular momentum recoupling 
theory (e.g. CGC) we have to apply the $W$-map in eqn \ref{Wmap} to all 
states in the SNF Hilbert space in order to justify to work in the angular 
momentum system Hilbert space. Therefore a matrix element of the usual 
flux operator is given by 
 \be
\begin{array}{l}
_{m'_{\sst 1},m'_{\sst 2}}\auf \beta^{\tilde{j}_{\sst 12}}\,
\wt{n}_{\sst 12}\,|\,\op{E}^Y_{k}(S)\,|\,\beta^{j_{\sst 12}}\,
n_{\sst 12}\zu_{m'_{\sst 1},m'_{\sst 2}}\nonumber\\
=\sum\limits_{m_{\sst 12},\wt{m}_{\sst 12}}\pi_{\tilde{j}_{\sst 
12}}(\epsilon^{-1})_{\tilde{n}_{\sst 12}\wt{m}_{\sst 12}}
\pi_{j_{\sst 12}}(\epsilon^{-1})_{n_{\sst 12}m_{\sst 12}}
\auf \beta^{\tilde{j}_{\sst 12}}\,\wt{m}_{\sst 12}\,;m'_{\sst 1}\,m'_{\sst 2}\,
|\,\op{E}^J_{k}(S)\,|\,\beta^{j_{\sst 12}}\,m_{\sst 12}\,;\,m'_{\sst 
1}\,\,m'_{\sst 2}\zu.
\end{array}
\ee
As has been 
pointed out before, this mapping is similar for the alternative and the 
usual flux operator. Therefore, we only will consider the matrix elements of $\op{E}_k$ in the abstract angular momentum system Hilbert space  here. Since the inverse of $\pi_{\ell}(\epsilon^{-1})$ exists, a possible difference between the usual and the alternative flux operator can only occur in the matrix element in the abstract angular momentum Hilbert space. Throughout this section we will neglect the additional indices $m'_{\sst 1},m'_{\sst 2}$ of the states $\bet{j_{\sst 12}}{m_{\sst 12}}$ as we did in the calculation of the alternative flux operator. Working in the abstract angular momentum system Hilbert space now, we can re-express the action of $\op{E}_k(S)$ in terms of angular momentum operators  of which the actions on the other hand are well known for states expressed in the tensor basis. Thus, it is suggestive to transform the recoupling states $\bet{0}{0}$ back into the tensor basis and afterwards apply  the operator $\op{E}_k(S)$ onto it. Thereafter we have to reformulate the result again in terms of the recoupling basis in order to be able to compare this result of the usual flux operator with the calculations of $\nflop{k}$ in the last sections.
\newline
The state $\bet{0}{0}$ transforms into the tensor basis according to the following linear combination
\ba
\bet{0}{0}&=& \sum\limits_{m=-j}^{+j}\auf j\, m, j\, -m\,|\, 0\, 0\,\zu |\,j\, m\, {\st ;}\, m_{e_{\sst 1}}'\zu_{e_{\sst 1}}\otimes|\,j\, m\, {\st ;}\, m_{e2}'\zu_{e_{\sst 2}} \\
                                          &=& \sum\limits_{m=-j}^{+j}\frac{(-1)^{j-m}}{\sqrt{2j+1}}|\,j\, m\, {\st ;}\, m_{e_{\sst 1}}'\zu_{e_{\sst 1}}\otimes |\,j\, -m\, {\st ;}\, m_{e_{\sst 2}}'\zu_{e_{\sst 2}},
\ea
where we have used the explicit expression for the CGC $\auf j\, m, j\, -m\,|\, 0\, 0\,\zu=\frac{(-1)^{j-m}}{\sqrt{2j+1}}$. Furthermore, the two edges $e_{\sst 1}$ and $e_{\sst 2}$ of our graph $\gamma$ couple to a resulting angular momentum $j_{\sst 12}=0$. Therefore, we have $\op{J}^k_{e_{\sst 1}}=-\op{J}^k_{e_{\sst 2}}$. Additionally, the tangent vectors  $\dot{e}_{\sst 1}(t)$ and $\dot{e}_{\sst 2}(t)$ have opposite orientations with respect to the surface $S$, from which follows that $\epsilon(e_{\sst 1},S)=-\epsilon(e_{\sst 2},S)=\sigma$, where $\sigma=+1$ for edges of type up and $\sigma=-1$ for type down edges. Hence, we obtain
\ba
\label{actEnorm}
\op{E}_k(S)\bet{0}{0}&=&-\frac{1}{2}\lp^2\left[\epsilon(e_{\sst 1},S)\op{J}^k_{e_{\sst 1}}+\epsilon(e_{\sst 2},S)\op{J}^k_{e_{\sst 2}}\right]\bet{0}{0}\nonumber\\
                                                  & = & -\frac{1}{2}\lp^2 \sum\limits_{m=-j}^{+j}\left[\epsilon(e_{\sst 1},S)\op{J}^k_{e_{\sst 1}}+\epsilon(e_{\sst 2},S)\op{J}^k_{e_{\sst 2}}\right]\frac{(-1)^{j-m}}{\sqrt{2j+1}}|\,j\, m\, {\st ;}\, m_{e_{\sst 1}}'\zu_{e_{\sst 1}}\otimes |\,j\, -m\, {\st ;}\, m_{e_{\sst 2}}'\zu_{e_{\sst 2}} \nonumber \\
                                                  & = &-\frac{\lp^2}{\sqrt{2j+1}}\sum\limits_{m=-j}^{+j}(-1)^{j-m}\left(\op{J}^k_{e_{\sst 1}}|\,j\, m\, {\st ;}\, m_{e_{\sst 1}}'\zu_{e_{\sst 1}}\right)\otimes |\,j\, -m\, {\st ;}\, m_{e_{\sst 2}}'\zu_{e_{\sst 2}}.
\ea
By applying eqn (\ref{actEnorm}), we calculate the action of $\op{E}_k(S)$ on our SNF for each $k=1,2,3$ separately with the case $k=1$ being the first one. From elementary quantum mechanics we know that we can introduce ladder angular momentum operators $\op{J}^+$ and $\op{J}^-$ defined by $\op{J}^+:=\op{J}^1+i\op{J}^2$ and $\op{J}^-:=\op{J}^1-i\op{J}^2$, respectively. Hence, we can express $\op{J}^1$ as $\op{J}^1=\frac{1}{2}(\op{J}^+ + \op{J}^-)$. The action of the ladder operators on a state in the abstract spin system $|j\,m\,{\st ;}\, m'\zu$ with spin $j$ and magnetic quantum number $m$ is given by
\ba
\label{J12eigen}
\op{J}^+|j\, m\, {\st ;}\,m'\zu &=& \sqrt{j(j+1)-m(m+1)}\,|j\, m+1\, {\st ;}\,m'\zu \nonumber \\
             && \\
\op{J}^-|j\, m\, {\st ;}\, m'\zu &=& \sqrt{j(j+1)-m(m-1)}\,|j\, m-1\, {\st ;}\,m' \zu. \nonumber
\ea
Therefore, by means of eqn (\ref{actEnorm}) we obtain for the $k=1$ component of the flux operator $\op{E}_1(S)$ acting on the SNF $\bet{0}{0}$ the following result
\ba
\op{E}_1(S)\bet{0}{0}&=&-\frac{\lp^2}{\sqrt{2j+1}}\sum\limits_{m=-j}^{+j}(-1)^{j-m}\left(\op{J}^1_{e_{\sst 1}}|\,j\, m\, {\st ;}\, m_{e_{\sst 1}}'\zu_{e_{\sst 1}}\right)\otimes |\,j\, -m\, {\st ;}\, m_{e_{\sst 2}}'\zu_{e_{\sst 2}} \nonumber \\
                                            &=&-\frac{\lp^2}{2\sqrt{2j+1}}\sum\limits_{m=-j}^{+j}(-1)^{j-m}\nonumber \\
                                            & & \hspace{2.2cm}\Big\{
+\sqrt{j(j+1)-m(m+1)}|\,j\, m+1\, {\st ;}\, m_{e_{\sst 1}}'\zu_{e_{\sst 1}}\otimes |\,j\, -m\, {\st ;}\, m_{e_{\sst 2}}'\zu_{e_{\sst 2}} \nonumber \\
                                            &&\hspace{2.2cm}\,\,\,\,\,
+\sqrt{j(j+1)-m(m-1)}|\,j\, m-1\, {\st ;}\, m_{e_{\sst 1}}'\zu_{e_{\sst 1}}\otimes |\,j\, -m\, {\st ;}\, m_{e_{\sst 2}}'\zu_{e_{\sst 2}}\Big\}.
\ea
We wish to express the final result in terms of recoupling states. Consequently, we have to transform the tensor product $|\,j\, m\pm1\, {\st ;}\, m_{e_{\sst 1}}'\zu_{e_{\sst 1}}\otimes |\,j\, -m\, {\st ;}\, m_{e_{\sst 2}}'\zu_{e_{\sst 2}}$ back into the recoupling basis.
\ba
\label{CGCE1}
|\,j\, m+1{\st ;}\, m'\zu_{e_{\sst 1}}\otimes\,|\,j\, -m{\st ;}\, -m'\zu_{e_{\sst 2}}&=&-(-1)^{j-m}\sqrt{\frac{3}{2}}\sqrt{\frac{j(j+1)-m(m+1)}{j(j+1)(2j+1)}}\bet{1}{1}\nonumber\\                                                    
&& + \sum\limits_{\wt{j}_{\sst 12}=2}^{2j}\auf \wt{j}_{\sst 12}\, \wt{m}_2=1\,|\,j\, m+1{\st ;}\, j\, -m\zu \bet{\wt{j}_{\sst 12}}{1}\nonumber \\
|\,j\, m-1{\st ;}\, m'\zu_{e_{\sst 1}}\otimes|\,j\, -m{\st ;}\, -m'\zu_{e_{\sst 2}}&=&(-1)^{j-m}\sqrt{\frac{3}{2}}\sqrt{\frac{j(j+1)-m(m-1)}{j(j+1)(2j+1)}}\bet{1}{-1} \nonumber\\
                                                     & &+ \sum\limits_{\wt{j}_{\sst 12}=2}^{2j}\auf \wt{j}_{\sst 12}\, \wt{m}_2=1\,|\,j\, m+1{\st ;}\, j\, -m\zu \bet{\wt{j}_{\sst 12}}{-1},
\ea
where we used the definition $\bet{j_{\sst 12}}{m_{\sst 12}}:=|\,a_{\sst 1}=j\,a_{\sst 2}=j_{\sst 12}\, m_{\sst 12}\,{\sst ;}\, m'_{e_{\sst 1}}\, m'_{e_{\sst 2}}\zu$ as we did during the whole calculation of the new flux operator.
We want to expand the action of $\op{E}_1(S)$ on $\bet{0}{0}$ in terms of the states $\bet{1}{m_{\sst 12}}$
\be
\label{ExpBet2}
\op{E}_{1}(S)\bet{0}{0}=\sum\limits_{\wt{m}_{\sst 12}=-1}^{+1}
\betc{1}{\wt{m}_{\sst 12}}\,\op{E}_1(S)\,\bet{0}{0}\bet{1}{\wt{m}_{\sst 12}}.
\ee
As the next step we insert eqn (\ref{CGCE1}) into eqn (\ref{ExpBet2}). As $j_{\sst 12}$ denotes the total angular momentum of the state $\bet{j_{\sst 12}}{m_{\sst 12}}$, we know that  two states with different values of $j_{\sst 12}$ and $m_{\sst 12}$ are orthogonal to each other, meaning $\auf\beta^{\wt{j}_{\sst 12}}\,{\wt{m}_{\sst 12}}\bet{j_{\sst 12}}{m_{\sst 12}}=\delta_{\wt{j}_{\sst 12},j_{\sst 12}}\delta_{\wt{m}_{\sst 12},m_{\sst 12}}$. Taking this into account, we obtain
\ba
\label{ResE1}
\op{E}_1(S)\bet{0}{0}&=&\frac{-\lp^2}{2\sqrt{2j+1}}\sum\limits_{m=-j}^{+j}\Big\{-(-1)^{2(j-m)}\sqrt{\frac{3}{2}}\sqrt{\frac{(j(j+1)-m(m+1))^2}{j(j+1)(2j+1)}}\bet{1}{1}\nonumber \\
                       & &\hspace{2.8cm}
+(-1)^{2(j-m)}\sqrt{\frac{3}{2}}\sqrt{\frac{(j(j+1)-m(m-1))^2}{j(j+1)(2j+1)}}\bet{1}{-1}\Big\}\nonumber \\
                       &=&-\frac{\lp^2}{\sqrt{6}}\sqrt{j(j+1)}\,\Big\{\bet{1}{-1}\,-\, \bet{1}{1}\Big\}.
\ea
Here we used $(-1)^{2(j-m)}=+1$, as $(j-m) \in \Zl$ and $\sum\limits_{m=-j}^{j}m^2=(1/3)j(j+1)(2j+1)$.\\
Analogous to $\op{J}^1$, we can formulate $\op{J}^2$ in terms of ladder operators $\op{J}^2=\frac{1}{2i}(\op{J}^+-J^-)$. Hence, the action of the $k=2$ component of the flux operator $\op{E}_2(S)$ acting on $\bet{0}{0}$ is given by
\ba
\op{E}_2(S)\bet{0}{0}&=&-\frac{\lp^2}{\sqrt{2j+1}}\sum\limits_{m=-j}^{+j}(-1)^{j-m}\left(\op{J}^2_{e_{\sst 1}}|j\, m\, {\st ;}\, m_{\sst 1}'\zu_{e_{\sst 1}}\right)\otimes |j\, -m\, {\st ;}\, m_2'\zu_{e_{\sst 2}} \nonumber \\
                                            &=&-\frac{\lp^2}{2i\sqrt{2j+1}}\sum\limits_{m=-j}^{+j}(-1)^{j-m}\nonumber \\
                                            &&\hspace{2.8cm} \Big\{+\sqrt{j(j+1)-m(m+1)}|j\, m+1\, {\st ;}\, m_{\sst 1}'\zu_{e_{\sst 1}}\otimes |j\, -m\, {\st ;}\, m_2'\zu_{e_{\sst 2}} \nonumber \\
                                            &&\,\,\,\,\,\hspace{2.8cm}-\sqrt{j(j+1)-m(m-1)}|j\, m-1\, {\st ;}\, m_{\sst 1}'\zu_{e_{\sst 1}}\otimes |j\, -m\, {\st ;}\, m_2'\zu_{e_{\sst 2}}\Big\}.
\ea
Again, we want to transform the appearing tensor product $|j\, m\pm 1\, {\st ;}\, m_{\sst 1}'\zu_{e_{\sst 1}}\otimes |j\, -m\, {\st ;}\, m_2'\zu_{e_{\sst 2}}$ into the recoupling basis by means of the necessary CGC that can be found in eqn (\ref{CGCE1}). Inserting eqn (\ref{CGCE1}) into the equation above and taking advantage of the orthogonality relation concerning different $m's$ and $j_{\sst 12}'s$, we get
\ba
\label{ResE2}
\op{E}_2(S)\bet{0}{0}&=&\frac{-\lp^2}{2i\sqrt{2j+1}}\sum\limits_{m=-j}^{+j}\Big\{-(-1)^{2(j-m)}\sqrt{\frac{3}{2}}\sqrt{\frac{(j(j+1)-m(m+1))^2}{j(j+1)(2j+1)}}\bet{1}{1}\nonumber \\
                       &&\hspace{2.8cm}-(-1)^{2(j-m)}\sqrt{\frac{3}{2}}\sqrt{\frac{(j(j+1)-m(m-1))^2}{j(j+1)(2j+1)}}\bet{1}{-1}\Big\}\nonumber \\
                       &=&-\frac{i\lp^2}{\sqrt{6}}\sqrt{j(j+1)}\,\Big\{\bet{1}{-1}\,+\,\bet{1}{1}\Big\},
\ea
where we again used  $(-1)^{2(j-m)}=+1$, as $(j-m) \in \Zl$ and $\sum\limits_{m=-j}^{j}m^2=(1/3)j(j+1)(2j+1)$.\\
It remains to calculate the $k=3$ component of $\op{E}_k(S)$. This case is easier than the other two components as  $\op{J}^3_{e_{\sst 1}}$ does not change the magnetic quantum number $m$. Rather $|j\,m\,{\st ;}\,m'\zu$ is already an eigenstate of $\op{J}^3_{e_{\sst 1}}$.
\be
\label{J3eigen}
\op{J}^3 |j\,m\zu = m\, |j\, m\zu.
\ee
Using the eigenvalue above, we can evaluate the action of $\op{E}_3(S)$ on the SNF $\bet{0}{0}$
\ba
\label{ResAE3}
\op{E}_3(S)\bet{0}{0}&=&-\frac{\lp^2}{\sqrt{2j+1}}\sum\limits_{m=-j}^{+j}(-1)^{j-m}\left(\op{J}^3_{e_{\sst 1}}|j\, m\, {\st ;}\, m_{\sst 1}'\zu_{e_{\sst 1}}\right)\otimes |j\, -m\, {\st ;}\, m_{\sst 2}'\zu_{e_{\sst 2}} \nonumber \\
                                            &=&-\frac{\lp^2}{\sqrt{2j+1}}\sum\limits_{m=-j}^{+j}(-1)^{j-m}\, m\,|j\, m\, {\st ;}\, m_{\sst 1}'\zu_{e_{\sst 1}}\otimes |j\, -m\, {\st ;}\, m_{\sst 2}'\zu_{e_{\sst 2}}     \ea 
As we have a different tensor product $|j\, m\, {\st ;}\, m_{\sst 1}'\zu_{e_{\sst 1}}\otimes |j\, -m\, {\st ;}\, m_{\sst 2}'\zu_{e_{\sst 2}}$ than in the $k=1,2$ component case, we will consequently have a different expansion in terms of the recoupling basis states, in particular different in terms of the appearing CGG.
\ba
\label{CGCE3}
|j\, m\, {\st ;}\, m_{\sst 1}'\zu_{e_{\sst 1}}\otimes |j\, -m\, {\st ;}\, m_{\sst 2}'\zu_{e_{\sst 2}} &=& +\frac{(-1)^{j-m}}{\sqrt{2j+1}}\,\bet{0}{0}\nonumber \\
                                                              && + \frac{(-1)^{j-m}m\sqrt{3}}{\sqrt{j(j+1)(2j+1)}}\,\bet{1}{0}\nonumber \\
                                                              && +\sum\limits_{\wt{j}_{\sst 12}=2}^{2j}\auf \wt{j}_{\sst 12}\, \wt{m}_{\sst 12}=0\,|\, j\, m\, {\st ;}\, j\, -m\zu\, \bet{\wt{j}_{\sst 12}}{0} 
\ea 
Here, we can neglect the first two summands in eqn (\ref{CGCE3}).As for the $k=1,2$ component, we will expand the final result in terms of the states $\bet{1}{m_{\sst 12}}$ 
(see also eqn (\ref{ExpBet2}) for this). Because $\bet{0}{0}$ and $\bet{1}{m_{\sst 12}}$ are orthogonal to each other, the scalar product 
$\auf\beta^{\st 1}\,m_{\sst 12}\,\bet{0}{0}$ vanishes. Additionally, the first summand in eqn (\ref{CGCE3}) leads to an expression proportional to $\sum\limits_{m=-j}^{+j}m=0$ 
when inserting it into eqn (\ref{ResAE3}). Therefore, we will just consider the second summand of eqn (\ref{CGCE3}) as all the other terms of the remaining sum vanish as well, because of the orthogonality relation concerning $\wt{j}_{\sst 12}$. Hence, we get
\ba
\label{ResE3}
\op{E}_3(S)\bet{0}{0}&=&-(-1)^{j-m}\sqrt{\frac{3}{2}}\sqrt{\frac{(j(j+1)-m(m-1))^2}{j(j+1)(2j+1)}}\bet{1}{0}\nonumber \\
                     &=&-\frac{\lp^2}{\sqrt{3}}\sqrt{j(j+1)}\,\bet{1}{0},
\ea
where we have taken advantage of the fact that $(-1)^{2(j-m)}=+1$, as $(j-m) \in \Zl$ and used $\sum\limits_{m=-j}^{j}m^2=(1/3)j(j+1)(2j+1)$.
Summarising the results of this section we can extract from eqn (\ref{ResE1}),(\ref{ResE2}) and (\ref{ResE3}) the following results for the three components of the flux operator $\op{E}_k(S)$
\ba
\label{EkRes}
\op{E}_1(S)\bet{0}{0} &=&-\frac{\lp^2}{\sqrt{6}}\sqrt{j(j+1)}\,\Big\{\bet{1}{-1}\quad-\quad \bet{1}{+1}\Big\} \nonumber\\
\op{E}_2(S)\bet{0}{0} &=&-\frac{i\lp^2}{\sqrt{6}}\sqrt{j(j+1)}\,\Big\{\bet{1}{-1}\quad+\quad \bet{1}{+1}\Big\} \nonumber\\
\op{E}_3(S)\bet{0}{0} &=&-\frac{\lp^2}{\sqrt{3}}\sqrt{j(j+1)}\,\bet{1}{0}
\ea
\section{Comparison of the Two Flux Operators}%
By comparing eqn (\ref{EkRes}) with the results for $\lL\op{\tilde{E}}^{\sst II,AL}_{k,{\sst tot}}(S_t)$ in eqn (\ref{NEkres}) and the results of $\lL\op{\tilde{E}}^{\sst II,RS}_{k,{\sst tot}}(S_t)$ shown in ewn (\ref{RSNEkres}) respectively, we can judge whether our new constructed flux operators $\lL\op{\tilde{E}}^{\sst II,AL}_{k,{\sst tot}}(S_t),\lL\op{\tilde{E}}^{\sst II,RS}_{k,{\sst tot}}(S_t)$ are consistent with the action of the usual 
one $\op{E}_k(S)$\footnote{The operators $\lL\op{\tilde{E}}^{\sst I,AL}_{k,{\sst tot}}(S_t)$ and $\lL\op{\tilde{E}}^{\sst I,RS}_{k,{\sst tot}}(S_t)$ have been ruled out before since they are the zero operator and not consistent with the usual flux operator $\op{E}_k(S)$.}. 
\newline
Let us first discuss the operator $\lL\op{\tilde{E}}^{\sst II,AL}_{k,{\sst tot}}(S_t)$.
It transpires that
\be
\label{CompAL}
\lL\op{\tilde{E}}^{\sst II,AL}_{k,{\sst tot}}(S)\bet{0}{0}=3!8C_{reg}\op{E}_k(S)\bet{0}{0}
\ee
Therefore the two operators differ only by a positive integer constant. As there is still the regularisation constant $C_{reg}$ in the above equation we can now fix it by requiring 
 that both operators do exactly agree with each other. In fact there is no other choice than exact agreement because the difference would be a global constant which does not
 decrease as we take the corresponding limit of large quantum numbers $j$. 
Thus, we can remove the regularisation ambiguity of the volume operator in this way and choose $C_{reg}$ to be $C_{reg}:=\frac{1}{3!8}=\frac{1}{48}$.
  \newline
 This is exactly the 
value of $C_{reg}$ that was obtained in \cite{4} by a completely different argument. Thus the geometrical interpretation of the value we have to choose for $C_{reg}$ is perfectly provided\footnote{The factor $8=2^3$ comes from the fact that during the regularisation one integrates a product of 3 $\delta-$distributions on $\Rl$ over $\Rl^+$ only. The factor $6=3!$ is due to the fact that one should sum over ordered triples of edges only.}
\newline
Note that the consistency check holds in the full theory and not only in the semiclassical sector.
Consequently, the operator $\lL\op{\tilde{E}}^{\sst II,AL}_{k,{\sst tot}}(S_t)$ is consistent with the usual flux operator.
\newline\newline
Now, considering the operator $\lL\op{\tilde{E}}^{\sst II,RS}_{k,{\sst tot}}(S_t)$ things look differently. Here, a  quantisation that is consistent with $\op{V}_{\sst RS}$ of the signum operator $\op{\cal S}$ cannot be found. Accordingly, we should stop here and draw the conclusion that $\lL\op{\tilde{E}}^{\sst II,RS}_{k,{\sst tot}}(S_t)$ is not consistent with $\op{E}_k(S)$. A way out of this problem is to use artificially $\op{V}_{\sst RS}\op{\cal S}_{\sst AL}\op{V}_{\sst RS}$ for $\lL\op{\tilde{E}}^{\sst II,RS}_{k,{\sst tot}}(S_t)$. In doing so, we obtain 
\be
\label{CompRS}
\lL\op{\tilde{E}}^{\sst II,RS}_{k,{\sst tot}}(S)\bet{0}{0}=C(j,\ell)C_{reg}\op{E}_k(S)\bet{0}{0},
\ee
whereby $C(j,\ell)\in\Rl$ is a constant depending on the spin labels $j,\ell$ in general. Precisely, the dependence on the spin label $j$ causes a discrepancy of $\lL\op{\tilde{E}}^{\sst II,RS}_{k,{\sst tot}}(S_t)$ with respect to $\op{E}_k(S)$. But since $C(j,\ell)\to C(\ell)$ semiclassically, i.e. in the limit of large $j$, which is shown  in appendix E and discussed in section 6.6.2 of of \cite{GT}, $\lL\op{\tilde{E}}^{\sst II,RS}_{k,{\sst tot}}(S_t)$ including the artificial operator $\op{V}_{\sst RS}\op{\cal S}_{\sst AL}\op{V}_{\sst RS}$ is consistent with $\op{E}_k(S)$ within the semiclassical regime of the theory if we choose $C_{reg}=1/C(\ell)$. Unfortunately, $C(\ell)$ has a non-trivial $\ell$ -dependence which is inacceptable because it is absent in the classical theory. Moreover, we do not see any geometrical interpretation available for the chosen value of $C_{reg}$ in this case. 
One could possibly get rid of the $\ell$-dependence by simply cancelling the linearly dependent triples by hand from the definition of $\op{V}_{\sst RS}$. But then the so modified $\op{V}^{\prime}_{\sst RS}$ and $\op{V}_{\sst AL}$ would  practically become identical on $3-$ and $4-$valent vertices and moreover $\op{V}^{\prime}_{\sst RS}$ now depends on the differentiable structure of $\Sigma$.
\section{Uniqueness of the Chosen Factor Ordering}%
Since the analysis here holds for $\op{V}_{\sst AL}$ as well as for $\op{V}_{\sst RS}$ we neglect the explicit labelling in this section.
Now, we want to discuss to which extent the factor ordering chosen by us in section \ref{factord} is unique.
For this purpose let us go back to eqn (\ref{clid}). 
 Instead of using the classical identity shown in that
equation we could have used the following identity
\ba
\label{clid2}
 \lL\wt{E}_k'(S_t)&=&\lim_{{\cal P}_t\to S_t}\sum\limits_{\sst 
 {\Box}\in{\cal P}_t}\epsilon_{kst}\frac{4}{\kappa^2}\left\{A^s_{\sst 
 3},V(R_{v({\sst \Box})})\right\}{\cal S}\left\{A^t_{\sst 4},V(R_{v({\sst 
\Box})})\right\}\\    
 &=&\lim_{{\cal P}_t\to S_t}\sum\limits_{\sst {\Box}\in{\cal 
 P}_t}\frac{16}{\kappa^2}\epsilon_{skt}
\frac{1}{\frac{4}{3}\ell(\ell+1)(2\ell+1)}\frac{1}{(2\ell+1)}\nonumber\\
&&\quad\quad\quad\quad
\tr\Big(\pi_{\ell}(\tau_s)\pi_{\ell}(h_{e_{\sst 3}})\left\{\pi_{\ell}(h^{-1}_{e_{\sst 3}}),V(R_{v({\sst\Box})})\right\}\Big)
\tr\Big(\fbox{$\cal S$}\mathbbm{1}_{\sst(2\ell+1)}\Big)\tr\Big(\pi_{\ell}(\tau_t)\pi_{\ell}(h_{e_{\sst 4}})\left\{\pi_{\ell}(h^{-1}_{e_{\sst
4}}),V(R_{v({\sst \Box})})\right\}\Big),\nonumber\\
\ea
where we used $\tr\big(\pi_{\ell}(\tau_s)\pi_{\ell}(\tau_s')\big)=-\frac{4}{3}2\ell(\ell+1)\ell(\ell+1)\delta_{s,s'}$.
Surely, the operator corresponding to eqn (\ref{clid}) would lead to a flux operator with a trivial action
so far, for the reason that only one edge is added to $\bet{0}{0}$ before $\op{V}$ acts. Nevertheless, as the
holonomies commute classically, and additionally the trace is invariant under
cyclic permutations, we are allowed to insert a well chosen unitary matrix in every trace.
\ba
\tr\Big(\pi_{\ell}(\tau_t)\pi_{\ell}(h_{e_{\sst 4}})\left\{\pi_{\ell}l(h^{-1}_{e_{\sst
4}}),V(R_{v({\sst \Box})})\right\}\Big)
&=&\tr\Big(\op{\pi}_{\ell}(\tau_t)\pi_{\ell}(h_{e_{\sst 4}})\left\{\pi_{\ell}(h^{-1}_{e_{\sst
4}}),V(R_{v({\sst \Box})})\right\}\pi_{\ell}(h_{e_{\sst 3}})\pi_{\ell}(h^{-1}_{e_{\sst 3}})\Big)\nonumber\\
&=&\tr\Big(\pi_{\ell}(h^{-1}_{e_{\sst 3}})\pi_{\ell}(\tau_t)\pi_{\ell}(h_{e_{\sst 4}})\left\{\pi_{\ell}(h^{-1}_{e_{\sst
4}}),V(R_{v({\sst \Box})})\right\}\pi_{\ell}(h_{e_{\sst 3}})\Big)
\ea
Considering the trace that includes the signum factor $\op{S}$, we note that we have to insert two unitary
matrices here, in order to avoid a trivial action of the corresponding operator. Accordingly, we end up with
\ba
\label{clid3}
\lL\wt{E}_k'(S_t)&=&\lim_{{\cal P}_t\to S_t}\sum\limits_{\sst {\Box}\in{\cal P}_t}\epsilon_{kst}\frac{4}{\kappa^2}\left\{A^s_{\sst 3},V(R_{v({\sst \Box})})\right\}{\cal S}\left\{A^t_{\sst 4},V(R_{v({\sst \Box})})\right\}\nonumber\\    
&=&\lim_{{\cal P}_t\to S_t}\sum\limits_{\sst {\Box}\in{\cal P}_t}\frac{16}{\kappa^2}\epsilon_{skt}\frac{1}{\frac{4}{3}\ell(\ell+1)(2\ell+1)}\frac{1}{(2\ell+1)}\nonumber \\
&&\quad\quad\quad\quad
\tr\Big(\pi_{\ell}(h^{-1}_{e_{\sst 4}})\pi_{\ell}(\tau_s)\pi_{\ell}(h_{e_{\sst 3}})\left\{\pi_{\ell}(h^{-1}_{e_{\sst 3}}),V(R_{v({\sst
\Box})})\right\}\pi_{\ell}(h_{e_{\sst 4}})\Big)\nonumber\\
&&\quad\quad\quad\quad
\tr\Big(\op{\pi}_{\ell}(h_{e_{\sst 4}})\pi_{\ell}(h^{-1}_{e_{\sst 3}})\fbox{$\cal S$}\mathbbm{1}_{\sst(2\ell+1)}\pi_{\ell}(h_{e_{\sst
3}})\pi_{\ell}(h^{-1}_{e_{\sst 4}})\Big)\nonumber\\
&&\quad\quad\quad\quad
\tr\Big(\pi_{\ell}(h^{-1}_{e_{\sst 3}})\pi_{\ell}(\tau_t)\pi_{\ell}(h_{e_{\sst 4}})\left\{\pi_{\ell}(h^{-1}_{e_{\sst
4}}),V(R_{v({\sst \Box})})\right\}\pi_{\ell}(h_{e_{\sst 3}})\Big),
\ea
When we apply the formalism of canonical quantisation now, we get an operator with a
different factor ordering than the one we used before
\ba
\label{Op2}
\lL\op{\wt{E}}_k'(S_t)    
&=&\lim_{{\cal P}_t\to S_t}\sum\limits_{\sst {\Box}\in{\cal P}_t}\epsilon_{skt}\frac{-4\lp^{-4}}{\frac{4}{3}\ell(\ell+1)(2\ell+1)}\frac{1}{(2\ell+1)}\nonumber \\
&&\quad\quad\quad\quad
\tr\Big(\op{\pi}_{\ell}(h^{-1}_{e_{\sst 4}})\pi_{\ell}(\tau_s)\op{\pi}_{\ell}(h_{e_{\sst
3}})\left[\op{V}(R_{v({\sst\Box})}),\op{\pi}_{\ell}(h^{-1}_{e_{\sst 3}})\right]\op{\pi}_{\ell}(h_{e_{\sst
4}})\Big)\nonumber\\
&&\quad\quad\quad\quad
\tr\Big(\op{\pi}_{\ell}(h_{e_{\sst 4}})\op{\pi}_{\ell}(h^{-1}_{e_{\sst 3}})\fbox{$\op{\cal S}$}\mathbbm{1}_{\sst(2\ell+1)}\op{\pi}_{\ell}(h_{e_{\sst
3}})\op{\pi}_{\ell}(h^{-1}_{e_{\sst 4}})\Big)\nonumber\\
&&\quad\quad\quad\quad
\tr\Big(\op{\pi}_{\ell}(h^{-1}_{e_{\sst 3}})\op{\pi}_{\ell}(\tau_t)\op{\pi}_{\ell}(h_{e_{\sst
4}})\left[\op{V}(R_{v({\sst \Box})},\op{\pi}_{\ell}(h^{-1}_{e_{\sst
4}}))\right]\op{\pi}_{\ell}(h_{e_{\sst 3}})\Big),
\ea
Hence, the matrix element of $\lL\op{\wt{E}}_k'(S_t)$ can be calculated in the following way
\ba
\label{ME2}
\betc{1}{\wt{m}_{\sst 12}}\,\lL\op{\wt{E}}_k'(S_t)\,\bet{0}{0}    
&=&\lim_{{\cal P}_t\to S_t}\sum\limits_{\sst {\Box}\in{\cal P}_t}\epsilon_{skt}\frac{-16\lp^{-4}}{\frac{4}{3}\ell(\ell+1)(2\ell+1)}\frac{1}{(2\ell+1)}\nonumber \\
&&\sum\limits_{\wt{m}'_{\sst 12}=-1}^{+1}
\betc{1}{\wt{m}_{\sst 12}}\,\tr\Big(\op{\pi}_{\ell}(h^{-1}_{e_{\sst 4}})\pi_{\ell}(\tau_s)\op{\pi}_{\ell}(h_{e_{\sst
3}})\left[\op{V}(R_{v({\sst\Box})}),\op{\pi}_{\ell}(h^{-1}_{e_{\sst 3}})\right]\op{\pi}_{\ell}(h_{e_{\sst
4}})\Big)\,\bet{0}{0}\nonumber\\
&&\quad\quad\quad
\betc{0}{0}\,\tr\Big(\op{\pi}_{\ell}(h_{e_{\sst 4}})\op{\pi}_{\ell}(h^{-1}_{e_{\sst 3}})\fbox{$\op{\cal S}$}\mathbbm{1}_{\sst(2\ell+1)}\op{\pi}_{\ell}(h_{e_{\sst
3}})\op{\pi}_{\ell}(h^{-1}_{e_{\sst 4}})\Big)\,\bet{1}{\wt{m}'_{\sst 12}}\nonumber\\
&&\quad\quad\quad
\betc{1}{\wt{m}'_{\sst 12}}\,\tr\Big(\op{\pi}_{\ell}(h^{-1}_{e_{\sst 3}})\op{\pi}_{\ell}(\tau_t)\op{\pi}_{\ell}(h_{e_{\sst
4}})\left[\op{V}(R_{v({\sst \Box})}),\op{\pi}_{\ell}(h^{-1}_{e_{\sst 4}})\right]\op{\pi}_{\ell}(h_{e_{\sst
3}})\Big)\,\bet{0}{0},\nonumber\\
\ea
In order to show why this factor ordering is not appropriate to construct an alternative flux operator, we take
a closer look at the trace terms, for instance the one on the rightmost side. Carrying out this trace
leads to
\ba
\lefteqn{\betc{1}{\wt{m}'_{\sst 12}}\,\tr\Big(\op{\pi}_{\ell}(h^{-1}_{e_{\sst 3}})\op{\pi}_{\ell}(\tau_t)\op{\pi}_{\ell}(h_{e_{\sst
4}})\left[\op{V}(R_{v({\sst \Box})}),\op{\pi}_{\ell}(h^{-1}_{e_{\sst 4}})\right]\op{\pi}_{\ell}(h_{e_{\sst
3}})\Big)\,\bet{0}{0}}\nonumber\\
&=&\lim_{{\cal P}_t\to S_t}\sum\limits_{\sst {\Box}\in{\cal P}_t}\frac{16\,\lp^{-4}(-1)^{2\ell}}{\frac{4}{3}\ell(\ell+1)(2\ell+1)}\pi_{\ell}(\tau_k)_{\sst CB}\nonumber \\
&&\betc{1}{\wt{m}'_{\sst 12}}\,\pi_{\ell}(\epsilon)_{\sst EI}\pi_{\ell}(\epsilon)_{\sst FC}\op{\pi}_{\ell}(h^{\dagger}_{e_{\sst 4}})_{\sst
FG}\op{\pi}_{\ell}(h^{\dagger}_{e_{\sst 3}})_{\sst BA}\;\,\op{V}\;\,\op{\pi}_{\ell}(h_{e_{\sst 4}})_{\sst
IG}\op{\pi}_{\ell}(h_{e_{\sst 3}})_{\sst EA}\,\bet{0}{0}.
\ea
However, this is exactly the expression of the former operator in eqn (\ref{Ektot}) with the small but important
difference that in this case the operator $\op{O}=\{\op{V}\op{S}\op{V}, \op{V}^2\}$ is replaced by the volume operator 
 $\op{V}$ itself. As $\op{V}^2$ and the operator $\op{V}$ have the same eigenvectors, we can conclude from the
 discussion about the case where $\op{O}=\op{V}^2$ in section \ref{OV2} and in appendix  \ref{CaseV2} that the matrix element is zero.
 Consequently, the whole flux operator $\lL\op{\wt{E}}_k'(S_t)$ has a trivial action. Therefore this factor
 ordering cannot be used. Moreover, one can show that the other trace terms vanish as well, so that the
 trivial action of $\lL\op{\wt{E}}_k'(S_t)$ is not only due to the disappearing of the matrix element which we
 took as an example.
 \newline
 Another idea could be to put an additional trace including additonal holonomies around the already existing traces.
 We did this for a trace including one more holonomy and calculated the case where all three edges that are
 added to $\bet{0}{0}$ carry a spin label of $\ell=\frac{1}{2}$ and it turned out that the result is zero, too.
\section{Conclusion}    %
\label{s11}             %
In contrast to our companion paper \cite{GT}, we focused in this paper on the technical and mathematical aspects of the consistency check. By following the technical details step by step we hope to have provided a possibility to present, among other things, the robustness of this consistency check. For instance the fact that case I where the densitised triad is given by $E^a_k=\det(e)e^a_k$ leads for $\op{V}_{\sst AL}$ as well as $\op{V}_{\sst RS}$ to an alternative flux operator that is the zero operator could not have been guessed from the outset. This seems to be caused by an abstract symmetry of the volume operator that we are not aware of up to now. We would appreciate if one could understand this issue from a more abstract perspective. Nevertheless, since the quantisation of the momentum operator $i\hbar\frac{d}{dx}$ on $L_2(\Rl^+,dx)$ is also not possible, the result that $E^a_k$ cannot be considered as a 2-form fits perfectly well.
\newline\newline
Quite unexpectedly, the quantisation of the signum operator becomes necessary in order to perform the consistency check. Furthermore, the explicit relation to the operator $\op{Q}^{\sst AL}_v$, namely, $\op{\cal S}=\sgn(\op{Q}^{\sst AL}_v)$ wich is equivalent to the operator identity $\op{Q}^{\sst AL}_v=\op{V}_{\sst AL}{\cal S}\op{V}_{\sst AL}$, provides us with i) the possibility to perform the check for arbitrary spin labels $\ell$  thanks to the techniques developed in \cite{14} and ii) to draw the conclusion that $\op{V}_{\sst RS}$ is not consistent with the usual flux operator, because there is no way to quantise a signum operator by using the regularisation that was taken when $\op{V}_{\sst RS}$ was defined. Even the artifical construction where one uses $\op{V}_{\sst RS}\op{\cal S}_{\sst AL}\op{V}_{\sst RS}$ leads to an alternative flux operator that differs from the usual one also semiclassically since it contains a regularisation constant still dependent on the spin label $\ell$.\newline
By comparing the detailed calculation of case I and II one realises that the signum operator $\op{\cal S}$ roughly speaking acts like a "switch" which either leads to cancellation or survival of terms in the eigenvector expansions.
\newline\newline
The regularisation of the alternative and the usual flux operator is based on the same method and it turns out that the classification of edges in types up, down, in and out that is sensible for the usual flux operator, is also meaningfull for the alternative one. Moreover the meaning of the limit as we remove the regulator and to define the alternative flux operator has to be understood in the same way as for the usual flux operator, otherwise the alternative flux operator is identical to zero. Moreover, without the additional smearing we would be missing a crucial factor of 1/2 and our $C_{reg}$ would be off the value found in \cite{4}.
\newline\newline
The correspondence between the Ashtekar-Lewandowski (${\cal H}_{\sst AL}$) and the abstract angular momentum system Hilbert space has to be taken into account and has a large impact on the final result. If we had not introduced the unitary map $W$ that allows us to transfrom between ${\cal H}_{\sst AL}$ and the abstract angular momentum Hilbert space the result of the alternative flux operator would differ from the result for the usual one.
\newline\newline
Finally, all the $\ell$-dependence cancels at the end. Since many $\ell$-dependent terms are involved in the calculation as for instance Clebsch-Gordan coefficients, tau-matrices and the matrix elements of $\op{Q}^{\sst AL}_v$, this is rather astonishing and demonstrates that all the ingredients of this consistency check  fit together harmonically.
\newline\newline
This paper along with our companion paper \cite{GT} is one of the first papers that tightens the mathematical 
structure of full LQG by using the kind of consistency argument that we 
used here. Many more such checks should be performed in the future to 
remove ambiguities of LQG and to make the theory more rigid, in particular 
those connected with the quantum dynamics.\\
\\
\\
{\large Acknowledgements}\\
\\
It is our pleasure to thank Johannes Brunnemann for countless discussions 
about the volume operator.
We also would like to thank Carlo Rovelli and, especially, Lee Smolin for illuminating discussions.\newline
K.G. thanks the Heinrich-B\"oll-Stiftung for financial support. 
This research project was supported in part by a grant from NSERC of 
Canada to the Perimeter Institute for Theoretical Physics.
\begin{appendix}
\section{Proof of Lemma \ref{lemma1} in section \ref{ExpAction}} %
\label{Beweis}                                                   %
In order to keep the proof comprehensible, we want to express the CGC in eqn (\ref{Lem1eqn}) in terms of Wigner-3j-symbols, because the symmetry properties of the  Wigner-3j-symbols are easier to handle than the one of the CGC itself\footnote{Notice that we already used the replacement $ I=-E$ in the lemma. We could have left both indices $E,I$ independent, but due to $\pi_{\ell}(\epsilon)_{\sst EI}=(-1)^{\ell{\st -E}}\delta_{E+I,0}$ all terms in which $I\not=-E$ will vanish anyway.}. 
The relation between the CGC and the corresponding 3j-symbol is given by
\be
\auf j_{\sst 12}\,m_{\sst 12}\, {\st ;}\,\ell\,{\sst E}\,|\,a_{\sst 3}\,m_{\sst 12}{\st +E}\zu=(-1)^{m_{\sst 12}{\sst +E+}j_{\sst 12}{\sst -}\ell}\sqrt{2a_{\sst 3}+1}\left(
\begin{array}{ccc}
j_{\sst 12}&\ell&a_{\sst 3}\\
m_{\sst 12}&{\st E}&{\st -}(m_{\sst 12}{\st +E})
\end{array}
\right)
\ee
Replacing the first CGC in eqn (\ref{lemma1}) by the corresponding 3j-symbol and using the definition of $\pi_{\ell}(\epsilon)_{\sst E-E}$, we get
\ba
\label{CGC3j}
\lefteqn{
\sum\limits_{E=-\ell}^{+\ell}\pi_{\ell}(\epsilon)_{\sst E-E}\auf j_{\sst 12}\,m_{\sst 12}\, {\st ;}\,\ell\,{\st E}\,|\,a_{\sst 3}\,m_{\sst 12}{\st +E}\zu\auf a_{\sst 3}\,m_{\sst 12}{\st +E}\, {\st ;}\, \ell {\st -E}\,|\,J\,m_{\sst 12}\zu}\nonumber\\
&=&\sum\limits_{E=-\ell}^{+\ell}
(-1)^{m_{\sst 12}+j_{\sst 12}}\sqrt{2a_{\sst 3}+1}\left(
\begin{array}{ccc}
j_{\sst 12}&\ell&a_{\sst 3}\\
m_{\sst 12}&{\st E}&{\st -}(m_{\sst 12}{\st +E})
\end{array}
\right)\auf a_{\sst 3}\,m_{\sst 12}{\st +E}\, {\st ;}\, \ell {\st -E}\,|\,J\,m_{\sst 12}\zu.
\ea
With the help of the symmetry properties of the 3j-symbol, we are able to show that $\left(\begin{array}{ccc}j_{\sst 12}&\ell&a_{\sst 3}\\m_{\sst 12}&{\st E}&{\st -}(m_{\sst 12}{\st +E})\end{array}\right)$ is proportional to the CGC $\auf a_{\sst 3}\,m_{\sst 12}{\st +E}\, {\st ;}\, \ell {\st -E}\,|\,j_{\st 12}\,m_{\sst 12}\zu$
\ba
\auf a_{\sst 3}\,m_{\sst 12}{\st +E}\, {\st ;}\, \ell {\st -E}\,|\,j_{\st 12}\,m_{\sst 12}\zu
&=&(-1)^{m_{\sst 12}{\st +}3a_{\sst 3}{\st +}\ell{\st +}2j_{\sst 12}}\sqrt{2j_{\sst 12}+1}
\left(
\begin{array}{ccc}
j_{\sst 12}&\ell&a_{\sst 3}\\
m_{\sst 12}&{\st E}&{\st -}(m_{\sst 12}{\st +E})
\end{array}
\right).
\ea
Hence, rearranging the equation above leads to the desired proportionality
\be
\label{3jCGC}
\left(
\begin{array}{ccc}
j_{\sst 12}&\ell&a_{\sst 3}\\
m_{\sst 12}&{\st E}&{\st -}(m_{\sst 12}{\st +E})
\end{array}
\right)=\frac{(-1)^{{\st -}m_{\sst 12}{\st -}3a_{\sst 3}{\st -}\ell{\st -}2j_{\sst 12}}}{\sqrt{2j_{\sst 12}+1}}\auf a_{\sst 3}\,m_{\sst 12}{\st +E}\, {\st ;}\, \ell {\st -E}\,|\,j_{\st 12}\,m_{\sst 12}\zu.
\ee
The next step will be to insert eqn (\ref{3jCGC}) into eqn (\ref{CGC3j}) in order to use the orthogonality relation of the CGC for the remaining two CGC of the rewritten version of eqn (\ref{CGC3j})
\be
\label{CGC3j2}
\begin{array}{l}
\sum\limits_{E=-\ell}^{+\ell}\pi_{\ell}(\epsilon)_{\sst E-E}\auf j_{\sst 12}\,m_{\sst 12}\, {\st ;}\,\ell\,{\st E}\,|\,a_{\sst 3}\,m_{\sst 12}{\st +E}\zu\auf a_{\sst 3}\,m_{\sst 12}{\st +E}\, {\st ;}\, \ell {\st -E}\,|\,J\,m_{\sst 12}\zu\\
=(-1)^{{\st -}\ell{\st -}3a_{\sst 3}{\st -}j_{\sst 12}}\frac{\sqrt{2a_{\sst 3}+1}}{\sqrt{2j_{\sst 12}+1}}
\sum\limits_{{\st E=-}\ell}^{\ell}\auf j_{\st 12}\,m_{\sst 12}\,|\,a_{\sst 3}\,m_{\sst 12}{\st +E}\, {\st ;}\, \ell {\st -E}\zu
\auf a_{\sst 3}\,m_{\sst 12}{\st +E}\, {\st ;}\, \ell {\st -E}\,|\,J\,m_{\sst 12}\zu,
\end{array}
\ee
where we utilised that the CGC are real by convention in the last step. Now, we can take advantage of the orthogonality relation of the CGC which is given by
\be
\label{orthCGC}
\sum\limits_{{\st E=-}\ell}^{\ell}\auf j_{\st 12}\,m_{\sst 12}\,|\,a_{\sst 3}\,m_{\sst 12}{\st +E}\, {\st ;}\, \ell {\st -E}\zu
\auf a_{\sst 3}\,m_{\sst 12}{\st +E}\, {\st ;}\, \ell {\st -E}\,|\,J\,m_{\sst 12}\zu=\delta_{J,j_{\sst 12}}\Big(\delta_{m_{\sst 12},-j_{\sst 12}}+\delta_{m_{\sst 12},-j_{\sst 12}+1}+...+\delta_{m_{\sst 12},j_{\sst 12}}\Big).
\ee
Replacing the sum in eqn (\ref{CGC3j2}) by the means of eqn (\ref{orthCGC}), we are able to show that lemma (\ref{lemma1}) is true
\ba
\label{truel1}
\lefteqn{\sum\limits_{E=-\ell}^{+\ell}\pi_{\ell}(\epsilon)_{\sst E-E}\auf j_{\sst 12}\,m_{\sst 12}\, {\st ;}\,\ell\,{\st E}\,|\,a_{\sst 3}\,m_{\sst 12}{\st +E}\zu\auf a_{\sst 3}\,m_{\sst 12}{\st +E}\, {\st ;}\, \ell {\st -E}\,|\,J\,m_{\sst 12}\zu}\nonumber\\
&=&(-1)^{{\st -}j_{\sst 12}{\st -}\ell{\st -}3a_{\sst 3}}\frac{\sqrt{2a_{\sst 3}+1}}{\sqrt{2j_{\sst 12}+1}}\delta_{J,j_{\sst 12}}\left(\delta_{m_{\sst 12},-j_{\sst 12}}+\delta_{m_{\sst 12},-j_{\sst 12}+1}+...+\delta_{m_{\sst 12},j_{\sst 12}}\right).\quad\quad\quad\quad\quad{\scriptstyle\blacksquare}
\ea
\section{Tau-Matrices in Arbitrary Representation with Weight $\ell$}%
\label{Taus}
In order to be able to define the alternative flux  $\lL\wt{E}_k(S)$ on the classical level, we need to derive the matrix elements $\pi_\ell(\tau_k)_{mn}$ for the three tau-matrices in an arbitrary representation with weight $\ell$. For this purpose, we will use a formula for the matrix elements suitable for general $SL(2,\Cl)$ matrices $h=\left( \begin{array}{cc} a & b \\
c & d  \end{array} \right)$ where $a,b,c,d \in \Cl$ and $det(h)=ad-bc=1$, given in \cite{Sexl}. 
\newline
 Let $\pi_\ell(h)$ be the $(2\ell+1)$-dimensional matrix for $h$ in a 
 particular representation with weight $\ell$. The entries of this 
 transformation matrix between totally symmetric spinors of rank $2\ell$. 
The $\pi_\ell(h)_{mn}$, where $m,n=\{-\ell,...,\ell\}$, are given by
\begin{equation}
\label{FormSexl}
\pi_\ell(h)_{mn}=\sum\limits_s\frac{\sqrt{(\ell+m)!(\ell-m)!(\ell+n)!(\ell-n)!}}{(\ell-m-s)!(\ell+n-s)!(m-n+s)!s!}\,a^{\ell+n-s}\,b^{\,m-n+s}\,c^{\,s}\,d^{\,\ell-m-s}.
\end{equation}
Here the sum has to be  taken over all integers $s$ that do not cause negative factorials. Using the definition of the matrix element of the tau-matrices in a particular representation with weight $\ell$
\begin{equation}
\pi_\ell(\tau_k)_{mn}=\frac{d}{dt}\Big|_{t=0}\pi_\ell(e^{t\tau_k})_{mn},
\end{equation}
where $\tau_k:=-i\sigma_k$, we can write down the three matrices $e^{t\tau_k}$ for $k=1,2,3$ that are shown in eqn (\ref{etau})
\begin{eqnarray}
\label{etau}
e^{t\tau_1} &=& \cos(t)1_1+\sin(t)\tau_1 \\
e^{t\tau_2} &=& \cos(t)1_2+\sin(t)\tau_2 \nonumber \\
e^{t\tau_3} &=& \cos(t)1_2+\sin(t)\tau_3 \nonumber 
\end{eqnarray}
Inserting the above matrices into the formula in eqn (\ref{FormSexl}) and taking the derivative at the point $t=0$, we achieve a general expression for the matrix elements of the three tau-matrices $\pi_\ell(\tau_k)$ in a particular representation with weight $\ell$
\begin{eqnarray}
\label{Taugeneral}
\pi_\ell(\tau_1)_{mn}&=&-i\sqrt{\ell(\ell+1)-m(m-1)}\,\delta_{m-n,1}-
i\sqrt{\ell(\ell+1)-m(m+1)}\,\delta_{m-n,-1}\\
\pi_\ell(\tau_2)_{mn}&=&\sqrt{\ell(\ell+1)-m(m+1)}\,\delta_{m-n,-1}-
\sqrt{\ell(\ell+1)-m(m-1)}\,\delta_{m-n,1} \nonumber\\
\pi_\ell(\tau_3)_{mn}&=&-2\,i\,m\,\delta_{m-n,0}. \nonumber
\end{eqnarray}
 During the derivation of the alternative flux $\lL\wt{E}_k(S)$, we  will 
need the following property of the tau-matrices $\pi_\ell(\tau_k)$.
\begin{Lemma}
\label{Taulemma}
 Let $\pi_\ell(\tau_k)$ be the $(2\ell+1)$-dimensional matrix for $\tau_k:=-i\sigma_k$ in a particular representation with weight $\ell$, then the following identity holds
\begin{equation}
\label{Trtauk}
\tr(\pi_\ell(\tau_k)\pi_\ell(\tau_{r})\pi_{\ell}(\tau_{s}))=-\frac{4}{3}\ell(\ell+1)(2\ell+1)\epsilon_{krs}.
\end{equation}
\end{Lemma}
We desist from writing the proof of lemma \ref{Taulemma}  here, since the lemma can be easily proven by using  basic algebraic tools and explicitly calculating the identity for the various cases.
\section{Derivation of the Formula for the Matrix Elements of $\op{q}_{\sst IJK}$}%
\label{Derqijk}                                                                                           %
In this section we will derive the explicit formulae for the matrix elements of $\op{q}_{\sst IJK}$, namely eqn (\ref{RSq134}), (\ref{RSq234}), (\ref{RSq123}) and (\ref{RSq124}), because it turned out \cite{CommJB} that these are two special cases in which the general formula in \cite{14} is not applicable. Therefore we have to start from the very beginning and use the defintion of $q_{\sst IJK}$ in eqn (\ref{qijkDef}). In the following we will adopt the notation introduced in \cite{14} and denote different recoupling schemes by $\vec{g}(IJ)$ where $I,J$ labels the momenta that are coupled together at first. Therefore, often  $\vec{g}(12)$ is called the standard recoupling scheme. The intermediate couplings of particular scheme $\vec{g}(IJ)$ will be called $g_i$, while the intermediate couplings of our states $\alp{J}{i}{M}$ and $\alp{J}{\tilde{i}}{M}$ are still $a_i$ and $\wt{a}_i$, respectively. Using eqn (\ref{qijkDef}) for the case of $I=1,J=3,K=4$, we obtain
\ba
\label{Derq134}
\lefteqn{\alpc{J}{i}{M}\,q_{\sst 134}\,\alp{J}{\tilde{i}}{M}}\nonumber\\
&=&\alpc{J}{i}{M}\,[(J_{\sst 13})^2,(J_{\sst 34})^2]\,\alp{J}{\tilde{i}}{M}\nonumber\\
&=&\alpc{J}{i}{M}\,(J_{\sst 13})^2(J_{\sst 34})^2\,\alp{J}{\tilde{i}}{M}
-\alpc{J}{i}{M}\,(J_{\sst 34})^2(J_{\sst 13})^2]\,\alp{J}{\tilde{i}}{M}\nonumber\\
&=&\sum\limits_{\sst \vec{g}''(12)}\Big\{\sum\limits_{\sst  \vec{g}(13),\vec{g}(34)}g_{\sst 2}(13)(g_{\sst 2}(13)+1)g_{\sst 2}(34)(g_{\sst 2}(34)+1)
\auf\vec{g}(13)\,|\,\vec{g}''(12)\zu\auf\vec{g}''(12)\,|\,\vec{g}(34)\zu\nonumber\\
&&\hspace{2.25cm}
\Big[\auf\vec{g}(13)\,|\,\alpha^{\st J}_{i}\,{\st M}\zu\auf\vec{g}(34)\,|\,\alpha^{J}_{\tilde{i}}\,{\st M}\zu-\auf\vec{g}(34)\,|\,\alpha^{\st J}_{i}\,{\st M}\zu\auf\vec{g}(13)\,|\,\alpha^{J}_{\tilde{i}}\,{\st M}\zu\Big]\nonumber\\
&=&\sum\limits_{\vec{g}''(12)}\Big\{\sum\limits_{\vec{g}(13)}g_{\sst 2}(13)(g_{\sst 2}(13)+1)\auf\vec{g}(13)\,|\,\vec{g}''(12)\zu\auf\vec{g}(13)\,|\,\alpha^{\st J}_{i}\,{\st M}\zu\times\nonumber\\
&&\hspace{1cm}
\times\sum\limits_{\vec{g}(34)}g_{\sst 2}(34)(g_{\sst 2}(34)+1)\auf\vec{g}(34)\,|\,\vec{g}''(12)\zu\auf\vec{g}(34)\,|\,\alpha^{\st J}_{\tilde{i}}\,{\st M}\zu\Big\}\nonumber\\
&& -\Big[\alp{J}{i}{M}\,\longleftrightarrow\,\alp{J}{\tilde{i}}{M}\Big]\Big\},
\ea
where the last term has to be understood as the analogon of the first term when the states $\alp{J}{i}{M}$ and $\alp{J}{i}{M}$ are interchanged. It was demonstrated in \cite{14} that by means of the Elliot-Biedenharn identity, one can actually carry out the sum over $\vec{g}(13)$ and $\vec{g}(34)$ in the above equation. Hence, we will take the result from \cite{14,CommJB} 
\ba
\label{Sumg13}
\lefteqn{\sum\limits_{\vec{g}(13)}g_{\sst 2}(13)(g_{\sst 2}(13)+1)\auf\vec{g}(13)\,|\,\vec{g}''(12)\zu\auf\vec{g}(13)\,|\,\alpha^{\st J}_{i}\,{\st M}\zu}\nonumber\\
&=&\Big[\frac{1}{2}(-1)^{-j_{\sst 1}-j_{\sst 2}+j_{\sst 3}+1}X(j_{\sst 1},j_{\sst 3})^{\frac{1}{2}}A(g_{\sst 2}'',a_{\sst 2})
\SixJ{j_{\sst 2}}{j_{\sst 1}}{g''_{\sst 2}}{1}{a_{\sst 2}}{j_{\sst 1}}(-1)^{a_{\sst 3}}
\SixJ{a_{\sst 3}}{j_{\sst 3}}{g''_{\sst 2}}{1}{a_{\sst 2}}{j_{\sst 3}}+C(j_{\sst 1},j_{\sst 3})\delta_{g''_{\sst 2},a_{\sst 2}}\Big]\delta_{g''_{\sst 3},a_{\sst 3}}\delta_{g''_{\sst 4},a_{\sst 4}}\nonumber\\
\ea
\ba
\label{Sumg34}
\lefteqn{\sum\limits_{\vec{g}(34)}g_{\sst 2}(34)(g_{\sst 2}(34)+1)\auf\vec{g}(34)\,|\,\vec{g}''(12)\zu\auf\vec{g}(34)\,|\,\alpha^{\st J}_{i}\,{\st M}\zu}\nonumber\\
&=&\Big[\frac{1}{2}(-1)^{-2(j_{\sst 1}+j_{\sst 2})+j_{\sst 4}-j_{\sst 3}}(-1)^{a_{\sst 2}+1}(-1)^{a_{\sst 3}-g''_{\sst 3}}
X(j_{\sst 3},j_{\sst 4})^{\frac{1}{2}}A(g_{\sst 3}'',a_{\sst 3})
\SixJ{a_{\sst 2}}{j_{\sst 3}}{g''_{\sst 3}}{1}{a_{\sst 3}}{j_{\sst 3}}(-1)^{a_{\sst 4}}
\SixJ{a_{\sst 4}}{j_{\sst 4}}{g''_{\sst 3}}{1}{a_{\sst 3}}{j_{\sst 4}}
+C(j_{\sst 3},j_{\sst 4})\prod\limits_{k=2}^3\delta_{g''_{\sst k},a_{\sst k}}\Big]\nonumber\\
&&\delta_{g''_{\sst 2},a_{\sst 2}}\delta_{g''_{\sst 4},a_{\sst 4}}.
\nonumber\\
\ea
In order to keep the equation comprehensible, we introduced the following abbreviations
\ba
\label{CXADef}
C(a,b)&:=&a(a+1)+b(b+1)\nonumber\\
X(a,b)&:=&2a(2a+1)(2a+2)2b(2b+1)(2b+2)\nonumber\\
A(a,b)&:=&\sqrt{(2a+1)(2b+1)}.
\ea
The next step is to insert eqn (\ref{Sumg13}) and (\ref{Sumg34}) back into eqn (\ref{Derq134}). By doing so, we recognise that the term containing $C(a,b)$ is symmetric under the interchange of $a_i\leftrightarrow \wt{a}_i$ and accordingly will be canceled, because we subtract the terms where $\alp{J}{i}{M}$ and $\alp{J}{\tilde{i}}{M}$ are interchanged from each other. Consequently only the first term of (\ref{Sumg13}) and (\ref{Sumg34}) survives and we end up with
%
\ba
\lefteqn{\alpc{J}{i}{M}\,q_{\sst 134}\,\alp{J}{\tilde{i}}{M}}\nonumber\\
&=&\sum\limits_{\vec{g}''(12)}
\Big\{+\frac{1}{4}(-1)^{-3(j_{\sst 1}+j_{\sst 2})+j_{\sst 4}+1}(-1)^{\wt{a}_{\sst 2}+1}(-1)^{\wt{a}_{\sst 3}-g''_{\sst 3}}(-1)^{a_{\sst 3}+\wt{a}_{\sst 4}}
X(j_{\sst 1},j_{\sst 3})^{\frac{1}{2}}X(j_{\sst 3},j_{\sst 4})^{\frac{1}{2}}A(g''_{\sst 2},a_{\sst 2})A(g''_{\sst 3},\wt{a}_{\sst 3})\nonumber\\
&&\hspace{1.15cm}
\SixJ{j_{\sst 2}}{j_{\sst 1}}{g''_{\sst 2}}{1}{a_{\sst 2}}{j_{\sst 1}}
\SixJ{a{\sst 3}}{j_{\sst 3}}{g''_{\sst 2}}{1}{a_{\sst 2}}{j_{\sst 3}}
\SixJ{\wt{a}_{\sst 2}}{j_{\sst 3}}{g''_{\sst 3}}{1}{\wt{a}_{\sst 3}}{j_{\sst 3}}
\SixJ{\wt{a}_{\sst 4}}{j_{\sst 4}}{g''_{\sst 3}}{1}{\wt{a}_{\sst 3}}{j_{\sst 4}}
\delta_{g''_{\sst 2},\wt{a}_{\sst 2}}\delta_{g''_{\sst 4},\wt{a}_{\sst 4}}
\delta_{g''_{\sst 3},a_{\sst 3}}\delta_{g''_{\sst 4},a_{\sst 4}}\nonumber\\
&&\hspace{1.1cm}
-\frac{1}{4}(-1)^{-3(j_{\sst 1}+j_{\sst 2})+j_{\sst 4}+1}(-1)^{a_{\sst 2}+1}(-1)^{a_{\sst 3}-g''_{\sst 3}}(-1)^{\wt{a}_{\sst 3}+a_{\sst 4}}
X(j_{\sst 1},j_{\sst 3})^{\frac{1}{2}}X(j_{\sst 3},j_{\sst 4})^{\frac{1}{2}}A(g''_{\sst 2},\wt{a}_{\sst 2})A(g''_{\sst 3},a_{\sst 3})\nonumber\\
&&\hspace{1.1cm}
\SixJ{j_{\sst 2}}{j_{\sst 1}}{g''_{\sst 2}}{1}{\wt{a}_{\sst 2}}{j_{\sst 1}}
\SixJ{\wt{a}{\sst 3}}{j_{\sst 3}}{g''_{\sst 2}}{1}{\wt{a}_{\sst 2}}{j_{\sst 3}}
\SixJ{a_{\sst 2}}{j_{\sst 3}}{g''_{\sst 3}}{1}{a_{\sst 3}}{j_{\sst 3}}
\SixJ{a_{\sst 4}}{j_{\sst 4}}{g''_{\sst 3}}{1}{a_{\sst 3}}{j_{\sst 4}}
\delta_{g''_{\sst 2},a_{\sst 2}}\delta_{g''_{\sst 4},a_{\sst 4}}
\delta_{g''_{\sst 3},\wt{a}_{\sst 3}}\delta_{g''_{\sst 4},\wt{a}_{\sst 4}}\Big\}\nonumber\\
&=&
+\frac{1}{4}(-1)^{-3(j_{\sst 1}+j_{\sst 2})+j_{\sst 4}+1}(-1)^{\wt{a}_{\sst 2}+1}(-1)^{\wt{a}_{\sst 3}-a_{\sst 3}}(-1)^{a_{\sst 3}+a_{\sst 4}}
X(j_{\sst 1},j_{\sst 3})^{\frac{1}{2}}X(j_{\sst 3},j_{\sst 4})^{\frac{1}{2}}A(\wt{a}_{\sst 2},a_{\sst 2})A(a_{\sst 3},\wt{a}_{\sst 3})\nonumber\\
&&\hspace{1.15cm}
\SixJ{j_{\sst 2}}{j_{\sst 1}}{\wt{a}_{\sst 2}}{1}{a_{\sst 2}}{j_{\sst 1}}
\SixJ{a{\sst 3}}{j_{\sst 3}}{\wt{a}_{\sst 2}}{1}{a_{\sst 2}}{j_{\sst 3}}
\SixJ{\wt{a}_{\sst 2}}{j_{\sst 3}}{a_{\sst 3}}{1}{\wt{a}_{\sst 3}}{j_{\sst 3}}
\SixJ{a_{\sst 4}}{j_{\sst 4}}{a_{\sst 3}}{1}{\wt{a}_{\sst 3}}{j_{\sst 4}}\nonumber\\
&&
-\frac{1}{4}(-1)^{-3(j_{\sst 1}+j_{\sst 2})+j_{\sst 4}+1}(-1)^{a_{\sst 2}+1}(-1)^{a_{\sst 3}-\wt{a}_{\sst 3}}(-1)^{\wt{a}_{\sst 3}+a_{\sst 4}}
X(j_{\sst 1},j_{\sst 3})^{\frac{1}{2}}X(j_{\sst 3},j_{\sst 4})^{\frac{1}{2}}A(a_{\sst 2},\wt{a}_{\sst 2})A(\wt{a}_{\sst 3},a_{\sst 3})\nonumber\\
&&\hspace{1.1cm}
\SixJ{j_{\sst 2}}{j_{\sst 1}}{a_{\sst 2}}{1}{\wt{a}_{\sst 2}}{j_{\sst 1}}
\SixJ{\wt{a}{\sst 3}}{j_{\sst 3}}{a_{\sst 2}}{1}{\wt{a}_{\sst 2}}{j_{\sst 3}}
\SixJ{a_{\sst 2}}{j_{\sst 3}}{\wt{a}_{\sst 3}}{1}{a_{\sst 3}}{j_{\sst 3}}
\SixJ{a_{\sst 4}}{j_{\sst 4}}{\wt{a}_{\sst 3}}{1}{a_{\sst 3}}{j_{\sst 4}}\nonumber\\
&=&+\frac{1}{4}(-1)^{-3(j_{\sst 1}+j_{\sst 2})+j_{\sst 4}+a_{\sst 4}}
X(j_{\sst 2},j_{\sst 3})^{\frac{1}{2}}X(j_{\sst 3},j_{\sst 4})^{\frac{1}{2}}
A(\wt{a}_{\sst 2},a_{\sst 2})A(a_{\sst 3},\wt{a}_{\sst 3})
\SixJ{j_{\sst 1}}{j_{\sst 2}}{a_{\sst 2}}{1}{\wt{a}_{\sst 2}}{j_{\sst 2}}
\SixJ{a_{\sst 4}}{j_{\sst 4}}{a_{\sst 3}}{1}{\wt{a}_{\sst 3}}{j_{\sst 4}}\nonumber\\
&&\left[(-1)^{\wt{a}_{\sst 2}+\wt{a}_{\sst 3}}
\SixJ{a{\sst 3}}{j_{\sst 3}}{a_{\sst 2}}{1}{\wt{a}_{\sst 2}}{j_{\sst 3}}
\SixJ{\wt{a}_{\sst 2}}{j_{\sst 3}}{a_{\sst 3}}{1}{\wt{a}_{\sst 3}}{j_{\sst 3}}
-(-1)^{a_{\sst 2}+a_{\sst 3}}
\SixJ{\wt{a}{\sst 3}}{j_{\sst 3}}{a_{\sst 2}}{1}{\wt{a}_{\sst 2}}{j_{\sst 3}}
\SixJ{a_{\sst 2}}{j_{\sst 3}}{a_{\sst 3}}{1}{\wt{a}_{\sst 3}}{j_{\sst 3}}\right].
\ea
In the last line we used the symmetry properties of the 6j-symbols, in particular the fact that $\SixJ{a}{b}{c}{d}{e}{b}=\SixJ{a}{b}{e}{d}{c}{b}$. Bringing back into our mind that for our SNF $\alp{J}{i}{M}$ we have  $j_{\sst 1}=j_{\sst 2}=j,j_{\sst 3}=j_{\sst 4}=\ell$ and $a_{\sst 4}=J$, the matrix element of $q_{\sst 134}$ can be expressed as
\ba
\lefteqn{\alpc{J}{i}{M}\,q_{\sst 134}\,\alp{J}{\tilde{i}}{M}}\nonumber\\
&=&+\frac{1}{4}(-1)^{+2j+\ell+{\st J}}
\sqrt{2j(2j+1)(2j+2)}[2\ell(2\ell+1)(2\ell+2)]^{\frac{3}{2}}
\sqrt{(2\wt{a}_{\sst 2}+1)(2a_{\sst 2}+1)}\sqrt{2(a_{\sst 3}+1)(2\wt{a}_{\sst 3}+1)}\nonumber\\
&&
\SixJ{j}{j}{a_{\sst 2}}{1}{\wt{a}_{\sst 2}}{j}
\SixJ{{\st J}}{\ell}{a_{\sst 3}}{1}{\wt{a}_{\sst 3}}{\ell}
\left[(-1)^{\wt{a}_{\sst 2}+\wt{a}_{\sst 3}}
\SixJ{a{\sst 3}}{\ell}{a_{\sst 2}}{1}{\wt{a}_{\sst 2}}{\ell}
\SixJ{\wt{a}_{\sst 2}}{\ell}{a_{\sst 3}}{1}{\wt{a}_{\sst 3}}{\ell}
-(-1)^{a_{\sst 2}+a_{\sst 3}}
\SixJ{\wt{a}{\sst 3}}{\ell}{a_{\sst 2}}{1}{\wt{a}_{\sst 2}}{\ell}
\SixJ{a_{\sst 2}}{\ell}{a_{\sst 3}}{1}{\wt{a}_{\sst 3}}{\ell}\right],\nonumber\\
\ea
where we used additionally that $2j\in\Zl$ and therefore $(-1)^{-6j}=(-1)^{-2j}=(-1)^{+2j}$. Now we have to repeat the whole calculation for the case $I=2,J=3,K=4$ in order to derive the formula for $q_{\sst 234}$. Since the intermediate steps in the calculation are analogous to $q_{\sst 134}$ not all details will be given.   Here eqn (\ref{qijkDef}) leads to
\ba
\label{Derq234}
\lefteqn{\alpc{J}{i}{M}\,q_{\sst 234}\,\alp{J}{\tilde{i}}{M}}\nonumber\\
&=&\alpc{J}{i}{M}\,[(J_{\sst 23})^2,(J_{\sst 34})^2]\,\alp{J}{\tilde{i}}{M}\nonumber\\
&=&\sum\limits_{\vec{g}''(12)}\Big\{\sum\limits_{\vec{g}(23)}g_{\sst 2}(23)(g_{\sst 2}(23)+1)\auf\vec{g}(23)\,|\,\vec{g}''(12)\zu\auf\vec{g}(23)\,|\,\alpha^{\st J}_{i}\,{\st M}\zu\times\nonumber\\
&&\hspace{1cm}
\times\sum\limits_{\vec{g}(34)}g_{\sst 2}(34)(g_{\sst 2}(34)+1)\auf\vec{g}(34)\,|\,\vec{g}''(12)\zu\auf\vec{g}(34)\,|\,\alpha^{\st J}_{\tilde{i}}\,{\st M}\zu\Big\}\nonumber\\
&& -\Big[\alp{J}{i}{M}\,\longleftrightarrow\,\alp{J}{\tilde{i}}{M}\Big].
\ea
Thus, we need to know the result of the summation over $\vec{g}(23)$ in this case. It is given by \cite{14,CommJB}
\ba
\label{Sumg23}
\lefteqn{\sum\limits_{\vec{g}(23)}g_{\sst 2}(23)(g_{\sst 2}(23)+1)\auf\vec{g}(23)\,|\,\vec{g}''(12)\zu\auf\vec{g}(23)\,|\,\alpha^{\st J}_{i}\,{\st M}\zu}\nonumber\\
&=&\Big[\frac{1}{2}(-1)^{-j_{\sst 1}-j_{\sst 2}+j_{\sst 3}+1}(-1)^{a_{\sst 2}-g''_{\sst 2}}X(j_{\sst 2},j_{\sst 3})^{\frac{1}{2}}A(g_{\sst 2}'',a_{\sst 2})
\SixJ{j_{\sst 1}}{j_{\sst 2}}{g''_{\sst 2}}{1}{a_{\sst 2}}{j_{\sst 2}}(-1)^{a_{\sst 3}}
\SixJ{a_{\sst 3}}{j_{\sst 3}}{g''_{\sst 2}}{1}{a_{\sst 2}}{j_{\sst 3}}+C(j_{\sst 2},j_{\sst 3})\delta_{g''_{\sst 2},a_{\sst 2}}\Big]\nonumber\\
&&\delta_{g''_{\sst 3},a_{\sst 3}}\delta_{g''_{\sst 4},a_{\sst 4}}.
\ea
Reinserting the equation above and the result of the summation over $\vec{g}(34)$ from eqn (\ref{Sumg34}) into eqn (\ref{Derq234}), we get
\ba
\lefteqn{\alpc{J}{i}{M}\,q_{\sst 234}\,\alp{J}{\tilde{i}}{M}}\nonumber\\
&=&\sum\limits_{\vec{g}''(12)}
\Big\{+\frac{1}{4}(-1)^{-3(j_{\sst 1}+j_{\sst 2})+j_{\sst 4}+1}(-1)^{a_{\sst 2}-g''_{\sst2}+\wt{a}_{\sst 2}+1}(-1)^{\wt{a}_{\sst 3}-g''_{\sst 3}}(-1)^{a_{\sst 3}+\wt{a}_{\sst 4}}
X(j_{\sst 2},j_{\sst 3})^{\frac{1}{2}}X(j_{\sst 3},j_{\sst 4})^{\frac{1}{2}}A(g''_{\sst 2},a_{\sst 2})A(g''_{\sst 3},\wt{a}_{\sst 3})\nonumber\\
&&\hspace{1.15cm}
\SixJ{j_{\sst 1}}{j_{\sst 2}}{g''_{\sst 2}}{1}{a_{\sst 2}}{j_{\sst 2}}
\SixJ{a{\sst 3}}{j_{\sst 3}}{g''_{\sst 2}}{1}{a_{\sst 2}}{j_{\sst 3}}
\SixJ{\wt{a}_{\sst 2}}{j_{\sst 3}}{g''_{\sst 3}}{1}{\wt{a}_{\sst 3}}{j_{\sst 3}}
\SixJ{\wt{a}_{\sst 4}}{j_{\sst 4}}{g''_{\sst 3}}{1}{\wt{a}_{\sst 3}}{j_{\sst 4}}
\delta_{g''_{\sst 2},\wt{a}_{\sst 2}}\delta_{g''_{\sst 4},\wt{a}_{\sst 4}}
\delta_{g''_{\sst 3},a_{\sst 3}}\delta_{g''_{\sst 4},a_{\sst 4}}\nonumber\\
&&\hspace{1.1cm}
-\frac{1}{4}(-1)^{-3(j_{\sst 1}+j_{\sst 2})+j_{\sst 4}+1}(-1)^{\wt{a}_{\sst 2}-g''_{\sst2}+a_{\sst 2}+1}(-1)^{a_{\sst 3}-g''_{\sst 3}}(-1)^{\wt{a}_{\sst 3}+a_{\sst 4}}
X(j_{\sst 2},j_{\sst 3})^{\frac{1}{2}}X(j_{\sst 3},j_{\sst 4})^{\frac{1}{2}}A(g''_{\sst 2},\wt{a}_{\sst 2})A(g''_{\sst 3},a_{\sst 3})\nonumber\\
&&\hspace{1.1cm}
\SixJ{j_{\sst 1}}{j_{\sst 2}}{g''_{\sst 2}}{1}{\wt{a}_{\sst 2}}{j_{\sst 2}}
\SixJ{\wt{a}{\sst 3}}{j_{\sst 3}}{g''_{\sst 2}}{1}{\wt{a}_{\sst 2}}{j_{\sst 3}}
\SixJ{a_{\sst 2}}{j_{\sst 3}}{g''_{\sst 3}}{1}{a_{\sst 3}}{j_{\sst 3}}
\SixJ{a_{\sst 4}}{j_{\sst 4}}{g''_{\sst 3}}{1}{a_{\sst 3}}{j_{\sst 4}}
\delta_{g''_{\sst 2},a_{\sst 2}}\delta_{g''_{\sst 4},a_{\sst 4}}
\delta_{g''_{\sst 3},\wt{a}_{\sst 3}}\delta_{g''_{\sst 4},\wt{a}_{\sst 4}}\Big\}\nonumber\\
&=&+\frac{1}{4}(-1)^{-3(j_{\sst 1}+j_{\sst 2})+j_{\sst 4}+a_{\sst 4}}
X(j_{\sst 2},j_{\sst 3})^{\frac{1}{2}}X(j_{\sst 3},j_{\sst 4})^{\frac{1}{2}}
A(\wt{a}_{\sst 2},a_{\sst 2})A(a_{\sst 3},\wt{a}_{\sst 3})
\SixJ{j_{\sst 1}}{j_{\sst 2}}{a_{\sst 2}}{1}{\wt{a}_{\sst 2}}{j_{\sst 2}}
\SixJ{a_{\sst 4}}{j_{\sst 4}}{a_{\sst 3}}{1}{\wt{a}_{\sst 3}}{j_{\sst 4}}\nonumber\\
&&\left[(-1)^{a_{\sst 2}+\wt{a}_{\sst 3}}
\SixJ{a{\sst 3}}{j_{\sst 3}}{a_{\sst 2}}{1}{\wt{a}_{\sst 2}}{j_{\sst 3}}
\SixJ{\wt{a}_{\sst 2}}{j_{\sst 3}}{a_{\sst 3}}{1}{\wt{a}_{\sst 3}}{j_{\sst 3}}
-(-1)^{\wt{a}_{\sst 2}+a_{\sst 3}}
\SixJ{\wt{a}{\sst 3}}{j_{\sst 3}}{a_{\sst 2}}{1}{\wt{a}_{\sst 2}}{j_{\sst 3}}
\SixJ{a_{\sst 2}}{j_{\sst 3}}{a_{\sst 3}}{1}{\wt{a}_{\sst 3}}{j_{\sst 3}}\right].
\ea
In the first step we used again the symmetry of the term containing $C(a,b)$ in eqn (\ref{Sumg34}) and (\ref{Sumg23}) under the interchange of $a_i\leftrightarrow\wt{a}_i$. Furthermore, as before, we took advantage of the symmetry properties of the appearing 6j-symbols in order to be able to write the equation more compactly in the last line. Let us as a last step implement the spin labels of our states $\alp{J}{i}{M}$, namely  $j_{\sst 1}=j_{\sst 2}=j,j_{\sst 3}=j_{\sst 4}=\ell$ and $a_{\sst 4}=J$. Moreover, as before, we rewrite $(-1)^{-6j}$ as $(-1)^{+2j}$. 
\newline
Considering all this we have our final result in eqn (\ref{finq234})
\ba
\label{finq234}
\lefteqn{\alpc{J}{i}{M}\,q_{\sst 234}\,\alp{J}{\tilde{i}}{M}}\nonumber\\
&=&+\frac{1}{4}(-1)^{+2j+\ell+{\st J}}
\sqrt{2j(2j+1)(2j+2)}[2\ell(2\ell+1)(2\ell+2)]^{\frac{3}{2}}
\sqrt{(2\wt{a}_{\sst 2}+1)(2a_{\sst 2}+1)}\sqrt{2(a_{\sst 3}+1)(2\wt{a}_{\sst 3}+1)}\nonumber\\
&&
\SixJ{j}{j}{a_{\sst 2}}{1}{\wt{a}_{\sst 2}}{j}
\SixJ{{\st J}}{\ell}{a_{\sst 3}}{1}{\wt{a}_{\sst 3}}{\ell}
\left[(-1)^{a_{\sst 2}+\wt{a}_{\sst 3}}
\SixJ{a{\sst 3}}{\ell}{a_{\sst 2}}{1}{\wt{a}_{\sst 2}}{\ell}
\SixJ{\wt{a}_{\sst 2}}{\ell}{a_{\sst 3}}{1}{\wt{a}_{\sst 3}}{\ell}
-(-1)^{\wt{a}_{\sst 2}+a_{\sst 3}}
\SixJ{\wt{a}{\sst 3}}{\ell}{a_{\sst 2}}{1}{\wt{a}_{\sst 2}}{\ell}
\SixJ{a_{\sst 2}}{\ell}{a_{\sst 3}}{1}{\wt{a}_{\sst 3}}{\ell}\right]\nonumber\\
\ea
In the case of $\op{q}_{\sst 123}$ the matrix element can be expressed as
\ba
\label{Derq123}
\lefteqn{\alpc{J}{i}{M}\,q_{\sst 123}\,\alp{J}{\tilde{i}}{M}}\nonumber\\
&=&\alpc{J}{i}{M}\,[(J_{\sst 12})^2,(J_{\sst 23})^2]\,\alp{J}{\tilde{i}}{M}\nonumber\\
&=&\sum\limits_{\vec{g}''(12)}\Big\{\sum\limits_{\vec{g}(12)}g_{\sst 2}(12)(g_{\sst 2}(12)+1)\auf\vec{g}(12)\,|\,\vec{g}''(12)\zu\auf\vec{g}(12)\,|\,\alpha^{\st J}_{i}\,{\st M}\zu\times\nonumber\\
&&\hspace{1cm}
\times\sum\limits_{\vec{g}(23)}g_{\sst 2}(23)(g_{\sst 2}(23)+1)\auf\vec{g}(23)\,|\,\vec{g}''(12)\zu\auf\vec{g}(23)\,|\,\alpha^{\st J}_{\tilde{i}}\,{\st M}\zu\Big\}\nonumber\\
&& -\Big[\alp{J}{i}{M}\,\longleftrightarrow\,\alp{J}{\tilde{i}}{M}\Big].
\ea
In order to perform the sums appearing in the equation above, we use eqn (\ref{Sumg23}) and take advantage of 
\ba
\label{Sumg12}
\sum\limits_{\vec{g}(12)}g_{\sst 2}(12)(g_{\sst 2}(12)+1)\auf\vec{g}(12)\,|\,\vec{g}''(12)\zu\auf\vec{g}(12)\,|\,\alpha^{\st J}_{i}\,{\st M}\zu
&=&a_{\sst 2}(a_{\sst 2}+1)\prod\limits_{k=2}^{4}\delta_{g''_k,a_k}.
\ea
Hence, we obtain
\ba
\lefteqn{\alpc{J}{i}{M}\,q_{\sst 123}\,\alp{J}{\tilde{i}}{M}}\nonumber\\
&=&\sum\limits_{\vec{g}''(12)}
\Big\{
+\frac{1}{2}(-1)^{-j_{\sst 1}-j_{\sst 2}+j_{\sst 3}+1}(-1)^{\wt{a}_{\sst 2}-g''_{\sst2}+\wt{a}_{\sst 3}}
X(j_{\sst 2},j_{\sst 3})^{\frac{1}{2}}A(g''_{\sst 2},\wt{a}_{\sst 2})\nonumber\\
&&\hspace{1.15cm}
\SixJ{j_{\sst 1}}{j_{\sst 2}}{g''_{\sst 2}}{1}{\wt{a}_{\sst 2}}{j_{\sst 2}}
\SixJ{\wt{a}{\sst 3}}{j_{\sst 3}}{g''_{\sst 2}}{1}{\wt{a}_{\sst 2}}{j_{\sst 3}}\Big[a_{\sst 2}(a_{\sst 2}-1)\Big]
\delta_{g''_{\sst 3},\wt{a}_{\sst 3}}\delta_{g''_{\sst 4},\wt{a}_{\sst 4}}
\delta_{g''_{\sst 2},a_{\sst 2}}\delta_{g''_{\sst 3},a_{\sst 3}}\delta_{g''_{\sst 4},a_{\sst 4}}\nonumber\\
&&\hspace{1.1cm}
-\frac{1}{2}(-1)^{-j_{\sst 1}-j_{\sst 2}+j_{\sst 3}+1}(-1)^{a_{\sst 2}-g''_{\sst2}+a_{\sst 3}}
X(j_{\sst 2},j_{\sst 3})^{\frac{1}{2}}A(g''_{\sst 2},a_{\sst 2})\nonumber\\
&&\hspace{1.15cm}
\SixJ{j_{\sst 1}}{j_{\sst 2}}{g''_{\sst 2}}{1}{a_{\sst 2}}{j_{\sst 2}}
\SixJ{a{\sst 3}}{j_{\sst 3}}{g''_{\sst 2}}{1}{a_{\sst 2}}{j_{\sst 3}}\Big[\wt{a}_{\sst 2}(\wt{a}_{\sst 2}-1)\Big]
\delta_{g''_{\sst 3},a_{\sst 3}}\delta_{g''_{\sst 4},a_{\sst 4}}
\delta_{g''_{\sst 2},\wt{a}_{\sst 2}}\delta_{g''_{\sst 3},\wt{a}_{\sst 3}}\delta_{g''_{\sst 4},\wt{a}_{\sst 4}}\nonumber\\
&=&+\frac{1}{2}(-1)^{-j_{\sst 1}-j_{\sst 2}+j_{\sst 3}+1}(-1)^{\wt{a}_{\sst 2}-a_{\sst2}+a_{\sst 3}}
X(j_{\sst 2},j_{\sst 3})^{\frac{1}{2}}A(a_{\sst 2},\wt{a}_{\sst 2})
\SixJ{j_{\sst 1}}{j_{\sst 2}}{a_{\sst 2}}{1}{\wt{a}_{\sst 2}}{j_{\sst 2}}
\SixJ{a_{\sst 3}}{j_{\sst 3}}{a_{\sst 2}}{1}{\wt{a}_{\sst 2}}{j_{\sst 3}}\nonumber\\
&&\hspace{0.15cm}
\Big[a_{\sst 2}(a_{\sst 2}-1) - \wt{a}_{\sst 2}(\wt{a}_{\sst 2}-1)\Big]\delta_{a_{\sst 3},\wt{a}_{\sst 3}},
\ea
where we used again that the term proportional to $C(a,b)$ in eqn (\ref{Sumg23}) is cancelled in the first step. As we did before, we omitted the delta-function $\delta_{a_{\sst 4},\wt{a}_{\sst 4}}$, because $a_{\sst 4}$ is equal to the total angular momentum $J$ of our states $\alp{J}{i}{M}$ and therefore we consider only cases where $a_{\sst 4}=\wt{a}_{\sst 4}$ anyway. However, this is different for the intermediate coupling $a_{\sst 3},\wt{a}_{\sst 3}$. for this reason we have to consider $\delta_{a_{\sst 3},\wt{a}_{\sst 3}}$ in the above equation. Applying the above equation to our particular case where $j_{\sst 1}=j_{\sst 2}=j$ and $j_{\sst 3}=\ell$ yields
\ba
\label{finq123}
\lefteqn{\alpc{J}{i}{M}\,q_{\sst 123}\,\alp{J}{\tilde{i}}{M}}\nonumber\\
&=&
+\frac{1}{2}(-1)^{+2j+\ell+1}(-1)^{\wt{a}_{\sst 2}-a_{\sst2}+a_{\sst 3}}
X(j,\ell)^{\frac{1}{2}}A(a_{\sst 2},\wt{a}_{\sst 2})
\SixJ{j}{j}{a_{\sst 2}}{1}{\wt{a}_{\sst 2}}{j}
\SixJ{a_{\sst 3}}{\ell}{a_{\sst 2}}{1}{\wt{a}_{\sst 2}}{\ell}
\Big[a_{\sst 2}(a_{\sst 2}-1) - \wt{a}_{\sst 2}(\wt{a}_{\sst 2}-1)\Big]\delta_{a_{\sst 3},\wt{a}_{\sst 3}}.\nonumber\\
\ea
Again, we rewrote $(-1)^{-2j}$ as $(-1)^{+2j}$ that is allowed due to $2j\in\Zl$ and used the symmetry properties of the 6j-symbol where appropriate. Now, we consider the last triple $q_{\sst 124}$. The corresponding matrix element is given by
\ba
\label{Derq124}
\lefteqn{\alpc{J}{i}{M}\,q_{\sst 124}\,\alp{J}{\tilde{i}}{M}}\nonumber\\
&=&\alpc{J}{i}{M}\,[(J_{\sst 12})^2,(J_{\sst 24})^2]\,\alp{J}{\tilde{i}}{M}\nonumber\\
&=&\sum\limits_{\vec{g}''(12)}\Big\{\sum\limits_{\vec{g}(12)}g_{\sst 2}(12)(g_{\sst 2}(12)+1)\auf\vec{g}(12)\,|\,\vec{g}''(12)\zu\auf\vec{g}(12)\,|\,\alpha^{\st J}_{i}\,{\st M}\zu\times\nonumber\\
&&\hspace{1cm}
\times\sum\limits_{\vec{g}(24)}g_{\sst 2}(24)(g_{\sst 2}(24)+1)\auf\vec{g}(24)\,|\,\vec{g}''(12)\zu\auf\vec{g}(24)\,|\,\alpha^{\st J}_{\tilde{i}}\,{\st M}\zu\Big\}\nonumber\\
&& -\Big[\alp{J}{i}{M}\,\longleftrightarrow\,\alp{J}{\tilde{i}}{M}\Big].
\ea
The summation including $\vec{g}(24)$ can be performed and leads to \cite{CommJB}
\ba
\label{Sumg24}
\lefteqn{\sum\limits_{\vec{g}(24)}g_{\sst 2}(24)(g_{\sst 2}(24)+1)\auf\vec{g}(24)\,|\,\vec{g}''(12)\zu\auf\vec{g}(24)\,|\,\alpha^{\st J}_{i}\,{\st M}\zu}\nonumber\\
&=&
\Big[\frac{1}{2}(-1)^{-j_{\sst 1}-j_{\sst 2}+a_{\sst 4}}(-1)^{a_{\sst 2}-g''_{\sst 2}}(-1)^{+g''_{\sst 2}+a_{\sst 2}}
X(j_{\sst 2},j_{\sst 4})^{\frac{1}{2}}A(g''_{\sst 2},a_{\sst 2})A(g''_{\sst 3},a_{\sst 3})\nonumber\\
&&\hspace{0.15cm}
\SixJ{j_{\sst 1}}{j_{\sst 2}}{g''_{\sst 2}}{1}{a_{\sst 2}}{j_{\sst 2}}\SixJ{j_{\sst 3}}{g''_{\sst 2}}{g''_{\sst 3}}{1}{a_{\sst 2}}{a_{\sst 3}}
\SixJ{a_{\sst 4}}{j_{\sst 4}}{g''_{\sst 3}}{1}{a_{\sst 3}}{j_{\sst 4}}+C(j_{\sst 2},j_{\sst 4})\delta_{g''_{\sst 2},a_{\sst 2}}\delta_{g''_{\sst 3},a_{\sst 3}}\Big]
\delta_{g''_{\sst 4},a_{\sst 4}}.
\ea
Inserting eqn (\ref{Sumg12}) and (\ref{Sumg23}) into eqn (\ref{Derq124}) and using again that the term proportional to $C(a,b)$ is antisymmetric under the interchange of $\alp{J}{i}{M}\,\leftrightarrow\,\alp{J}{\tilde{i}}{M}$, we get
\ba
\lefteqn{\alpc{J}{i}{M}\,q_{\sst 124}\,\alp{J}{\tilde{i}}{M}}\nonumber\\
&=&\sum\limits_{\vec{g}''(12)}
\Big\{
+\frac{1}{2}(-1)^{-j_{\sst 1}-j_{\sst 2}+\wt{a}_{\sst 4}}(-1)^{\wt{a}_{\sst 2}-g''_{\sst 2}}(-1)^{+g''_{\sst 2}+\wt{a}_{\sst 2}}
X(j_{\sst 2},j_{\sst 4})^{\frac{1}{2}}A(g''_{\sst 2},\wt{a}_{\sst 2})A(g''_{\sst 3},\wt{a}_{\sst 3})\nonumber\\
&&\hspace{1.15cm}
\SixJ{j_{\sst 1}}{j_{\sst 2}}{g''_{\sst 2}}{1}{\wt{a}_{\sst 2}}{j_{\sst 2}}\SixJ{j_{\sst 3}}{g''_{\sst 2}}{g''_{\sst 3}}{1}{\wt{a}_{\sst 2}}{\wt{a}_{\sst 3}}
\SixJ{\wt{a}_{\sst 4}}{j_{\sst 4}}{g''_{\sst 3}}{1}{\wt{a}_{\sst 3}}{j_{\sst 4}}\Big[a_{\sst 2}(a_{\sst 2}-1)\Big]
\delta_{g''_{\sst 2},a_{\sst 2}}\delta_{g''_{\sst 3},a_{\sst 3}}\delta_{g''_{\sst 4},a_{\sst 4}}
\delta_{g''_{\sst 4},\wt{a}_{\sst 4}}\nonumber\\
&&\hspace{1.1cm}
-\frac{1}{2}(-1)^{-j_{\sst 1}-j_{\sst 2}+a_{\sst 4}}(-1)^{a_{\sst 2}-g''_{\sst 2}}(-1)^{+g''_{\sst 2}+a_{\sst 2}}
X(j_{\sst 2},j_{\sst 4})^{\frac{1}{2}}A(g''_{\sst 2},a_{\sst 2})A(g''_{\sst 3},a_{\sst 3})\nonumber\\
&&\hspace{1.15cm}
\SixJ{j_{\sst 1}}{j_{\sst 2}}{g''_{\sst 2}}{1}{a_{\sst 2}}{j_{\sst 2}}\SixJ{j_{\sst 3}}{g''_{\sst 2}}{g''_{\sst 3}}{1}{a_{\sst 2}}{a_{\sst 3}}
\SixJ{a_{\sst 4}}{j_{\sst 4}}{g''_{\sst 3}}{1}{a_{\sst 3}}{j_{\sst 4}}\Big[\wt{a}_{\sst 2}(\wt{a}_{\sst 2}-1)\Big]
\delta_{g''_{\sst 2},\wt{a}_{\sst 2}}\delta_{g''_{\sst 3},\wt{a}_{\sst 3}}\delta_{g''_{\sst 4},\wt{a}_{\sst 4}}
\delta_{g''_{\sst 4},a_{\sst 4}}\nonumber\\
&=&
+\frac{1}{2}(-1)^{-j_{\sst 1}-j_{\sst 2}+a_{\sst 4}}X(j_{\sst 2},j_{\sst 4})^{\frac{1}{2}}A(a_{\sst 2},\wt{a}_{\sst 2})A(a_{\sst 3},\wt{a}_{\sst 3})\nonumber\\
&&\hspace{1.15cm}
\SixJ{j_{\sst 1}}{j_{\sst 2}}{a_{\sst 2}}{1}{\wt{a}_{\sst 2}}{j_{\sst 2}}\SixJ{j_{\sst 3}}{a_{\sst 2}}{a_{\sst 3}}{1}{\wt{a}_{\sst 2}}{\wt{a}_{\sst 3}}
\SixJ{a_{\sst 4}}{j_{\sst 4}}{a_{\sst 3}}{1}{\wt{a}_{\sst 3}}{j_{\sst 4}}\Big[(-1)^{+2\wt{a}_{\sst 2}}a_{\sst 2}(a_{\sst 2}-1)-(-1)^{+2a_{\sst 2}}\wt{a}_{\sst 2}(\wt{a}_{\sst 2}+1)\Big].
\ea
In the last line we took advantage of the symmetry properties of the 6j-symbol and moreover used that $(-1)^{\wt{a}_{\sst 2}-a_{\sst 2}}=(-1)^{a_{\sst 2}-\wt{a}_{\sst 2}}$ as $\wt{a}_{\sst 2}-a_{\sst 2}\in\Zl$. In our special situation where $j_{\sst 1}=j_{\sst 2}=j$ the value of the intermediate coupling $a_{\sst 2}$ and $\wt{a}_{\sst 2}$ can only be an integer and thus $(-1)^{+2\wt{a}_{\sst 2}}=(-1)^{+2a_{\sst 2}}=+1$. Accordingly, we can completely neglect these factors and obtain
\ba
\label{finq124}
\lefteqn{\alpc{J}{i}{M}\,q_{\sst 124}\,\alp{J}{\tilde{i}}{M}}\nonumber\\
&=&
+\frac{1}{2}(-1)^{+2j+{\st J}}X(j,\ell)^{\frac{1}{2}}A(a_{\sst 2},\wt{a}_{\sst 2})A(a_{\sst 3},\wt{a}_{\sst 3})
\SixJ{j}{j}{a_{\sst 2}}{1}{\wt{a}_{\sst 2}}{j}\SixJ{\ell}{a_{\sst 2}}{a_{\sst 3}}{1}{\wt{a}_{\sst 2}}{\wt{a}_{\sst 3}}
\SixJ{a_{\sst 4}}{\ell}{a_{\sst 3}}{1}{\wt{a}_{\sst 3}}{\ell}\Big[a_{\sst 2}(a_{\sst 2}-1)-\wt{a}_{\sst 2}(\wt{a}_{\sst 2}+1)\Big],\nonumber\\
\ea
where we additionally inserted $j_{\sst 3}=j_{\sst 4}=\ell$ and $a_{\sst 4}=J$ in the equation above.
\section{Case $\lL\op{\tilde{E}}^{\sst I,AL}_{k,{\sst tot}}(S_t)$: \newline
Detailed Calculation of the Matrix Elements of $\op{O}_{1}=\op{O}_2=V^2_{\sst AL}$}%
\label{CaseV2}
The aim of this section is to calculate the four matrix elements  $\alpc{0}{2}{0}\,\op{V}_{\sst AL}^2\,\alp{0}{1}{0}$ and $\alpc{1}{i}{M}\,\op{V}_{\sst AL}^2\,\alp{1}{1}{M}$ with $i=2,3,4$. As has been mentioned before, in order to calculate matrix elements of $\op{V}_{\sst AL}^2$ we first have to calculate the matrix elements of $\op{Q}^{\sst AL}_v$ and derive the eigenvalues und eigenvectors for $\op{Q}^{\sst AL}_v$. If $\lambda_{Q}$ is an eigenvalue of $\op{Q}^{\sst AL}_v$ with corresponding eigenvector $|\phi\zu$, then $|\lambda_{Q}|$ is an eigenvalue of $V^2$ with the same eigenvector. Consequently, we have to calculate all possible matrix elements of $\op{Q}^{\sst AL}_v$ for each fixed total angular momentum $J$. For this reason, we are not able to evaluate matrix elements for the case of arbitrary spin representation, as we could do in the case of $\op{O}_{1}=\op{O}_2=\op{V }_{\sst AL}\op{\cal S}\op{V}_{\sst AL}=\op{Q}^{\sst AL}_v$. In that case we needed only particular matrix elements of $\op{Q}^{\sst AL}$ but not the knowledge of the spectrum of  $\op{Q}^{\sst AL}_v$ itself. However, we will calculate the matrix elements of $\op{V}_{\sst AL}^2$ for the case of $\ell=\frac{1}{2},1$ here. Fortunately, they already show the major difference between the case of $\op{O}_{1}=\op{O}_2=\op{Q}^{\sst AL}_v$ and $\op{O}_{1}=\op{O}_2=\op{V}_{\sst AL}^2$.
\newline
When we calculated the five necessary matrix elements $\alpc{0}{2}{0}\,\op{Q}^{\sst AL}_v\,\alp{0}{1}{0},\alpc{1}{2}{M}\,\op{Q}^{\sst AL}_v\,\alp{1}{1}{M}$ and $\alpc{1}{i}{M}\,\op{Q}^{\sst AL}_v\,\alp{1}{1}{M}$ in the case of $\op{O}_{\sst 1}=\op{O}_2=\op{Q}^{\sst AL}_v$ with $i$ being $2,3,4$ in section \ref{QVSV}, we took advantage of the fact that the intermediate coupling $a_{\sst 2}$ of $\alp{1}{1}{M}$ is identical to zero and therefore we have $J_{e_{\sst 1}}=-J_{e_{\sst 2}}$. Furthermore the orientations of the two triples $\{e_{\sst 1},e_{\sst 3},e_{\sst 4}\},\{e_{\sst 2},e_{\sst 3},e_{\sst 4}\}$ were exactly opposite to each other. Accordingly, we only needed to consider one triple, e.g. $\op{q}_{\st 134}$, and multiply the result by a factor of 2, because the second triple had exactly the same contribution as the first one. Now, in contrast, we will also have to consider matrix elements where the incoming state has intermediate couplings $a_{\sst 2}$ different from zero. Consequently, in these cases we will have to consider the contribution of the second triple exactly, as it might not just be a trivial factor of 2.
\newline
By comparing the formulae for general matrix elments of $\op{q}_{\sst 134}$ in eqn (\ref{RSq134}) and of $\op{q}_{\sst 234}$ in eqn (\ref{RSq234}) respectively, we note that the only difference between these two formulae is due to different pre-factors in  the square bracket in front of the 6j-symbols. Before we can actually calculate the matrix elements, we have to know how the corresponding Hilbert space looks like. 
\subsection{Matrix Elements for the Case of a Spin-$\frac{1}{2}$-Representation}%
Let us begin with the case $\ell=\frac{1}{2}$ and a total angular momentum $J=0$. From eqn (\ref{HJ0}) we can easily extract the basis states of this Hilbert space
\ba
\label{Hj05J0}
\alp{0}{1}{0}&:=&|\,a_{\sst 1}=j\,a_{\sst 2}=0\,a_{\sst 3}=\frac{1}{2}\,J=0\zu\nonumber\\
\alp{0}{2}{0}&:=&|\,a_{\sst 1}=j\,a_{\sst 2}=1\,a_{\sst 3}=\frac{1}{2}\,J=0\zu.
\ea 
With the Hilbert space being only two-dimensional and the fact that $\op{Q}^{\sst AL}_v$ and consequently $\op{q}_{\sst 134},\op{q}_{\sst 234}$ are anti-symmetric, we know that $\alpc{0}{2}{0}\,\op{Q}^{\sst AL}_v\alp{0}{1}{0}=-\alpc{0}{1}{0}\,\op{Q}^{\sst AL}_v\,\alp{0}{2}{0}$ are the only non-vanishing matrix elements. Moreover, we have  $\epsilon(e_{\sst 1},e_{\sst 3},e_{\sst 4})\alpc{0}{2}{0}\,\op{q}_{\sst 134}\,\alp{0}{1}{0}=\epsilon(e_{\sst 2},e_{\sst 3},e_{\sst 4})\alpc{0}{2}{0}\,\op{q}_{\sst 234}\alp{0}{1}{0}$, because the intermediate coupling $a_{\sst 2}$ of $\alp{0}{1}{0}$ is zero. Therefore, we obtain the following matrix structure for $\op{Q}^{\sst AL}_v$
\be
\op{Q}^{\st J=0}_{\sst AL}=\left(\begin{array}{cc} 0& -i a\\ i a & 0\end{array}\right),
\ee
where $a:=(\lp^6\frac{3!}{2}C_{reg})\frac{1}{2}\sqrt{j(j+1)}$. We labelled the rows of $\op{Q}^{\st J=0}_{\sst AL}$ by $\alp{J}{i}{M}$ and columns by $\alp{J}{\tilde{i}}{M}$. The eigenvalues of $\op{Q}^{\sst AL}_v$ are given by $\lambda_{1}=-a,\lambda_2=+a$ with corresponding eigenvectors $\vec{v}_1=(i,1),\vec{v}_2=(-i,1)$. Hence, for $\op{V}_{\sst AL}^2$ we have one degenerated eigenvalue $\lambda=|a|$ and the two corresponding eigenvector components $\vec{v}_1,\vec{v}_2$ in the basis $\{\alp{0}{1}{0},\alp{0}{2}{0}\}$. The matrix element $\alpc{0}{2}{0}\,\op{V}_{\sst AL}^2\,\alp{0}{1}{0}$ is thus given by
\ba
\alpc{0}{2}{0}\,\op{V}_{\sst AL}^2\alp{0}{1}{0}&=&\sum\limits_{k=1}^{2}\alpc{0}{2}{0}\,\op{V}_{\sst AL}^2\,|\,\vec{e}_k\zu\auf \vec{e}_k\,\alp{0}{1}{0}=0.\\
 \ea
 Here the vectors $\vec{e}_k$ denote the corresponding normed eigenvectors of $\op{V}_{\sst AL}^2$.
The surprising issue is that in contrast to the matrix element of $\op{Q}^{\sst AL}_v$, the analogous matrix element of $\op{V}_{\sst AL}^2$ vanishes. In this special situation where we chose $\ell=\frac{1}{2}$ and the total angular momentum $J$ to be zero, we realise that  $\op{Q}^{\sst AL}_v$ purely consists of off-diagonal entries, while $\op{V}_{\sst AL}^2$ is a diagonal matrix. In order to calculate the  remaining matrix elements, we have to consider the Hilbert space for a total angular momentum of $J=1$ in the case of $\ell=\frac{1}{2}$. Because we consider the special case of $\ell=\frac{1}{2}$ the intermediate coupling $a_{\sst 3}=\ell-1$ is not sensible here. Therefore the matrix element  $\alpc{1}{2}{M}\,\op{V}_{\sst AL}^2\,\alp{1}{1}{M}$ does not exist. Thus, we will only have two remaining matrix elements. Using eqn (\ref{HJ1}) for this purpose, we end up with a $4\times 3$-dimensional Hilbert space. 
\ba
\label{Hj05J1}
\alp{1}{1}{M}&:=&|\,a_{\sst 1}=j\,a_{\sst 2}=0\,a_{\sst 3}=\frac{1}{2}\,J=1\zu\nonumber\\
\alp{1}{3}{M}&:=&|\,a_{\sst 1}=j\,a_{\sst 2}=1\,a_{\sst 3}=\frac{1}{2}\,J=1\zu\nonumber\\
\alp{1}{4}{M}&:=&|\,a_{\sst 1}=j\,a_{\sst 2}=1\,a_{\sst 3}=\frac{3}{2}\,J=1\zu\nonumber\\
\alp{1}{5}{M}&:=&|\,a_{\sst 1}=j\,a_{\sst 2}=2\,a_{\sst 3}=\frac{3}{2}\,J=1\zu,
\ea
where we skipped the number two in labelling the states in order to keep our notation consistent with the former calculations.
With the states $\alp{1}{1}{M}$ being orthogonal for different values of the magnetic quantum number $M$ and the knowledge that $\op{Q}^{\sst AL}_v$ does not change the magnetic quantun number, we can treat the calculation  separately for each value of $M=\{-1,0,1\}$. Furthermore, we know that the result is equal for each value of $M$.
\newline
Thus, we have a $4\times 4$-matrix, but as $\op{Q}^{\sst AL}_v$ is anti-symmetric, its diagonal entries are zero and $(\op{Q}^{\sst AL}_v)_{\sst AB}=-(\op{Q}^{\sst AL}_v)_{\sst BA}$. Hence, we only have to calculate 6 different matrix elements. Two out of these 6 matrix elements have already been calculated before and can be extracted from eqn (\ref{Resq134}) if we set $\ell=\frac{1}{2}$. For both matrix elements, we have $\epsilon(e_{\sst 1},e_{\sst 3},e_{\sst 4})\op{q}_{\sst 134}=\epsilon(e_{\sst 2},e_{\sst 3},e_{\sst 4})\op{q}_{\sst 234}$, so that the contribution coming from the second triple is again only a factor of 2. So, we are left with four matrix elements that have to be evaluated, namely $\alpc{1}{5}{M}\,\op{Q}^{\sst AL}_v\,\alp{1}{1}{M},\alpc{1}{4}{M}\,\op{Q}^{\sst AL}_v\,\alp{1}{3}{M},\alpc{1}{5}{M}\,\op{Q}^{\sst AL}_v\,\alp{1}{3}{M}$ and $\alpc{1}{5}{M}\,\op{Q}^{\sst AL}_v\,\alp{1}{4}{M}$. By simply looking at eqn (\ref{RSq234}) and (\ref{RSq134}) we see that  $\alpc{1}{5}{M}\,\op{Q}^{\sst AL}_v\,\alp{1}{1}{M}$ vanishes, because $\wt{a}_{\sst 2}-a_{\sst 2}=2$ here and therefore the 6j-symbols  in the corresponding square brackets will be zero. Consequently, the matrix elements of $\op{q}_{\sst 134}$ and $\op{q}_{\sst 234}$ are zero and thus the corresponding matrix elements of $\op{Q}^{\sst AL}_v$ disappear. In the case of $\alpc{1}{4}{M}\,\op{Q}^{\sst AL}_v\,\alp{1}{3}{M}$ we have $\wt{a}_{\sst 2}=a_{\sst 2}$. Implementing this condition into eqn (\ref{RSq234}) and (\ref{RSq134}), we realise that both equations become identical. Accordingly, we get  $\epsilon(e_{\sst 1},e_{\sst 3},e_{\sst 4})\op{q}_{\sst 134}=-\epsilon(e_{\sst 2},e_{\sst 3},e_{\sst 4})\op{q}_{\sst 234}$ which leads to a vanishing matrix element for $\op{Q}^{\sst AL}_v$. With all this in mind, we end up with the following expression for $\op{Q}^{\sst AL}_v$
\be
\op{Q}^{\st J=1}_{\sst AL}=\left(\begin{array}{cccc} 0 & +i a & -i\sqrt{2}a & 0 \\ -ia & 0 & 0 & +\frac{ib}{\sqrt{2}}\\
                                 +i\sqrt{2}a & 0 & 0 & -ib\\ 0 & -\frac{ib}{\sqrt{2}} & +ib & 0  \end{array}\right),
\ee
where we defined $a:=(\lp^6\frac{3!}{2}C_{reg})\frac{2}{3}\sqrt{j(j+1)}$ and $b:=(\lp^6\frac{3!}{2}C_{reg})\frac{2}{3}\sqrt{4j(j+1)-3}$. In this case the eigenvalues of $\op{Q}^{\sst AL}_v$ are $\lambda_1=\lambda_2=0$ and  $\lambda_3=-\lambda_4=-\sqrt{\frac{3}{2}}\sqrt{2a^2+b^2}=:-\lambda$. The corresponding eigenvectors are given by
\ba
\vec{v}_1&=&(\frac{b}{\sqrt{2a}},0,0,1)\nonumber\\
\vec{v}_3&=&(0,\sqrt{2},0,1)\nonumber\\
\vec{v}_4&=&\frac{1}{b}(-\sqrt{2}a,-i\frac{\sqrt{2}}{3}\lambda,+i\lambda,b)\nonumber\\
\vec{v}_5&=&\frac{1}{b}(-\sqrt{2}a,+i\frac{\sqrt{2}}{3}\lambda,-i\lambda,b).
\ea
With the first two eigenvalues being identical to zero, we do not have to consider them when we calculate the matrix element of $\op{V}^2$. Hence, we obtain
\ba
\alpc{1}{3}{M}\,\op{V}_{\sst AL}^2\alp{1}{1}{M}&=&\sum\limits_{k=3}^{4}\alpc{1}{3}{M}\,\op{V}_{\sst AL}^2\,|\,\vec{e}_k\zu\auf \vec{e}_k\,\alp{1}{1}{M}=0
\ea
\ba
\alpc{1}{4}{M}\,\op{V}_{\sst AL}^2\alp{1}{1}{M}&=&\sum\limits_{k=3}^{4}\alpc{1}{4}{M}\,\op{V}_{\sst AL}^2\,|\,\vec{e}_k\zu\auf \vec{e}_k\,\alp{1}{1}{M}=0.                                     
\ea
As in the case of $J=0$ all matrix elements that occur in the action of the operator $^{\frac{1}{2}}\op{\tilde{E}}^{\sst I,AL}_{k,{\sst tot}}(S_t)$ vanish. Consequently the whole matrix element $\betc{\wt{j}_{\sst 12}}{\wt{m}_{\sst 12}}\,^{\frac{1}{2}}\op{\tilde{E}}^{\sst I,AL}_{k,{\sst tot}}(S_t)\,\bet{j_{\sst 12}}{m_{\sst 12}}$ vanishes. 
\newline
In order to see whether the vanishing of  $\betc{\wt{j}_{\sst 12}}{\wt{m}_{\sst 12}}\,^{\frac{1}{2}}\op{\tilde{E}}^{\sst I,AL}_{k,{\sst tot}}(S_t)\,\bet{j_{\sst 12}}{m_{\sst 12}}$ is somehow connected with the fact 
 that we chose the most easiest case where $\ell=\frac{1}{2}$, we will investigate the matrix elements for the
  case of $\ell=1$ as well.
\subsection{Matrix Elements for the Case of a Spin-1-Representation}%
In the case that both additional edges carry a spin-1-representation ($\ell=1$), the Hilbert space belonging to a total angular momentum $J=0$ is 3-dimensional
\ba
\label{Hj1J0}
\alp{0}{1}{0}&:=&|\,a_{\sst 1}=j\,a_{\sst 2}=0\,a_{\sst 3}=1\,J=0\zu\nonumber\\
\alp{0}{2}{M}&:=&|\,a_{\sst 1}=j\,a_{\sst 2}=1\,a_{\sst 3}=1\,J=0\zu\nonumber\\
\alp{0}{3}{M}&:=&|\,a_{\sst 1}=j\,a_{\sst 2}=2\,a_{\sst 3}=1\,J=0\zu.
\ea 
Again, the matrix element $\alpc{0}{3}{0}\,\op{Q}\,\alp{0}{1}{0}$ vanishes, because $\Delta a_{\sst 2}:=|\wt{a}_{\sst 2}-a_{\sst 2}|>1$. Consequently the 6j-symbols including $\wt{a}_{\sst 2}$ and $a_{\sst 2}$ becomes zero. Thus the whole matrix element is zero. Considering the matrix element  $\alpc{0}{3}{0}\,\op{Q}^{\sst AL}_v\,\alp{0}{2}{0}$ we see that we have $\wt{a}_{\sst 3}=a_{\sst 3}$ and $\wt{a}_{\sst 2}=a_{\sst 2}+1$ here. Inserting this into eqn (\ref{RSq234}) and (\ref{RSq134}), we get $\alpc{0}{3}{0}\,\op{q}_{\sst 134}\,\alp{0}{1}{0}=-\alpc{0}{3}{0}\,\op{q}_{\sst 234}\,\alp{0}{1}{0}$, so that we only have to calculate one of the triples and multiply it by two. Hence, the operator $\op{Q}^{\sst AL}_v$ can be described by the following matrix
\be
\op{Q}^{\st J=0}=\left(\begin{array}{ccc} 0& -i a & 0\\ +i a & 0 & -ib \\ 0 & -ib & 0\end{array}\right),
\ee
where $a:=(\lp^6\frac{3!}{2}C_{reg})\frac{4}{\sqrt{3}}\sqrt{2}\sqrt{j(j+1)}$ and $b:=(\lp^6\frac{3!}{2}C_{reg})\frac{4}{\sqrt{3}}\sqrt{4j(j+1)-3}$. The eigenvalues are given by $\lambda_1=0,\lambda_2=-\sqrt{a^2+b^2},\lambda_3=+\sqrt{a^2+b^2}=:\lambda$. The corresponding eigenvectors can be found in the equation below
\ba
\vec{v}_1&=&(\frac{b}{\sqrt{a}},0,1)\nonumber\\
\vec{v}_2&=&(-\frac{a}{b},\frac{i}{b}\lambda,1)\nonumber\\
\vec{v}_3&=&(-\frac{a}{b},-\frac{i}{b}\lambda,1).
\ea
With the eigenvectors again having only purely real and purely imaginary entries  we can again guess that the matrix elements of $\op{V}_{\sst AL}^2$ will vanish. This is indeed the case, as can be seen in the following lines
\ba
\alpc{0}{2}{0}\,\op{V}_{\sst AL}^2\,\alp{0}{1}{0}&=&\sum\limits_{k=2}^{3}\alpc{0}{2}{0}\,\op{V}_{\sst AL}^2\,|\,\vec{e}_k\zu\auf \vec{e}_k\,\alp{0}{1}{0}=0.
\ea
Thus, as long as we have eigenvectors that do have only purely imaginary and purely real components and we are furthermore forced to consider matrix element of $\alpc{J}{i}{M}\,\op{V}_{\sst AL}^2\,\alp{J}{1}{M}$ such that one of the states has an imaginary expansion coefficient in terms of the eigenvectors, while the other has a real expansion coefficient, we will obtain a vanishing matrix element for $\op{V}_{\sst AL}^2$. Note that this is not the case for the operator $\op{Q}^{\sst AL}_v$, because there the eigenvectors have different eigenvalues $+\lambda$ and $-\lambda$. Accordingly, the corresponding terms would not be canceled by each other, but would just be added up.
\newline
Let us consider a total angular momentum of $J=1$ now and investigate whether we will get the same behaviour of the eigenvectors as well. For $J=1$ the associated Hilbert space is already $(7\times 3)$-dimensional.
\ba
\label{Hj1J1}
\alp{1}{1}{M}&:=&|\,a_{\sst 1}=j\,a_{\sst 2}=0\,a_{\sst 3}=1\,J=1\zu\nonumber\\
\alp{1}{2}{M}&:=&|\,a_{\sst 1}=j\,a_{\sst 2}=1\,a_{\sst 3}=0\,J=1\zu\nonumber\\
\alp{1}{3}{M}&:=&|\,a_{\sst 1}=j\,a_{\sst 2}=1\,a_{\sst 3}=1\,J=1\zu\nonumber\\
\alp{1}{4}{M}&:=&|\,a_{\sst 1}=j\,a_{\sst 2}=1\,a_{\sst 3}=2\,J=1\zu\nonumber\\
\alp{1}{5}{M}&:=&|\,a_{\sst 1}=j\,a_{\sst 2}=2\,a_{\sst 3}=1\,J=1\zu\nonumber\\
\alp{1}{6}{M}&:=&|\,a_{\sst 1}=j\,a_{\sst 2}=2\,a_{\sst 3}=2\,J=1\zu\nonumber\\
\alp{1}{7}{M}&:=&|\,a_{\sst 1}=j\,a_{\sst 2}=3\,a_{\sst 3}=2\,J=1\zu.
\ea
In order to minimise the amount of computation, we will discuss some particular matrix elements in advance, especially those for which we can read off the result easily from the eqn (\ref{RSq234}) and (\ref{RSq134}).With $a_{\sst 2}$ being zero for $\alp{1}{1}{M}$, we know that for all matrix elements 
\be
\epsilon(e_{\sst 1},e_{\sst 3},e_{\sst 4})\alpc{1}{i}{M}\,\op{q}_{\sst 134}\,\alp{1}{1}{M}=\epsilon(e_{\sst 2},e_{\sst 2},e_{\sst 3})\alpc{1}{i}{M}\,\op{q}_{\sst 234}\,\alp{1}{1}{M},\quad i=2,..,7.
\ee
Furthermore, we have $\alpc{1}{i}{M}\,\op{q}_{\sst 134}\,\alp{1}{1}{M}=0$ for $i>4$, because then $\Delta a_{\sst 2}=|\wt{a}_{\sst 2}-a_{\sst 2}|>1$. For the same reason the matrix elements $\alpc{1}{7}{M}\,\op{Q}^{\sst AL}_v\,\alp{1}{i}{M}=0$ with $i=1,..,4$ vanish. As $\Delta a_{\sst 3}=|\wt{a}_{\sst 3}-a_{\sst 3}|>1$ for $\alpc{1}{i}{M}\,\op{Q}^{\sst AL}_v\,\alp{1}{2}{M}=0$ for $i=4,6,7$ these matrix elements are zero as well. Then there can be found several matrix elements where $\wt{a}_{\sst 2}=a_{\sst 2}$ and $\wt{a}_{\sst 3}=a_{\sst 3}+1$. In this case we get $\epsilon(e_{\sst 1},e_{\sst 3},e_{\sst 4})\alpc{1}{i}{M}\,\op{q}_{\sst 134}\,\alp{1}{i+1}{M}=-\epsilon(e_{\sst 2},e_{\sst 2},e_{\sst 3})\alpc{1}{i+1}{M}\,\op{q}_{\sst 234}\,\alp{1}{i}{M}$ with $i=2,3,5$ and consequently the matrix element of $\op{Q}^{\sst AL}_v$ is zero for this particular combination of states $\alp{1}{i}{M}$. Considering these arguments, we obtain the following kind of matrix for $\op{Q}^{\sst AL}_v$
\be
\op{Q}^{\st J=1}_{\sst AL}=\left(\begin{array}{ccccccc} 
0 & -i\frac{8}{3}\sqrt{2}a& -i2\sqrt{\frac{2}{3}}a  & -i\frac{2}{3}\sqrt{10}a & 0 & 0 & 0\\
+i\frac{8}{3}\sqrt{2}a& 0 & 0 & 0& +i\frac{4}{3}b & 0 & 0\\
+i2\sqrt{\frac{2}{3}}a & 0 & 0 & 0& -i\frac{2}{\sqrt{3}}b & 0 & 0\\
+i\frac{2}{3}\sqrt{10}a& 0 & 0 & 0& -i\frac{4}{3\sqrt{5}}b & -i2\sqrt{\frac{3}{5}}b & 0\\
0 & -i\frac{4}{3}b & +\frac{2}{\sqrt{3}}b & +i\frac{4}{3\sqrt{5}}b & 0 & 0 & +i2\sqrt{\frac{6}{5}}c\\
0 & 0 & 0 & +i2\sqrt{\frac{3}{5}}b & 0 & 0 & -i6\sqrt{\frac{2}{5}}c \\
0 & 0 & 0 & 0 & -i2\sqrt{\frac{6}{5}}c & +i6\sqrt{\frac{2}{5}}c & 0
\end{array}\right),
\ee
where we introduced 
\be
a:=(\lp^6\frac{3!}{2}C_{reg})\sqrt{j(j+1)}\quad b:=(\lp^6\frac{3!}{2}C_{reg})\sqrt{4j(j+1)-3}\quad c:=(\lp^6\frac{3!}{2}C_{reg})\sqrt{j(j+1)-2}.
\ee
The seven eigenvalues of $\op{Q}^{\st J=1}_{\sst AL}$ are 
\ba
\lambda_1&=&0\nonumber\\
\lambda_2&=&-2(\lp^6\frac{3!}{2}C_{reg})\sqrt{4j(j+1)-3}=-\lambda_3\nonumber\\
\lambda_4&=&-(\lp^6\frac{3!}{2}C_{reg})\sqrt{24j(j+1)-2(11+\sqrt{121+8j(j+1)(2j(j+1)-5))}}=-\lambda_5\nonumber\\
\lambda_6&=&-(\lp^6\frac{3!}{2}C_{reg})\sqrt{2}\sqrt{12j(j+1)-11+\sqrt{121+8j(j+1)(2j(j+1)-5)}}=-\lambda_7.
\ea 
The corresponding eigenvectors can be given in the following form
\ba
\vec{v}_1&=&(0,0,\gamma_1,\delta_1,0,0,1)\nonumber\\
\vec{v}_2&=&(0,-i\beta_2,+i,+i\delta_2,\epsilon_2,1,0)\nonumber\\
\vec{v}_3&=&(0,+i\beta_2,-i,-i\delta_2,\epsilon_2,1,0)\nonumber\\
\vec{v}_4&=&(+i\alpha_3,\beta_3,\gamma_3,\delta_3,-i\epsilon_3,+i\phi,1)\nonumber\\
\vec{v}_5&=&(-i\alpha_3,\beta_3,\gamma_3,\delta_3,+i\epsilon_3,-i\phi,1)\nonumber\\
\vec{v}_6&=&(+i\alpha_4,\beta_4,\gamma_4,\delta_4,-i\epsilon_4,+i\phi,1)\nonumber\\
\vec{v}_7&=&(-i\alpha_4,\beta_4,\gamma_4,\delta_4,+i\epsilon_4,-i\phi,1).
\ea
All Greek letters appearing in the equation above are supposed to represent real numbers. Our aim is to calculate the matrix elements $\alpc{1}{i}{M}\,\op{V}_{\sst AL}^2\,\alp{1}{1}{M}$ where $i=2,3,4$. Hence, similar to the calculations before we have to expand the states $\alp{1}{i}{M}$ in terms of the eigenvectors $\vec{e}_k$ that are the normed versions of the vectors $\vec{v}_k$. But only by looking at the structure of the eigenvectors $\vec{e}_k$ we can already read off that the three matrix elements will vanish for the following reasons: First of all, the first 3 eigenvectors do not contribute to the matrix elements at all, because their expansion coefficient for $\alp{1}{1}{M}$ is zero. Additionally, we have $\vec{e}_4^*=\vec{e}_5$ and $\vec{e}_6^*=\vec{e}_7$. As the expansion coefficient for $\alp{1}{1}{M}$ is purely imaginary, while the one for $\alp{1}{i}{M}$ with $i=2,3,4$ is real and moreover the two eigenvectors have  the same eigenvalue, namely $|\lambda_3|$ and $|\lambda_5|$ respectively, the contribution of $\vec{e}_4$ cancels the contribution of $\vec{e}_5$. The same is true for $\vec{e}_6$ and $\vec{e}_7$. Accordingly, we get
\be
\alpc{1}{2}{M}\,\op{V}_{\sst AL}^2\,\alp{1}{1}{M}=\alpc{1}{3}{M}\,\op{V}_{\sst AL}^2\,\alp{1}{1}{M}=\alpc{1}{4}{M}\,\op{V}_{\sst AL}^2\,\alp{1}{1}{M}=0.
\ee
Unfortunately, we are not able to repeat the analysis for arbitrary spin representation $\ell$, because the matrices representing $\op{Q}$ cannot be solved analytically anymore. Nevertheless, as the structure of the basis states in the Hilbert spaces stays the same, only the amount of states is changed, we would guess that the eigenvalues and eigenvectors look analogous also in the general case. Hence, we would expect a vanishing of the matrix elements of $\op{V}_{\sst AL}^2$ that are contained in $\betc{\wt{j}_{\sst 12}}{\wt{m}_{\sst 12}}\,^{1}\op{\tilde{E}}^{\sst I,AL}_{k,{\sst tot}}(S_t)\,\bet{j_{\sst 12}}{m_{\sst12}}$ and therefore expect that the result of $\betc{\wt{j}_{\sst 12}}{\wt{m}_{\sst 12}}\,^{1}\op{\tilde{E}}^{\sst I,AL}_{k,{\sst tot}}(S_t)\,\bet{j_{\sst 12}}{m_{\sst12}}$ is zero. In any case since the choice of $\ell$ should not be important in the semiclassical limit of large $j$, we can rule out the choice $\op{V}_{\sst AL}^2$ already based on the result of the present section.
\section{Detailed Calculation in the Case of the Volume Operator 
$\op{V}_{\sst RS}$ Introduced by Rovelli and Smolin} %
\label{RSOp}
As a first step we will derive the explicit expressions of the operators $\op{O}_1$ and $\op{O}_2$  in the case of $\lL\op{\tilde{E}}^{\sst I,RS}_{k,{\sst tot}}(S_t)$ and $\lL\op{\tilde{E}}^{\sst II,RS}_{k,{\sst tot}}(S_t)$, respectively shown in eqn (\ref{C1RSO1O2}) and in eqn (\ref{C2RSO1O2}), respectively.
Let us begin with $\lL\op{\tilde{E}}^{\sst I,RS}_{k,{\sst tot}}(S_t)$. Apart from some prefactors including numbers, that are not important for our argument, the precise expression of $\lL\op{\tilde{E}}^{\sst I,RS}_{k,{\sst tot}}(S_t)$ is given by a sum consisting of 8 terms
\ba
\label{Ek1eins}
\lL\op{\tilde{E}}^{\sst I,RS}_{k,{\sst tot}}(S_t) &\propto&\pi_{\ell}(\tau_k)_{\sst BC}\pi_{\ell}(\epsilon)_{\sst EI}\Big[ \nonumber\\
&&
+\pi_{\ell}(\epsilon)_{\sst FC}\Big(
+\op{\pi}_{\ell}(h^{\dagger}_{e_{\sst 4}})_{\sst FG}\op{\pi}_{\ell}(h^{\dagger}_{e_{\sst 3}})_{\sst BA }\op{V}_{\sst RS}\op{V}_{\sst RS}\op{\pi}_{\ell}(h_{e_{\sst 4}})_{\sst IG }\op{\pi}_{\ell}(h_{e_{\sst 3}})_{\sst EA }\nonumber\\
&&\hspace{1.7cm}
+\op{\pi}_{\ell}(h^{\dagger}_{e_{\sst 4}})_{\sst FG}\op{V}_{\sst RS}\op{\pi}_{\ell}(h^{\dagger}_{e_{\sst 3}})_{\sst BA }\op{\pi}_{\ell}(h_{e_{\sst 4}})_{\sst IG }\op{V}_{\sst RS}\op{\pi}_{\ell}(h_{e_{\sst 3}})_{\sst EA }\nonumber\\
&&\hspace{1.7cm}
-\op{\pi}_{\ell}(h^{\dagger}_{e_{\sst 4}})_{\sst FG}\op{V}_{\sst RS}\op{\pi}_{\ell}(h^{\dagger}_{e_{\sst 3}})_{\sst BA }\op{V}_{\sst RS}\op{\pi}_{\ell}(h_{e_{\sst 4}})_{\sst IG }\op{\pi}_{\ell}(h_{e_{\sst 3}})_{\sst EA }\nonumber\\
&&\hspace{1.7cm}
-\op{\pi}_{\ell}(h^{\dagger}_{e_{\sst 4}})_{\sst FG}\op{\pi}_{\ell}(h^{\dagger}_{e_{\sst 3}})_{\sst BA }\op{V}_{\sst RS}\op{\pi}_{\ell}(h_{e_{\sst 4}})_{\sst IG}\op{V}_{\sst RS}\op{\pi}_{\ell}(h_{e_{\sst 3}})_{\sst EA }\Big)\nonumber\\
&&
-\pi_{\ell}(\epsilon)_{\sst FB}\Big(
+\op{\pi}_{\ell}(h^{\dagger}_{e_{\sst 4}})_{\sst IG}\op{\pi}_{\ell}(h^{\dagger}_{e_{\sst 3}})_{\sst EA }\op{V}_{\sst RS}\op{V}_{\sst RS}\op{\pi}_{\ell}(h_{e_{\sst 4}})_{\sst FG }\op{\pi}_{\ell}(h_{e_{\sst 3}})_{\sst CA }\nonumber\\
&&\hspace{1.7cm}
+\op{\pi}_{\ell}(h^{\dagger}_{e_{\sst 4}})_{\sst IG}\op{V}_{\sst RS}\op{\pi}_{\ell}(h^{\dagger}_{e_{\sst 3}})_{\sst EA }\op{\pi}_{\ell}(h_{e_{\sst 4}})_{\sst FG }\op{V}_{\sst RS}\op{\pi}_{\ell}(h_{e_{\sst 3}})_{\sst CA }\nonumber\\
&&\hspace{1.7cm}
-\op{\pi}_{\ell}(h^{\dagger}_{e_{\sst 4}})_{\sst IG}\op{V}_{\sst RS}\op{\pi}_{\ell}(h^{\dagger}_{e_{\sst 3}})_{\sst EA }\op{V}_{\sst RS}\op{\pi}_{\ell}(h_{e_{\sst 4}})_{\sst FG }\op{\pi}_{\ell}(h_{e_{\sst 3}})_{\sst CA }\nonumber\\
&&\hspace{1.7cm}
-\op{\pi}_{\ell}(h^{\dagger}_{e_{\sst 4}})_{\sst IG}\op{\pi}_{\ell}(h^{\dagger}_{e_{\sst 3}})_{\sst EA }\op{V}_{\sst RS}\op{\pi}_{\ell}(h_{e_{\sst 4}})_{\sst FG}\op{V}_{\sst RS}\op{\pi}_{\ell}(h_{e_{\sst 3}})_{\sst CA }\Big)\Big]
\ea
As mentioned before, the volume operator $\op{V}_{\sst RS}$ is the the sum of the contributing triples
\be
\label{Vsum}
\op{V}_{\sst RS}=\op{V}_{q_{\sst 134}}+\op{V}_{q_{\sst 234}}+\op{V}_{q_{\sst 123}}+\op{V}_{q_{\sst 124}}
\ee
Recall that the SNF $\bet{j_{\sst 12}}{m_{\sst 1}}$ consists of two edges $e_{\sst 1},e_{\sst 2}$ only. Therefore, if $\op{V}_{\sst RS}$ acts before for instance $\op{\pi}_{\ell}(h_{e_{\sst 4}})$ acts the only non-vanishing contribution in $\op{V}_{\sst RS}$ is due to $\op{V}_{q_{\sst 123}}$, because for $\op{V}_{q_{\sst 134}}$,$\op{V}_{q_{\sst 234}}$ and $\op{V}_{q_{\sst 124}}$ the edge $e_{\sst 4}$ is missing. Analogous, only $\op{V}_{q_{\sst 124}}$ contributes to $\op{V}_{\sst RS}$ when the latter is applied to $\bet{j_{\sst 12}}{m_{\sst 1}}$ before $\op{\pi}_{\ell}(h_{e_{\sst 3}})$ has acted. Consequently, eqn (\ref{Ek1eins}) reduces to
\ba
\label{Ek1zwei}
\lL\op{\tilde{E}}^{\sst I,RS}_{k,{\sst tot}}(S_t) &\propto&\pi_{\ell}(\tau_k)_{\sst BC}\pi_{\ell}(\epsilon)_{\sst EI}\Big[ \nonumber\\
&&
+\pi_{\ell}(\epsilon)_{\sst FC}\Big(
+\op{\pi}_{\ell}(h^{\dagger}_{e_{\sst 4}})_{\sst FG}\op{\pi}_{\ell}(h^{\dagger}_{e_{\sst 3}})_{\sst BA }\op{V}_{\sst RS}\op{V}_{\sst RS}\op{\pi}_{\ell}(h_{e_{\sst 4}})_{\sst IG }\op{\pi}_{\ell}(h_{e_{\sst 3}})_{\sst EA }\nonumber\\
&&\hspace{1.7cm}
+\op{\pi}_{\ell}(h^{\dagger}_{e_{\sst 4}})_{\sst FG}\op{V}_{q_{\sst 124}}\op{\pi}_{\ell}(h^{\dagger}_{e_{\sst 3}})_{\sst BA }\op{\pi}_{\ell}(h_{e_{\sst 4}})_{\sst IG }\op{V}_{q_{\sst 123}}\op{\pi}_{\ell}(h_{e_{\sst 3}})_{\sst EA }\nonumber\\
&&\hspace{1.7cm}
-\op{\pi}_{\ell}(h^{\dagger}_{e_{\sst 4}})_{\sst FG}\op{V}_{q_{\sst 124}}\op{\pi}_{\ell}(h^{\dagger}_{e_{\sst 3}})_{\sst BA }\op{V}_{\sst RS}\op{\pi}_{\ell}(h_{e_{\sst 4}})_{\sst IG }\op{\pi}_{\ell}(h_{e_{\sst 3}})_{\sst EA }\nonumber\\
&&\hspace{1.7cm}
-\op{\pi}_{\ell}(h^{\dagger}_{e_{\sst 4}})_{\sst FG}\op{\pi}_{\ell}(h^{\dagger}_{e_{\sst 3}})_{\sst BA }\op{V}_{\sst RS}\op{\pi}_{\ell}(h_{e_{\sst 4}})_{\sst IG}\op{V}_{q_{\sst 123}}\op{\pi}_{\ell}(h_{e_{\sst 3}})_{\sst EA }\Big)\nonumber\\
&&
-\pi_{\ell}(\epsilon)_{\sst FB}\Big(
+\op{\pi}_{\ell}(h^{\dagger}_{e_{\sst 3}})_{\sst IG}\op{\pi}_{\ell}(h^{\dagger}_{e_{\sst 4}})_{\sst EA }\op{V}_{\sst RS}\op{V}_{\sst RS}\op{\pi}_{\ell}(h_{e_{\sst 3}})_{\sst FG }\op{\pi}_{\ell}(h_{e_{\sst 4}})_{\sst CA }\nonumber\\
&&\hspace{1.7cm}
+\op{\pi}_{\ell}(h^{\dagger}_{e_{\sst 3}})_{\sst IG}\op{V}_{q_{\sst 123}}\op{\pi}_{\ell}(h^{\dagger}_{e_{\sst 4}})_{\sst EA }\op{\pi}_{\ell}(h_{e_{\sst 3}})_{\sst FG }\op{V}_{q_{\sst 124}}\op{\pi}_{\ell}(h_{e_{\sst 4}})_{\sst CA }\nonumber\\
&&\hspace{1.7cm}
-\op{\pi}_{\ell}(h^{\dagger}_{e_{\sst 3}})_{\sst IG}\op{V}_{q_{\sst 123}}\op{\pi}_{\ell}(h^{\dagger}_{e_{\sst 4}})_{\sst EA }\op{V}_{\sst RS}\op{\pi}_{\ell}(h_{e_{\sst 3}})_{\sst FG }\op{\pi}_{\ell}(h_{e_{\sst 4}})_{\sst CA }\nonumber\\
&&\hspace{1.7cm}
-\op{\pi}_{\ell}(h^{\dagger}_{e_{\sst 3}})_{\sst IG}\op{\pi}_{\ell}(h^{\dagger}_{e_{\sst 4}})_{\sst EA }\op{V}_{\sst RS}\op{\pi}_{\ell}(h_{e_{\sst 3}})_{\sst FG}\op{V}_{q_{\sst 124}}\op{\pi}_{\ell}(h_{e_{\sst 4}})_{\sst CA }\Big)\Big]
\ea
Futhermore, $\op{V}_{q_{\sst 123}}$ commutes with $\op{\pi}_{\ell}(h_{e_{\sst 4}})$ as well as $\op{V}_{q_{\sst 124}}$ commutes with $\op{\pi}_{\ell}(h_{e_{\sst 3}})$.Using this, we get
\ba
\lL\op{\tilde{E}}^{\sst I,RS}_{k,{\sst tot}}(S_t) &\propto&\pi_{\ell}(\tau_k)_{\sst BC}\pi_{\ell}(\epsilon)_{\sst EI}\Big[ \nonumber\\
&&
+\pi_{\ell}(\epsilon)_{\sst FC}\Big(
+\op{\pi}_{\ell}(h^{\dagger}_{e_{\sst 4}})_{\sst FG}\op{\pi}_{\ell}(h^{\dagger}_{e_{\sst 3}})_{\sst BA }\op{V}_{\sst RS}\op{V}_{\sst RS}\op{\pi}_{\ell}(h_{e_{\sst 4}})_{\sst IG }\op{\pi}_{\ell}(h_{e_{\sst 3}})_{\sst EA }\nonumber\\
&&\hspace{1.7cm}
+\op{\pi}_{\ell}(h^{\dagger}_{e_{\sst 4}})_{\sst FG}\op{\pi}_{\ell}(h^{\dagger}_{e_{\sst 3}})_{\sst BA }\op{V}_{q_{\sst 124}}\op{V}_{q_{\sst 123}}\op{\pi}_{\ell}(h_{e_{\sst 4}})_{\sst IG }\op{\pi}_{\ell}(h_{e_{\sst 3}})_{\sst EA }\nonumber\\
&&\hspace{1.7cm}
-\op{\pi}_{\ell}(h^{\dagger}_{e_{\sst 4}})_{\sst FG}\op{\pi}_{\ell}(h^{\dagger}_{e_{\sst 3}})_{\sst BA }\op{V}_{q_{\sst 124}}\op{V}_{\sst RS}\op{\pi}_{\ell}(h_{e_{\sst 4}})_{\sst IG }\op{\pi}_{\ell}(h_{e_{\sst 3}})_{\sst EA }\nonumber\\
&&\hspace{1.7cm}
-\op{\pi}_{\ell}(h^{\dagger}_{e_{\sst 4}})_{\sst FG}\op{\pi}_{\ell}(h^{\dagger}_{e_{\sst 3}})_{\sst BA }\op{V}_{\sst RS}\op{V}_{q_{\sst 123}}\op{\pi}_{\ell}(h_{e_{\sst 4}})_{\sst IG}\op{\pi}_{\ell}(h_{e_{\sst 3}})_{\sst EA }\Big)\nonumber\\
&&
-\pi_{\ell}(\epsilon)_{\sst FB}\Big(
+\op{\pi}_{\ell}(h^{\dagger}_{e_{\sst 3}})_{\sst IG}\op{\pi}_{\ell}(h^{\dagger}_{e_{\sst 4}})_{\sst EA }\op{V}_{\sst RS}\op{V}_{\sst RS}\op{\pi}_{\ell}(h_{e_{\sst 3}})_{\sst FG }\op{\pi}_{\ell}(h_{e_{\sst 4}})_{\sst CA }\nonumber\\
&&\hspace{1.7cm}
+\op{\pi}_{\ell}(h^{\dagger}_{e_{\sst 3}})_{\sst IG}\op{\pi}_{\ell}(h^{\dagger}_{e_{\sst 4}})_{\sst EA }\op{V}_{q_{\sst 123}}\op{V}_{q_{\sst 124}}\op{\pi}_{\ell}(h_{e_{\sst 3}})_{\sst FG }\op{\pi}_{\ell}(h_{e_{\sst 4}})_{\sst CA }\nonumber\\
&&\hspace{1.7cm}
-\op{\pi}_{\ell}(h^{\dagger}_{e_{\sst 3}})_{\sst IG}\op{\pi}_{\ell}(h^{\dagger}_{e_{\sst 4}})_{\sst EA }\op{V}_{q_{\sst 123}}\op{V}_{\sst RS}\op{\pi}_{\ell}(h_{e_{\sst 3}})_{\sst FG }\op{\pi}_{\ell}(h_{e_{\sst 4}})_{\sst CA }\nonumber\\
&&\hspace{1.7cm}
-\op{\pi}_{\ell}(h^{\dagger}_{e_{\sst 3}})_{\sst IG}\op{\pi}_{\ell}(h^{\dagger}_{e_{\sst 4}})_{\sst EA }\op{V}_{\sst RS}\op{V}_{q_{\sst 124}}\op{\pi}_{\ell}(h_{e_{\sst 3}})_{\sst FG}\op{\pi}_{\ell}(h_{e_{\sst 4}})_{\sst CA }\Big)\Big]\nonumber\\
&=&\pi_{\ell}(\tau_k)_{\sst BC}\pi_{\ell}(\epsilon)_{\sst EI}\Big[ \nonumber\\
&&
+\pi_{\ell}(\epsilon)_{\sst FC} 
\op{\pi}_{\ell}(h^{\dagger}_{e_{\sst 4}})_{\sst FG}\op{\pi}_{\ell}(h^{\dagger}_{e_{\sst 3}})_{\sst BA }
(\op{V}_{\sst RS}\op{V}_{\sst RS}+\op{V}_{q_{\sst 124}}\op{V}_{q_{\sst 123}}-\op{V}_{q_{\sst 124}}\op{V}_{\sst RS}-\op{V}_{\sst RS}\op{V}_{q_{\sst 123}})
\op{\pi}_{\ell}(h_{e_{\sst 4}})_{\sst IG }\op{\pi}_{\ell}(h_{e_{\sst 3}})_{\sst EA }\nonumber\\
&&
-\pi_{\ell}(\epsilon)_{\sst FB}
\op{\pi}_{\ell}(h^{\dagger}_{e_{\sst 3}})_{\sst IG}\op{\pi}_{\ell}(h^{\dagger}_{e_{\sst 4}})_{\sst EA }
(\op{V}_{\sst RS}\op{V}_{\sst RS}+\op{V}_{q_{\sst 123}}\op{V}_{q_{\sst 124}}-\op{V}_{q_{\sst 123}}\op{V}_{\sst RS}-\op{V}_{\sst RS}\op{V}_{q_{\sst 124}})
\op{\pi}_{\ell}(h_{e_{\sst 3}})_{\sst FG }\op{\pi}_{\ell}(h_{e_{\sst 4}})_{\sst CA }\Big]\nonumber\\
&=&\pi_{\ell}(\tau_k)_{\sst BC}\pi_{\ell}(\epsilon)_{\sst EI}\Big[ 
+\pi_{\ell}(\epsilon)_{\sst FC}
\op{\pi}_{\ell}(h^{\dagger}_{e_{\sst 4}})_{\sst FG}\op{\pi}_{\ell}(h^{\dagger}_{e_{\sst 3}})_{\sst BA }
\,\op{O}^{\sst I,RS }_1\,
\op{\pi}_{\ell}(h_{e_{\sst 4}})_{\sst IG }\op{\pi}_{\ell}(h_{e_{\sst 3}})_{\sst EA }\nonumber\\
&&\hspace{2.7cm}
-\pi_{\ell}(\epsilon)_{\sst FB}
\op{\pi}_{\ell}(h^{\dagger}_{e_{\sst 3}})_{\sst IG}\op{\pi}_{\ell}(h^{\dagger}_{e_{\sst 4}})_{\sst EA }
\,\op{O}^{\sst I,RS}_2\,
\op{\pi}_{\ell}(h_{e_{\sst 3}})_{\sst FG }\op{\pi}_{\ell}(h_{e_{\sst 4}})_{\sst CA }\Big],
\ea
whereby we used the definition of $\op{O}^{\sst I,RS }_1$ and $\op{O}^{\sst I,RS}_2$ from eqn (\ref{C1RSO1O2}) as well as the definition of $\op{V}_{\sst RS}$ in eqn (\ref{Vsum}) in the last step. The calculation for $\lL\op{\tilde{E}}^{\sst II,RS}_{k,{\sst tot}}(S_t)$ is similar with the small difference that the signum operator $\op{\cal S}$ is sandwiched between the two volume operators $\op{V}_{\sst RS}$. Hence, we end up with
\ba
\lL\op{\tilde{E}}^{\sst II,RS}_{k,{\sst tot}}(S_t) &\propto&
\pi_{\ell}(\tau_k)_{\sst BC}\pi_{\ell}(\epsilon)_{\sst EI}\Big[ 
+\pi_{\ell}(\epsilon)_{\sst FC}
\op{\pi}_{\ell}(h^{\dagger}_{e_{\sst 4}})_{\sst FG}\op{\pi}_{\ell}(h^{\dagger}_{e_{\sst 3}})_{\sst BA }
\,\op{O}^{\sst II,RS }_1\,
\op{\pi}_{\ell}(h_{e_{\sst 4}})_{\sst IG }\op{\pi}_{\ell}(h_{e_{\sst 3}})_{\sst EA }\nonumber\\
&&\hspace{2.7cm}
-\pi_{\ell}(\epsilon)_{\sst FB}
\op{\pi}_{\ell}(h^{\dagger}_{e_{\sst 3}})_{\sst IG}\op{\pi}_{\ell}(h^{\dagger}_{e_{\sst 4}})_{\sst EA }
\,\op{O}^{\sst II,RS}_2\,
\op{\pi}_{\ell}(h_{e_{\sst 3}})_{\sst FG }\op{\pi}_{\ell}(h_{e_{\sst 4}})_{\sst CA }\Big].
\ea
Here we used the expressions for $\op{O}^{\sst II,RS}_1$ and $\op{O}^{\sst II,RS}_2$ in eqn (\ref{C2RSO1O2}).
\subsection{Case $\lL\op{\tilde{E}}^{\sst I,RS}_{k,{\sst tot}}(S_t)$:\newline
Detailed Calculation of the Matrix Elements of $\op{O}^{\sst I,RS}_{1}$ and $\op{O}^{\sst I,RS}_2$}
As in section \ref{CaseV2} we will investigate the case of a spin-representation $\ell=\frac{1}{2},1$ for the
reason that these are the two easiest cases where the matrices of $\op{Q}^{\sst RS}_{v,{\sst IJK}}$ can still  be solved
analytically. Here, we want to keep the discussion succinct and mainly present our results, for the reason
that section \ref{CaseV2}  already explains in a quite detailed way how matrix elements of of the volume operator are actually calculated.
\subsubsection{Matrix Elements for the Case of a Spin-$\frac{1}{2}$-Representation}%
With $\ell=\frac{1}{2}$, the matrix elements $\alpc{0}{M}{0}\,\op{O}^{\sst I,RS}_1\,\alp{0}{1}{0}$ and 
$\alpc{1}{i}{M}\,\op{O}^{\sst I,RS}_2\,\alp{1}{1}{M}$, where $i=3,4$, contribute to the matrix element of $\lL\op{\tilde{E}}^{\sst I,RS}_{k,{\sst tot}}(S_t)$. The matrix elements are given by
\ba
\label{RSMEJ}
\alpc{\st 0}{2}{\st M}\,\op{O}_1^{I,\sst RS}\,\alp{\st 0}{1}{\st M}
&=&+\alpc{\st 0}{2}{\st M}\,\op{V}_{q_{\sst 134}}^2\,\alp{\st 0}{1}{\st M}
   +\alpc{\st 0}{2}{\st M}\,\op{V}_{q_{\sst 234}}^2\,\alp{\st 0}{1}{\st M}\nonumber\\
&&+\alpc{\st 0}{2}{\st M}\,\op{V}_{q_{\sst 134}}\op{V}_{q_{\sst 234}}\,\alp{\st 0}{1}{\st M}
  +\alpc{\st 0}{2}{\st M}\,\op{V}_{q_{\sst 234}}\op{V}_{q_{\sst 134}}\,\alp{\st 0}{1}{\st M}\nonumber\\
&&+\alpc{\st 0}{2}{\st M}\,\op{V}_{q_{\sst 134}}\op{V}_{q_{\sst 123}}\,\alp{\st 0}{1}{\st M}
  +\alpc{\st 0}{2}{\st M}\,\op{V}_{q_{\sst 124}}\op{V}_{q_{\sst 134}}\,\alp{\st 0}{1}{\st M}\nonumber\\
&&+\alpc{\st 0}{2}{\st M}\,\op{V}_{q_{\sst 234}}\op{V}_{q_{\sst 123}}\,\alp{\st 0}{1}{\st M}
  +\alpc{\st 0}{2}{\st M}\,\op{V}_{q_{\sst 124}}\op{V}_{q_{\sst 234}}\,\alp{\st 0}{1}{\st M}\nonumber\\
&&+\alpc{\st 0}{2}{\st M}\,\op{V}_{q_{\sst 124}}\op{V}_{q_{\sst 123}}\,\alp{\st 0}{1}{\st M}
\ea
and with $i=3,4$
\ba
\alpc{\st 1}{i}{\st M}\,\op{O}_2^{I,\sst RS}\,\alp{\st 1}{1}{\st M}
&=&+\alpc{\st 1}{i}{\st M}\,\op{V}_{q_{\sst 134}}^2\,\alp{\st 1}{1}{\st M}
   +\alpc{\st 1}{i}{\st M}\,\op{V}_{q_{\sst 234}}^2\,\alp{\st 1}{1}{\st M}\nonumber\\
&&+\alpc{\st 1}{i}{\st M}\,\op{V}_{q_{\sst 234}}\op{V}_{q_{\sst 134}}\,\alp{\st 1}{1}{\st M}
  +\alpc{\st 1}{i}{\st M}\,\op{V}_{q_{\sst 134}}\op{V}_{q_{\sst 234}}\,\alp{\st 1}{1}{\st M}\nonumber\\
&&+\alpc{\st 1}{i}{\st M}\,\op{V}_{q_{\sst 123}}\op{V}_{q_{\sst 134}}\,\alp{\st 1}{1}{\st M}
  +\alpc{\st 1}{i}{\st M}\,\op{V}_{q_{\sst 134}}\op{V}_{q_{\sst 124}}\,\alp{\st 1}{1}{\st M}\nonumber\\
&&+\alpc{\st 1}{i}{\st M}\,\op{V}_{q_{\sst 123}}\op{V}_{q_{\sst 234}}\,\alp{\st 1}{1}{\st M}
  +\alpc{\st 1}{i}{\st M}\,\op{V}_{q_{\sst 234}}\op{V}_{q_{\sst 124}}\,\alp{\st 1}{1}{\st M}\nonumber\\
&&+\alpc{\st 1}{i}{\st M}\,\op{V}_{q_{\sst 123}}\op{V}_{q_{\sst 124}}\,\alp{\st 1}{1}{\st M}
\ea
Here we used the definitions of the operators $\op{O}_1^{I,\sst RS},\op{O}_2^{I,\sst RS}$ in eqn (\ref{C1RSO1O2}).
These matrix elements for $\op{O}_1^{I,\sst RS}$ consist of the sum of the matrix elements with the following structure
\be
\label{expalp}
\alpc{0}{2}{M}\,\op{V}_{q_{\sst IJK}}\op{V}_{q_{\sst \tilde{I}\tilde{J}\tilde{K}}}\,\alp{0}{1}{M}=\sum\limits_{|\alpha^{\prime}\zu}\alpc{0}{2}{M}\,\op{V}_{ q_{\sst IJK}}\,|\alpha^{\prime}\zu\auf\alpha^{\prime}\,|\,\op{V}_{ q_{\sst \tilde{I}\tilde{J}\tilde{K}}}\,\alp{0}{1}{M},
\ee
where we expanded the matrix element in terms of basis vectors $|\alpha^{\prime}\zu$ of the Hilbert space ${\cal H}^{\st J=0}$.
\newline
Now each $\alpc{0}{2}{M}\,\op{V}_{ q_{\sst IJK}}\,|\alpha^{\prime}\zu$ can be caculated through an eigenvector expansion
\ba
\label{expv}
\alpc{0}{2}{M}\,\op{V}_{ q_{\sst IJK}}\,|\alpha^{\prime}\zu&=&\sum\limits_{k}\alpc{0}{2}{M}\,\op{V}_{ q_{\sst IJK}}\,|\vec{e}_k\zu\auf\vec{e}_k\,|\,\alpha^{\prime}\zu\nonumber\\
\auf\alpha^{\prime}\,|\,\op{V}_{ q_{\sst \tilde{I}\tilde{J}\tilde{K}}}\,\alp{0}{1}{M}&=&\sum\limits_{k}\auf\alpha^{\prime}\,|\,\op{V}_{ q_{\sst \tilde{I}\tilde{J}\tilde{K}}}\,|\,\vec{e}_k\zu\auf\vec{e}_k\,\alp{0}{1}{M}
\ea
 where $\vec{e}_k$ denotes the normed eigenvectors of $\op{V}_{q_{\sst IJK}}$. As in section \ref{CaseV2}, the 
 eigenvectors of $\op{V}_{q_{\sst IJK}}$ are equal to the eigenvectors of $\op{Q}^{\sst RS}_{v,{\sst IJK}}$ and  $\sqrt{|\lambda|}$ is an eigenvalue of $\op{V}_{q_{\sst IJK}}$ assuming that the corresponding eigenvalue of 
 $\op{Q}^{\sst RS}_{v,{\sst IJK}}$ is $\lambda$. (See also the discussion in 
section \ref{TwoVs}.)
\newline
 The states $\alp{0}{i}{M}$ that have
to be taken into account in order to derive the matrix of $\op{Q}^{\sst RS}_{v,{\sst IJK}}$, can be found in eqn (\ref{Hj05J0}). Using eqn (\ref{RSq134}), (\ref{RSq234}),
(\ref{RSq123}) and (\ref{RSq124}) we get the following matrices, eigenvalues and eigenvectors for a total angular momentum $J=0$ 
\ba
\label{QJ0IJK}
\op{Q}_{\sst RS,134}^{\sst J=0}=\left(\begin{array}{cc} 0& -2i a \\ +2i a & 0 \end{array}\right),&&\quad\lambda_1=-2a=-\lambda_2,\quad \vec{v}_1=(=+i,1),\quad \vec{v}_2=(-i,1)\nonumber\\
\op{Q}_{\sst RS,234}^{\sst J=0}=\op{Q}_{\sst RS,123}^{\sst J=0}=\op{Q}_{\sst RS,124}^{\sst J=0}=\left(\begin{array}{cc} 0& +2i a \\ -2i a & 0 \end{array}\right),&&\quad\lambda_1=-2a=-\lambda_2,\quad \vec{v}_1=(=-i,1),\quad \vec{v}_2=(+i,1),
\ea
where we defined $a:=(\lp^6\frac{3!}{4}C_{reg})\sqrt{j(j+1)}$ and labelled the rows by $\alp{J}{i}{M}$, whereas columns are labelled by $\alp{J}{\tilde{i}}{M}$. Inserting these eigenvectors above into eqn (\ref{expv}), yields to vanishing off-diagonal matrix elements of $\op{V}_{q_{\sst IJK}}$
\ba
\alpc{0}{2}{0}\,\op{V}_{q_{\sst 134}}\,\alp{0}{1}{0}=\alpc{0}{2}{0}\,
\op{V}_{q_{\sst 234}}\,\alp{0}{1}{0}=\alpc{0}{2}{0}\,\op{V}_{q_{\sst 123}}\,\alp{0}{1}{0}=\alpc{0}{2}{0}\,\op{V}_{q_{\sst 124}}\,\alp{0}{1}{0}&=&0\nonumber\\
\alpc{0}{1}{0}\,\op{V}_{q_{\sst 134}}\,\alp{0}{2}{0}=\alpc{0}{1}{0}\,\op{V}_{q_{\sst 234}}\,\alp{0}{2}{0}=\alpc{0}{1}{0}\,\op{V}_{q_{\sst 123}}\,\alp{0}{2}{0}=\alpc{0}{1}{0}\,
\op{V}_{q_{\sst 124}}\,\alp{0}{2}{0}&=&0
\ea
for the reason that the expansion coefficient of $\alp{0}{1}{0}$ is purely imaginary, whereas the one of $\alp{0}{2}{M}$ is real and therefore the terms appearing in the sum of the expansion will cancel each other (see also the explicit calculations done in section \ref{CaseV2} for this).
\newline
Accordingly, if we sum over $|\alpha^{\prime}\zu$ in eqn (\ref{expalp}) either the first or the second matrix element of $\op{V}_{q_ {\sst IJK}}$ in the product is zero. Thus each $\alpc{0}{2}{M}\,\op{V}_{q_{\sst IJK}}\op{V}_{q_{\sst \tilde{I}\tilde{J}\tilde{K}}}\,\alp{0}{1}{M}=0$ and therefore the whole sum in eqn (\ref{RSMEJ}) is identical to zero and we have
\be
\alpc{0}{2}{M}\,\op{O}^{\sst I,RS}_1\,\alp{0}{1}{M}=0
\ee
Let us discuss the matrix element $\alpc{\st 1}{i}{\st M}\,\op{O}_2^{I,\sst RS}\,\alp{\st 1}{1}{\st M}$ from eqn (\ref{RSMEJ}) now.
The four states $|\alpha^{\st 1}{i}{M}\zu$ that are included in the Hilbert space belonging to a total angular momentum of $J=1$ are written down in eqn (\ref{Hj05J1}). Inserting them into eqn (\ref{RSq134}) leads to  
\be
\op{Q}_{\sst RS,134}^{\sst J=1}=\left(\begin{array}{cccc} 0 & +i\frac{2}{3}a & -i\frac{2}{3}\sqrt{2}a & 0 \\ -i\frac{2}{3}a & 0 & -i\sqrt{2} & +i\frac{1}{3}\sqrt{2}b\\
                                 +i\frac{2}{3}\sqrt{2}a & +i\sqrt{2} & 0 & -i\frac{2}{3}b\\ 0 & -i\frac{1}{3}\sqrt{2}b & +i\frac{2}{3}b & 0  \end{array}\right),\quad \lambda_1=0=\lambda_2,\quad \lambda_3=-\sqrt{\frac{2}{3}}\sqrt{3+2a^2+b^2}=-\lambda_4,
\ee
with $a:=(\lp^6\frac{3!}{4}C_{reg})\sqrt{j(j+1)}$, $b:=(\lp^6\frac{3!}{4}C_{reg})\sqrt{4j(j+1)-3}$, while the corresponding eigenvectors are given by
\ba
\vec{v}_1&=&(+\frac{b}{\sqrt{2}a},0,0,1)\nonumber\\
\vec{v}_3&=&(-\frac{3}{\sqrt{2}a},\sqrt{2},1,0)\nonumber\\
\vec{v}_4&=&(-\frac{\sqrt{2}a}{b},\frac{-3\sqrt{2}-i\sqrt{3}\sqrt{3+2a^2+b^2}}{3b},\frac{-3+i\sqrt{6}\sqrt{3+2a^2+b^2}}{3b},1)\nonumber\\
\vec{v}_5&=&(-\frac{\sqrt{2}a}{b},\frac{-3\sqrt{2}+i\sqrt{3}\sqrt{3+2a^2+b^2}}{3b},\frac{-3-i\sqrt{6}\sqrt{3+2a^2+b^2}}{3b},1).
\ea
In contrast to the case of a total angular momentum $J=0$ the expansion coefficients of $\alp{1}{i}{M}$, where
$i=3,4$, have a real and an imaginary part, while the one of $\alp{1}{1}{M}$ is real. Consequently, we get a result different from zero here.  Additionally, we show the result of $\alpc{1}{i}{M}\,\op{V}_{\sst q_{134}}\,\alp{1}{5}{M}$ with $i=3,4$, because we need these matrix elements later when we expand $\alpc{\st 1}{i}{\st M}\,\op{O}_2^{I,\sst RS}\,\alp{\st 1}{1}{\st M}$ in terms of $\alp{1}{i}{M}$.
\ba
\label{RSME1341}
\alpc{1}{3}{M}\,\op{V}_{\sst q_{134}}\,\alp{1}{1}{M}=\sum\limits_{k}\alpc{1}{3}{M}\,\op{V}_{\sst q_{134}}\,|\vec{e}_k\zu\auf\vec{e}_k\,\alp{1}{1}{M}
&=&+\sqrt{|\lambda_3|}\frac{12a}{|\lambda_3|^2}\nonumber\\
\alpc{1}{4}{M}\,\op{V}_{\sst q_{134}}\,\alp{1}{1}{M}=\sum\limits_{k}\alpc{1}{4}{M}\,\op{V}_{\sst q_{134}}\,|\vec{e}_k\zu\auf\vec{e}_k\,\alp{1}{1}{M}
&=&+\sqrt{|\lambda_3|}\frac{6\sqrt{2}a}{|\lambda_3|^2}\nonumber\\
\alpc{1}{3}{M}\,\op{V}_{\sst q_{134}}\,\alp{1}{5}{M}=\sum\limits_{k}\alpc{1}{3}{M}\,\op{V}_{\sst q_{134}}\,|\vec{e}_k\zu\auf\vec{e}_k\,\alp{1}{5}{M}&=&
-\sqrt{|\lambda_3|}\frac{6\sqrt{2}b}{|\lambda_3|^2}\nonumber\\
\alpc{1}{4}{M}\,\op{V}_{\sst q_{134}}\,\alp{1}{5}{M}=\sum\limits_{k}\alpc{1}{4}{M}\,\op{V}_{\sst q_{134}}\,|\vec{e}_k\zu\auf\vec{e}_k\,\alp{1}{5}{M}&=&
-\sqrt{|\lambda_3|}\frac{6b}{|\lambda_3|^2},
\ea
whereby $\vec{e}_k$ are the normed eigenvectors of $\op{Q}^{\sst J=1}_{RS,{\sst 134}}$.
The same is true for the triple $\op{q}_{\sst 234}$ in which case we have the following matrix and eigenvalues
\be
\op{Q}_{\sst RS,234}^{\sst J=1}=\left(\begin{array}{cccc} 0 & -i\frac{2}{3}a & +i\frac{2}{3}\sqrt{2}a & 0 \\ +i\frac{2}{3}a & 0 & -i\sqrt{2} & -i\frac{1}{3}\sqrt{2}b\\
                                 -i\frac{2}{3}\sqrt{2}a & +i\sqrt{2} & 0 & +i\frac{2}{3}b\\ 0 & +i\frac{1}{3}\sqrt{2}b & -i\frac{2}{3}b & 0  \end{array}\right),\quad \lambda_1=0=\lambda_2,\quad \lambda_3=-\sqrt{\frac{2}{3}}\sqrt{3+2a^2+b^2}=-\lambda_4
\ee
and the corresponding eigenvectors
\ba
\vec{v}_1&=&(+\frac{b}{\sqrt{2}a},0,0,1)\nonumber\\
\vec{v}_3&=&(+\frac{3}{\sqrt{2}a},\sqrt{2},1,0)\nonumber\\
\vec{v}_4&=&(-\frac{\sqrt{2}a}{b},\frac{3\sqrt{2}+i\sqrt{3}\sqrt{3+2a^2+b^2}}{3b},\frac{3-i\sqrt{6}\sqrt{3+2a^2+b^2}}{3b},1)\nonumber\\
\vec{v}_5&=&(-\frac{\sqrt{2}a}{b},\frac{3\sqrt{2}-i\sqrt{3}\sqrt{3+2a^2+b^2}}{3b},\frac{3+i\sqrt{6}\sqrt{3+2a^2+b^2}}{3b},1).
\ea
Accordingly, these eigenvectors yield a non-vanishing matrix element for the states $\alp{1}{i}{M}$ where $i=3,4$ as well
\ba
\label{RSME2341}
\alpc{1}{3}{M}\,\op{V}_{\sst q_{234}}\,\alp{1}{1}{M}=\sum\limits_{k}\alpc{1}{3}{M}\,\op{V}_{\sst q_{234}}\,|\vec{e}_k\zu\auf\vec{e}_k\,\alp{1}{1}{M}
&=&-\sqrt{|\lambda_3|}\frac{12a}{|\lambda_3|^2}\nonumber\\
\alpc{1}{4}{M}\,\op{V}_{\sst q_{234}}\,\alp{1}{1}{M}=\sum\limits_{k}\alpc{1}{4}{M}\,\op{V}_{\sst q_{234}}\,|\vec{e}_k\zu\auf\vec{e}_k\,\alp{1}{1}{M}
&=&-\sqrt{|\lambda_3|}\frac{6\sqrt{2}a}{|\lambda_3|^2}\nonumber\\
\alpc{1}{3}{M}\,\op{V}_{\sst q_{234}}\,\alp{1}{5}{M}=\sum\limits_{k}\alpc{1}{3}{M}\,\op{V}_{\sst q_{234}}\,|\vec{e}_k\zu\auf\vec{e}_k\,\alp{1}{5}{M}&=&
+\sqrt{|\lambda_3|}\frac{6\sqrt{2}b}{|\lambda_3|^2}\nonumber\\
\alpc{1}{4}{M}\,\op{V}_{\sst q_{234}}\,\alp{1}{5}{M}=\sum\limits_{k}\alpc{1}{4}{M}\,\op{V}_{\sst q_{234}}\,|\vec{e}_k\zu\auf\vec{e}_k\,\alp{1}{5}{M}&=&
+\sqrt{|\lambda_3|}\frac{6b}{|\lambda_3|^2}.
\ea
The situation is different if we consider the triple $\op{q}_{\sst 123}$. In this case the matrix obtained from eqn (\ref{RSq123}) includes more entries that are zero due to the $\delta_{\wt{a}_3,a_3}$ in eqn (\ref{RSq123})
\be
\op{Q}_{\sst RS,123}^{\sst J=1}=\left(\begin{array}{cccc} 0 & -i2a & 0 & 0 \\ +i2a & 0 & 0 & 0\\
                                 0 & 0 & 0 & -i2b\\ 0 & 0 & -i2b & 0  \end{array}\right),\quad \lambda_1=-2a=\lambda_2,\quad \lambda_3=-2b=-\lambda_4.
\ee
Therefore, the eigenvectors look much simpler 
\be
\vec{v}_1=(-i,1,0,0),\quad
\vec{v}_3=(+i,1,0,0), \quad
\vec{v}_4=(0,0,-i,1), \quad
\vec{v}_5=(0,0,+i,1)
\ee
and we can easily extract from them 
\ba
\label{RSME1231}
\alpc{1}{i}{M}\,\op{V}_{\sst q_{123}}\,\alp{1}{1}{M}=\sum\limits_{k}\alpc{1}{i}{M}\,\op{V}_{\sst q_{123}}\,|\vec{e}_k\zu\auf\vec{e}_k\,\alp{1}{1}{M}&=&0\nonumber\\
\alpc{1}{i}{M}\,\op{V}_{\sst q_{123}}\,\alp{1}{5}{M}=\sum\limits_{k}\alpc{1}{i}{M}\,\op{V}_{\sst q_{123}}\,|\vec{e}_k\zu\auf\vec{e}_k\,\alp{1}{5}{M}&=&0\nonumber\\
\alpc{1}{3}{M}\,\op{V}_{\sst q_{123}}\,\alp{1}{3}{M}=\sum\limits_{k}\alpc{1}{3}{M}\,\op{V}_{\sst q_{123}}\,|\vec{e}_k\zu\auf\vec{e}_k\,\alp{1}{3}{M}&=&+\sqrt{2a}\nonumber\\
\alpc{1}{3}{M}\,\op{V}_{\sst q_{123}}\,\alp{1}{4}{M}=\sum\limits_{k}\alpc{1}{3}{M}\,\op{V}_{\sst q_{123}}\,|\vec{e}_k\zu\auf\vec{e}_k\,\alp{1}{4}{M}&=&0\nonumber\\
\alpc{1}{4}{M}\,\op{V}_{\sst q_{123}}\,\alp{1}{3}{M}=\sum\limits_{k}\alpc{1}{4}{M}\,\op{V}_{\sst q_{123}}\,|\vec{e}_k\zu\auf\vec{e}_k\,\alp{1}{3}{M}&=&0\nonumber\\
\alpc{1}{4}{M}\,\op{V}_{\sst q_{123}}\,\alp{1}{4}{M}=\sum\limits_{k}\alpc{1}{4}{M}\,\op{V}_{\sst q_{123}}\,|\vec{e}_k\zu\auf\vec{e}_k\,\alp{1}{4}{M}&=&+\sqrt{2b}
\ea
Here $i=3,4$. The matrix of the last triple can be evaluated by using eqn (\ref{RSq124}). The matrix itself and its eigenvalues can be found in the equation below 
\be
\op{Q}_{\sst RS,124}^{\sst J=1}=\left(\begin{array}{cccc} 0 & -i\frac{2}{3}a & -i\frac{4}{3}\sqrt{2}a & 0 \\ +i\frac{2}{3}a & 0 & 0 & -i\frac{4}{3}\sqrt{2}b\\
                                 +i\frac{4}{3}\sqrt{2}a & 0 & 0 & +i\frac{2}{3}b\\ 0 & +i\frac{4}{3}\sqrt{2}b & -i\frac{2}{3}b & 0  \end{array}\right),\quad \lambda_1=0=\lambda_2,\quad \lambda_3=-\sqrt{\frac{2}{3}}\sqrt{3+2a^2+b^2}=-\lambda_4.
\ee
The corresponding eigenvectors are given by
\be
\vec{v}_1=(+i\frac{3}{2\sqrt{2}},\frac{1}{2\sqrt{2}},1,0),\quad
\vec{v}_3=(-\frac{3}{2\sqrt{2}},\frac{1}{2\sqrt{2}},1,0),\quad
\vec{v}_4=(0,+i\frac{2\sqrt{2}}{3},-\frac{i}{3},1),\quad
\vec{v}_5=(0,-i\frac{2\sqrt{2}}{3},+\frac{i}{3},1).
\ee
In this case either the eigenvalues or the expansion coefficient of $\alp{1}{1}{M}$ and $\alp{1}{5}{M}$ are zero, so that we also end up with only trivial matrix elements. Only the diagonal matrix elements of $\alp{1}{3}{M}$,$\alp{1}{4}{M}$ are non-vanishing.
\ba
\label{RSME1241}
\alpc{1}{i}{M}\,\op{V}_{\sst q_{124}}\,\alp{1}{1}{M}=\sum\limits_{k}\alpc{1}{i}{M}\,\op{V}_{\sst q_{124}}\,|\vec{e}_k\zu\auf\vec{e}_k\,\alp{1}{1}{M}&=&0\nonumber\\
\alpc{1}{i}{M}\,\op{V}_{\sst q_{124}}\,\alp{1}{5}{M}=\sum\limits_{k}\alpc{1}{i}{M}\,\op{V}_{\sst q_{124}}\,|\vec{e}_k\zu\auf\vec{e}_k\,\alp{1}{5}{M}&=&0\nonumber\\
\alpc{1}{3}{M}\,\op{V}_{\sst q_{124}}\,\alp{1}{3}{M}=\sum\limits_{k}\alpc{1}{3}{M}\,\op{V}_{\sst q_{124}}\,|\vec{e}_k\zu\auf\vec{e}_k\,\alp{1}{3}{M}&=&+\sqrt{2a}\nonumber\\
\alpc{1}{3}{M}\,\op{V}_{\sst q_{124}}\,\alp{1}{4}{M}=\sum\limits_{k}\alpc{1}{3}{M}\,\op{V}_{\sst q_{124}}\,|\vec{e}_k\zu\auf\vec{e}_k\,\alp{1}{4}{M}&=&0\nonumber\\
\alpc{1}{4}{M}\,\op{V}_{\sst q_{124}}\,\alp{1}{3}{M}=\sum\limits_{k}\alpc{1}{4}{M}\,\op{V}_{\sst q_{124}}\,|\vec{e}_k\zu\auf\vec{e}_k\,\alp{1}{3}{M}&=&0\nonumber\\
\alpc{1}{4}{M}\,\op{V}_{\sst q_{124}}\,\alp{1}{4}{M}=\sum\limits_{k}\alpc{1}{4}{M}\,\op{V}_{\sst q_{124}}\,|\vec{e}_k\zu\auf\vec{e}_k\,\alp{1}{4}{M}&=&+\sqrt{2b}\nonumber\\
\alpc{1}{5}{M}\,\op{V}_{\sst q_{124}}\,\alp{1}{1}{M}=\sum\limits_{k}\alpc{1}{5}{M}\,\op{V}_{\sst q_{124}}\,|\vec{e}_k\zu\auf\vec{e}_k\,\alp{1}{1}{M}&=&0\nonumber\\
\ea
After having calculated all the necessary matrix elements of $\op{V}_ {q_{\sst IJK}}$, we will expand each matrix element \\
$\alpc{1}{i}{M}\,\op{V}_{\sst q_{IJK}}\op{V}_{\sst q_{\tilde{I}\tilde{J}\tilde{K}}}\,\alp{1}{1}{M}$ included in $\alpc{\st 1}{i}{\st M}\,\op{O}_2^{I,\sst RS}\,\alp{\st 1}{1}{\st M}$ in terms of the basis states $\alp{1}{j}{M}$
\be
\label{1Expa}
\alpc{1}{i}{M}\,\op{V}_{q_{\sst IJK}}\op{V}_{q_{\sst \tilde{I}\tilde{J}\tilde{K}}}\,\alp{1}{1}{M}=\sum\limits_{|\alpha^{\prime}\zu}\alpc{1}{i}{M}\,\op{V}_{ q_{\sst IJK}}\,|\alpha^{\prime}\zu\auf\alpha^{\prime}\,|\,\op{V}_{ q_{\sst \tilde{I}\tilde{J}\tilde{K}}}\,\alp{1}{1}{M}
\ee
Considering the operator $\op{V}_{q_{\sst 123}}\op{V}_{q_{\sst IJK}}$, where $IJK\in\{134,234\}$. The expansion is given by
\ba
\alpc{1}{i}{M}\,\op{V}_{q_{\sst 123}}\op{V}_{q_{\sst IJK}}\,\alp{1}{1}{M}&=&
+\alpc{1}{i}{M}\,\op{V}_{ q_{\sst 123}}\,\alp{1}{1}{M}\alpc{1}{1}{M}\,\op{V}_{ q_{\sst IJK}}\,\alp{1}{1}{M}\nonumber\\
&&+\alpc{1}{i}{M}\,\op{V}_{ q_{\sst 123}}\,\alp{1}{3}{M}\alpc{1}{3}{M}\,\op{V}_{ q_{\sst IJK}}\,\alp{1}{1}{M}\nonumber\\
&&+\alpc{1}{i}{M}\,\op{V}_{ q_{\sst 123}}\,\alp{1}{4}{M}\alpc{1}{4}{M}\,\op{V}_{ q_{\sst IJK}}\,\alp{1}{1}{M}\nonumber\\
&&+\alpc{1}{i}{M}\,\op{V}_{ q_{\sst 123}}\,\alp{1}{5}{M}\alpc{1}{5}{M}\,\op{V}_{ q_{\sst IJK}}\,\alp{1}{1}{M}
\ea
We can read off from eqn (\ref{RSME1231}) 
\be
\alpc{1}{i}{M}\,\op{V}_{\sst q_{123}}\,\alp{1}{1}{M}=\alpc{1}{i}{M}\,\op{V}_{\sst q_{123}}\,\alp{1}{5}{M}=\alpc{1}{3}{M}\,\op{V}_{\sst q_{123}}\,\alp{1}{4}{M}=\alpc{1}{4}{M}\,\op{V}_{\sst q_{123}}\,\alp{1}{3}{M}=0
\ee
Therefore the expansion reduces to
\ba
\alpc{1}{i}{M}\,\op{V}_{q_{\sst 123}}\op{V}_{q_{\sst IJK}}\,\alp{1}{1}{M}&=&
+\alpc{1}{3}{M}\,\op{V}_{ q_{\sst 123}}\,\alp{1}{3}{M}\alpc{1}{3}{M}\,\op{V}_{ q_{\sst IJK}}\,\alp{1}{3}{M}\nonumber\\
&&+\alpc{1}{4}{M}\,\op{V}_{ q_{\sst 123}}\,\alp{1}{4}{M}\alpc{1}{4}{M}\,\op{V}_{ q_{\sst IJK}}\,\alp{1}{1}{M}
\ea
If we choose $\tilde{I}\tilde{J}\tilde{K}=124$ we see due to $\alpc{1}{i}{M}\,\op{V}_{ q_{\sst 124}}\,\alp{1}{1}{M}=0$ (see eqn (\ref{RSME1241})) that the matrix element of $\op{V}_{q_{\sst 123}}\op{V}_{q_{\sst 124}}$ vanishes. In the case of $\tilde{I}\tilde{J}\tilde{K}=134,234$ by comparing eqn (\ref{RSME1341}) with  eqn (\ref{RSME2341}) we realise $\alpc{1}{i}{M}\,\op{V}_{ q_{\sst 134}}\,\alp{1}{1}{M}=-\alpc{1}{i}{M}\,\op{V}_{ q_{\sst 234}}\,\alp{1}{1}{M}=0$
. Thus, the the two contributions cancel each other. Consequently,
\be
\label{R123IJK}
\alpc{1}{i}{M}\,\op{V}_{q_{\sst 123}}\left(\op{V}_{q_{\sst 134}}+\op{V}_{q_{\sst 234}}+\op{V}_{q_{\sst 124}}\right)\,\alp{1}{1}{M}
=0
\ee
In the case of  the operator $\op{V}_{q_{\sst IJK}}\op{V}_{q_{\sst 124}}$, where $IJK\in\{134,234\}$, the expansion in terms of the basis states $\alp{\st 1}{i}{M}$ can be written as
\ba
\alpc{1}{i}{M}\,\op{V}_{q_{\sst IJK}}\op{V}_{q_{\sst 124}}\,\alp{1}{1}{M}&=&
+\alpc{1}{i}{M}\,\op{V}_{ q_{\sst IJK}}\,\alp{1}{1}{M}\alpc{1}{1}{M}\,\op{V}_{ q_{\sst 124}}\,\alp{1}{1}{M}\nonumber\\
&&+\alpc{1}{i}{M}\,\op{V}_{ q_{\sst IJK}}\,\alp{1}{3}{M}\alpc{1}{3}{M}\,\op{V}_{ q_{\sst 124}}\,\alp{1}{1}{M}\nonumber\\
&&+\alpc{1}{i}{M}\,\op{V}_{ q_{\sst IJK}}\,\alp{1}{4}{M}\alpc{1}{4}{M}\,\op{V}_{ q_{\sst 124}}\,\alp{1}{1}{M}\nonumber\\
&&+\alpc{1}{i}{M}\,\op{V}_{ q_{\sst IJK}}\,\alp{1}{5}{M}\alpc{1}{5}{M}\,\op{V}_{ q_{\sst 124}}\,\alp{1}{1}{M}
\ea
The matrix elements $\alpc{1}{j}{M}\,\op{V}_{ q_{\sst 124}}\,\alp{1}{1}{M}$ with $j=3,4,5$ are identical to zero, as can be seen from eqn (\ref{RSME1241}). Hence, only the first term in the expansion survives. Moreover, if we choose for $IJK=\{134,234\}$, we have \\
$\alpc{1}{i}{M}\,\op{V}_{ q_{\sst 134}}\,\alp{1}{1}{M}=-\alpc{1}{i}{M}\,\op{V}_{ q_{\sst 234}}\,\alp{1}{1}{M}$, as can be seen in eqn (\ref{RSME1341}) and eqn (\ref{RSME2341}), so that the non-vanishung contributions get canceled by each other. Accordingly, we obtain
\be
\label{RIJK124}
\alpc{1}{i}{M}\,\left(\op{V}_{q_{\sst 134}}+\op{V}_{q_{\sst 234}}\right)\op{V}_{q_{\sst 124}}\,\alp{1}{1}{M}
=0
\ee
Analysing the operator $\op{V}_{q_ {\sst 134}}\op{V}_{q_{\sst IJK}}$ with $IJK\in\{134,234\}$, we get
\ba
\alpc{1}{i}{M}\,\op{V}_{q_{\sst 134}}\op{V}_{q_{\sst IJK}}\,\alp{1}{1}{M}&=&
+\alpc{1}{i}{M}\,\op{V}_{ q_{\sst 134}}\,\alp{1}{1}{M}\alpc{1}{1}{M}\,\op{V}_{ q_{\sst IJK}}\,\alp{1}{1}{M}\nonumber\\
&&+\alpc{1}{i}{M}\,\op{V}_{ q_{\sst 134}}\,\alp{1}{3}{M}\alpc{1}{3}{M}\,\op{V}_{ q_{\sst IJK}}\,\alp{1}{1}{M}\nonumber\\
&&+\alpc{1}{i}{M}\,\op{V}_{ q_{\sst 134}}\,\alp{1}{4}{M}\alpc{1}{4}{M}\,\op{V}_{ q_{\sst IJK}}\,\alp{1}{1}{M}\nonumber\\
&&+\alpc{1}{i}{M}\,\op{V}_{ q_{\sst 134}}\,\alp{1}{5}{M}\alpc{1}{5}{M}\,\op{V}_{ q_{\sst IJK}}\,\alp{1}{1}{M}
\ea
From eqn (\ref{RSME1341}) and eqn (\ref{RSME2341}) we can extract for $i=3,4$ $\alpc{1}{i}{M}\,\op{V}_{ q_{\sst 134}}\,\alp{1}{1}{M}=-\alpc{1}{i}{M}\,\op{V}_{ q_{\sst 234}}\,\alp{1}{1}{M}$. Hence, the expansion above yields
\ba
\alpc{1}{i}{M}\,\op{V}_{q_{\sst 134}}\left(\op{V}_{q_{\sst 134}}+\op{V}_{q_{\sst 234}}\right)\,\alp{1}{1}{M}&=&
+\alpc{1}{i}{M}\,\op{V}_{ q_{\sst 134}}\,\alp{1}{1}{M}\alpc{1}{1}{M}\,\left(\op{V}_{q_{\sst 134}}+\op{V}_{q_{\sst 234}}\right)\,\alp{1}{1}{M}\nonumber\\
&&+\alpc{1}{i}{M}\,\op{V}_{ q_{\sst 134}}\,\alp{1}{5}{M}\alpc{1}{5}{M}\,\left(\op{V}_{q_{\sst 134}}+\op{V}_{q_{\sst 234}}\right)\,\alp{1}{1}{M}
\ea
The same argument applies to the operator $\op{V}_{q_ {\sst 234}}\op{V}_{q_{\sst IJK}}$ with $IJK\in\{134,234\}$. Terefore we obtain here 
\ba
\alpc{1}{i}{M}\,\op{V}_{q_{\sst 234}}\left(\op{V}_{q_{\sst 134}}+\op{V}_{q_{\sst 234}}\right)\,\alp{1}{1}{M}&=&
+\alpc{1}{i}{M}\,\op{V}_{ q_{\sst 234}}\,\alp{1}{1}{M}\alpc{1}{1}{M}\,\left(\op{V}_{q_{\sst 134}}+\op{V}_{q_{\sst 234}}\right)\,\alp{1}{1}{M}\nonumber\\
&&+\alpc{1}{i}{M}\,\op{V}_{ q_{\sst 234}}\,\alp{1}{5}{M}\alpc{1}{5}{M}\,\left(\op{V}_{q_{\sst 134}}+\op{V}_{q_{\sst 234}}\right)\,\alp{1}{1}{M}
\ea
Using the fact that $\alpc{1}{i}{M}\,\op{V}_{q_{\sst 134}}\alp{1}{1}{M}=-\alpc{1}{i}{M}\,\op{V}_{q_{\sst 234}}\alp{1}{1}{M}$ and $\alpc{1}{i}{M}\,\op{V}_{q_{\sst 134}}\alp{1}{5}{M}=-\alpc{1}{i}{M}\,\op{V}_{q_{\sst 234}}\alp{1}{5}{M}$ that can be seen by comparing eqn (\ref{RSME1341}) with eqn (\ref{RSME2341}), we get
\be
\label{R134234}
\alpc{1}{i}{M}\,\left(\op{V}_{q_{\sst 134}}+\op{V}_{q_{\sst 234}}\right)\left(\op{V}_{q_{\sst 134}}+\op{V}_{q_{\sst 234}}\right)\,\alp{1}{1}{M}
=0
\ee
Now, we add eqn (\ref{R123IJK}), eqn (\ref{RIJK124}) and eqn (\ref{R134234}) and note that the sum is precisely the operator $\op{O}^{\sst I,RS}_2$. Accordingly, 
\be
\alpc{1}{i}{M}\,\op{O}^{\sst I,RS}_2\,\alp{1}{1}{M}=0\quad\quad i=3,4
\ee
Since the matrix element of $\op{O}^{\sst I,RS}_1$ as well as the matrix element of $\op{O}^{\sst I,RS}_2$ vanishes and exactly these matrix elements are the only one that contribute to $^{\frac{1}{2}}\op{\tilde{E}}^{\sst I,RS}_{k,{\sst tot}}(S_t)$, the operator $^{\frac{1}{2}}\op{\tilde{E}}^{\sst I,RS}_{k,{\sst tot}}(S_t)$ becomes the zero operator.
\subsubsection{Matrix Elements for the Case of a Spin-$1$-Representation}%
In this subsection we will repeat the calculation of the last subsection for the case of a spin-representation $\ell=1$. The matrix elements that are included in the calculations of the alternative flux operator $^{1}\op{\tilde{E}}^{\sst I,RS}_{k,{\sst tot}}(S_t)$ are  $\alpc{0}{2}{M}\,\op{O}_1^{\sst I,RS}\,\alp{0}{1}{M}$ and  $\alpc{1}{i}{M}\,\op{O}_2^{\sst I,RS}\,\alp{1}{1}{M}$ where $i=2,3,4$. The definition of $\op{O}_1^{\sst I,RS}$ and $\op{O}_2^{\sst I,RS}$ can be found in eqn (\ref{C1RSO1O2}), whereas we have to, as before, derive the value of the matrix element of each single triple.
\newline\newline
A basis of the Hilbert space associated with a total angular momentum $J=0$ can be found in eqn (\ref{Hj1J0}) and consists of four states. Hence, for every single triple $\op{Q}_{v,{\sst IJK}}$ we obtain a $4\times 4$-matrix. Starting with $\op{Q}_{\sst RS,134}^{\sst J=0}$ we have 
\be
\op{Q}_{\sst RS,134}^{\sst J=0}=\left(\begin{array}{ccc} 0 & -i4\sqrt{\frac{2}{3}}a &  0 \\ +i4\sqrt{\frac{2}{3}}a & 0 & -i4\sqrt{\frac{1}{3}}b \\
                                 0 & +i4\sqrt{\frac{1}{3}}b & 0 \end{array}\right),\quad \lambda_1=0,\quad \lambda_2=-\frac{4}{\sqrt{3}}\sqrt{2a^2+b^2}=-\lambda_3,
\ee
where we used eqn (\ref{RSq134}) in order to obtain the matrix. The corresponding four eigenvectors are given by
\be
\vec{v}_1=(+i\frac{b}{\sqrt{2}a},0,1),\quad
\vec{v}_2=(-\frac{\sqrt{2}a}{b},+i\frac{\sqrt{2a^2+b^2}}{b},1)\,\quad
\vec{v}_3=(-\frac{\sqrt{2}a}{b},-i\frac{\sqrt{2a^2+b^2}}{b},1).
\ee
As in the situation of $\ell=\frac{1}{2}$ for  $J=0$ the remaining three triples are identical and  moreover just the negative of the matrix of $\op{Q}_{\sst RS,234}^{\sst J=0}$
\be
\op{Q}_{\sst RS,234}^{\sst J=0}=\op{Q}_{\sst RS,123}^{\sst J=0}=\op{Q}_{\sst RS,124}^{\sst J=0}=\left(\begin{array}{ccc} 0 & +i4\sqrt{\frac{2}{3}}a &  0 \\ -i4\sqrt{\frac{2}{3}}a & 0 & +i4\sqrt{\frac{1}{3}}b \\
                                 0 & -i4\sqrt{\frac{1}{3}}b & 0 \end{array}\right),\quad \lambda_1=0,\quad \lambda_2=-\frac{4}{\sqrt{3}}\sqrt{2a^2+b^2}=-\lambda_3.
\ee
while the corresponding eigenvectors are
\be
\vec{v}_1=(-i\frac{b}{\sqrt{2}a},0,1),\quad
\vec{v}_2=(-\frac{\sqrt{2}a}{b},-i\frac{\sqrt{2a^2+b^2}}{b},1),\quad
\vec{v}_3=(-\frac{\sqrt{2}a}{b},+i\frac{\sqrt{2a^2+b^2}}{b},1).
\ee
Now, we expand each operator $\op{V}_{q_{\sst IJK}}\op{V}_{q_{\sst \tilde{I}\tilde{J}\tilde{K}}}$ included in $\alpc{0}{2}{M}\,\op{O}_1^{\sst I,RS}\,\alp{0}{1}{M}$ with the help of the states $\alp{0}{j}{M}$
\ba
\label{alpEx}
\alpc{0}{2}{M}\,\op{V}_{q_{\sst IJK}}\op{V}_{q_{\sst \tilde{I}\tilde{J}\tilde{K}}}\,\alp{0}{1}{M}&=&
+\alpc{1}{2}{M}\,\op{V}_{ q_{\sst IJK}}\,\alp{\st 1}{1}{M}\alpc{\st 1}{1}{M}\,\op{V}_{ q_{\sst \tilde{I}\tilde{J}\tilde{K}}}\,\alp{1}{1}{M}\nonumber\\
&&+\alpc{1}{2}{M}\,\op{V}_{ q_{\sst IJK}}\,\alp{\st 1}{2}{M}\alpc{\st 1}{2}{M}\,\op{V}_{ q_{\sst \tilde{I}\tilde{J}\tilde{K}}}\,\alp{1}{1}{M}\nonumber\\
&&+\alpc{1}{2}{M}\,\op{V}_{ q_{\sst IJK}}\,\alp{\st 1}{3}{M}\alpc{\st 1}{3}{M}\,\op{V}_{ q_{\sst \tilde{I}\tilde{J}\tilde{K}}}\,\alp{1}{1}{M}
\ea
From the eigenvector expansion, we get 
\ba
\alpc{0}{2}{M}\,\op{V}_{ q_{\sst IJK}}\,\alp{1}{1}{M}&=&\sum\limits_{k}\alpc{0}{2}{M}\,\op{V}_{ q_{\sst IJK}}\,|\vec{e}_k\zu\auf\vec{e}_k\,\alp{1}{1}{M}=0\nonumber\\
\alpc{0}{2}{M}\,\op{V}_{ q_{\sst IJK}}\,\alp{1}{3}{M}&=&\sum\limits_{k}\alpc{0}{2}{M}\,\op{V}_{ q_{\sst IJK}}\,|\vec{e}_k\zu\auf\vec{e}_k\,\alp{1}{3}{M}=0,
\ea
whereby $\vec{e}_1,\vec{e}_2,\vec{e}_3$ result from $\vec{v}_1,\vec{v}_2,\vec{v}_3$ by just dividing each vector by its norm. Consequently, each term in the expansion in eqn (\ref{alpEx}) vanishes separately an we obtain as in the case of $\ell=0.5$
\be
\alpc{0}{2}{M}\,\op{O}_1^{\sst I,RS}\,\alp{0}{1}{M}=0
\ee
If we consider a total angular momentum of $J=1$ eqn (\ref{Hj1J1}) tells us that we have already to deal with seven states and consequently get a $7\times 7$-matrix for each triple $\op{q}_{\sst IJK}$. 
Again we have to discuss the matrix elements $\alpc{1}{i}{M}\,\op{O}^{\sst I,RS}_2\,\alp{1}{1}{M}$ with $i=2,3,4$. Starting with $\op{Q}^{\sst RS}_{v,{\sst 134}}$ and using eqn (\ref{RSq134}) yields
\be
\op{Q}^{\sst J=1}_{\sst RS,134}=\left(\begin{array}{ccccccc} 
0 & -i\frac{8}{3}\sqrt{2}\tilde{a}& -i2\sqrt{\frac{2}{3}}\tilde{a}  & -i\frac{2}{3}\sqrt{10}\tilde{a} & 0 & 0 & 0\\
+i\frac{8}{3}\sqrt{2}\tilde{a}& 0 & -i\frac{4}{\sqrt{3}} & 0& +i\frac{4}{3}\tilde{b} & 0 & 0\\
+i2\sqrt{\frac{2}{3}}\tilde{a} & +i\frac{4}{\sqrt{3}} & 0 & -i2\sqrt{\frac{5}{3}}& -i\frac{2}{\sqrt{3}}\tilde{b} & 0 & 0\\
+i\frac{2}{3}\sqrt{10}\tilde{a}& 0 & +i2\sqrt{\frac{5}{3}} & 0& -i\frac{4}{3\sqrt{5}}\tilde{b} & -i2\sqrt{\frac{3}{5}}\tilde{b} & 0\\
0 & -i\frac{4}{3}\tilde{b} & +i\frac{2}{\sqrt{3}}\tilde{b} & +i\frac{4}{3\sqrt{5}}\tilde{b} & 0 & -i2\sqrt{3} & +i2\sqrt{\frac{6}{5}}\tilde{c}\\
0 & 0 & 0 & +i2\sqrt{\frac{3}{5}}\tilde{b} & +i2\sqrt{3} & 0 & -i6\sqrt{\frac{2}{5}}\tilde{c} \\
0 & 0 & 0 & 0 & -i2\sqrt{\frac{6}{5}}\tilde{c} & +i6\sqrt{\frac{2}{5}}\tilde{c} & 0
\end{array}\right),
\ee
where we introduced 
\be
\tilde{a}:=(\lp^6\frac{3!}{4}C_{reg})\sqrt{j(j+1)}\quad \tilde{b}:=(\lp^6\frac{3!}{4}C_{reg})\sqrt{4j(j+1)-3}\quad \tilde{c}:=(\lp^6\frac{3!}{4}C_{reg})\sqrt{j(j+1)-2}.
\ee
The seven eigenvalues of $\op{Q}^{\sst RS}_{v,{\sst 134}}$ are 
\ba
\lambda_1&=&0\nonumber\\
\lambda_2&=&-4(\lp^6\frac{3!}{4}C_{reg})\sqrt{j(j-1)}=-\lambda_3\nonumber\\
\lambda_4&=&-4(\lp^6\frac{3!}{4}C_{reg})\sqrt{j(j+2)}=-\lambda_5\nonumber\\
\lambda_6&=&-4(\lp^6\frac{3!}{4}C_{reg})\sqrt{2j(j+1)-1}=-\lambda_7.
\ea 
The corresponding eigenvectors can be written in the following form
\ba
\vec{v}_1&=&(0,0,-\sqrt{15}a''',+3a''',-\sqrt{3}b''',-b''',1)\nonumber\\
\vec{v}_2&=&(+ic,a-ib,d+ie,f+ig,h-i\ell,m+in,1)\nonumber\\
\vec{v}_3&=&(-ic,a+ib,d-ie,f-ig,h+i\ell,m-in,1)\nonumber\\
\vec{v}_4&=&(+ic',a'+ib',d'-ie',f'-ig',-h'-i\ell',-m'+in',1)\nonumber\\
\vec{v}_5&=&(-ic',a'-ib',d'+ie',f'+ig',-h'+i\ell',-m'-in',1)\nonumber\\
\vec{v}_6&=&(-ic'',-a''+ib'',d''-ie'',-f''-ig'',-h''-i\ell'',-m''+in'',1)\nonumber\\
\vec{v}_7&=&(+ic'',-a''-ib'',d''+ie'',-f''+ig'',-h''+i\ell'',-m''-in'',1).
\ea
Here all letters $\{a,...,n''\}$ denote real numbers which depend on the chosen value for the spin label $j$ that is attached to the edges $e_{\sst 1},e_{\sst 2}$. Using the expansion in terms of eigenvectors in eqn (\ref{expv}), we get
\ba
\label{RS1ME134}
\alpc{1}{2}{M}\,\op{V}_{q_{\sst 134}}\,\alp{1}{1}{M}=\sum\limits_{k=1}^4\alpc{1}{2}{M}\,\op{V}_{\sst q_{134}}\,|\vec{e}_k\zu\auf\vec{e}_k\,\alp{1}{1}{M}&=&
-2c_2bc + 2c_4b'c' - 2c_6b''c''\nonumber\\ 
\alpc{1}{3}{M}\,\op{V}_{\sst q_{134}}\,\alp{1}{1}{M}=\sum\limits_{k=1}^4\alpc{1}{3}{M}\,\op{V}_{\sst q_{134}}\,|\vec{e}_k\zu\auf\vec{e}_k\,\alp{1}{1}{M}&=&
+ 2c_2ce - 2c_4c'e' + 2c_6c''e''\nonumber\\ 
\alpc{1}{4}{M}\,\op{V}_{\sst q_{134}}\,\alp{1}{1}{M}=\sum\limits_{k=1}^4\alpc{1}{4}{M}\,\op{V}_{\sst q_{134}}\,|\vec{e}_k\zu\auf\vec{e}_k\,\alp{1}{1}{M}&=&
 + 2c_2cg - 2c_4c'g' - 2c_6c''g''\nonumber\\
 \alpc{1}{5}{M}\,\op{V}_{\sst q_{134}}\,\alp{1}{1}{M}=\sum\limits_{k=1}^4\alpc{1}{5}{M}\,\op{V}_{\sst q_{134}}\,|\vec{e}_k\zu\auf\vec{e}_k\,\alp{1}{1}{M}&=&
-2c_2c\ell - 2c_4c'\ell' + 2c_6c''\ell''\nonumber\\ 
\alpc{1}{6}{M}\,\op{V}_{\sst q_{134}}\,\alp{1}{1}{M}=\sum\limits_{k=1}^4\alpc{1}{6}{M}\,\op{V}_{\sst q_{134}}\,|\vec{e}_k\zu\auf\vec{e}_k\,\alp{1}{1}{M}&=&
 + 2c_2cn + 2c_4c'n' - 2c_6c''n''\nonumber\\
 \alpc{1}{7}{M}\,\op{V}_{\sst q_{134}}\,\alp{1}{1}{M}=\sum\limits_{k=1}^4\alpc{1}{7}{M}\,\op{V}_{\sst q_{134}}\,|\vec{e}_k\zu\auf\vec{e}_k\,\alp{1}{1}{M}&=&0\nonumber\\
 \alpc{1}{2}{M}\,\op{V}_{q_{\sst 134}}\,\alp{1}{5}{M}=\sum\limits_{k=1}^4\alpc{1}{2}{M}\,\op{V}_{\sst q_{134}}\,|\vec{e}_k\zu\auf\vec{e}_k\,\alp{1}{5}{M}&=&
+2c_2(ah+b\ell) - 2c_4(a'h'+b'\ell') + 2c_6(a''h''-b''\ell'')\nonumber\\ 
\alpc{1}{3}{M}\,\op{V}_{\sst q_{134}}\,\alp{1}{5}{M}=\sum\limits_{k=1}^4\alpc{1}{3}{M}\,\op{V}_{\sst q_{134}}\,|\vec{e}_k\zu\auf\vec{e}_k\,\alp{1}{5}{M}&=&
+ 2c_2(dh-e\ell) - 2c_4(d'h'-e'\ell') - 2c_6(d''h''-e''\ell'')\nonumber\\ 
\alpc{1}{4}{M}\,\op{V}_{\sst q_{134}}\,\alp{1}{5}{M}=\sum\limits_{k=1}^4\alpc{1}{4}{M}\,\op{V}_{\sst q_{134}}\,|\vec{e}_k\zu\auf\vec{e}_k\,\alp{1}{5}{M}&=&
 + 2c_2(fh-g\ell) - 2c_4(f'h'-g'\ell') + 2c_6(f''h''+g''\ell'')\nonumber\\
 \alpc{1}{2}{M}\,\op{V}_{q_{\sst 134}}\,\alp{1}{6}{M}=\sum\limits_{k=1}^4\alpc{1}{2}{M}\,\op{V}_{\sst q_{134}}\,|\vec{e}_k\zu\auf\vec{e}_k\,\alp{1}{6}{M}&=&
+2c_2(am-bn) - 2c_4(a'm'-b'n') + 2c_6(a''m''+b''n'')\nonumber\\ 
\alpc{1}{3}{M}\,\op{V}_{\sst q_{134}}\,\alp{1}{6}{M}=\sum\limits_{k=1}^4\alpc{1}{3}{M}\,\op{V}_{\sst q_{134}}\,|\vec{e}_k\zu\auf\vec{e}_k\,\alp{1}{6}{M}&=&
+ 2c_2(dm-en) - 2c_4(d'm'+e'n') - 2c_6(d''m''+e''n'')\nonumber\\ 
\alpc{1}{4}{M}\,\op{V}_{\sst q_{134}}\,\alp{1}{6}{M}=\sum\limits_{k=1}^4\alpc{1}{4}{M}\,\op{V}_{\sst q_{134}}\,|\vec{e}_k\zu\auf\vec{e}_k\,\alp{1}{6}{M}&=&
 + 2c_2(fm-gn) - 2c_4(f'm'-g'n') + 2c_6(f''m''+g''n'')\nonumber\\
\ea
Here we introduced the constants $c_1,c_2,c_3$ that are defined by
\ba
c_2:&=&\frac{\sqrt{|\lambda_2|}}{1+a^2+b^2+c^2+d^2+e^2+f^2+g^2+h^2+\ell^2+m^2+n^2}\nonumber\\
c_4:&=&\frac{\sqrt{|\lambda_4|}}{1+a'^2+b'^2+c'^2+d'^2+e'^2+f'^2+g'^2+h'^2+\ell'^2+m'^2+n'^2}\nonumber\\
c_6:&=&\frac{\sqrt{|\lambda_6|}}{1+a''^2+b''^2+c''^2+d''^2+e''^2+f''^2+g''^2+h''^2+\ell''^2+m''^2+n''^2}.
\ea
The non-vanishing of these matrix elements is as in the case of $\ell=\frac{1}{2}$ caused by the fact that the
expansion coefficients of $\alp{1}{i}{M}$, where $i=2,3,4$, have real as well as imaginary parts. When we apply eqn (\ref{RSq234}) to the states in eqn (\ref{Hj1J1}) we obtain the following matrix
\be
\op{Q}^{\sst J=1}_{\sst RS,234}=\left(\begin{array}{ccccccc} 
0 & +i\frac{8}{3}\sqrt{2}\tilde{a}& +i2\sqrt{\frac{2}{3}}\tilde{a}  & +i\frac{2}{3}\sqrt{10}\tilde{a} & 0 & 0 & 0\\
-i\frac{8}{3}\sqrt{2}\tilde{a}& 0 & -i\frac{4}{\sqrt{3}} & 0& -i\frac{4}{3}\tilde{b} & 0 & 0\\
-i2\sqrt{\frac{2}{3}}\tilde{a} &+i\frac{4}{\sqrt{3}} & 0 & -i2\sqrt{\frac{5}{3}}& +i\frac{2}{\sqrt{3}}\tilde{b} & 0 & 0\\
-i\frac{2}{3}\sqrt{10}\tilde{a}& 0 & +i2\sqrt{\frac{5}{3}} & 0& +i\frac{4}{3\sqrt{5}}\tilde{b} & +i2\sqrt{\frac{3}{5}}\tilde{b} & 0\\
0 & +i\frac{4}{3}\tilde{b} & -i\frac{2}{\sqrt{3}}\tilde{b} & -i\frac{4}{3\sqrt{5}}\tilde{b} & 0 & -i2\sqrt{3} & -i2\sqrt{\frac{6}{5}}\tilde{c}\\
0 & 0 & 0 & -i2\sqrt{\frac{3}{5}}\tilde{b} & +i2\sqrt{3} & 0 & +i6\sqrt{\frac{2}{5}}\tilde{c} \\
0 & 0 & 0 & 0 & +i2\sqrt{\frac{6}{5}}\tilde{c} & -i6\sqrt{\frac{2}{5}}\tilde{c} & 0
\end{array}\right),
\ee
where we used the abbreviations
\be
\tilde{a}:=(\lp^6\frac{3!}{4}C_{reg})\sqrt{j(j+1)}\quad \tilde{b}:=(\lp^6\frac{3!}{4}C_{reg})\sqrt{4j(j+1)-3}\quad \tilde{c}:=(\lp^6\frac{3!}{2}C_{reg})\sqrt{j(j+1)-2}.
\ee
The eigenvalues are similar to the one of $\op{q}_{\sst 134}$  
\ba
\lambda_1&=&0\nonumber\\
\lambda_2&=&-4(\lp^6\frac{3!}{4}C_{reg})\sqrt{j(j-1)}=-\lambda_3\nonumber\\
\lambda_4&=&-4(\lp^6\frac{3!}{4}C_{reg})\sqrt{j(j+2)}=-\lambda_5\nonumber\\
\lambda_6&=&-4(\lp^6\frac{3!}{4}C_{reg})\sqrt{2j(j+1)-1}=-\lambda_7.
\ea 
and can be used to derive the corresponding eigenvectors
\ba
\vec{v}_1&=&(0,0,-\sqrt{15}a''',+3a''',-\sqrt{3}b''',-b''',1)\nonumber\\
\vec{v}_2&=&(-ic,a-ib,d+ie,f+ig,-h+i\ell,-m-in,1)\nonumber\\
\vec{v}_3&=&(+ic,a+ib,d-ie,f-ig,-h-i\ell,-m+in,1)\nonumber\\
\vec{v}_4&=&(-ic',a'+ib',d'-ie',f'-ig',h'+i\ell',m'-in',1)\nonumber\\
\vec{v}_5&=&(+ic',a'-ib',d'+ie',f'+ig',h'-i\ell',m'+in',1)\nonumber\\
\vec{v}_6&=&(+ic'',-a''+ib'',d''-ie'',-f''-ig'',h''+i\ell'',m''-in'',1)\nonumber\\
\vec{v}_7&=&(-ic'',-a''-ib'',d''+ie'',-f''+ig'',h''-i\ell'',m''+in'',1)\nonumber\\
\ea
where we again suppose that $\{a,...,n''\}$ denote real numbers. Thus the desired matrix elements are
\ba
\label{RS1ME234}
\alpc{1}{2}{M}\,\op{V}_{q_{\sst 234}}\,\alp{1}{1}{M}=\sum\limits_{k=1}^4\alpc{1}{2}{M}\,\op{V}_{\sst q_{234}}\,|\vec{e}_k\zu\auf\vec{e}_k\,\alp{1}{1}{M}&=&
+2c_2bc - 2c_4b'c' + 2c_6b''c''\nonumber\\ 
\alpc{1}{3}{M}\,\op{V}_{\sst q_{234}}\,\alp{1}{1}{M}=\sum\limits_{k=1}^4\alpc{1}{3}{M}\,\op{V}_{\sst q_{234}}\,|\vec{e}_k\zu\auf\vec{e}_k\,\alp{1}{1}{M}&=&
- 2c_2ce + 2c_4c'e' - 2c_6c''e''\nonumber\\ 
\alpc{1}{4}{M}\,\op{V}_{\sst q_{234}}\,\alp{1}{1}{M}=\sum\limits_{k=1}^4\alpc{1}{4}{M}\,\op{V}_{\sst q_{234}}\,|\vec{e}_k\zu\auf\vec{e}_k\,\alp{1}{1}{M}&=&
 - 2c_2cg + 2c_4c'g' + 2c_6c''g''\nonumber\\
 \alpc{1}{5}{M}\,\op{V}_{\sst q_{234}}\,\alp{1}{1}{M}=\sum\limits_{k=1}^4\alpc{1}{5}{M}\,\op{V}_{\sst q_{234}}\,|\vec{e}_k\zu\auf\vec{e}_k\,\alp{1}{1}{M}&=&
-2c_2c\ell - 2c_4c'\ell' + 2c_6c''\ell''\nonumber\\ 
\alpc{1}{6}{M}\,\op{V}_{\sst q_{234}}\,\alp{1}{1}{M}=\sum\limits_{k=1}^4\alpc{1}{6}{M}\,\op{V}_{\sst q_{234}}\,|\vec{e}_k\zu\auf\vec{e}_k\,\alp{1}{1}{M}&=&
 + 2c_2cn + 2c_4c'n' - 2c_6c''n''\nonumber\\
 \alpc{1}{7}{M}\,\op{V}_{\sst q_{234}}\,\alp{1}{1}{M}=\sum\limits_{k=1}^4\alpc{1}{7}{M}\,\op{V}_{\sst q_{234}}\,|\vec{e}_k\zu\auf\vec{e}_k\,\alp{1}{1}{M}&=&0\nonumber\\
 \alpc{1}{2}{M}\,\op{V}_{q_{\sst 234}}\,\alp{1}{5}{M}=\sum\limits_{k=1}^4\alpc{1}{2}{M}\,\op{V}_{\sst q_{234}}\,|\vec{e}_k\zu\auf\vec{e}_k\,\alp{1}{5}{M}&=&
-2c_2(ah+b\ell) + 2c_4(a'h'+b'\ell') - 2c_6(a''h''-b''\ell'')\nonumber\\ 
\alpc{1}{3}{M}\,\op{V}_{\sst q_{234}}\,\alp{1}{5}{M}=\sum\limits_{k=1}^4\alpc{1}{3}{M}\,\op{V}_{\sst q_{234}}\,|\vec{e}_k\zu\auf\vec{e}_k\,\alp{1}{5}{M}&=&
- 2c_2(dh-e\ell) + 2c_4(d'h'-e'\ell') + 2c_6(d''h''-e''\ell'')\nonumber\\ 
\alpc{1}{4}{M}\,\op{V}_{\sst q_{234}}\,\alp{1}{5}{M}=\sum\limits_{k=1}^4\alpc{1}{4}{M}\,\op{V}_{\sst q_{234}}\,|\vec{e}_k\zu\auf\vec{e}_k\,\alp{1}{5}{M}&=&
 - 2c_2(fh-g\ell) + 2c_4(f'h'-g'\ell') - 2c_6(f''h''+g''\ell'')\nonumber\\
 \alpc{1}{2}{M}\,\op{V}_{q_{\sst 234}}\,\alp{1}{6}{M}=\sum\limits_{k=1}^4\alpc{1}{2}{M}\,\op{V}_{\sst q_{234}}\,|\vec{e}_k\zu\auf\vec{e}_k\,\alp{1}{6}{M}&=&
-2c_2(am-bn) + 2c_4(a'm'-b'n') - 2c_6(a''m''+b''n'')\nonumber\\ 
\alpc{1}{3}{M}\,\op{V}_{\sst q_{234}}\,\alp{1}{6}{M}=\sum\limits_{k=1}^4\alpc{1}{3}{M}\,\op{V}_{\sst q_{234}}\,|\vec{e}_k\zu\auf\vec{e}_k\,\alp{1}{6}{M}&=&
- 2c_2(dm-en) + 2c_4(d'm'+e'n') + 2c_6(d''m''+e''n'')\nonumber\\ 
\alpc{1}{4}{M}\,\op{V}_{\sst q_{234}}\,\alp{1}{6}{M}=\sum\limits_{k=1}^4\alpc{1}{4}{M}\,\op{V}_{\sst q_{234}}\,|\vec{e}_k\zu\auf\vec{e}_k\,\alp{1}{6}{M}&=&
 - 2c_2(fm-gn) + 2c_4(f'm'-g'n') - 2c_6(f''m''+g''n'')\nonumber\\
\ea
In the case of $\op{Q}^{\sst RS}_{v,{\sst 123}}$ eqn (\ref{RSq123}) leads to a matrix that looks less complicated
\be
\op{Q}^{\sst J=1}_{\sst RS,123}=\left(\begin{array}{ccccccc} 
0 & 0 & =i\sqrt{\frac{2}{3}}a  & 0 & 0 & 0 & 0\\
0 & 0 & 0 & 0& 0 & 0 & 0\\
-i\sqrt{\frac{2}{3}}a & 0 & 0 & 0& +i\frac{4}{\sqrt{3}}b & 0 & 0\\
0 & 0 & 0 & 0& 0 & +i4\sqrt{\frac{3}{5}}b & 0\\
0 & 0 & -i\frac{4}{\sqrt{3}}b & 0 & 0 & 0 & 0\\
0 & 0 & 0 & -i4\sqrt{\frac{3}{5}}b & 0 & 0 & +i12\sqrt{\frac{2}{5}}c \\
0 & 0 & 0 & 0 & 0 & -i12\sqrt{\frac{2}{5}}c & 0
\end{array}\right),
\ee
with 
\be
a:=(\lp^6\frac{3!}{4}C_{reg})\sqrt{j(j+1)}\quad b:=(\lp^6\frac{3!}{4}C_{reg})\sqrt{4j(j+1)-3}\quad c:=(\lp^6\frac{3!}{2}C_{reg})\sqrt{j(j+1)-2}.
\ee
and the corresponding eigenvalues 
\ba
\lambda_1&=&0=\lambda_2=\lambda_3\nonumber\\
\lambda_4&=&-4(\lp^6\frac{3!}{4}C_{reg})\sqrt{3}\sqrt{2j(j+1)-3}=-\lambda_5\nonumber\\
\lambda_6&=&-4(\lp^6\frac{3!}{4}C_{reg})\sqrt{2j(j+1)-1}=-\lambda_7.
\ea 
Unsurprisingly, the eigenvectors are simpler as well and are shown below
\ba
\vec{v}_1&=&(0,0,0,\frac{\sqrt{6}}{\alpha},0,0,1),\quad
\vec{v}_2+(\frac{1}{\sqrt{2}\beta},0,0,0,1,0,0),\quad
\vec{v}_3=(0,1,0,0,0,0,0)\nonumber\\
\vec{v}_4&=&(0,0,0,-\frac{1}{\sqrt{6}}\alpha,0,-i\sqrt{\frac{5}{6}}\gamma,1),\quad
\vec{v}_5=(0,0,0,-\frac{1}{\sqrt{6}}\alpha,0,+i\sqrt{\frac{5}{6}}\gamma,1)\nonumber\\
\vec{v}_6&=&(-\sqrt{2}\beta,0,-i\delta,0,1,0,0),\quad
\vec{v}_7=(-\sqrt{2}\beta,0,+i\delta,0,1,0,0).
\ea
Here, the dependence on $j$ of the components of $\vec{v}_k$ is less tricky and therefore we mention them
explicitly for those interested
\be
\alpha:=\frac{b}{c},\quad\beta:=\frac{a}{b},\quad \gamma:=\frac{\sqrt{2a^2-3}}{c},\quad \delta:=\frac{\sqrt{7a^2-3}}{b}.
\ee
These eigenvectors demonstrate that all matrix elements $\alpc{1}{i}{M}\,\op{V}^2_{\sst
q_{123}}\,\alp{1}{1}{M}$, where $i=2,3,4$, are zero
\ba
\label{RS1ME123}
\alpc{1}{i}{M}\,\op{V}_{\sst q_{123}}\,\alp{1}{1}{M}=\sum\limits_{k=1}^4\alpc{1}{i}{M}\,\op{V}_{\sst q_{123}}\,|\vec{e}_k\zu\auf\vec{e}_k\,\alp{1}{1}{M}&=&0
\nonumber\\ 
\alpc{1}{i}{M}\,\op{V}_{\sst q_{123}}\,\alp{1}{5}{M}=\sum\limits_{k=1}^4\alpc{1}{i}{M}\,\op{V}_{\sst q_{123}}\,|\vec{e}_k\zu\auf\vec{e}_k\,\alp{1}{5}{M}&=&0
\nonumber\\ 
\alpc{1}{i}{M}\,\op{V}_{\sst q_{123}}\,\alp{1}{6}{M}=\sum\limits_{k=1}^4\alpc{1}{i}{M}\,\op{V}_{\sst q_{123}}\,|\vec{e}_k\zu\auf\vec{e}_k\,\alp{1}{1}{M}&=&0,
\ea
as we have either an expansion coefficient equal to zero or the combination of a real and a purely imaginary expansion coefficient.
The last triple that has to be discussed is $\op{Q}^{\sst RS}_{v,{\sst 124}}$. Considering eqn (\ref{RSq124}) we get
\be
\op{Q}^{\sst J=1}_{\sst RS,124}=\left(\begin{array}{ccccccc} 
0 & +i\frac{4}{3}\sqrt{2}a& +i2\sqrt{\frac{2}{3}}a  & -i\frac{2}{3}\sqrt{10}a & 0 & 0 & 0\\
-i\frac{4}{3}\sqrt{2}a& 0 & 0 & 0& -i\frac{8}{3}b & 0 & 0\\
-i2\sqrt{\frac{2}{3}}a & 0 & 0 & 0& +i\frac{2}{\sqrt{3}}b & -2ib & 0\\
+i\frac{2}{3}\sqrt{10}a& 0 & 0 & 0& +i\frac{2}{3\sqrt{5}}b & +i2\sqrt{\frac{3}{5}}b & 0\\
0 & +i\frac{8}{3}b & -i\frac{2}{\sqrt{3}}b & -i\frac{4}{3\sqrt{5}}b & 0 & 0 & -i6\sqrt{\frac{6}{5}}c\\
0 & 0 & +2ib & -i2\sqrt{\frac{3}{5}}b & 0 & 0 & +i6\sqrt{\frac{2}{5}}c \\
0 & 0 & 0 & 0 & +i6\sqrt{\frac{6}{5}}c & -i6\sqrt{\frac{2}{5}}c & 0
\end{array}\right),
\ee
where we introduced 
\be
a:=(\lp^6\frac{3!}{4}C_{reg})\sqrt{j(j+1)}\quad b:=(\lp^6\frac{3!}{4}C_{reg})\sqrt{4j(j+1)-3}\quad c:=(\lp^6\frac{3!}{4}C_{reg})\sqrt{j(j+1)-2}.
\ee
The seven eigenvalues of $\op{q}_{\sst 124}$ are 
\ba
\lambda_1&=&0=\lambda_2=\lambda_3\nonumber\\
\lambda_4&=&-4(\lp^6\frac{3!}{4}C_{reg})\sqrt{3}\sqrt{2j(j+1)-3}=-\lambda_5\nonumber\\
\lambda_6&=&-4(\lp^6\frac{3!}{4}C_{reg})\sqrt{2j(j+1)-1}=-\lambda_7.
\ea 
and the corresponding eigenvectors can be expressed as
\ba
\vec{v}_1&=&(0,3\sqrt{\frac{3}{10}}\frac{1}{\alpha},-3\sqrt{\frac{2}{5}}\frac{1}{\alpha},0,0,0,1),\quad
\vec{v}_2=(-\sqrt{\frac{2}{3}}\frac{1}{\beta},0,0,0,\frac{1}{\sqrt{3}},1,0),\quad
\vec{v}_3=(0,\frac{1}{\sqrt{5}},\sqrt{\frac{3}{5}},1,0,0,0)\nonumber\\
\vec{v}_4&=&(0,-\frac{1}{3}\sqrt{\frac{5}{6}}\alpha,\sqrt{\frac{5}{2}}\frac{1}{6}\alpha,-\frac{1}{6\sqrt{6}}\alpha,+i\sqrt{\frac{5}{2}}\gamma,-i\sqrt{\frac{5}{6}}\gamma,1)\nonumber\\
\vec{v}_5&=&(0,-\frac{1}{3}\sqrt{\frac{5}{6}}\alpha,\sqrt{\frac{5}{2}}\frac{1}{6}\alpha,-\frac{1}{2\sqrt{6}}\alpha,-i\sqrt{\frac{5}{2}}\gamma,+i\sqrt{\frac{5}{6}}\gamma,1)\nonumber\\
\vec{v}_6&=&(2\sqrt{\frac{2}{3}}\beta,+i\frac{2}{\sqrt{3}}\delta,+i\delta,-i\sqrt{\frac{5}{3}}\delta,\frac{1}{\sqrt{3}},1,0)\nonumber\\
\vec{v}_7&=&(2\sqrt{\frac{2}{3}}\beta,-i\frac{2}{\sqrt{3}}\delta,-i\delta,+i\sqrt{\frac{5}{3}}\delta,\frac{1}{\sqrt{3}},1,0).
\ea
with the following abbreviations 
\be
\alpha:=\frac{b}{c},\quad\beta:=\frac{a}{b},\quad \gamma:=\frac{\sqrt{2a^2-3}}{2c},\quad \delta:=\frac{\sqrt{2a^2-1}}{b}.
\ee
For this particular triple the matrix elements disappear as well, because the first three eigenvalues are zero, the eigenvectors $\vec{v}_4,\vec{v}_5$ have an expansion coefficient for $\alp{1}{1}{M}$ which is zero and the vectors $\vec{v}_6,\vec{v}_7$ have a real expansion coefficient for $\alp{1}{1}{M}$, while the one for the states $\alp{1}{i}{M}$ with $i$ being $2,3,4$ is purely imaginary. Consequently, we have
\ba
\label{RS1ME124}
\alpc{1}{i}{M}\,\op{V}_{\sst q_{124}}\,\alp{1}{1}{M}=\sum\limits_{k=1}^4\alpc{1}{i}{M}\,\op{V}_{\sst q_{124}}\,|\vec{e}_k\zu\auf\vec{e}_k\,\alp{1}{1}{M}&=&0
\nonumber\\ 
\alpc{1}{7}{M}\,\op{V}_{\sst q_{124}}\,\alp{1}{1}{M}=\sum\limits_{k=1}^4\alpc{1}{7}{M}\,\op{V}_{\sst q_{124}}\,|\vec{e}_k\zu\auf\vec{e}_k\,\alp{1}{1}{M}&=&0.
\ea
The expansion of each operator $\op{V}_{q_ {\sst IJK}}\op{V}_{q_{\sst \tilde{I}\tilde{J}\tilde{K}}}$ that
occurs in the operator $\op{O}^{\sst I,RS}_2$ is given by
\be
\alpc{1}{i}{M}\,\op{V}_{q_{\sst IJK}}\op{V}_{q_{\sst
\tilde{I}\tilde{J}\tilde{K}}}\,\alp{1}{1}{M}=
\sum\limits_{k=1}^7 \alpc{1}{i}{M}\,\op{V}_{q_{\sst IJK}}\,\alp{1}{k}{M}\alpc{1}{k}{M}\,\op{V}_{q_{\sst
\tilde{I}\tilde{J}\tilde{K}}}\,\alp{1}{1}{M}
\ee
Considering the operator $\op{V}_{q_ {\sst 123}}\op{V}_{q_{\sst IJK}}$ where $IJK\in\{134,234,124\}$, the expansion above leads to
\ba
\alpc{1}{i}{M}\,\op{V}_{q_{\sst 123}}\op{V}_{q_{\sst
IJK}}\,\alp{1}{1}{M}&=&
+\alpc{1}{i}{M}\,\op{V}_{q_{\sst 123}}\,\alp{1}{1}{M}\alpc{1}{1}{M}\,\op{V}_{q_{\sst
IJK}}\,\alp{1}{1}{M}\nonumber\\
&&+\alpc{1}{i}{M}\,\op{V}_{q_{\sst 123}}\,\alp{1}{2}{M}\alpc{1}{2}{M}\,\op{V}_{q_{\sst
IJK}}\,\alp{1}{1}{M}\nonumber\\
&&+\alpc{1}{i}{M}\,\op{V}_{q_{\sst 123}}\,\alp{1}{3}{M}\alpc{1}{3}{M}\,\op{V}_{q_{\sst
IJK}}\,\alp{1}{1}{M}\nonumber\\
&&+\alpc{1}{i}{M}\,\op{V}_{q_{\sst 123}}\,\alp{1}{4}{M}\alpc{1}{4}{M}\,\op{V}_{q_{\sst
IJK}}\,\alp{1}{1}{M}\nonumber\\
&&+\alpc{1}{i}{M}\,\op{V}_{q_{\sst 123}}\,\alp{1}{5}{M}\alpc{1}{5}{M}\,\op{V}_{q_{\sst
IJK}}\,\alp{1}{1}{M}\nonumber\\
&&+\alpc{1}{i}{M}\,\op{V}_{q_{\sst 123}}\,\alp{1}{6}{M}\alpc{1}{6}{M}\,\op{V}_{q_{\sst
IJK}}\,\alp{1}{1}{M}\nonumber\\
&&+\alpc{1}{i}{M}\,\op{V}_{q_{\sst 123}}\,\alp{1}{7}{M}\alpc{1}{7}{M}\,\op{V}_{q_{\sst
IJK}}\,\alp{1}{1}{M}
\ea
We can read off from eqn (\ref{RS1ME123}) $\alpc{1}{i}{M}\,\op{V}_{q_{\sst 123}}\,\alp{1}{1}{M}=0$ with $i=2,3,4$ and $j=1,5,6$. Consequently, the expansion reduces to
\ba
\alpc{1}{i}{M}\,\op{V}_{q_{\sst 123}}\op{V}_{q_{\sst
IJK}}\,\alp{1}{1}{M}&=&
+\alpc{1}{i}{M}\,\op{V}_{q_{\sst 123}}\,\alp{1}{2}{M}\alpc{1}{2}{M}\,\op{V}_{q_{\sst
IJK}}\,\alp{1}{1}{M}\nonumber\\
&&+\alpc{1}{i}{M}\,\op{V}_{q_{\sst 123}}\,\alp{1}{3}{M}\alpc{1}{3}{M}\,\op{V}_{q_{\sst
IJK}}\,\alp{1}{1}{M}\nonumber\\
&&+\alpc{1}{i}{M}\,\op{V}_{q_{\sst 123}}\,\alp{1}{4}{M}\alpc{1}{4}{M}\,\op{V}_{q_{\sst
IJK}}\,\alp{1}{1}{M}\nonumber\\
&&+\alpc{1}{i}{M}\,\op{V}_{q_{\sst 123}}\,\alp{1}{7}{M}\alpc{1}{7}{M}\,\op{V}_{q_{\sst
IJK}}\,\alp{1}{1}{M}
\ea
Since $\alpc{1}{7}{M}\,\op{V}_{q_{\sst IJK}}\,\alp{1}{1}{M}=0$ for $IJK\in\{134,234,124\}$ as can be seen in eqn (\ref{RS1ME134}), eqn (\ref{RS1ME234}) and eqn (\ref{RS1ME124}), the last term in the sum drops out. Furthermore, $\alpc{1}{i}{M}\,\op{V}_{q_{\sst
124}}\,\alp{1}{1}{M}=0$, whereas $\alpc{1}{i}{M}\,\op{V}_{q_{\sst
134}}\,\alp{1}{1}{M}=-\alpc{1}{i}{M}\,\op{V}_{q_{\sst
234}}\,\alp{1}{1}{M}$. Accordingly, the non-vanishing contributions of the triples $\{e_{\sst 1},e_{\sst 3},e_{\sst 4}\}$ and $\{e_{\sst 2},e_{\sst 3},e_{\sst 4}\}$ cancel each other. Hence, we get
\be
\label{R1123IJK}
\alpc{1}{i}{M}\,\op{V}_{q_{\sst 123}}\left(\op{V}_{q_{\sst 134}}+\op{V}_{q_{\sst 234}}\right)\,\alp{1}{1}{M}=0
\ee
In the case of the operator $\op{V}_{q_ {\sst IJK}}\op{V}_{q_{\sst 124}}$ with $IJK\in\{134,234\}$, we can expand the matrix elements as
\ba
\alpc{1}{i}{M}\,\op{V}_{q_{\sst IJK}}\op{V}_{q_{\sst
124}}\,\alp{1}{1}{M}&=&
+\alpc{1}{i}{M}\,\op{V}_{q_{\sst IJK}}\,\alp{1}{1}{M}\alpc{1}{1}{M}\,\op{V}_{q_{\sst
124}}\,\alp{1}{1}{M}\nonumber\\
&&+\alpc{1}{i}{M}\,\op{V}_{q_{\sst IJK}}\,\alp{1}{2}{M}\alpc{1}{2}{M}\,\op{V}_{q_{\sst
124}}\,\alp{1}{1}{M}\nonumber\\
&&+\alpc{1}{i}{M}\,\op{V}_{q_{\sst IJK}}\,\alp{1}{3}{M}\alpc{1}{3}{M}\,\op{V}_{q_{\sst
124}}\,\alp{1}{1}{M}\nonumber\\
&&+\alpc{1}{i}{M}\,\op{V}_{q_{\sst IJK}}\,\alp{1}{4}{M}\alpc{1}{4}{M}\,\op{V}_{q_{\sst
124}}\,\alp{1}{1}{M}\nonumber\\
&&+\alpc{1}{i}{M}\,\op{V}_{q_{\sst IJK}}\,\alp{1}{5}{M}\alpc{1}{5}{M}\,\op{V}_{q_{\sst
124}}\,\alp{1}{1}{M}\nonumber\\
&&+\alpc{1}{i}{M}\,\op{V}_{q_{\sst IJK}}\,\alp{1}{6}{M}\alpc{1}{6}{M}\,\op{V}_{q_{\sst
124}}\,\alp{1}{1}{M}\nonumber\\
&&+\alpc{1}{i}{M}\,\op{V}_{q_{\sst IJK}}\,\alp{1}{7}{M}\alpc{1}{7}{M}\,\op{V}_{q_{\sst
124}}\,\alp{1}{1}{M}
\ea
In eqn (\ref{RS1ME124}) is shown $\alpc{1}{j}{M}\,\op{V}_{q_{\sst
124}}\,\alp{1}{1}{M}=0$ with $j=2,3,4,7$. Therefore, we can neglect four terms in the sum above and get
\ba
\alpc{1}{i}{M}\,\op{V}_{q_{\sst IJK}}\op{V}_{q_{\sst
124}}\,\alp{1}{1}{M}&=&
+\alpc{1}{i}{M}\,\op{V}_{q_{\sst IJK}}\,\alp{1}{1}{M}\alpc{1}{1}{M}\,\op{V}_{q_{\sst
124}}\,\alp{1}{1}{M}\nonumber\\
&&+\alpc{1}{i}{M}\,\op{V}_{q_{\sst IJK}}\,\alp{1}{5}{M}\alpc{1}{5}{M}\,\op{V}_{q_{\sst
124}}\,\alp{1}{1}{M}\nonumber\\
&&+\alpc{1}{i}{M}\,\op{V}_{q_{\sst IJK}}\,\alp{1}{6}{M}\alpc{1}{6}{M}\,\op{V}_{q_{\sst
124}}\,\alp{1}{1}{M}.
\ea
By comparing the results in eqn (\ref{RS1ME134}) with the one in eqn (\ref{RS1ME134}), we note that
$\alpc{1}{i}{M}\,\op{V}_{q_{\sst 134}}\,\alp{1}{j}{M}=-\alpc{1}{i}{M}\,\op{V}_{q_{\sst 134}}\,\alp{1}{j}{M}$ whereby $i=2,3,4$ and $j=1,5,6$. Accordingly, this yields
\be
\label{R1IJK124}
\alpc{1}{i}{M}\,\left(\op{V}_{q_{\sst 134}}+\op{V}_{q_{\sst 234}}\right)\op{V}_{q_{\sst 124}}\,\alp{1}{1}{M}=0.
\ee
The expansion in terms of $\alp{1}{k}{M}$ of the operator $\op{V}_{q_{\sst 134}}\op{V}_{q_{\sst IJK}}$ whereby $IJK\in\{134,234\}$ can be found below
\ba
\alpc{1}{i}{M}\,\op{V}_{q_{\sst 134}}\op{V}_{q_{\sst
IJK}}\,\alp{1}{1}{M}&=&
+\alpc{1}{i}{M}\,\op{V}_{q_{\sst 134}}\,\alp{1}{1}{M}\alpc{1}{1}{M}\,\op{V}_{q_{\sst
IJK}}\,\alp{1}{1}{M}\nonumber\\
&&+\alpc{1}{i}{M}\,\op{V}_{q_{\sst 134}}\,\alp{1}{2}{M}\alpc{1}{2}{M}\,\op{V}_{q_{\sst
IJK}}\,\alp{1}{1}{M}\nonumber\\
&&+\alpc{1}{i}{M}\,\op{V}_{q_{\sst 134}}\,\alp{1}{3}{M}\alpc{1}{3}{M}\,\op{V}_{q_{\sst
IJK}}\,\alp{1}{1}{M}\nonumber\\
&&+\alpc{1}{i}{M}\,\op{V}_{q_{\sst 134}}\,\alp{1}{4}{M}\alpc{1}{4}{M}\,\op{V}_{q_{\sst
IJK}}\,\alp{1}{1}{M}\nonumber\\
&&+\alpc{1}{i}{M}\,\op{V}_{q_{\sst 134}}\,\alp{1}{5}{M}\alpc{1}{5}{M}\,\op{V}_{q_{\sst
IJK}}\,\alp{1}{1}{M}\nonumber\\
&&+\alpc{1}{i}{M}\,\op{V}_{q_{\sst 134}}\,\alp{1}{6}{M}\alpc{1}{6}{M}\,\op{V}_{q_{\sst
IJK}}\,\alp{1}{1}{M}\nonumber\\
&&+\alpc{1}{i}{M}\,\op{V}_{q_{\sst 134}}\,\alp{1}{7}{M}\alpc{1}{7}{M}\,\op{V}_{q_{\sst
IJK}}\,\alp{1}{1}{M}
\ea
If we compare eqn (\ref{RS1ME134}) with eqn (\ref{RS1ME234}), we realise that $\alpc{1}{i}{M}\,\op{V}_{q_{\sst 134}}\,\alp{1}{j}{M}=-\alpc{1}{i}{M}\,\op{V}_{q_{\sst 234}}\,\alp{1}{1}{M}$ with $i=2,3,4$ and $\alp{1}{7}{M}\alpc{1}{7}{M}\,\op{V}_{q_{\sst
IJK}}\,\alp{1}{1}{M}=0$ Therefore, we have
\ba
\alpc{1}{i}{M}\,\op{V}_{q_{\sst 134}}\left(\op{V}_{q_{\sst
134}}+\op{V}_{q_{\sst
234}}\right)\,\alp{1}{1}{M}&=&
+\alpc{1}{i}{M}\,\op{V}_{q_{\sst 134}}\,\alp{1}{1}{M}\alpc{1}{1}{M}\,\left(\op{V}_{q_{\sst
134}}+\op{V}_{q_{\sst
234}}\right)\,\alp{1}{1}{M}\nonumber\\
&&+\alpc{1}{i}{M}\,\op{V}_{q_{\sst 134}}\,\alp{1}{5}{M}\alpc{1}{5}{M}\,\left(\op{V}_{q_{\sst
134}}+\op{V}_{q_{\sst
234}}\right)\,\alp{1}{1}{M}\nonumber\\
&&+\alpc{1}{i}{M}\,\op{V}_{q_{\sst 134}}\,\alp{1}{6}{M}\alpc{1}{6}{M}\,\left(\op{V}_{q_{\sst
134}}+\op{V}_{q_{\sst
234}}\right)\,\alp{1}{1}{M}.
\ea
The same argument applies to the operator $\op{V}_{q_{\sst 234}}\op{V}_{q_{\sst IJK}}$ whereby $IJK\in\{134,234\}$, so that its expansion is given by
\ba
\alpc{1}{i}{M}\,\op{V}_{q_{\sst 234}}\left(\op{V}_{q_{\sst
134}}+\op{V}_{q_{\sst
234}}\right)\,\alp{1}{1}{M}&=&
+\alpc{1}{i}{M}\,\op{V}_{q_{\sst 234}}\,\alp{1}{1}{M}\alpc{1}{1}{M}\,\left(\op{V}_{q_{\sst
134}}+\op{V}_{q_{\sst
234}}\right)\,\alp{1}{1}{M}\nonumber\\
&&+\alpc{1}{i}{M}\,\op{V}_{q_{\sst 234}}\,\alp{1}{5}{M}\alpc{1}{5}{M}\,\left(\op{V}_{q_{\sst
134}}+\op{V}_{q_{\sst
234}}\right)\,\alp{1}{1}{M}\nonumber\\
&&+\alpc{1}{i}{M}\,\op{V}_{q_{\sst 234}}\,\alp{1}{6}{M}\alpc{1}{6}{M}\,\left(\op{V}_{q_{\sst
134}}+\op{V}_{q_{\sst
234}}\right)\,\alp{1}{1}{M}.
\ea
By using $\alpc{1}{i}{M}\,\op{V}_{q_{\sst 134}}\,\alp{1}{j}{M}=-\alpc{1}{i}{M}\,\op{V}_{q_{\sst 134}}\,\alp{1}{j}{M}$ where $i=2,3,4$ and $j=1,5,6$ which can be easily extracted from eqn (\ref{RS1ME134}) and eqn (\ref{RS1ME234}), we obtain
\be
\label{R1134IJK}
\alpc{1}{i}{M}\,\left(\op{V}_{q_{\sst
134}}+\op{V}_{q_{\sst
234}}\right)\left(\op{V}_{q_{\sst 134}}+\op{V}_{q_{\sst 234}}+\op{V}_{q_{\sst 124}}\right)\,\alp{1}{1}{M}=0
\ee
If we add up eqn (\ref{R1123IJK}), eqn (\ref{R1IJK124}) and eqn (\ref{R1134IJK}) the occuring operators $\op{V}_{q_{\sst IJK}}\op{V}_{q_{\sst \tilde{I}\tilde{J}\tilde{K}}}$ add up to the operator $\op{O}^{\sst I,RS}_2$. Hence, we can conclude
\be
\alpc{1}{i}{M}\,\op{O}^{\sst I,RS}_2\,\alp{1}{1}{M}=0\quad\quad i=2,3,4
\ee
Since, the matrix elements of $\op{O}^{\sst I,RS}_1$ and $\op{O}^{\sst I,RS}_2$ also vanish in the case of a spin label $\ell=1$, the operator $^{1}\op{\tilde{E}}^{\sst I,RS}_{k,{\sst tot}}(S_t)$ is the zero operator as well.
\subsection{Case $\lL\op{\tilde{E}}^{\sst II,RS}_{k,{\sst tot}}(S_t)$:\newline
Detailed Calculation of the Matrix Elements of $\op{O}^{\sst II,RS}_{1}$ and $\op{O}^{\sst II,RS}_2$}
In this section we discuss the matrix elements $\alpc{0}{2}{M}\,\op{O}^{\sst II,RS}_1\,\alp{0}{1}{M}$ and $\alpc{0}{2}{M}\,\op{O}^{\sst II,RS}_2\,\alp{0}{1}{M}$ that contribute to the matrix element of the alternative flux operator $\lL\op{\tilde{E}}^{\sst II,RS}_{k,{\sst tot}}(S_t)$. As discussed in section \ref{Problems}, from our point of view the operator $\lL\op{\tilde{E}}^{\sst II,RS}_{k,{\sst tot}}(S_t)$ including the combination $\op{V}_{\sst RS}\op{\cal S}_{\sst AL}\op{V}_{\sst RS}$ is highly artificial. Nevertheless, we investigate this operator in detail for a spin label $\ell=0.5$ here.
\subsubsection{Matrix Elements for the Case of a Spin-$\frac{1}{2}$-Representation}%
In order to calculate the matrix element of $\lL\op{\tilde{E}}^{\sst II,RS}_{k,{\sst tot}}(S_t)$, we have to know the matrix elements $\alpc{0}{2}{M}\,\op{O}^{\sst II,RS}_1\,\alp{0}{1}{M}$ and $\alpc{1}{i}{M}\,\op{O}^{\sst II,RS}_2\,\alp{0}{1}{M}$. This can be seen in eqn (\ref{act7Ek}). The explicit definition of the operators $\op{O}^{\sst II,RS}_1,\op{O}^{\sst II,RS}_2$ are shown in eqn (\ref{C2RSO1O2}). Here, the calculation for $\op{O}^{\sst II,RS}_1,\op{O}^{\sst II,RS}_2$ differs from the discussion of $\op{O}^{\sst I,RS}_1,\op{O}^{\sst I,RS}_2$ in the last section, because now additionally the signum operator $\op{\cal S}$ occurs sandwiched between the two volume operators $\op{V}_{\sst RS}$. Since the matrices of the operators $\op{Q}^{\sst RS}_{v,{\sst IJK}}$ and their corresbonding eigenvectors and eigenvalues are already given in the last section, we will not show them here again, but only refer to the results of the last section.
\newline
The expansion of each operator $\op{V}_{q_{\sst IJK}}\op{\cal S}\op{V}_{q_{\sst \tilde{I}\tilde{J}\tilde{K}}}$ that contributes to $\op{O}^{\sst II,RS}_2$ is given by
\ba
\alpc{0}{2}{M}\,\op{V}_{q_{\sst IJK}}\op{V}_{q_{\sst \tilde{I}\tilde{J}\tilde{K}}}\,\alp{0}{1}{M}&=&\sum\limits_{i,j=1}^2\alpc{0}{2}{M}\,\op{V}_{ q_{\sst IJK}}\,\alp{0}{i}{M}\alpc{0}{i}{M}\,\op{S}\,\alp{0}{j}{M}\alpc{0}{j}{M}\,\op{V}_{ q_{\sst \tilde{I}\tilde{J}\tilde{K}}}\,\alp{0}{1}{M}\nonumber\\
&=&\alpc{0}{2}{M}\,\op{V}_{ q_{\sst IJK}}\,\alp{0}{2}{M}\alpc{0}{2}{M}\,\op{S}\,\alp{0}{1}{M}\alpc{0}{1}{M}\,\op{V}_{ q_{\sst \tilde{I}\tilde{J}\tilde{K}}}\,\alp{0}{1}{M},
\ea
because $\op{V}_{q_{\sst RS}}$  is diagonal in this case.\newline
The Matrix elements of the signum operator $\op{\cal S}$ can be calculated by
\ba
\alpc{0}{i}{M}\,\op{S}\,\alp{0}{j}{M}&=&\sum\limits_{k}\alpc{0}{i}{M}\,\op{S}\,|\,\vec{e}_k\zu\auf\vec{e}_k\,\alp{0}{j}{M}=\sum\limits_{k}\sgn(\lambda_k^{\sst Q})\alpc{0}{i}{M}\,\vec{e}_k\zu\auf\vec{e}_k\,\alp{0}{j}{M},
\ea
whereby $\lambda_k^{\sst Q}$ denotes the eigenvalue of the operator $\op{Q}^{\st J=0}_{v,AL}$ associated with the eigenvector $\vec{e}_k$. Using the results of $\op{Q}^{\st J=0}_{\sst RS,IJK}$ in eqn (\ref{QJ0IJK}), we end up with
\ba
\alpc{0}{2}{M}\,\op{V}_{ q_{\sst IJK}}\,\alp{0}{2}{M}&=&\alpc{0}{1}{M}\,\op{V}_{ q_{\sst IJK}}\,\alp{0}{1}{M}=\sqrt{2a}\quad\quad IJK\in\{134,234,123,124\}\nonumber\\
\alpc{0}{2}{M}\,\op{S}\,\alp{0}{1}{M}&=&+i
\ea
Considering the definition of $\op{O}^{\sst II,RS}_1$ in eqn (\ref{C2RSO1O2}) and the results above, we obtain
\be
\label{C1}
\alpc{0}{2}{M}\,\op{O}^{\sst II,RS}_1\,\alp{0}{1}{M}=+i18a=+i9a_{\sst AL}=9\alpc{0}{2}{M}\,\op{O}^{\sst II,AL}_1\,\alp{0}{1}{M}=:C_1(\ell)\alpc{0}{2}{M}\,\op{O}^{\sst II,AL}_1\,\alp{0}{1}{M}
\ee
Here we used $a=\frac{1}{2}a_{\sst AL}$ that can be found by comparing the matrix entries of $op{Q}^{\st J=0}_{v,AL}$ with the one of $\op{Q}^{\st J=0}_{\sst RS,IJK}$. We want to express everything in terms of the $AL-$paramaters here in order to compare the results of $\lL\op{\tilde{E}}^{\sst II,RS}_{k,{\sst tot}}(S_t)$ and $\lL\op{\tilde{E}}^{\sst II,AL}_{k,{\sst tot}}(S_t)$ directly.
\newline
For the operator $\op{O}^{\sst II,RS}_2$, we have to consider the case of a total angular momentum $J=1$. The
 expansion of $\op{V}_{q_{\sst IJK}}\op{V}_{q_{\sst \tilde{I}\tilde{J}\tilde{K}}}$ in terms of the basis states $\alp{1}{k}{M}$ of ${\cal H}^{J=1}$ is shown below
\ba
\alpc{1}{i}{M}\,\op{V}_{q_{\sst IJK}}\op{V}_{q_{\sst \tilde{I}\tilde{J}\tilde{K}}}\,\alp{1}{1}{M}&=&\sum\limits_{j,k}\alpc{1}{i}{M}\,\op{V}_{ q_{\sst IJK}}\,\alp{1}{j}{M}\alpc{1}{j}{M}\,\op{S}\,\alp{1}{k}{M}\alpc{1}{k}{M}\,\op{V}_{ q_{\sst \tilde{I}\tilde{J}\tilde{K}}}\,\alp{1}{1}{M}.
\ea
In this case the matrix elements of the signum operator $\op{\cal S}$ are given by
\ba
\alpc{1}{j}{M}\,\op{S}\,\alp{1}{k}{M}&=&\sum\limits_{k'}\alpc{1}{j}{M}\,\op{S}\,|\,\vec{e}_{k'}\zu\auf\vec{e}_{k'}\,\alp{1}{k}{M}=\sum\limits_{k'}\sgn(\lambda_{k'}^{\sst Q})\alpc{1}{j}{M}\,\vec{e}_{k'}\zu\auf\vec{e}_{k'}\,\alp{1}{k}{M}.
\ea
By using the results shown in section \ref{QVSV}, we obtain
\ba
\label{MES}
\alpc{1}{j}{M}\,\op{S}\,\alp{1}{j}{M}&=&\sum\limits_{k}\alpc{1}{j}{M}\,\op{S}\,|\,\vec{e}_k\zu\auf\vec{e}_k\,\alp{1}{j}{M}=0\nonumber\\
\alpc{1}{j}{M}\,\op{S}\,\alp{1}{i}{M}&=&\sum\limits_{k}\alpc{1}{j}{M}\,\op{S}\,|\,\vec{e}_k\zu\auf\vec{e}_k\,\alp{1}{i}{M}=-\alpc{1}{i}{M}\,\op{S}\,\alp{1}{j}{M}\nonumber\\
\alpc{1}{3}{M}\,\op{S}\,\alp{1}{1}{M}&=&\sum\limits_{k}\alpc{1}{3}{M}\,\op{S}\,|\,\vec{e}_k\zu\auf\vec{e}_k\,\alp{1}{1}{M}=-i\frac{a_{\sst AL}}{\lambda_{\sst AL}}\nonumber\\
\alpc{1}{4}{M}\,\op{S}\,\alp{1}{1}{M}&=&\sum\limits_{k}\alpc{1}{4}{M}\,\op{S}\,|\,\vec{e}_k\zu\auf\vec{e}_k\,\alp{1}{1}{M}=-i\frac{a_{\sst AL}}{\lambda_{\sst AL}}\nonumber\\
\alpc{1}{5}{M}\,\op{S}\,\alp{1}{1}{M}&=&\sum\limits_{k}\alpc{1}{5}{M}\,\op{S}\,|\,\vec{e}_k\zu\auf\vec{e}_k\,\alp{1}{1}{M}=0\nonumber\\
\alpc{1}{3}{M}\,\op{S}\,\alp{1}{5}{M}&=&\sum\limits_{k}\alpc{1}{3}{M}\,\op{S}\,|\,\vec{e}_k\zu\auf\vec{e}_k\,\alp{1}{5}{M}=-i\frac{b_{\sst AL}}{\sqrt{2}\lambda_{\sst AL}}\nonumber\\
\alpc{1}{4}{M}\,\op{S}\,\alp{1}{3}{M}&=&\sum\limits_{k}\alpc{1}{4}{M}\,\op{S}\,|\,\vec{e}_k\zu\auf\vec{e}_k\,\alp{1}{3}{M}=0\nonumber\\
\alpc{1}{4}{M}\,\op{S}\,\alp{1}{5}{M}&=&\sum\limits_{k}\alpc{1}{4}{M}\,\op{S}\,|\,\vec{e}_k\zu\auf\vec{e}_k\,\alp{1}{5}{M}=-i\frac{b_{\sst AL}}{\lambda_{\sst AL}}.\nonumber\\
\ea
Here we explicitly labelled the constants $a_{\sst AL},b_{\sst AL}$ by AL, because the differ from the constants $a,b$ used in the case of $\op{V}_{\sst RS}$. The relation between these two constants is for $J=1$  only a factor of $2/3$, namely $a_{\sst AL}=(\frac{2}{3})a$ and $b_{\sst AL}=(\frac{2}{3})b$. This can be easily seen by comparing the matrix entries of $\op{Q}^{\st J=1}_{v,{\sst AL}}$ with the one of $\op{Q}_{\sst RS;IJK}^{\st J=1}$. Additionally, we labelled the eigenvalue $\lambda_{\sst AL}$ by AL, because $\op{Q}^{\st J=1}_{v,{\sst AL}}$ and $\op{Q}_{\sst RS;IJK}^{\st J=1}$ have different eigenvalues.
\newline
Starting with the operator $\op{V}_{q_{\sst IJK}}\op{V}_{q_{\sst 124}}$ with $IJK\in\{134,234,123\}$ and taking into account the vanishing of certain matrix elements of $\op{\cal S}$ shown in eqn (\ref{MES}), we get the following expansion
\ba
\alpc{1}{i}{M}\,\op{V}_{q_{\sst IJK}}\op{V}_{q_{\sst 124}}\,\alp{1}{1}{M}
&=&+\alpc{1}{i}{M}\,\op{V}_{ q_{\sst IJK}}\,\alp{1}{3}{M}\alpc{1}{3}{M}\,\op{S}\,\alp{1}{1}{M}\alpc{1}{1}{M}\,\op{V}_{ q_{\sst 124}}\,\alp{1}{1}{M}\nonumber\\
&&+\alpc{1}{i}{M}\,\op{V}_{ q_{\sst IJK}}\,\alp{1}{4}{M}\alpc{1}{4}{M}\,\op{S}\,\alp{1}{1}{M}\alpc{1}{1}{M}\,\op{V}_{ q_{\sst 124}}\,\alp{1}{1}{M}\nonumber\\
&&+\alpc{1}{i}{M}\,\op{V}_{ q_{\sst IJK}}\,\alp{1}{1}{M}\alpc{1}{1}{M}\,\op{S}\,\alp{1}{3}{M}\alpc{1}{3}{M}\,\op{V}_{ q_{\sst 124}}\,\alp{1}{1}{M}\nonumber\\
&&+\alpc{1}{i}{M}\,\op{V}_{ q_{\sst IJK}}\,\alp{1}{5}{M}\alpc{1}{5}{M}\,\op{S}\,\alp{1}{3}{M}\alpc{1}{3}{M}\,\op{V}_{ q_{\sst 124}}\,\alp{1}{1}{M}\nonumber\\
&&+\alpc{1}{i}{M}\,\op{V}_{ q_{\sst IJK}}\,\alp{1}{1}{M}\alpc{1}{1}{M}\,\op{S}\,\alp{1}{4}{M}\alpc{1}{4}{M}\,\op{V}_{ q_{\sst 124}}\,\alp{1}{1}{M}\nonumber\\
&&+\alpc{1}{i}{M}\,\op{V}_{ q_{\sst IJK}}\,\alp{1}{5}{M}\alpc{1}{5}{M}\,\op{S}\,\alp{1}{4}{M}\alpc{1}{4}{M}\,\op{V}_{ q_{\sst 124}}\,\alp{1}{1}{M}\nonumber\\
&&+\alpc{1}{i}{M}\,\op{V}_{ q_{\sst IJK}}\,\alp{1}{3}{M}\alpc{1}{3}{M}\,\op{S}\,\alp{1}{5}{M}\alpc{1}{5}{M}\,\op{V}_{ q_{\sst 124}}\,\alp{1}{1}{M}\nonumber\\
&&+\alpc{1}{i}{M}\,\op{V}_{ q_{\sst IJK}}\,\alp{1}{4}{M}\alpc{1}{4}{M}\,\op{S}\,\alp{1}{5}{M}\alpc{1}{5}{M}\,\op{V}_{ q_{\sst 124}}\,\alp{1}{1}{M}
\ea
From eqn (\ref{RSME1241}) we can read off $\alpc{1}{i}{M}\,\op{V}_{ q_{\sst 124}}\,\alp{1}{1}{M}=0$ with $i=3,4,5$. Hence, only the first two terms of the sum are not zero
\ba
\alpc{1}{i}{M}\,\op{V}_{q_{\sst IJK}}\op{V}_{q_{\sst 124}}\,\alp{1}{1}{M}
&=&+\alpc{1}{i}{M}\,\op{V}_{ q_{\sst IJK}}\,\alp{1}{3}{M}\alpc{1}{3}{M}\,\op{S}\,\alp{1}{1}{M}\alpc{1}{1}{M}\,\op{V}_{ q_{\sst 124}}\,\alp{1}{1}{M}\nonumber\\
&=&+\alpc{1}{i}{M}\,\op{V}_{ q_{\sst IJK}}\,\alp{1}{4}{M}\alpc{1}{4}{M}\,\op{S}\,\alp{1}{1}{M}\alpc{1}{1}{M}\,\op{V}_{ q_{\sst 124}}\,\alp{1}{1}{M}.
\ea
The matrix elements that are necessary to know in order to calculate the matrix element of $\op{V}_{q_{\sst IJK}}\op{V}_{q_{\sst 124}}$ explicitly are given below
\ba
\alpc{1}{1}{M}\,\op{V}_{\sst q_{124}}\,\alp{1}{1}{M}=\sum\limits_{k}\alpc{1}{1}{M}\,\op{V}_{\sst q_{124}}\,|\vec{e}_k\zu\auf\vec{e}_k\,\alp{1}{1}{M}&=&\sqrt{2a}\nonumber\\
\alpc{1}{3}{M}\,\op{V}_{\sst q_{134}}\,\alp{1}{3}{M}=\alpc{1}{3}{M}\,\op{V}_{\sst q_{234}}\,\alp{1}{3}{M}\sum\limits_{k}\alpc{1}{3}{M}\,\op{V}_{\sst q_{124}}\,|\vec{e}_k\zu\auf\vec{e}_k\,\alp{1}{3}{M}&=&\left(3+\frac{12}{\lambda^2}\right)\sqrt{\lambda}\nonumber\\
\alpc{1}{3}{M}\,\op{V}_{\sst q_{134}}\,\alp{1}{4}{M}=\alpc{1}{3}{M}\,\op{V}_{\sst q_{234}}\,\alp{1}{4}{M}\sum\limits_{k}\alpc{1}{3}{M}\,\op{V}_{\sst q_{124}}\,|\vec{e}_k\zu\auf\vec{e}_k\,\alp{1}{4}{M}&=&\left(\frac{6\sqrt{2}}{\lambda^2}-3\sqrt{2}\right)\sqrt{\lambda}\nonumber\\
\alpc{1}{4}{M}\,\op{V}_{\sst q_{134}}\,\alp{1}{3}{M}=\alpc{1}{4}{M}\,\op{V}_{\sst q_{234}}\,\alp{1}{3}{M}\sum\limits_{k}\alpc{1}{4}{M}\,\op{V}_{\sst q_{124}}\,|\vec{e}_k\zu\auf\vec{e}_k\,\alp{1}{3}{M}&=&\left(\frac{6\sqrt{2}}{\lambda^2}-3\sqrt{2}\right)\sqrt{\lambda}\nonumber\\
\alpc{1}{4}{M}\,\op{V}_{\sst q_{134}}\,\alp{1}{4}{M}=\alpc{1}{4}{M}\,\op{V}_{\sst q_{234}}\,\alp{1}{4}{M}\sum\limits_{k}\alpc{1}{4}{M}\,\op{V}_{\sst q_{124}}\,|\vec{e}_k\zu\auf\vec{e}_k\,\alp{1}{4}{M}&=&\left(6+\frac{6}{\lambda^2}\right)\sqrt{\lambda}\nonumber\\
\alpc{1}{3}{M}\,\op{V}_{\sst q_{123}}\,\alp{1}{3}{M}=\sum\limits_{k}\alpc{1}{3}{M}\,\op{V}_{\sst q_{124}}\,|\vec{e}_k\zu\auf\vec{e}_k\,\alp{1}{3}{M}&=&\sqrt{2a}\nonumber\\
\alpc{1}{4}{M}\,\op{V}_{\sst q_{123}}\,\alp{1}{4}{M}=\sum\limits_{k}\alpc{1}{4}{M}\,\op{V}_{\sst q_{124}}\,|\vec{e}_k\zu\auf\vec{e}_k\,\alp{1}{4}{M}&=&\sqrt{2b}\nonumber\\
\alpc{1}{3}{M}\,\op{V}_{\sst q_{123}}\,\alp{1}{4}{M}=\alpc{1}{4}{M}\,\op{V}_{\sst q_{123}}\,\alp{1}{3}{M}=\sum\limits_{k}\alpc{1}{3}{M}\,\op{V}_{\sst q_{124}}\,|\vec{e}_k\zu\auf\vec{e}_k\,\alp{1}{4}{M}&=&0.
\ea
Thus, we obtain
\ba
\label{SIJK124}
\alpc{1}{3}{M}\,\left(\op{V}_{q_{\sst 134}}+\op{V}_{q_{\sst 234}}+\op{V}_{q_{\sst 123}}\right)\op{V}_{q_{\sst 124}}\,\alp{1}{1}{M}&=&-i12a_{\sst AL}\frac{\sqrt{\lambda}\sqrt{2a}}{\lambda_{\sst AL}}-i2a_{\sst AL}\frac{a}{\lambda_{\sst AL}}\nonumber\\
\alpc{1}{4}{M}\,\left(\op{V}_{q_{\sst 134}}+\op{V}_{q_{\sst 234}}+\op{V}_{q_{\sst 123}}\right)\op{V}_{q_{\sst 123}}\,\alp{1}{1}{M}&=&+i\sqrt{2}18a_{\sst AL}\frac{\sqrt{\lambda}\sqrt{2a}}{\lambda_{\sst AL}}+i\sqrt{2}a_{\sst AL}\frac{2\sqrt{ab}}{\lambda_{\sst AL}}\nonumber\\
\ea
The operator $\op{V}_{q_{\sst 123}}\op{V}_{q_{\sst IJK}}$ where $IJK\in\{134,234\}$ can be expanded as
\ba
\alpc{1}{i}{M}\,\op{V}_{q_{\sst 123}}\op{V}_{q_{\sst IJK}}\,\alp{1}{1}{M}
&=&+\alpc{1}{i}{M}\,\op{V}_{ q_{\sst 123}}\,\alp{1}{3}{M}\alpc{1}{3}{M}\,\op{S}\,\alp{1}{1}{M}\alpc{1}{1}{M}\,\op{V}_{ q_{\sst IJK}}\,\alp{1}{1}{M}\nonumber\\
&&+\alpc{1}{i}{M}\,\op{V}_{ q_{\sst 123}}\,\alp{1}{4}{M}\alpc{1}{4}{M}\,\op{S}\,\alp{1}{1}{M}\alpc{1}{1}{M}\,\op{V}_{ q_{\sst IJK}}\,\alp{1}{1}{M}\nonumber\\
&&+\alpc{1}{i}{M}\,\op{V}_{ q_{\sst 123}}\,\alp{1}{1}{M}\alpc{1}{1}{M}\,\op{S}\,\alp{1}{3}{M}\alpc{1}{3}{M}\,\op{V}_{ q_{\sst IJK}}\,\alp{1}{1}{M}\nonumber\\
&&+\alpc{1}{i}{M}\,\op{V}_{ q_{\sst 123}}\,\alp{1}{5}{M}\alpc{1}{5}{M}\,\op{S}\,\alp{1}{3}{M}\alpc{1}{3}{M}\,\op{V}_{ q_{\sst IJK}}\,\alp{1}{1}{M}\nonumber\\
&&+\alpc{1}{i}{M}\,\op{V}_{ q_{\sst 123}}\,\alp{1}{1}{M}\alpc{1}{1}{M}\,\op{S}\,\alp{1}{4}{M}\alpc{1}{4}{M}\,\op{V}_{ q_{\sst IJK}}\,\alp{1}{1}{M}\nonumber\\
&&+\alpc{1}{i}{M}\,\op{V}_{ q_{\sst 123}}\,\alp{1}{5}{M}\alpc{1}{5}{M}\,\op{S}\,\alp{1}{4}{M}\alpc{1}{4}{M}\,\op{V}_{ q_{\sst IJK}}\,\alp{1}{1}{M}\nonumber\\
&&+\alpc{1}{i}{M}\,\op{V}_{ q_{\sst 123}}\,\alp{1}{3}{M}\alpc{1}{3}{M}\,\op{S}\,\alp{1}{5}{M}\alpc{1}{5}{M}\,\op{V}_{ q_{\sst IJK}}\,\alp{1}{1}{M}\nonumber\\
&&+\alpc{1}{i}{M}\,\op{V}_{ q_{\sst 123}}\,\alp{1}{4}{M}\alpc{1}{4}{M}\,\op{S}\,\alp{1}{5}{M}\alpc{1}{5}{M}\,\op{V}_{ q_{\sst IJK}}\,\alp{1}{1}{M}
\ea
The results in eqn (\ref{RSME1231}) show $\alpc{1}{3}{M}\,\op{V}_{q_{\sst 123}}\,\alp{1}{4}{M}=\alpc{1}{3}{M}\,\op{V}_{q_{\sst 123}}\,\alp{1}{4}{M}=0$. Moreover, by comparing eqn (\ref{RSME1341}) with eqn (\ref{RSME2341}), we note $\alpc{1}{i}{M}\,\op{V}_{q_{\sst 134}}\,\alp{1}{1}{M}=-
\alpc{1}{i}{M}\,\op{V}_{q_{\sst 234}}\,\alp{1}{1}{M}$ where $i=3,4$. Consequently, we have
\ba
\alpc{1}{i}{M}\,\op{V}_{q_{\sst 123}}\left(\op{V}_{q_{\sst 134}}+\op{V}_{q_{\sst 234}}\right)\,\alp{1}{1}{M}
&=&+\alpc{1}{i}{M}\,\op{V}_{ q_{\sst 123}}\,\alp{1}{3}{M}\alpc{1}{3}{M}\,\op{S}\,\alp{1}{1}{M}\alpc{1}{1}{M}\,
\left(\op{V}_{q_{\sst 134}}+\op{V}_{q_{\sst 234}}\right)\,\alp{1}{1}{M}\nonumber\\
&&+\alpc{1}{i}{M}\,\op{V}_{ q_{\sst 123}}\,\alp{1}{4}{M}\alpc{1}{4}{M}\,\op{S}\,\alp{1}{5}{M}\alpc{1}{5}{M}\,\left(\op{V}_{q_{\sst 134}}+\op{V}_{q_{\sst 234}}\right)\,\alp{1}{1}{M}.\nonumber\\
\ea
The particular matrix elements that contribute to the expansion above are
\ba
\alpc{1}{3}{M}\,\op{V}_{\sst q_{123}}\,\alp{1}{3}{M}=\sum\limits_{k}\alpc{1}{3}{M}\,\op{V}_{\sst q_{123}}\,|\vec{e}_k\zu\auf\vec{e}_k\,\alp{1}{3}{M}&=&\sqrt{2a}\nonumber\\
\alpc{1}{4}{M}\,\op{V}_{\sst q_{123}}\,\alp{1}{4}{M}=\sum\limits_{k}\alpc{1}{4}{M}\,\op{V}_{\sst q_{123}}\,|\vec{e}_k\zu\auf\vec{e}_k\,\alp{1}{4}{M}&=&\sqrt{2b}\nonumber\\
\alpc{1}{1}{M}\,\op{V}_{\sst q_{134}}\,\alp{1}{1}{M}=\alpc{1}{1}{M}\,\op{V}_{\sst q_{234}}\,\alp{1}{1}{M}=\sum\limits_{k}\alpc{1}{1}{M}\,\op{V}_{\sst q_{134}}\,|\vec{e}_k\zu\auf\vec{e}_k\,\alp{1}{1}{M}&=&+\frac{12a^2}{\lambda^2}\sqrt{\lambda}\nonumber\\
\alpc{1}{5}{M}\,\op{V}_{\sst q_{134}}\,\alp{1}{1}{M}=\alpc{1}{5}{M}\,\op{V}_{\sst q_{234}}\,\alp{1}{1}{M}=\sum\limits_{k}\alpc{1}{5}{M}\,\op{V}_{\sst q_{134}}\,|\vec{e}_k\zu\auf\vec{e}_k\,\alp{1}{1}{M}&=&-\frac{6\sqrt{2}ab}{\lambda^2}\sqrt{\lambda}.
\ea
Using the results above, we obtain
\ba
\label{S123IJK}
\alpc{1}{3}{M}\,\op{V}_{q_{\sst 123}}\left(\op{V}_{q_{\sst 134}}+\op{V}_{q_{\sst 234}}\right)\,\alp{1}{1}{M}&=&-i12a_{\sst AL}\frac{(2a^2+b^2)\sqrt{2a}\sqrt{\lambda}}{\lambda_{\sst AL}\lambda^2}\nonumber\\
\alpc{1}{4}{M}\,\op{V}_{q_{\sst 123}}\left(\op{V}_{q_{\sst 134}}+\op{V}_{q_{\sst 234}}\right)\,\alp{1}{1}{M}&=&+i\sqrt{2}12a_{\sst AL}\frac{(2a^2+b^2)\sqrt{2b}\sqrt{\lambda}}{\lambda_{\sst AL}\lambda^2},
\ea
whereby we used $a=(3/2)a_{\sst AL}$ and $b_{\sst AL}=(2/3)b$. Expanding the operator $\op{V}_{q_{\sst 134}}\op{V}_{q_{\sst IJK}}$ where $IJK\in\{134,234\}$ yields
\ba
\alpc{1}{i}{M}\,\op{V}_{q_{\sst 134}}\op{V}_{q_{\sst IJK}}\,\alp{1}{1}{M}
&=&+\alpc{1}{i}{M}\,\op{V}_{ q_{\sst 134}}\,\alp{1}{3}{M}\alpc{1}{3}{M}\,\op{S}\,\alp{1}{1}{M}\alpc{1}{1}{M}\,\op{V}_{ q_{\sst IJK}}\,\alp{1}{1}{M}\nonumber\\
&&+\alpc{1}{i}{M}\,\op{V}_{ q_{\sst 134}}\,\alp{1}{4}{M}\alpc{1}{4}{M}\,\op{S}\,\alp{1}{1}{M}\alpc{1}{1}{M}\,\op{V}_{ q_{\sst IJK}}\,\alp{1}{1}{M}\nonumber\\
&&+\alpc{1}{i}{M}\,\op{V}_{ q_{\sst 134}}\,\alp{1}{1}{M}\alpc{1}{1}{M}\,\op{S}\,\alp{1}{3}{M}\alpc{1}{3}{M}\,\op{V}_{ q_{\sst IJK}}\,\alp{1}{1}{M}\nonumber\\
&&+\alpc{1}{i}{M}\,\op{V}_{ q_{\sst 134}}\,\alp{1}{5}{M}\alpc{1}{5}{M}\,\op{S}\,\alp{1}{3}{M}\alpc{1}{3}{M}\,\op{V}_{ q_{\sst IJK}}\,\alp{1}{1}{M}\nonumber\\
&&+\alpc{1}{i}{M}\,\op{V}_{ q_{\sst 134}}\,\alp{1}{1}{M}\alpc{1}{1}{M}\,\op{S}\,\alp{1}{4}{M}\alpc{1}{4}{M}\,\op{V}_{ q_{\sst IJK}}\,\alp{1}{1}{M}\nonumber\\
&&+\alpc{1}{i}{M}\,\op{V}_{ q_{\sst 134}}\,\alp{1}{5}{M}\alpc{1}{5}{M}\,\op{S}\,\alp{1}{4}{M}\alpc{1}{4}{M}\,\op{V}_{ q_{\sst IJK}}\,\alp{1}{1}{M}\nonumber\\
&&+\alpc{1}{i}{M}\,\op{V}_{ q_{\sst 134}}\,\alp{1}{3}{M}\alpc{1}{3}{M}\,\op{S}\,\alp{1}{5}{M}\alpc{1}{5}{M}\,\op{V}_{ q_{\sst IJK}}\,\alp{1}{1}{M}\nonumber\\
&=&+\alpc{1}{i}{M}\,\op{V}_{ q_{\sst 134}}\,\alp{1}{4}{M}\alpc{1}{4}{M}\,\op{S}\,\alp{1}{5}{M}\alpc{1}{5}{M}\,\op{V}_{ q_{\sst IJK}}\,\alp{1}{1}{M}
\ea
As before by using $\alpc{1}{i}{M}\,\op{V}_{q_{\sst 134}}\,\alp{1}{1}{M}=-
\alpc{1}{i}{M}\,\op{V}_{q_{\sst 234}}\,\alp{1}{1}{M}$ where $i=3,4$, the expansion reduces to
\ba
\alpc{1}{i}{M}\,\op{V}_{q_{\sst 134}}\left(\op{V}_{q_{\sst 134}}+\op{V}_{q_{\sst 234}}\right)\,\alp{1}{1}{M}
&=&+\alpc{1}{i}{M}\,\op{V}_{ q_{\sst 134}}\,\alp{1}{3}{M}\alpc{1}{3}{M}\,\op{S}\,\alp{1}{1}{M}\alpc{1}{1}{M}\,
\left(\op{V}_{q_{\sst 134}}+\op{V}_{q_{\sst 234}}\right)\,\alp{1}{1}{M}\nonumber\\
&&+\alpc{1}{i}{M}\,\op{V}_{ q_{\sst 134}}\,\alp{1}{4}{M}\alpc{1}{4}{M}\,\op{S}\,\alp{1}{1}{M}\alpc{1}{1}{M}\,
\left(\op{V}_{q_{\sst 134}}+\op{V}_{q_{\sst 234}}\right)\,\alp{1}{1}{M}\nonumber\\
&&+\alpc{1}{i}{M}\,\op{V}_{ q_{\sst 134}}\,\alp{1}{3}{M}\alpc{1}{3}{M}\,\op{S}\,\alp{1}{5}{M}\alpc{1}{5}{M}\,
\left(\op{V}_{q_{\sst 134}}+\op{V}_{q_{\sst 234}}\right)\,\alp{1}{1}{M}\nonumber\\
&&+\alpc{1}{i}{M}\,\op{V}_{ q_{\sst 134}}\,\alp{1}{4}{M}\alpc{1}{4}{M}\,\op{S}\,\alp{1}{5}{M}\alpc{1}{5}{M}\,
\left(\op{V}_{q_{\sst 134}}+\op{V}_{q_{\sst 234}}\right)\,\alp{1}{1}{M}.\nonumber\\
\ea
The same is true for $\op{V}_{q_ {\sst 234}}\op{V}_{q_{\sst IJK}}$, thus
\ba
\alpc{1}{i}{M}\,\op{V}_{q_{\sst 234}}\left(\op{V}_{q_{\sst 134}}+\op{V}_{q_{\sst 234}}\right)\,\alp{1}{1}{M}
&=&+\alpc{1}{i}{M}\,\op{V}_{ q_{\sst 234}}\,\alp{1}{3}{M}\alpc{1}{3}{M}\,\op{S}\,\alp{1}{1}{M}\alpc{1}{1}{M}\,
\left(\op{V}_{q_{\sst 134}}+\op{V}_{q_{\sst 234}}\right)\,\alp{1}{1}{M}\nonumber\\
&&+\alpc{1}{i}{M}\,\op{V}_{ q_{\sst 234}}\,\alp{1}{4}{M}\alpc{1}{4}{M}\,\op{S}\,\alp{1}{1}{M}\alpc{1}{1}{M}\,
\left(\op{V}_{q_{\sst 134}}+\op{V}_{q_{\sst 234}}\right)\,\alp{1}{1}{M}\nonumber\\
&&+\alpc{1}{i}{M}\,\op{V}_{ q_{\sst 234}}\,\alp{1}{3}{M}\alpc{1}{3}{M}\,\op{S}\,\alp{1}{5}{M}\alpc{1}{5}{M}\,
\left(\op{V}_{q_{\sst 134}}+\op{V}_{q_{\sst 234}}\right)\,\alp{1}{1}{M}\nonumber\\
&&+\alpc{1}{i}{M}\,\op{V}_{ q_{\sst 234}}\,\alp{1}{4}{M}\alpc{1}{4}{M}\,\op{S}\,\alp{1}{5}{M}\alpc{1}{5}{M}\,
\left(\op{V}_{q_{\sst 134}}+\op{V}_{q_{\sst 234}}\right)\,\alp{1}{1}{M}.\nonumber\\
\ea
Inserting the explicit results of the matrix elements of $\op{V}_{q_{\sst 134}}$ and $\op{V}_{q_{\sst 234}}$, we get
\ba
\label{S134IJK}
\alpc{1}{3}{M}\,\left(\op{V}_{q_{\sst 134}}+\op{V}_{q_{\sst 234}}\right)\left(\op{V}_{q_{\sst 134}}+\op{V}_{q_{\sst 234}}\right)\,\alp{1}{1}{M}&=&-i12a_{\sst AL}18\left(\frac{2a^2+b^2}{\lambda_{\sst AL}\lambda}\right)\nonumber\\
\alpc{1}{4}{M}\,\left(\op{V}_{q_{\sst 134}}+\op{V}_{q_{\sst 234}}\right)\left(\op{V}_{q_{\sst 134}}+\op{V}_{q_{\sst 234}}\right)\,\alp{1}{1}{M}&=&+i12\sqrt{2}a_{\sst AL}18\left(\frac{2a^2+b^2}{\lambda_{\sst AL}\lambda}\right),
\ea
where for the latter matrix element we used $a=(3/2)a_{\sst AL}$ and $b_{\sst AL}=(2/3)b$. By summing the results in eqn (\ref{SIJK124}), eqn (\ref{S123IJK}) and eqn (\ref{S134IJK}), we obtain the result of the operator $\op{O}^{\sst II,RS}_2$, because the separated operators $\op{V}_{q_{\sst IJK}}\op{V}_{q_{\sst \tilde{I}\tilde{J}\tilde{K}}}$ exactly add up to $\op{O}^{\sst II,RS}_2$.
\ba
\alpc{1}{3}{M}\,\op{O}^{\sst II,RS}_2\,\alp{1}{1}{M}&=&-ia_{\sst AL}\left(12\frac{\sqrt{\lambda}\sqrt{2a}}{\lambda_{\sst AL}}+12\frac{a}{\lambda_{\sst AL}}+12\frac{(2a^2+b^2)\sqrt{2a}\sqrt{\lambda}}{\lambda_{\sst AL}\lambda^2}+12\cdot18\left(\frac{2a^2+b^2}{\lambda_{\sst AL}\lambda}\right)\right)\nonumber\\
\alpc{1}{4}{M}\,\op{O}^{\sst II,RS}_2\,\alp{1}{1}{M}&=&+i\sqrt{2}a_{\sst AL}\left(18\frac{\sqrt{\lambda}\sqrt{2a}}{\lambda_{\sst AL}}+\frac{2\sqrt{ab}}{\lambda_{\sst AL}}+12\frac{(2a^2+b^2)\sqrt{2b}\sqrt{\lambda}}{\lambda_{\sst AL}\lambda^2}+12\cdot18\left(\frac{2a^2+b^2}{\lambda_{\sst AL}\lambda}\right)\right)
\ea
Since the eigenvalues
\be
\lambda_{\sst AL}=\sqrt{\frac{3}{2}}\sqrt{2a_{\sst AL}^2+b_{\sst AL}^2}\quad\mbox{and}\quad\lambda=\sqrt{\frac{2}{3}}\sqrt{2a^2+b^2+3},
\ee
the matrix elements of $\op{O}^{\sst II,RS}_2$ will depend on the spin label $j$ in general.
The relation between the matrix elements of $\op{O}^{\sst II,RS}_2$ and $\op{O}^{\sst II,AL}_2$ is given by
\ba
\alpc{1}{3}{M}\,\op{O}^{\sst II,RS}_2\,\alp{1}{1}{M}&=&C_3(j,\frac{1}{2})\alpc{1}{3}{M}\,\op{O}^{\sst II,AL}_2\,\alp{1}{1}{M}\nonumber\\
\alpc{1}{4}{M}\,\op{O}^{\sst II,RS}_2\,\alp{1}{1}{M}&=&C_4(j,\frac{1}{2})\alpc{1}{4}{M}\,\op{O}^{\sst II,AL}_2\,\alp{1}{1}{M},
\ea
whereby
\ba
\label{ResC1C2}
C_3(j,\frac{1}{2})&=&\left(12\frac{\sqrt{\lambda}\sqrt{2a}}{\lambda_{\sst AL}}+12\frac{a}{\lambda_{\sst AL}}+12\frac{(2a^2+b^2)\sqrt{2a}\sqrt{\lambda}}{\lambda_{\sst AL}\lambda^2}+12\cdot18\left(\frac{2a^2+b^2}{\lambda_{\sst AL}\lambda}\right)\right)\nonumber\\
C_4(j,\frac{1}{2})&=&\left(18\frac{\sqrt{\lambda}\sqrt{2a}}{\lambda_{\sst AL}}+\frac{2\sqrt{ab}}{\lambda_{\sst AL}}+12\frac{(2a^2+b^2)\sqrt{2b}\sqrt{\lambda}}{\lambda_{\sst AL}\lambda^2}+12\cdot18\left(\frac{2a^2+b^2}{\lambda_{\sst AL}\lambda}\right)\right)
\ea
In order to see whether this dependence vanishes in the semiclassical regime of the theory, i.e. in the limit of large $j$, we will analyse this limit now.
\subsubsection{Semiclassical Limit of the Matrix Elements of $\op{O}^{\sst II,RS}_2$}
First, let us investigate the semiclassical behaviour of the eigenvalues $\lambda_{\sst AL}$ and $\lambda$.
\ba
\lambda_{\sst AL}&=&\sqrt{\frac{3}{2}}\sqrt{2a_{\sst AL}^2+b_{\sst AL}^2}\nonumber\\
\lambda&=&\sqrt{\frac{2}{3}}\sqrt{2a^2+b^2+3}=\sqrt{\frac{3}{2}}\sqrt{2a_{\sst AL}^2+b_{\sst AL}^2+\frac{4}{3}},
\ea
whereby we used $a_{\sst AL}=(2/3)a$ and $b_{\sst AL}=(2/3)b$. Hence, semiclassically, we get $\lambda\to\lambda_{\sst AL}$. The constants $a_{\sst AL},b_{\sst AL}$ are given by 
\be
a_{\sst AL}:=(\lp^6\frac{3!}{2}C_{reg})\frac{2}{3}\sqrt{j(j+1)}\quad\quad
b_{\sst AL}:=(\lp^6\frac{3!}{2}C_{reg})\frac{2}{3}\sqrt{4j(j+1)-3}
\ee
Accordingly, in the semiclassical limit $b_{\sst AL}\to 2a_{\sst AL}$.\newline
Summarsising, in the semiclassical sector of the theory, we have
\be
\lambda\to\lambda_{\sst AL},\quad b_{\sst AL}\to 2_{\sst AL}\quad\Rightarrow\quad\lambda_{\sst AL}\to 3a_{\sst AL}.
\ee
If we express all $a,b$ occuring in $C_3(j,\frac{1}{2}),C_4(j,\frac{1}{2})$ in terms of $a_{\sst AL}$ and $b_{\sst AL}$, and afterwards take the semiclassical limit, we end up with
\be
\label{C3C4}
C_3(j,\frac{1}{2})\to C_3(\frac{1}{2})=9\cdot 42\quad\quad C_4(j,\frac{1}{2})\to C_4(\frac{1}{2})=(18+1)(18+\sqrt{2})
\ee
It is precisely due to the linearly dependent triples that the akward $\sqrt{2}$ term appears which certainly lacks any combinatorial or geometrical interpretation.
\end{appendix}


\begin{thebibliography}{99}
\parskip -6pt
\bibitem{1} C. Rovelli, ``Quantum Gravity'', Cambridge University
Press, Cambridge, 2004\\
T. Thiemann, ``Modern Canonical Quantum General Relativity'', Cambridge
University Press, 2005; gr-qc/0110034
\bibitem{1a} 
C. Rovelli, ``Loop Quantum Gravity", Living Rev. Rel. {\bf 1} (1998) 1,
gr-qc/9710008\\
T. Thiemann,``Lectures on Loop Quantum Gravity'', Lecture Notes in
Physics, {\bf 631} (2003) 41 -- 135, gr-qc/0210094\\
A. Ashtekar, J. Lewandowski, ``Background Independent Quantum Gravity:
A Status Report'', Class. Quant. Grav. {\bf 21} (2004) R53;
[gr-qc/0404018]\\
L. Smolin, ``An Invitation to Loop Quantum Gravity'', hep-th/0408048
\bibitem{2} 
T. Thiemann, ``Anomaly-free Formulation of non-perturbative,
four-dimensional Lorentzian Quantum Gravity", Physics Letters {\bf B380}
(1996) 257-264, [gr-qc/9606088]\\
T. Thiemann, ``Quantum Spin Dynamics (QSD)",
Class. Quantum Grav. {\bf 15} (1998) 839-73, [gr-qc/9606089];
``II. The Kernel of the Wheeler-DeWitt
Constraint Operator",
Class. Quantum Grav. {\bf 15} (1998) 875-905, [gr-qc/9606090];
``III.
Quantum Constraint Algebra and Physical Scalar Product in Quantum General
Relativity", Class. Quantum Grav. {\bf 15} (1998) 1207-1247,
[gr-qc/9705017];
``IV. 2+1 Euclidean Quantum Gravity as a model to test 3+1
Lorentzian Quantum Gravity", Class. Quantum Grav. {\bf 15} (1998)
1249-1280, [gr-qc/9705018]; ``V. Quantum Gravity as the Natural Regulator 
of 
the Hamiltonian Constraint of Matter Quantum Field Theories",
Class. Quantum Grav. {\bf 15} (1998) 1281-1314, [gr-qc/9705019]
\bibitem{2a} 
 T. Thiemann, ``The Phoenix Project: Master Constraint
Programme for Loop Quantum Gravity'', gr-qc/0305080\\
B. Dittrich, T. Thiemann, ``Testing the Master Constraint Programme for
Loop Quantum Gravity
I. General Framework'', gr-qc/0411138;
``II. Finite Dimensional Systems'', gr-qc/0411139;
``III. SL(2,R) Models'', gr-qc/0411140;
``IV. Free Field Theories'', gr-qc/0411141;
``V. Interacting Field Theories'', gr-qc/0411142
\bibitem{3}  C. Rovelli, L. Smolin,
``Discreteness of volume and area in quantum gravity",
Nucl. Phys. {\bf B442} (1995) 593, Erratum : Nucl. Phys. {\bf B456}
(1995) 734
\bibitem{4} A. Ashtekar, J. Lewandowski, ``Quantum Theory of Geometry II :
Volume Operators", Adv. Theo. Math. Phys. {\bf 1} (1997) 388-429 
\bibitem{5}  T. Thiemann, ``A Length Operator for Canonical Quantum
Gravity'', J. Math. Phys. {\bf 39} (1998) 3372-3392; [gr-qc/9606092]
\bibitem{6} T. Thiemann, ``Quantum Spin Dynamics (QSD): VII.
Symplectic Structures and Continuum Lattice Formulations of
Gauge Field Theories", Class.Quant.Grav.18:3293-3338,2001,
[hep-th/0005232]; ``Gauge Field Theory Coherent States (GCS): I.
General Properties", Class.Quant.Grav.18:2025-2064,2001,
[hep-th/0005233]\\
T. Thiemann, O. Winkler, ``Gauge Field Theory Coherent States
(GCS): II. Peakedness Properties", Class.Quant.Grav.18:2561-2636,2001,
[hep-th/0005237]; ``III. Ehrenfest Theorems",
Class. Quantum Grav. {\bf 18} (2001) 4629-4681, [hep-th/0005234];
``IV. Infinite Tensor Product and Thermodynamic Limit",
Class. Quantum Grav. {\bf 18} (2001) 4997-5033, [hep-th/0005235]\\
H. Sahlmann, T. Thiemann, O. Winkler, ``Coherent States for
Canonical Quantum General Relativity and the Infinite Tensor Product
Extension", Nucl.Phys.B606:401-440,2001;
[gr-qc/0102038]\\
T. Thiemann, ``Complexifier Coherent States for Canonical
Quantum General Relativity", gr-qc/0206037
\bibitem{7}  H. Sahlmann, T. Thiemann, ``Towards the QFT on
Curved Spacetime Limit of QGR. 1. A General Scheme'', [gr-qc/0207030];
``2. A Concrete Implementation", [gr-qc/0207031]
\bibitem{8}  H. Sahlmann, ``When do Measures on the Space of Connections
Support the Triad Operators of Loop Quantum Gravity?'', gr-qc/0207112;
``Some Comments on the Representation Theory of the Algebra Underlying
Loop Quantum Gravity'', gr-qc/0207111\\
H. Sahlmann, T. Thiemann, ``On the Superselection Theory of
the Weyl Algebra for Diffeomorphism Invariant Quantum Gauge Theories'',
gr-qc/0302090;
``Irreducibility of the Ashtekar-Isham-Lewandowski Representation'',
gr-qc/0303074\\
A. Okolow, J. Lewandowski, ``Diffeomorphism Covariant
Representations of the Holonomy Flux Algebra'', gr-qc/0302059\\
C. Fleischhack, ``Representations of the Weyl Algebra in Quantum
Geometry'', math-ph/0407006\\
J. Lewandowski, A. Okolow, H. Sahlmann, T. Thiemann, ``Uniqueness of
Diffeomorphism Invariant States on Holonomy -- Flux
Algebras'', to appear
\bibitem{9} A. Ashtekar, C.J. Isham, ``Representations of the Holonomy
Algebras of Gravity and Non-Abelean Gauge Theories",
Class. Quantum Grav. {\bf 9} (1992) 1433, [hep-th/9202053]\\
A. Ashtekar, J. Lewandowski, ``Representation
theory of analytic Holonomy $C^\star$ algebras", in ``Knots and
Quantum Gravity", J. Baez (ed.), Oxford University Press, Oxford 1994
                                                                                
\bibitem{10}  T. Thiemann, ``Closed Formula for the Matrix Elements of the
Volume Operator in Canonical Quantum Gravity'',
J. Math. Phys. {\bf 39} (1998) 3347-3371; gr-qc/9606091
%
\bibitem{11}
R.~Loll,
``Imposing det E > 0 in discrete quantum gravity,''
Phys.\ Lett.\ B {\bf 399} (1997) 227
[arXiv:gr-qc/9703033].
%
\bibitem{12} 
G. Immirzi,
``Quantum Gravity and Regge Calculus", Nucl. Phys. Proc. Suppl. {\bf 57} 
(1997) 65; [gr-qc/9701052]\\
C. Rovelli, T. Thiemann, ``The Immirzi Parameter in Quantum General 
Relativity'', Phys. Rev. {\bf D57} (1998) 1009-14; [gr-qc/9705059]

\bibitem{13} A. Ashtekar, J. C. Baez, K. Krasnov,
``Quantum Geometry of Isolated Horizons and Black Hole Entropy",
Adv.Theor.Math.Phys.4:1-94,2001, [gr-qc/0005126]\\
K. Meissner , ``Black Hole Entropy in Loop Quantum Gravity'',
Class. Quant. Grav. {\bf 21} (2004) 5245-5252; [gr-qc/0407052]\\
M. Domagala, J. Lewandowski, ``Black 
Hole Entropy from Quantum Geometry'', 
Class.Quant.Grav. {\bf 21} (2004) 5233-5244, 2004; [gr-qc/0407051]
%
\bibitem{14} J. Brunnemann, T. Thiemann, ``Simplification of the Spectral 
Analysis
of the Volume Operator in Loop Quantum Gravity'', gr-qc/0405060
\bibitem{15} L. Smolin,  ``Recent Developments in Non-Perturbative Quantum Gravity``, hep-th/9202022                               
%
\bibitem{Sexl}
R.~Sexl, H.~Urbantke, 
``Relativit\"at, Gruppen und Teilchen'', Springer Verlag, Berlin, 1984.
%
\bibitem{CommJB}
J. Brunnemann, private communication 
\bibitem{GT}
K.~Giesel, T.~Thiemann, ``Consistency  Check on Volume and Triad Operator Quantisation in Loop Quantum Gravity I'', gr-qc/0507036 
%
\bibitem{FairRov}
W.~Fairbairn and C.~Rovelli,
``Separable Hilbert space in loop quantum gravity,''
J.\ Math.\ Phys.\  {\bf 45} (2004) 2802
[arXiv:gr-qc/0403047].
%
\bibitem{GrotRov}
N.~Grot and C.~Rovelli,
``Moduli-space structure of knots with intersections,''
J.\ Math.\ Phys.\  {\bf 37}, 3014 (1996)
[arXiv:gr-qc/9604010].
\end{thebibliography}
\end{document}